\documentclass[a4paper,12pt, notitlepage]{article}

\usepackage{amssymb,amsmath,amsfonts,eurosym,geometry,ulem,setspace,subfigure,comment,footmisc,pdflscape,array,dcolumn}
\usepackage[utf8]{inputenc}
\usepackage[dvipsnames]{xcolor}
\usepackage[pdftex]{hyperref}
\usepackage{bookmark}
\usepackage{booktabs} % top/bottom rule
\usepackage{makecell} % make cell command
\usepackage{colortbl} % grey hlines in table
\usepackage{appendix}
\usepackage[english]{babel}
\usepackage{float}
\usepackage{enumitem} % for itemlist
\usepackage{adjustbox} % adjusting table and figure size
\usepackage{verbatim} % comment out entire section
\usepackage{threeparttable}
\usepackage[justification=centering]{caption}
\usepackage{graphicx}
\usepackage{tabularx}       % If using tabularx
\usepackage{makecell}
\usepackage{longtable}
\usepackage{tikz}
\usepackage{array} 
\usepackage{tabularray}
\usepackage{rotating}          % If you want sideways tables (optional)
\usepackage{csquotes}
\usepackage[backend=biber, style=apa, natbib=true]{biblatex}
\usepackage{chngcntr}
\apptocmd{\bibsetup}{\raggedright}{}{}  % Allow ragged right for better line breaks
\newcommand{\floatnote}[1]{\par\smallskip\noindent\begin{minipage}{\linewidth}\footnotesize\textit{Note:} #1\end{minipage}\par}
\newcommand{\tabnote}[1]{\floatnote{#1}}
\newcommand{\fignote}[1]{\floatnote{#1}}

\begin{document}

\title{Digital State Capacity\thanks{We are grateful to Satya Borgohain for his extensive contributions to developing the large-language-model methods used in this paper.}}
\author{Patrick Healy$^{a,d}$\thanks{Corresponding author: \href{mailto:Patrick.Healy2@monash.edu}{Patrick.Healy2@monash.edu}}, Simon D Angus$^{a,d}$, Paul A Raschky$^{a,d}$,\\
Klaus Ackermann$^{b,d}$, Nathan Lane$^{c,d}$,\\
Weijia Li$^{a,d}$ and Cynthia Huang$^{b,d}$\\[2ex]
\small\itshape $^{a}$Department of Economics, Monash University\\
\small\itshape $^{b}$Department of Econometrics and Business Statistics, Monash University\\
\small\itshape $^{c}$London School of Economics\\
\small\itshape $^{d}$SoDa Laboratories, Monash University}
\date{\today}
\maketitle

\section*{Abstract}

Digital State Capacity is the ability of governments to deploy ICT infrastructure and information systems to implement policy. This paper introduces a new measure of government ICT capacity based on an observable stock of deployable public-sector network infrastructure: public IPv4 address space held by government organisations. These address holdings are key inputs into digital administration because they support internet-facing systems, networked information exchange, and coordination across agencies and functions. The core panel covers approximately 150,000 country-entity records classified as government across more than 150 countries from 2019 to 2024 and can be disaggregated by administrative level and government function. In the 2019 to 2024 Admin-1 panel, government IP holdings are observed in 1,681 subnational regions across all years. We validate the measure at the cross-country and subnational levels and apply it to government tasks related to corruption control and vaccination rollout. In illustrative country-year analysis, higher Digital State Capacity is associated with higher-quality governance and public-service outcomes in the expected directions, including lower measured corruption and higher vaccination coverage. These associations are descriptive; they demonstrate the empirical relevance of the measure and are not causal estimates. 
\pagebreak

\section{Introduction}
The concept of state capacity, the quality and competence of governments in implementing policies and providing public services, is a central topic in political economy \parencite{hanson2021leviathan, Besley2022BureaucracyDevelopment}. State capacity fundamentally depends on effectively utilising information about the population \parencite{Scott1998}. Standard institutions, including census collection \parencite{LeeZhang2017StateCapacity}, cadastral mapping \parencite{d2024cadasters}, and statistical agencies \parencite{brambor2020lay}, form the informational foundations of the pre-modern state by rendering society legible to government and transforming local knowledge into administratively actionable formats that enable fiscal development and public good provision \parencite{LeeZhang2017StateCapacity, vom2023no, Scott1998}. Just as censuses and cadastral systems were crucial for the pre-modern state, ICT infrastructure is a key input into modern state capacity. Modern states have a more complex set of functions and policies, which require granular, high-frequency information \parencite[e.g.][]{Cingolani2023CovidApps, Muralidharan2016BiometricIndia}. More generally, building institutions requires governments to elicit and aggregate dispersed local knowledge rather than rely only on imported blueprints \parencite{rodrik2000institutions}. In contemporary administration, inadequate ICT capacity limits the granularity and timeliness with which information can be processed and used \parencite{Fetzer2024Informational, Scott1998}. Whether these informational resources improve implementation depends on the organisational processes that translate capacity into performance \parencite{williams2021beyond}. The capacity of governments to gather, process, and deploy data through digital networks allows for organisational coordination, real-time information exchange, and targeted policy interventions \parencite{Fetzer2024Informational}.

Despite its importance, measuring government ICT capacity presents significant challenges: the digital state is multifaceted, encompassing websites, databases, communication networks, and more recently machine learning and artificial intelligence systems that utilise real-time inputs. Yet precisely because ICT infrastructure spans the breadth and depth of modern government activity, measuring it provides new knowledge about the operations of the state. Existing measures of government ICT capacity face fundamental limitations. First, subnational coverage is limited: leading digital-government measures are usually national, or observe selected local portals rather than a systematic Admin-1 panel, which is insufficient for within-country analysis \parencite{worldbank2022gtmi, UN2022EGovernmentSurvey, UNDESA2024LocalEgov}. Second, existing measures aggregate across government functions: they bundle sectors into a single national or subnational index, so digitisation across distinct functions of government, such as health, taxation, or statistical administration, cannot be separated when relating digital capacity to outcomes. This aggregation problem leaves existing work unable to distinguish whether outcomes track digitisation of specific sectors or the breadth of digitisation across levels and functions of government, a distinction central to recent debates on the patchwork structure of state capacity \parencite{mcdonnell2017patchwork}. The missing measurement is therefore not another national score of digital-government quality, but a government-specific, owner-level trace of digital infrastructure that can be aggregated across agencies, geography, levels of government, and functions.

% \parencite{mcdonnell2017patchwork, besley2011pillars}. besley2011pillars missing!

Digital State Capacity (DSC) is the ability of governments to use digital infrastructure and information systems for administration, monitoring, enforcement, coordination, and service delivery. Our empirical measure captures an observable input into this capacity: public IPv4 address space associated with government organisations and linked to locations, administrative levels, and functions. Public IPv4 resources provide the addressing infrastructure required for internet-facing systems, network administration, and information exchange across organisations \parencite{hanson2021leviathan, brambor2020lay, Vaccaro2023StateCapacity}. We therefore interpret their scale and distribution as a measure of public-sector network infrastructure, the core network component of the ICT capital available to government. This input-based approach follows work that measures state capacity through the resources, public investments, and institutional presence available for policy implementation, rather than through the outcomes the state produces
\parencite{BesleyPersson2011Pillars,acemoglu2015state,brambor2020lay,
Soifer2008StateInfrastructuralPower}.

The distinction between available infrastructure and its use is important. Infrastructure does not translate mechanically into implementation or performance \parencite{williams2021beyond}. The same systems that support coordination, targeting, and service delivery can also support surveillance, censorship, exclusion, or manipulation
\parencite{heldring2023problematizing}. This distinction follows V-Dem's separation of internet filtering capacity from internet filtering practice \parencite{Vdemcodebook2023} and matters for interpreting the empirical applications below: associations between Digital State Capacity and outcomes such as corruption, vaccination coverage, mortality, or information controls should be read as describing patterns where the measured infrastructure stock is larger.

In this paper, we address these limitations by introducing a new measure of ICT capacity based on approximately 150,000 country-entity records classified as government across more than 150 countries from 2019 to 2024. We term this Digital State Capacity.\footnote{A separate historical panel contains approximately 30,000 matched government entities and covers the period from 2008 to 2018.} Our measure provides broader subnational coverage than existing digital-government measures. In the 2019 to 2024 Admin-1 panel, government IP holdings are observed in 1,681 subnational regions in every year. The measure can be hierarchically disaggregated by government function (health, taxation, public order, social protection), administrative level, and geography. More broadly, our measure addresses three core gaps in the state capacity literature. First, \textcite{mcdonnell2017patchwork} argues that state capacity should be reconceptualised as comprising sub-units that may vary in organisational capacity. We address this gap by measuring government information and ICT infrastructure capacity across a broad range of government sectors and operations. Second, subnational research highlights the difficulty of selecting comparable units and measuring state infrastructure below the national level \parencite{luna2017capturing, soifer2019}. We provide a comparable measure of government digital infrastructure recorded in the DB-IP database \parencite{DBIP2021} that can be assigned to subnational units across countries. Third, state-capacity measurement research emphasises matching indicators to the particular dimensions of capacity they are intended to capture \parencite{Soifer2008StateInfrastructuralPower, Vaccaro2023StateCapacity}. Our measurement is registry-based and government-specific at the national level, and can be disaggregated into subnational, functional, and agency-level units where the source data support those assignments.

We introduce this new dataset to the literature and show that it is objective, practical to collect, and suitable for cross-country and subnational analysis of government ICT capacity. We create our measure of Digital State Capacity by leveraging an institutional feature of internet governance: the formal allocation of Internet Protocol (IP) addresses. IP addressing is necessary for communication within internal networks and with internet-facing systems. Public IPv4 address space is an observable input into internet-facing communication and network administration. Government-held public IPv4 address space therefore provides a proxy for a stock of usable public-sector network infrastructure. Requests for directly registered address space require an identifiable organisation and authorised contacts, and registries may request documentation of the organisation, need, and network plans \parencite{rfc2050,housley2013rfc,APNIC2024PolicyResources}. Meeting these requirements demonstrates baseline ICT capacity to plan and administer network resources. Regional Internet Registry and RDAP records identify registered resource holders, address ranges, and allocation events, while DB-IP provides the owner and geolocation fields used in the monthly panel. Just as \textcite{acemoglu2015state} measure state infrastructural reach by counting post offices and police stations for a single country, we measure a digital-infrastructure component of state capacity by counting public IPv4 address space held by government organisations. This yields a broad, government-specific dataset with approximately 150,000 country-entity records classified as government in the monthly data and approximately 30,000 matched government entities in the historical panel.

Existing measurements of informational capacity and digital government have additional limitations that we address. First, established measures of informational capacity \parencite{brambor2020lay, d2024cadasters, LeeZhang2017StateCapacity} are informative for understanding pre-modern states and, to some extent, contemporary governments. For example, tax administration draws on census data, public health surveillance integrates vital statistics, and environmental agencies utilise cadastral systems. However, these analogue-era informational systems, together with the indices constructed from them, are increasingly incomplete proxies for modern government ICT capacity in the context of rapid advances in government computing capacity and data collection methods \parencite{dang2024country}. Having an effective statistical agency or analogue cadastral system does not necessarily translate into effective ICT use within that agency or across other sectors of government. While there are efforts to update these measurements to better capture the use of modern ICT systems and data \parencite[e.g.][]{dang2024country, worldbank_spi}, they assess particular government functions, rely on highly costly methodologies, and are specialised for a single purpose. They therefore cannot be easily used to study different dimensions of government, such as function, administrative level, spatial distribution, or subnational variation.

Second, while existing measures of digital government and e-government support cross-country comparison, they typically focus on subsets of government capabilities, such as national government portals and online services \parencite{UN2022EGovernmentSurvey}, the performance of national statistical systems \parencite{worldbank_spi}, or expert surveys prone to measurement bias \parencite{luna2017capturing}, and rarely extend to the subnational level where substantial variation exists \parencite{Mann1984AutonomousPowerState, mcdonnell2017patchwork, Soifer2008StateInfrastructuralPower}. Several existing measures incorporate economy-wide ICT indicators that do not isolate government infrastructure from private-sector infrastructure \parencite{UN2020EGDI, dener2021govtech}.

Existing measures capture service quality, policy maturity, digital ID coverage, expert assessments of government performance, and specific informational systems, but they are generally not designed to provide objective, government-specific, organisation-level, function-level, and subnationally assignable measures of digital infrastructure.
DSC instead captures government IPs recorded in DB-IP that can be assigned to owners, locations, administrative levels, and functions. This distinction is especially important for empirical work: e-government indices measure visible service quality or policy maturity, while DSC measures the underlying public-sector ICT infrastructure stock that can be disaggregated across agencies, functions, and subnational regions. Three features make this possible: (1) registry-based identification of government-assigned IP addresses, which isolates the digital footprint of public institutions from economy-wide ICT activity; (2) subnational and organisational granularity, which captures intra-country variation across levels and functions of government; and (3) verifiability against registry records and a historical extension to 2008 and a common monthly construction from 2019 onward, which permits agency-level disaggregation, longitudinal analysis, and regular updating at relatively low cost.

We identify the internet addresses of government entities using the \textcite{DBIP2021} database, which reports owner and geographic-location fields for public IP address ranges. To classify organisations as government entities, we use the organisational name associated with each IP range. Requests for directly registered address space require an identifiable organisation and authorised contacts, and registries may request documentation of the organisation, need, and network plans \parencite{rfc2050,housley2013rfc,APNIC2024PolicyResources}. These requirements make the organisation associated with directly registered address space materially different from an informal web label. We use large language models (LLMs) to classify organisations as government or non-government entities. We treat these classifications as scalable first-pass labels rather than ground truth; Appendix~\ref{app:classification_validation} evaluates benchmark agreement, structured error patterns, and open-source reproducibility checks. 

In the monthly DB-IP data \parencite{DBIP2021}, DSC observes government IPs where public IPv4 ranges can be assigned to government owners and geolocated. Table~\ref{tab:dsc_regional_coverage} reports a 2020 benchmark cross-section for the population-covered subnational analysis sample, defined as countries and territories with usable 2020 population data from \textcite{CIESIN2016}.\footnote{Country and region counts differ across figures because the plotted samples use different denominators. The entity-location map counts geolocated government entities in the monthly DB-IP data. The country maps count countries with positive government IPs and the denominator needed for the displayed transformation: population for the per-capita map and public-employment data for the per-worker map. The subnational coverage table is restricted to the population-covered Admin-1 sample using \textcite{CIESIN2016}; its population-coverage column is population-weighted. The raw appendix map counts countries with positive raw government IP counts and does not require either population or public-employment denominators.} In this 2020 benchmark, government IPs are observed in 217 of 237 countries and territories. At the Admin-1 level, positive government IP counts are observed in 2,041 of 3,643 population-covered regions, covering 86.7 percent of the population. DB-IP records at least one public IPv4 range of any ownership type in 2,921 regions, of which 880 have no government IPs recorded in DB-IP. In the separate annual 2019 to 2024 panel, the number of regions with positive government IP counts is highest in 2022 at 2,224. Annual population coverage reaches 88.9 percent in both 2022 and 2023 when rounded to one decimal place. The stricter balanced panel contains 1,681 regions with positive government IP counts in every year, covering 80.7 percent of the population. These coverage statistics measure government IPs recorded in DB-IP and should not be interpreted mechanically as evidence that government digital infrastructure is absent elsewhere. The main regional limitation is Africa, where DB-IP often records IPv4 ranges more coarsely at the country or capital-city level rather than assigning them across the full set of Admin-1 units.

We begin by cataloguing and mapping the number of government organisations with an IP address at the country and subnational levels between 2019 and 2023. We then extend our data to 2008 using historical databases. Next, we present the main features of the data and highlight stylised facts both across countries at the national level and within countries at the subnational level. To assess the validity of our Digital State Capacity measure, we perform validation tests on the number of IP addresses per capita and per public-sector worker, following methods standard in the literature \parencite[e.g.][]{hanson2021leviathan}. When interpreting our analysis, we consider that an increase in digital infrastructure capacity is unlikely to result in a one-to-one increase in public-sector performance. Variation in outcomes across government organisations depends not only on available infrastructure but also on institutional barriers, including political constraints, human capital, and bureaucratic factors \parencite{mcdonnell2017patchwork}. Following \textcite{williams2021beyond}, we distinguish the public-sector ICT capital stock available to government organisations from the complementary technical and organisational processes that translate that stock into performance across tasks and policy domains. 
We use three transformations. The first measure is the raw count of public IPv4 addresses attributed to government organisations in a country, region, agency, administrative level, or government function. Per-capita values describe government IP intensity relative to the population, allowing comparisons across territorial units and subnational regions. Per-public-sector-worker values describe the digital capital intensity of the bureaucracy, subject to the availability and comparability of public-employment data.

We validate the usefulness of DSC as a measure of modern informational state capacity at both the national and subnational levels using correlations with existing measures. Digital State Capacity is positively correlated with measures of e-government and state capacity, while remaining conceptually distinct from them. To demonstrate the usefulness of our measure to researchers, we examine the relationship between our measure and several other outcomes, including measured corruption, tax revenue, public-service outcomes, vaccination rates, and infant death rates. These are essential government functions across all countries, and prior work commonly links effective ICT and e-government use to performance in these domains \parencite[e.g.][]{andersen2009government, odone2021use}. Our primary analysis is at the country-year level, comparing level differences between countries. We find that higher Digital State Capacity is associated with lower perceived corruption, higher vaccination rates, and lower infant mortality. The results are robust to inclusion of a number of existing measures of state capacity. These applications illustrate that the measure is related to outcomes and institutional environments where information-processing capacity is likely to matter. They do not imply that digital state capacity is necessarily used for welfare-improving purposes; the same infrastructure may support service delivery, monitoring, enforcement, censorship, or other state objectives.

The paper is structured as follows: Section~\ref{sec:dsc} introduces the concept of digital state capacity. Section~\ref{sec:datacreation} outlines our methods for creating the measure. Section~\ref{sec:comparison} situates our approach relative to existing measures of informational and digital capacity. Then, in Section~\ref{sec:val}, we provide stylised facts and validate the measure through correlations with related state-capacity, informational-capacity, and digital-government measures. Next, in Section~\ref{sec:nat_outcomes}, we explore relationships with several outcomes and conduct heterogeneity analysis to describe where these associations are stronger or weaker.  

\clearpage

\section{Digital State Capacity}
\label{sec:dsc}

We argue that public IPv4 address space held by government organisations is an observable input into internet-facing communication and network administration, and a proxy for the stock of ICT infrastructure available to government.\footnote{Private IPv4 addresses support communication within internal networks, while public IPv4 addresses provide globally unique addressing for internet-facing systems \parencite{rfc1918}.} The allocation and recording process is what makes this input observable. RIR and RDAP registration records identify registered resource holders, address ranges, and allocation events, while DB-IP provides the owner and geolocation fields used to construct the monthly panel. Requests for directly registered address space require an identifiable organisation and authorised contacts, and registries may request documentation of the organisation, need, and network plans \parencite{rfc2050,housley2013rfc,APNIC2024PolicyResources}. This produces an observable trace of which organisations hold network resources that is independent of self-report, expert judgement, or survey participation. Government IP allocations therefore reflect an input into digital infrastructure that governments have chosen to acquire and maintain, rather than a measure of the outcomes that infrastructure is used to produce.\footnote{RFC 791 (September 1981) standardised the Internet Protocol (IP) as a U.S. Department of Defense specification, developed through DARPA's research programmes \url{https://datatracker.ietf.org/doc/html/rfc791}. Developed within the Defense Advanced Research Projects Agency of the U.S. Department of Defense, the Internet Protocol is a way to interconnect administered networks \parencite{clark1988design}. This protocol was designed to be cost effective and to accommodate a variety of networks and communication services. The most important goal is that communication between two administered networks should be able to continue if any part of the infrastructure allowing them to communicate is damaged or interrupted. ``In March 1982, the US Department of Defense declared TCP/IP as the standard for all military computer networking.''\url{https://tcpip.training/history-of-tcp-ip/}.}

While IP addresses provide the technical foundation for digital operations, acquiring and administering them reflects complementary organisational capacity in government. The digitisation of government functions requires far more than simply obtaining IP addresses: it demands ICT infrastructure, technical expertise, and institutional coordination.\footnote{RIR allocation policies generally require applicants to document need and planned utilisation, demonstrating baseline ICT capacity to plan and administer network resources \parencite{rfc2050}.} The digitisation of a single government agency or function, such as the statistical agency or cadastral records, is a large undertaking. While RIRs do not audit ICT deployment across every office, planning, deploying, and managing digitised systems across the relevant organisational units requires substantial coordination and sustained technical capacity.\footnote{For example, plans to digitise and modernise Indian cadastral records were developed in 2008 and detail the investments needed to build capacity for this government function \parencite{DLR2008NLRMP}. The programme included plans to: (i) digitise all textual records; (ii) digitise cadastral maps and integrate them with textual records such as ownership information; and (iii) ensure GIS compatibility for modern survey methods, including aerial and satellite scans. For centralised management, real-time data updates, and citizen service provision, the plans specified hardware (computers, printers, smart card readers), secure network infrastructure connecting tehsil, district, and state-level land record offices and sub-registrars' offices, computer centres with local area networks, and state-level data centres for data aggregation and backup.} Given this complexity, the ability of government organisations to secure and manage public IPv4 address space across levels and functions of government serves as a revealing indicator of digital state capacity.

Recent literature has identified key standards for measuring state capacity effectively. First, measures should capture subnational variation, as state reach is often uneven across territory and the performance of public-sector organisations varies substantially within countries, with weak states potentially containing strong regional governments or agencies \parencite{Soifer2008StateInfrastructuralPower, mcdonnell2017patchwork}. Second, measures should focus on direct state provision of infrastructure rather than economy-wide indicators that conflate state and private-sector contributions. Third, measures should distinguish state capabilities from outcome-based indicators, such as tax ratios or education levels, which are influenced by factors beyond state capacity \parencite{Soifer2008StateInfrastructuralPower}. Fourth, to better understand the building blocks of state capacity, measures should extend to the micro level \parencite{Besley2022BureaucracyDevelopment}. Measures should also distinguish the resources available to bureaucratic organisations from the processes through which those resources produce performance across policy domains \parencite{williams2021beyond}. Fifth, measures should enable both cross-country and within-country comparison and longitudinal analysis, as measuring individual government institutions at the local level has been shown to be effective but has rarely been extended geographically or temporally beyond single-country studies \parencite{acemoglu2015state}.

We argue that aggregating government-held public IPv4 address space provides an observable proxy for the public-sector ICT capital stock available to government organisations. Our Digital State Capacity measure addresses the five criteria outlined above. First, by disaggregating IP address allocations to individual government organisations, we capture subnational variation across territory, levels, and functions of government \parencite{Soifer2008StateInfrastructuralPower, mcdonnell2017patchwork}. Second, we measure direct state provision by identifying government-owned infrastructure rather than economy-wide ICT indicators that conflate state and private-sector contributions. Third, our measurement is registry-based, using observed ICT resource allocations rather than outcome-based indicators \parencite{Soifer2008StateInfrastructuralPower, Besley2022BureaucracyDevelopment}. Fourth, the data extend to the micro level, with observations at the national, subnational, organisational, and functional levels. This structure allows future work to examine how organisational resources relate to task-specific bureaucratic performance \parencite{Besley2022BureaucracyDevelopment, williams2021beyond}. Fifth, our dataset generalises the local-institution measurement approach of \textcite{acemoglu2015state} beyond a single country, enabling cross-country and subnational comparison. Repeated monthly DB-IP data support temporal comparisons from 2019 onward under a common construction method, while the historical allocation data extend the national panel to 2008 for supplementary evidence. In addition, we greatly expand on \textcite{dang2024country}, who assess the capacity of statistical agencies to use technology for function-specific tasks. More broadly, we capture governments' ability to use technology to collect, store, process, and disseminate information for policy implementation and public service delivery \parencite{LeeZhang2017StateCapacity, Fetzer2024Informational} across all dimensions of government.

\section{Dataset creation}
\label{sec:datacreation}

We construct our dataset by identifying public IPv4 ranges registered to government organisations. The raw data contain the IP range, the number of addresses in the range, the registered owner name, and geolocation from proprietary DB-IP data. We use the owner name and country as context for an LLM to classify whether the range is registered to a government organisation, and enrich these records with information from Google search results where available. We then use historical allocation records from the Regional Internet Registries to recover assignment dates and extend the measure backward in time to 2008.

Appendix Table~\ref{tab:dsc_process_overview} provides a guide to the data-construction pipeline, linking the raw public IPv4 range data, search-enriched classification steps, historical allocation matching, and normalised DSC measures to the appendix sections where each step is documented.

The primary source is the DB-IP monthly database \textcite{DBIP2021}, which reports ownership and geolocation fields for public IPv4 address ranges. For each public IPv4 range, DB-IP reports the start and end address, owner name, internet service provider, country, and geolocation fields.\footnote{Table~\ref{tab:ip-details} provides an example for the White House, with randomly altered IP addresses and geolocation data. A single owner may have multiple IP ranges, including within the same location.} DB-IP is updated monthly to reflect changes in ownership and location information \parencite{ipinfo_geolocation_changes}. We use these records to identify the stock of public IPv4 resources allocated to government organisations. In the main analysis, we aggregate the data at the country and Admin-1 levels.

We classify probable government entities using the DB-IP owner-name field, which identifies the organisation associated with each public IP range. Requests for directly registered address space require an identifiable organisation and authorised contacts, and generally require supporting organisational records \parencite{rfc2050,housley2013rfc,APNIC2024PolicyResources}. To determine whether an IP range belongs to a government organisation, we classify the resource-holder name in its country context. The distinction is between organisations that are part of the public sector, such as ministries, departments, municipalities, courts, regulators, police agencies, or tax authorities, and non-state organisations, such as private firms, industry associations, charities, or commercial service providers. In many cases this is straightforward, but simple keyword rules are not sufficient: terms such as “authority,” “council,” “state,” “national,” or “public” can appear in both government and non-government names, and the same acronym can refer to different types of organisations across countries. The DB-IP database contains more than six million unique owners across 67 monthly files from January 2019 through July 2024 \parencite{DBIP2021}. Owner names are also inconsistently formatted. They can contain acronyms, missing spaces, spelling errors, translations, and multilingual terms. Standard text-classification methods are therefore poorly suited to the government classification task. We use a stepwise large-language-model classification procedure to identify government organisations and then classify their level and function of government. Appendix~\ref{app:prompt1} reports the prompt design and model comparisons. The prompt requires the models to return a short reason for each prediction, creating an intermediate output that can be reviewed against the input evidence. This use of an explicit verification step follows the broader logic of checking model-generated claims against supporting evidence \parencite{Dhuliawala2023ChainOfVerification}; it does not make the classifications self-validating.\footnote{Because government status, level, and function are classified rather than directly observed for many entities, coverage statistics should be read alongside the classification validation and boundary-sensitivity exercises below.}

We use search results to provide additional information about candidate organisations. For each candidate, Serper \textcite{serper_dev_2023} returns the top Google result, including fields such as the title, URL, snippet, and sitelinks where available (Appendix~\ref{app:serper}). These fields help the LLM classifier distinguish similarly named organisations and assign more detailed entity types and government functions. We then use a final LLM classification step to integrate the DB-IP owner name, country context, initial model predictions, and search-enriched organisational information. This final step classifies government status, level of government, and function of government using the Classification of the Functions of Government (COFOG) reported in the International Monetary Fund's Government Finance Statistics Manual \parencite{IMF2014GFSM} and OpenAI's GPT-4o mini model \parencite{OpenAI2024GPT4oMini}. Appendix~\ref{app:gov_examples_level} provides examples of government entities categorised by level. After additional processing detailed in Appendix~\ref{app:data_cleaning}, we identify approximately 150,000 country-entity records classified as government with one or more IP ranges between 2019 and 2024.\footnote{We exclude education entities from the main measure because owner names and search results are often insufficient to distinguish public, private, and mixed education providers consistently across countries. We report the exclusion rules in Appendix~\ref{app:data_cleaning}.} A single organisation can hold multiple IP ranges and may appear in more than one geographic location. For the analysis, we aggregate ranges by country, subnational region, level of government, and function depending on the research question.

To extend the data historically, we use RDAP to query standardised registration records and draw on the RIR system's historical allocation and assignment records \parencite{APNIC_RDAP, apnicHistoryInternet}. We use these records to recover allocation dates for matched owner-name and IP-range pairs. These historical matches allow us to construct country-year measures before the DB-IP monthly panel, but coverage is lower than in the post-2019 data. We match approximately 30,000 of the final government entities to historical allocation records. These matches are useful for cross-country analysis before 2019, but researchers should account for the lower coverage of the historical sample (Appendix~\ref{app:historical}). We set 2008 as the starting year for our observations to deal with limitations of these databases.\footnote{Historical queries extend to the early 2000s for Europe and Latin America and the 1990s for North America. APNIC's Whowas service provides prior states of APNIC registration records from 2008 onward \parencite{APNIC_Whowas}. We use \textcite{apnic-rex}, which aggregates RIR data dating back to the 1980s. Some delegations pre-date the RIRs themselves and represent resources the RIR currently has authority over (known as legacy resource holders).} Validity checks discussed in Appendix~\ref{app:historical_validation} confirm that the historical matches cover only a subset of entities and understate pre-2019 holdings. We therefore interpret the historical extension as a lower-bound rather than as an equivalent panel. The measure is most informative for cross-sectional differences in government IPs recorded in DB-IP and for countries where public IPv4 remains central to government network architecture. It is less informative for tracking digital infrastructure growth in countries undergoing rapid IPv6 transition. This limitation is most relevant for within-country growth in countries where digital expansion occurred during rapid IPv6 adoption; it is less problematic for cross-sectional comparisons of government-held public IPv4 resources.

\section{Comparison with Existing Measures}
\label{sec:comparison}

This section situates DSC relative to two adjacent measurement literatures. The first measures general state capacity and the informational foundations of the state: censuses, statistical systems, cadastral records, bureaucratic quality, and latent state-capacity indices. The second measures digital government, digital public infrastructure, and state activity in the online information environment. The goal is not to argue that DSC supersedes these measures. Rather, the comparison clarifies the object each measure observes. DSC is a complementary measure of modern informational capacity: an objective, government-specific, organisation-level, function-level, and geographically assignable measure of public ICT infrastructure.

The first set of measures locates DSC within the informational-capacity tradition. Census accuracy, statistical agencies, cadastral records, and information-capacity indices measure the state's ability to make society legible and convert local information into administratively usable data \parencite{LeeZhang2017StateCapacity, brambor2020lay, d2024cadasters, Scott1998}. Broad state-capacity measures, including \textcite{hanson2021leviathan}, aggregate across several dimensions of administrative, fiscal, coercive, and informational capacity. These measures provide useful points of comparison, but their unit of observation is usually national, restricted to a specific government function, or inferred from latent indices. DSC extends the informational-capacity logic to the modern public ICT apparatus by observing government-held digital infrastructure across organisations, functions, levels of government, and subnational regions. This distinction is central to the patchwork view of state capacity \parencite{mcdonnell2017patchwork}: if capacities vary across agencies, functions, and territories, then a single national score cannot reveal where digital informational capacity is concentrated or absent.

\clearpage

\begingroup
\small
\setlength{\tabcolsep}{3pt}
\renewcommand{\arraystretch}{1.15}
\begin{longtable}{@{}
>{\raggedright\arraybackslash}p{0.20\linewidth}
>{\raggedright\arraybackslash}p{0.17\linewidth}
>{\raggedright\arraybackslash}p{0.22\linewidth}
>{\raggedright\arraybackslash}p{0.16\linewidth}
>{\raggedright\arraybackslash}p{0.16\linewidth}
@{}}
  \caption{State-Capacity and Informational-Capacity Measures Related to DSC}
  \label{tab:state_info_measures} \\
  \toprule
  \textbf{Measure} & \textbf{Observed construct} & \textbf{Measurement basis} & \textbf{Coverage} & \textbf{Relation to DSC} \\
  \midrule
  \endfirsthead

  \multicolumn{5}{c}%
  {{\tablename\ \thetable{} continued from previous page}} \\
  \toprule
  \textbf{Measure} & \textbf{Observed construct} & \textbf{Measurement basis} & \textbf{Coverage} & \textbf{Relation to DSC} \\
  \midrule
  \endhead

  \midrule
  \multicolumn{5}{r}{{Continued on next page}} \\
  \endfoot

  \bottomrule
  \endlastfoot

  Hanson–Sigman State Capacity Index \parencite{hanson2021leviathan} &
  General state capacity &
  Latent index from administrative, fiscal, coercive, and informational indicators &
  Country-year; 1960–2015 &
  National latent benchmark; not digital, organisational, functional, or subnational \\
  \addlinespace

  V-Dem and WGI administrative and governance measures
  \parencite{VDemV13, Vdemcodebook2023, Kaufmann2011WorldwideGovernance} &
  Administrative quality and government effectiveness &
  Expert assessments and perception-based governance indicators &
  Country-year; broad cross-national coverage &
  Validation benchmarks, but not direct measures of public ICT infrastructure \\
  \addlinespace

  Census accuracy, information capacity, and cadasters
  \parencite{LeeZhang2017StateCapacity, brambor2020lay, d2024cadasters} &
  Informational foundations of the state &
  Census records, statistical yearbooks, population registers, and cadastral systems &
  National; country and period coverage varies by source &
  Closest conceptual comparators; generally historical or function-specific \\
  \addlinespace

  World Bank Statistical Capacity and Statistical Performance measures
  \parencite{worldbank_spi, dang2024country} &
  National statistical-system performance &
  Objective indicators of statistical data use, services, products, sources, and infrastructure &
  Country-year; national statistical systems &
  Close comparator, but focused on the statistical function rather than the broader public sector \\

\end{longtable}
\endgroup

The second set of measures concerns a more proximate empirical domain for DSC: digital government and ICT-enabled administration. These measures capture service quality, policy maturity, digital identity, online service provision, and digital control \parencite{UN2022EGovernmentSurvey, UN2020EGDI, EU2020eGov}. The main difference is the object observed. E-government and maturity indices often assess portals, policies, services, or national readiness; DSC measures the public ICT infrastructure stock held by government organisations. This distinction is especially important because the infrastructure components of major e-government indices often rely on economy-wide ICT indicators, such as internet use, affordability, or human capital, rather than digital infrastructure held by public organisations. These indicators measure national digital readiness, but they can conflate government capacity with household access, private-sector infrastructure, and broader economic development.

\begingroup
\small
\setlength{\tabcolsep}{3pt}
\renewcommand{\arraystretch}{1.15}
\begin{longtable}{@{}
>{\raggedright\arraybackslash}p{0.20\linewidth}
>{\raggedright\arraybackslash}p{0.17\linewidth}
>{\raggedright\arraybackslash}p{0.22\linewidth}
>{\raggedright\arraybackslash}p{0.16\linewidth}
>{\raggedright\arraybackslash}p{0.16\linewidth}
@{}}
  \caption{Digital-Government and Digital-Infrastructure Measures Related to DSC}
  \label{tab:existing_measures} \\
  \toprule
  \textbf{Measure} & \textbf{Observed construct} & \textbf{Measurement basis} & \textbf{Coverage} & \textbf{Relation to DSC} \\
  \midrule
  \endfirsthead

  \multicolumn{5}{c}%
  {{\tablename\ \thetable{} continued from previous page}} \\
  \toprule
  \textbf{Measure} & \textbf{Observed construct} & \textbf{Measurement basis} & \textbf{Coverage} & \textbf{Relation to DSC} \\
  \midrule
  \endhead

  \midrule
  \multicolumn{5}{r}{{Continued on next page}} \\
  \endfoot

  \bottomrule
  \endlastfoot

  UN E-Government Development Index
  \parencite{UN2022EGovernmentSurvey, UN2020EGDI} &
  Citizen-facing e-government readiness &
  Online services, telecommunications infrastructure, and human capital &
  National; 193 UN Member States; biennial 2003–2024 &
  Service and readiness measure; infrastructure inputs are partly economy-wide rather than government-specific \\
  \addlinespace

  UN Local Online Services Index \parencite{UNDESA2024LocalEgov} &
  Local online-service provision &
  Assessment of local government portals &
  Largest or selected city portal in each UN Member State; periodic since 2018 &
  Local portal measure; not full regional, agency, functional, or infrastructure coverage \\
  \addlinespace

  World Bank Government Technology Maturity Index
  \parencite{dener2021govtech, worldbank2022gtmi, worldbank2025gtmi} &
  GovTech maturity &
  Core government systems, public service delivery, digital citizen engagement, and GovTech enablers &
  National economies; 2020, 2022, and 2025 updates &
  Maturity and readiness benchmark; not owner-level or subnational infrastructure stock \\
  \addlinespace

  OECD Digital Government Index \parencite{oecd2024digital} &
  Digital-government policy maturity &
  Survey-based indicators of digital-by-design, data-driven, platform, open, user-driven, and proactive government &
  National; 38 countries in the 2023 edition &
  Detailed policy benchmark; limited global and subnational coverage \\
  \addlinespace

  World Bank ID4D \parencite{WorldBank2018ID4D, WorldBank2021ID4D} &
  Identification and digital-verification coverage &
  Foundational ID ownership, digital ID availability, and ID-system characteristics &
  National; global and periodic &
  One foundational DPI pillar, not whole-of-government ICT capacity \\
  \addlinespace

  World Bank Digital Adoption Index \parencite{WorldBank2016DAI} &
  Digital adoption by government, people, and business &
  Composite index with government, people, and business subindices &
  National; 180 countries; 2014 and 2016 &
  Includes a government subindex, but discontinued and not organisation-level or subnational \\
  \addlinespace

  World Bank Digital Progress and Trends Report \parencite{WorldBank2024DigitalProgress} &
  Economy-wide digitisation and digital public infrastructure &
  Report and dashboards on infrastructure, adoption, digital sectors, DPI, and emerging technologies &
  National and global; current report series &
  Broad digital-development benchmark, not a single government digital-capacity score \\
  \addlinespace

  V-Dem Digital Society and internet-control indicators
  \parencite{VDemV13, Vdemcodebook2023} &
  Online information control capacity and practice &
  Expert-coded indicators of internet filtering, shutdowns, censorship, cybersecurity, and online-content regulation &
  Country-year; internet-era coverage &
  Discriminant comparator; captures control capacity or practice, not general public ICT infrastructure \\
  \midrule

  Digital State Capacity, this paper &
  Public-sector ICT infrastructure stock &
  Government-held public IPv4 address space linked to owners, locations, administrative levels, and functions &
  Country, Admin-1, organisation, level, and function; annual 2008–2023; see Table~\ref{tab:dsc_coverage_margins} &
  Objective government-specific public ICT stock, assignable by owner, level, function, and geography \\

\end{longtable}
\endgroup

The coverage distinction is central to the comparison. Table~\ref{tab:dsc_coverage_margins} summarises the main coverage margins opened by DSC and separates raw source visibility from denominator-specific analysis samples.

\begin{table}[htbp]
\centering
\begingroup
\scriptsize
\setlength{\tabcolsep}{3pt}
\renewcommand{\arraystretch}{1.08}
\caption{Coverage Margins of Digital State Capacity}
\label{tab:dsc_coverage_margins}
\begin{tabular}{@{}
>{\raggedright\arraybackslash}p{0.18\linewidth}
>{\raggedright\arraybackslash}p{0.37\linewidth}
>{\raggedright\arraybackslash}p{0.36\linewidth}
@{}}
\toprule
\textbf{Margin} & \textbf{DSC coverage} & \textbf{Comparison and use} \\
\midrule
Owner-level source data &
Approximately 150,000 country-entity records classified as government in the monthly DB-IP data from 2019 to 2024; approximately 30,000 matched government entities in the historical extension to 2008. &
Unlike country-year indices or selected portal evaluations, DSC can be aggregated by owner, level, function, and location. \\
\addlinespace

National country coverage &
219 countries and territories with positive raw government IP holdings in the 2019–2023 raw-count appendix panel; 202 countries in the 2019–2023 per-capita map and 179 in the 2019–2023 per-worker map. &
Provides broad national reach comparable to leading global measures, but with a government-specific infrastructure numerator for raw maps and normalised national comparisons. \\
\addlinespace

Subnational geography &
Population-covered 2020 benchmark: 217 of 237 countries and territories and 2,041 of 3,643 Admin-1 regions with positive government IPs, covering 86.7 percent of the Admin-1 population in that denominator. The annual region count is highest in 2022 at 2,224, and annual population coverage reaches 88.9 percent in 2022 and 2023. The balanced panel contains 1,681 regions with positive government IPs in every year from 2019 to 2024, covering 80.7 percent of the population. &
Extends beyond national measures and selected-city portal evaluations to systematic Admin-1 coverage. Broader any-DB-IP source visibility is treated as a diagnostic comparison, not as the headline DSC contribution. \\
\addlinespace

Administrative level and function &
Government-entity records can be aggregated by federal/local level and COFOG function where classification is available. &
Captures within-state heterogeneity usually collapsed in composite indices; used in the level/function descriptives and function-specific outcome applications. \\
\bottomrule
\end{tabular}
\tabnote{Counts use different denominators by design. Raw country coverage requires only positive government IP holdings. Per-capita and per-worker country counts require the corresponding denominator data. The 2020 Admin-1 benchmark is restricted to the population-covered subnational sample in Table~\ref{tab:dsc_regional_coverage}. The 2022 maximum is an annual count, while the 1,681-region figure is the stricter balanced panel with positive government IP counts in every year. These figures measure government IPs recorded in DB-IP, not the absence of public ICT infrastructure where no government IP is observed.}
\endgroup
\end{table}

Taken together, these measures are best understood as complements rather than substitutes. Existing state-capacity measures provide broad benchmarks, while digital-government measures capture service quality, maturity, digital identity, online control, or economy-wide digitisation. DSC measures a different object: the stock of public ICT infrastructure held by government organisations. Development proxies such as GDP and nighttime lights \parencite{luna2017capturing, Fetzer2024Informational}, and single-country studies of biometric identification, digital payments, tax technology, and administrative information systems \parencite{Muralidharan2016BiometricIndia, Cingolani2023CovidApps}, remain important for validation and interpretation, but they are not cross-national measurement systems for government ICT infrastructure. This distinction motivates the validation exercises below, which ask whether DSC converges with related measures while retaining non-redundant variation across countries, functions, and subnational regions.

\section{Validation}
\label{sec:val}
The literature on state infrastructural power and subnational reach motivates examining variation below the national level \parencite{Mann1984AutonomousPowerState, Soifer2008StateInfrastructuralPower, soifer2019}. We expect three patterns. First, we expect substantial subnational variation in digital infrastructure, with capital cities and economically dominant regions concentrating resources while peripheral territories exhibit systematically lower state presence. Second, drawing on \textcite{mcdonnell2017patchwork}, we expect significant variation across government functions and administrative levels, driven by differential access to labour pools with requisite technical skills and divergent organisational management capacity. Third, we anticipate that agencies with specialised mandates, autonomous budgets, and higher skill requirements demonstrate elevated digital state capacity relative to general administrative units. To examine these patterns, we present three data transformations: first, raw government IP counts measuring absolute digital infrastructure stock at national and subnational scales; second, per-capita normalisation measuring government IP intensity relative to population; and third, per-public-sector-worker ratios capturing capital intensity available to civil servants. 

\subsection{Coverage and Data Construction}

\paragraph{Data construction.}
The DSC series combines two components constructed from different source records. The first is the DB-IP monthly ownership and geolocation panel. For 2019–2023, we classify government-owned IPv4 ranges in each monthly DB-IP file, aggregate government IPs across organisations within each country, region, level of government, or function, and then take the average across months within the calendar year. Annual averaging reduces dependence on any single monthly database release and treats the country-year measure as the average number of government IPs recorded in DB-IP during that year.

The second component extends the national panel to 2008 using historical allocation records. For 2008–2018, we use RDAP and Regional Internet Registry records to recover allocation dates for matched owner-name and IP-range pairs. We then construct cumulative country-year stocks from the matched historical allocation dates. This historical measure records whether a matched government IP range had been allocated by a given year; it does not reproduce the monthly ownership and geolocation update process available in DB-IP from 2019 onward.

The historical extension has lower coverage because it is available only for entities and ranges that can be matched to historical allocation records. After cleaning, the monthly DB-IP data contain approximately 150,000 country-entity records classified as government, while the historical data contain approximately 30,000 matched government entities. Because unmatched entities are omitted, the historical component generally records fewer government IPs than the monthly DB-IP data and should be interpreted as a lower-bound estimate.
The panel is most informative for cross-country differences in levels and less informative about within-country changes over time.\footnote{IPv4 has a fixed number of addresses that can be allocated in each region.} We therefore do not interpret changes from 2018 to 2019 as growth in DSC. In the main text, we treat 2019–2023 results based only on monthly DB-IP observations as the preferred validation sample where possible, and use 2008–2023 results as extended evidence.

Appendix~\ref{app:historical_validation} evaluates the change in data source and construction directly. The Pearson correlation between reconstructed 2018 log counts and DB-IP-based 2019 log counts is 0.81. Although the lower-bound historical measure omits many observations, this correlation indicates that it recovers substantial cross-country variation in levels. It should not be read as evidence that the two components recover identical levels or within-country changes. By contrast, the 2019–2020 comparison within the monthly DB-IP data is much tighter, with a correlation of 0.983 (Figure~\ref{fig:govip_2019_2020}). Figure~\ref{fig:2019_break_ip_trend} illustrates the corresponding level shift for selected countries.

\subsection{National Level Correlations}
We examine two normalised measures of DSC: government IPs per capita and government IPs per public-sector worker. The per-capita transformation captures the visible territorial reach of government digital infrastructure, while the per-worker transformation captures the digital capital intensity of the public bureaucracy. The per-worker sample is smaller because public-employment estimates are not available for all countries. Observation counts also vary across validation tables because each comparator measure has its own country and temporal coverage; the Year column reports the overlap period used for each comparison.

In our empirical analysis, we use a panel dataset spanning 2008 to 2023 to examine two measures of Digital State Capacity: the number of government IPs per capita\footnote{Using population data from the World Bank.}, and the number of government IPs per public-sector worker\footnote{Using modelled estimates of the public-sector workforce from the International Labour Organization \parencite{bescond2024global}.}. Throughout the empirical analysis, $\ln(\cdot)$ denotes the natural logarithm and $\log_{10}(\cdot)$ denotes the base-10 logarithm.

\begin{figure}
    \centering
    \caption{Locations of Government Entities, 2019--2023}
    \includegraphics[width=\linewidth]{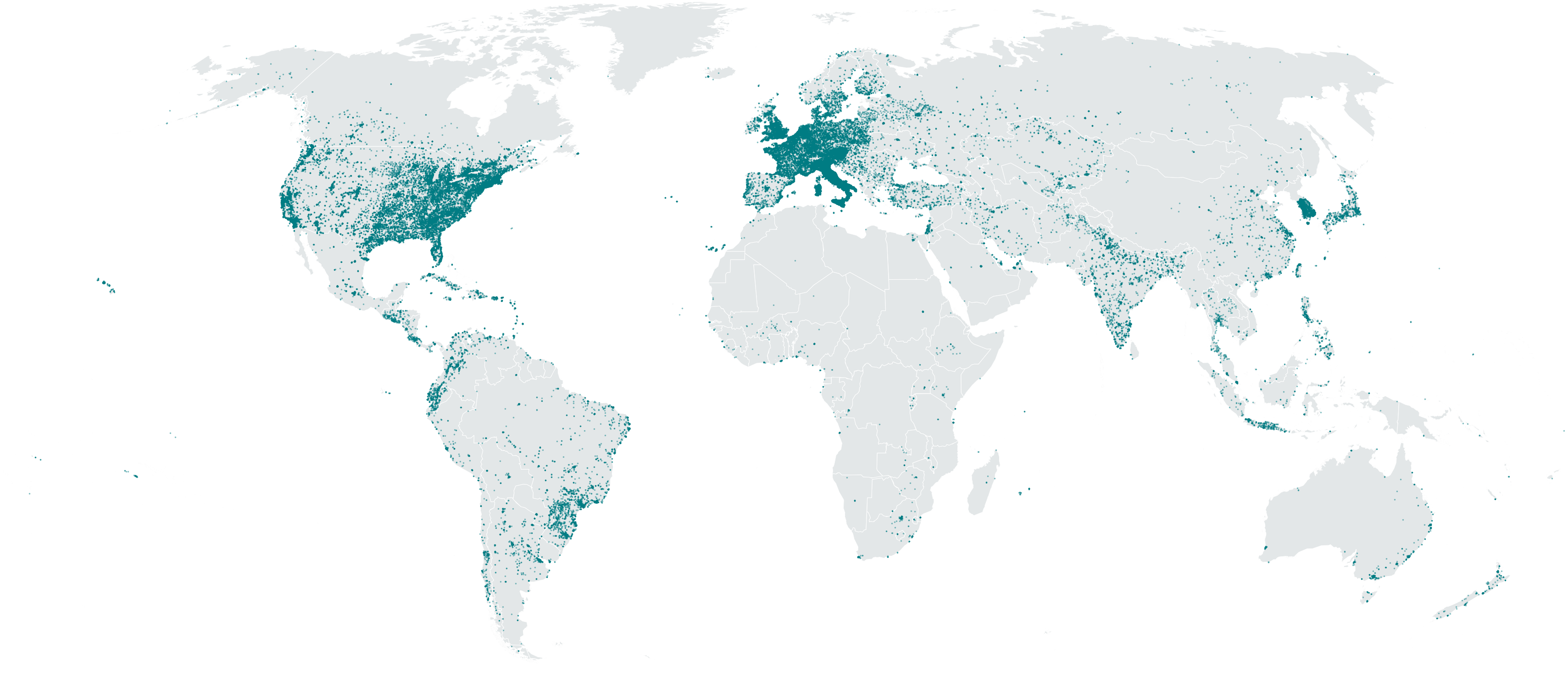}
    \fignote{Each point marks a distinct geocoded location at which at least one government entity is observed during 2019--2023. Locations are plotted once and are not weighted by repeated monthly observations, the number of entities at the same coordinate, government IP counts, population, or public-sector employment. Overlapping points appear darker.}
    \label{fig:dbipmap}
\end{figure}

\begin{figure}[htp]
    \centering
    \caption{Subnational Log Government IPs per Capita, 2020}
        \centering
        \includegraphics[width=\linewidth]{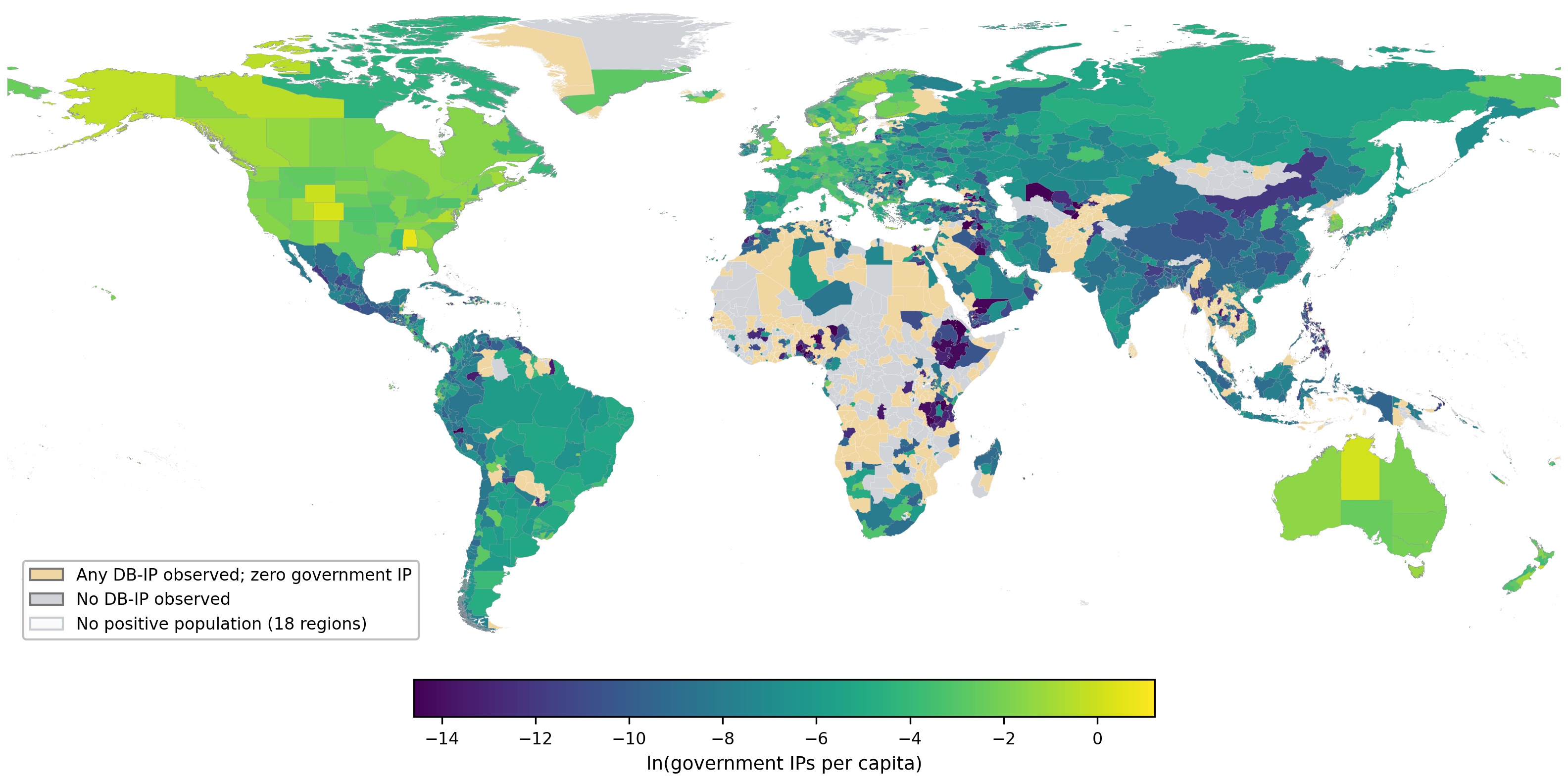}
        \fignote{The map reports Admin-1 variation in the natural logarithm of government IPs per capita in 2020. Among regions with positive government IPs, lighter colours indicate larger values; the continuous colour scale is clipped at the 1st and 99th percentiles. Tan regions are observed in the DB-IP source data but have no government IPs, while grey regions are not observed in the DB-IP source. Eighteen mostly uninhabited regions with no positive recorded population are excluded from the denominator and left white. Table~\ref{tab:dsc_regional_coverage} reports the corresponding coverage counts and population-weighted coverage shares.}
        \label{fig:total_gov_ip_map}
    
\end{figure}

\begin{figure}
    \centering
    \caption{Country-Level Log Total Government IPs per Public-Sector Worker, 2002--2023}
    \includegraphics[width=\linewidth]{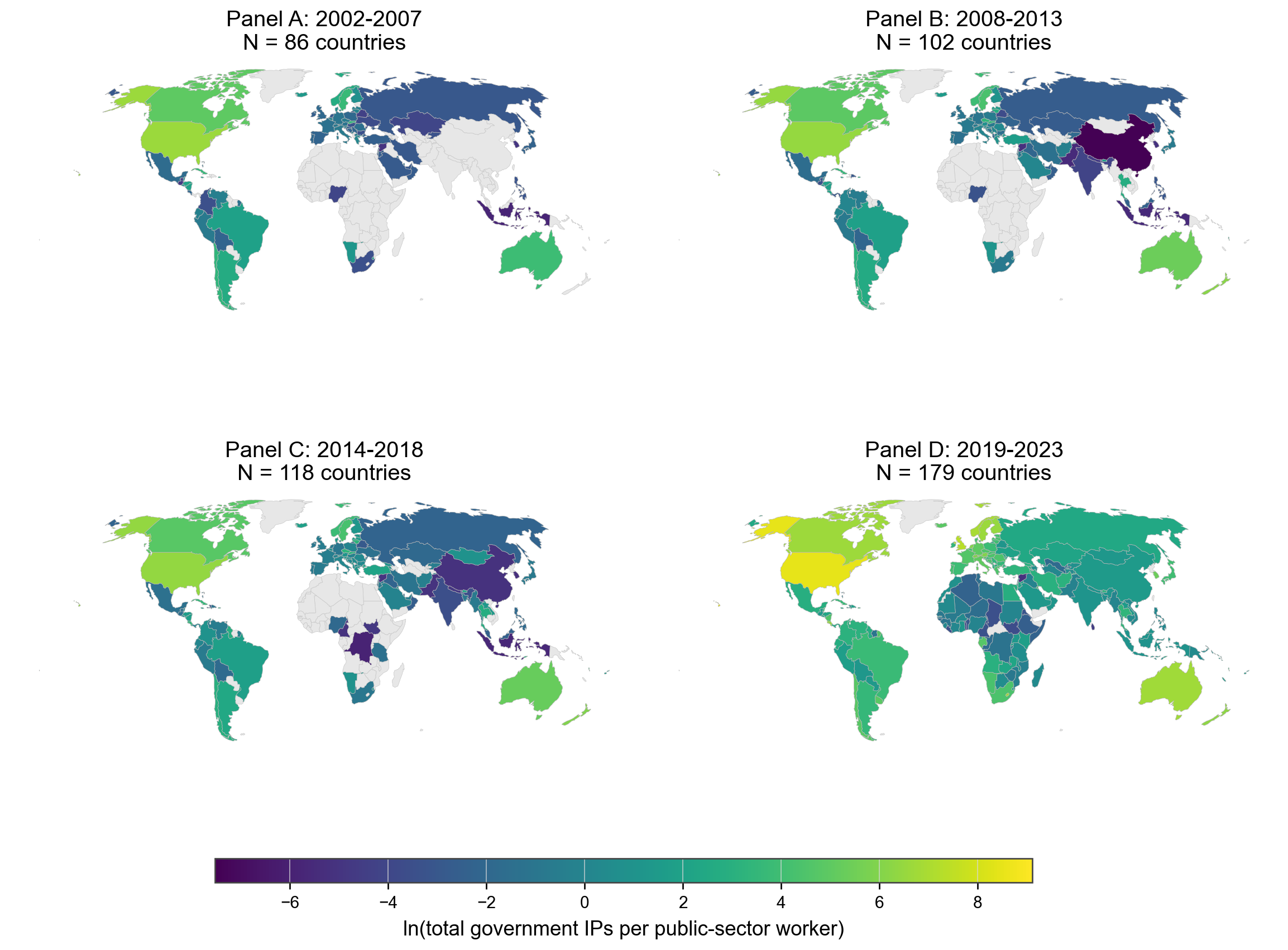}
    \fignote{Each country is shaded by the arithmetic mean, over the years shown, of its annual natural logarithm of total government IPs per public-sector worker. A country is included if it has positive government IPs and observed positive public-sector employment in at least one year of the period. Panels A--D contain 86, 102, 118, and 179 countries, respectively. Grey countries lack a defined value for the relevant period. Values before 2008 are shown descriptively; the main analysis sample begins in 2008.}
    
    \label{fig:nat_gov_ip_pubemp}
\end{figure}

\begin{figure}
    \centering
        \caption{Country-Level Log Total Government IPs per Capita, 2002--2023}
        \includegraphics[width=\linewidth]{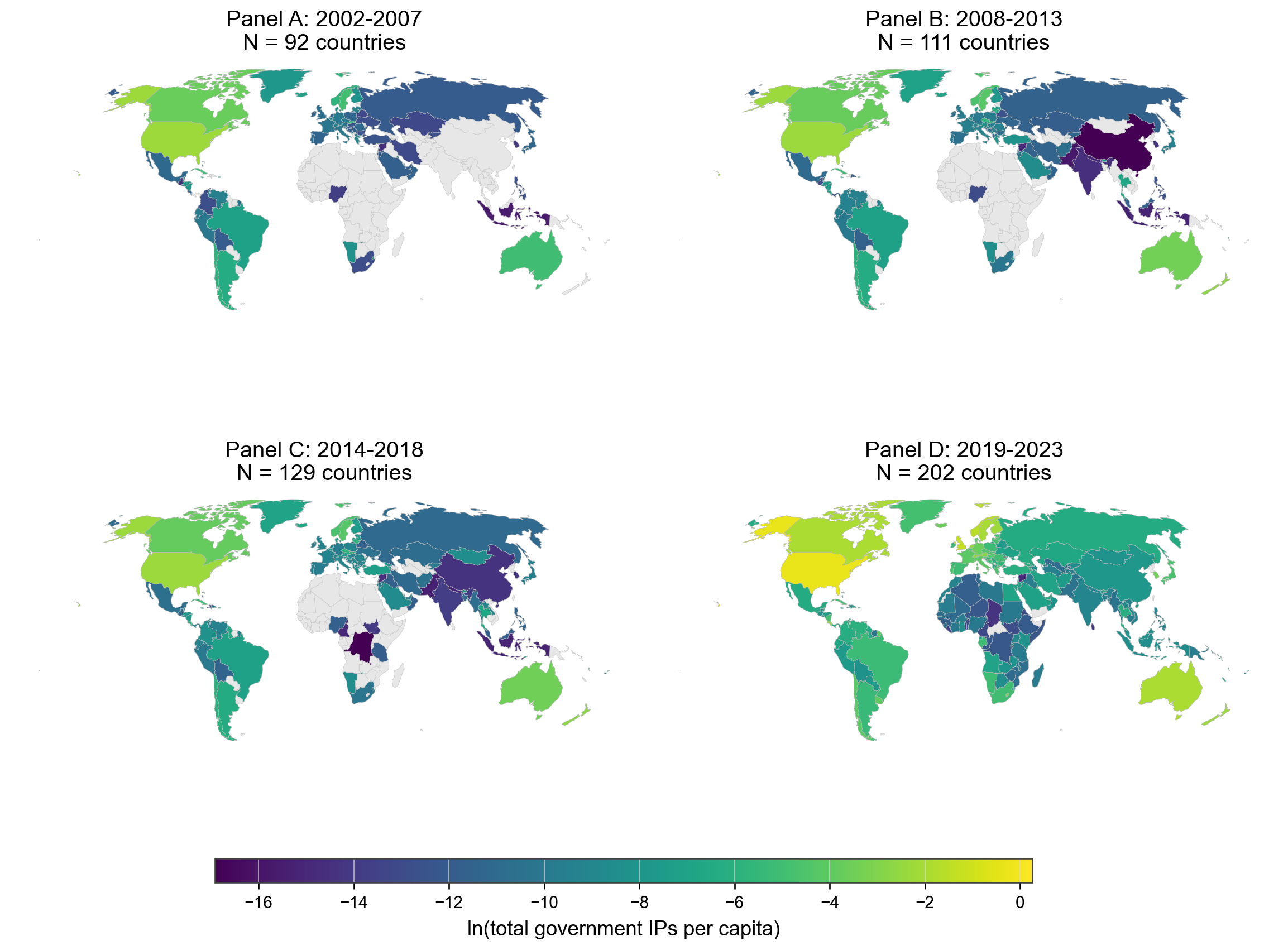}
        \fignote{Each country is shaded by the arithmetic mean, over the years shown, of its annual natural logarithm of total government IPs per capita. A country is included if it has positive government IPs and observed positive population in at least one year of the period. Panels A--D contain 92, 111, 129, and 202 countries, respectively. Grey countries lack a defined value for the relevant period. The main analysis sample begins in 2008.}
    \label{fig:nat_gov_ip_cap}
\end{figure}

For the country-level government IP count in 2019–2023, we aggregate IP addresses assigned to government-classified organisations in each month to the country level and take the average for the calendar year. Figure~\ref{fig:dbipmap} maps the government entities observed in the monthly DB-IP data, while Figure~\ref{fig:total_gov_ip_map} shows the corresponding Admin-1 distribution of government ICT infrastructure intensity, measured as the natural logarithm of government IPs per capita in 2020. For 2008–2018, we use historical registration dates and aggregate cumulative government IP allocations at the country-year level. Because the historical source underestimates pre-2019 holdings (Appendix~\ref{app:historical_validation}), the analysis focuses on level differences between countries rather than within-country changes over time. The denominator-specific coverage maps in Figures~\ref{fig:nat_gov_ip_pubemp} and~\ref{fig:nat_gov_ip_cap} illustrate the change from the historical panel to monthly DB-IP data: observed countries increase from 129 to 202 in the per-capita figure and from 118 to 179 in the per-worker figure between the 2014–2018 and 2019–2023 panels, with many additions in Africa. These counts differ because each transformation depends on denominator-specific data availability.

\paragraph{Validation framework.}
We evaluate DSC through measurement-validity exercises and a set of outcome applications. Convergent validation asks whether DSC correlates with related constructs, including e-government, statistical capacity, and state capacity. Discriminant validation asks whether DSC remains distinct from neighbouring but conceptually different constructs, such as GDP per capita, household internet access, and expert-coded internet-control capacity or practice. Conditional convergent associations examine whether these relationships remain similar after accounting for basic country characteristics. Finally, the outcome applications examine whether DSC is associated with selected government outcomes in expected directions, without interpreting these associations causally.

\textcite{Vaccaro2023StateCapacity} shows that state-capacity measures can have high convergent validity while remaining non-interchangeable. We therefore interpret positive relationships between DSC and existing state-capacity or digital-government measures as evidence that government IP holdings capture a related public-sector ICT-capacity dimension, not as evidence that DSC is equivalent to broader latent state-capacity indices. Following \textcite{McMann2022DataQuality}, \textcite{Seawright2014RivalStrategies}, \textcite{hanson2021leviathan}, \textcite{LeeZhang2017StateCapacity}, and \textcite{chong2014letter}, we first compare DSC with related and unrelated measures of e-government and state capacity. We then perform a similar OLS correlation test on the related measures controlling for population and GDP. Finally, we use our measures as independent variables in OLS regressions for selected outcomes that are theoretically related to Digital State Capacity.

\paragraph{Comparison Measures}
\label{sec:comparisons}

We compare both digital state capacity measures to existing measures of state capacity, informational capacity, digital government, and government quality. The World Bank Statistical Performance Indicators \parencite[SPI;][]{worldbank_spi} measure the performance of national statistical systems across data use, data services, data products, data sources, and data infrastructure. In the controlled validation tables, we use the current World Bank SPI release and cite the 2025 update where those values are discussed \parencite{WorldBankSPI2025Update}. The United Nations E-Government Survey \parencite{UN2022EGovernmentSurvey} evaluates national government portals and online-service provision. The authors emphasise that it is not intended as an absolute measure of e-government capability. The E-Government Development Index combines the Online Service Index, Telecommunications Infrastructure Index, and Human Capital Index; the E-Participation Index is reported separately. Broad measures of government and bureaucratic quality from the Varieties of Democracy (V-Dem) dataset \parencite{VDemV13, Vdemcodebook2023} aggregate expert judgements on government quality. We focus on Rigorous and Impartial Public Administration, State Fiscal Capacity, and State Authority over Territory. Of these, our measure correlates most strongly with Impartial Public Administration. In addition, we consider the State Capacity Index from \textcite{hanson2021leviathan}, which uses high-dimensional methods to create a latent measure of state capacity derived from many standard indicators, and the Information Capacity Index from \textcite{brambor2020lay}, which measures the frequency of conducting a census and the historical start of census taking in a country. Finally, we examine the Worldwide Governance Indicators \parencite{Kaufmann2011WorldwideGovernance}, which reflect expert opinions and survey responses on government quality across several dimensions: voice and accountability, political stability and the absence of violence or terrorism, government effectiveness, regulatory quality, rule of law, and control of corruption. We do not run comparisons on the \textcite{LeeZhang2017StateCapacity} legibility dataset, which measures bunching of age in census records as an objective measure of informational state capacity, because most observations occur before our observation period and the years differ across countries.

If DSC measures a dimension of modern informational state capacity, it should be positively correlated with measures of digital government, statistical capacity, government effectiveness, and broader state capacity, while not being identical to any one of them. The first set of validation tables presents government IPs per capita and per public-sector worker across a range of standard measures, with the Year column indicating the period during which DSC overlaps with each reference measure. The main validation tables use monthly DB-IP observations from 2019 to 2023. Table~\ref{tab:pearson_correlation_per_capita_2019} reports pairwise correlations for government IPs per capita, and Table~\ref{tab:pearson_correlation_per_worker_2019} reports the corresponding correlations for government IPs per public-sector worker. Table~\ref{tab:country_sum} provides variable-specific summary statistics for the extended 2008 to 2023 national DSC panel; the preceding validation tables use only monthly DB-IP observations from 2019 onward. Extended-sample correlations from 2008 to 2023 are reported in Appendix~\ref{app:extended_validation_tables}; these combine the lower-coverage historical data through 2018 with monthly DB-IP observations from 2019 onward and are interpreted as supplementary evidence rather than the preferred DB-IP-only validation sample. The descriptive statistics for the raw government IP counts are in Appendix~\ref{app:raw_ip}.

The correlations from 2019 to 2023 are generally stronger and use only monthly DB-IP observations. In Tables~\ref{tab:pearson_correlation_per_capita_2019} and~\ref{tab:pearson_correlation_per_worker_2019}, DSC is positively correlated with the World Bank statistical system measures, the U.N. E-Government indicators, and the Worldwide Governance Indicators. Across these measures, the correlations are strongest for the log transformations. In the extended-sample tables in Appendix~\ref{app:extended_validation_tables}, the raw per-capita DSC measure is negatively correlated with the Information Capacity Index from \textcite{brambor2020lay}, while the log transformation is close to zero. This is consistent with the index's historical and relatively time-invariant construction, limited overlap with the DSC sample, and the fact that several technically advanced states do not maintain centralised population registers. The discontinued World Bank Doing Business Report registration measures \parencite{business2020world} are weakly negatively correlated with DSC, suggesting that countries with higher digital state capacity tend to have lower business-registration costs. The State Capacity Index from \textcite{hanson2021leviathan}, which uses high-dimensional methods to create a latent measure of state capacity from standard indicators, is positively correlated with DSC. The per-worker overlap is limited around 2010 because the denominator requires labour statistics coverage. The World Bank statistical system measures are also positive, with the strongest subindicator correlations for Data Sources and weaker correlations for Data Use and Data Products. Correlations with digital government indicators are strongest for the composite index and the telecommunications infrastructure index. Taken together, the log-transformed correlations are in a similar range to the convergent-validity analysis in \textcite{chong2014letter}, which reports coefficients between 0.24 and 0.67; they are below the strongest correlations reported for the aggregate state-capacity indicator in \textcite{hanson2021leviathan}, where the strongest correlations range between 0.55 and 0.85. Federal and local government IPs show similar results in Table~\ref{tab:pearson_fed} and Table~\ref{tab:pearson_local}, with correlations in a similar range to \textcite{LeeZhang2017StateCapacity}. Appendix Table~\ref{tab:app_spi_country_universe_sensitivity} compares the SPI correlations with results for all SPI-covered ISO areas. The aggregate and most pillar correlations are similar, while local Data Use and Data Products are sensitive to the inclusion of small territories and non-member areas.

Following \textcite{hanson2021leviathan}, we also incorporate measures of civil society participation, including the involvement of women in civil society, overall civil society participation, the presence of multiparty elections, and the occurrence of consecutive presidential elections. From the World Bank's Sustainable Development Goals data, we include economic and infrastructure indicators, such as access to electricity, birth registration completeness, exports as a percentage of GDP, inflation rates, oil rents, population size, GDP per capita, and internet access. We also include the Polity V democracy score, with values after 2018 imputed from the 2018 value, and variables from the Varieties of Democracy dataset \parencite{VDemV13, Vdemcodebook2023} that capture expert opinions on the government's technical capacity to control and regulate the internet. These include measures of internet filtering, shutdown capacity, cybersecurity, and the ability to regulate online content. Discriminant validation compares DSC with related but conceptually distinct constructs. A valid measure should be correlated with economic development and economy-wide ICT diffusion, but not collapse into GDP per capita, household internet access, electricity access, democracy, civil-society participation, or V-Dem measures of internet control. Near-perfect correlations with GDP or internet access would suggest that DSC is simply a proxy for development or national ICT diffusion rather than a government-specific infrastructure measure \parencite{hanson2021leviathan}. Discriminant validity therefore requires DSC to retain variation that is distinct from these adjacent constructs. Appendix Table~\ref{tab:app_pearson_convergent_per_capita_2008_2023} reports the per-capita measure of Digital State Capacity. Appendix Table~\ref{tab:app_pearson_convergent_per_worker_2008_2023} reports the per-worker measure of Digital State Capacity.

In the broad extended-sample discriminant tables, the strongest correlations are with GDP per capita and the percent of the population using the internet. This pattern shows that DSC is higher in richer countries with more developed internet access, but the correlations are not high enough to suggest that DSC is simply a development or economy-wide ICT proxy. The V-Dem measures of cyber capacity are significant but relatively small for filtering and regulating the internet, consistent with those expert-coded measures capturing the joint legal and technical capability of the central government, while DSC measures digitisation across levels and functions of government. The lowest correlations are with export growth, consecutive presidential elections, and oil rents as a percent of GDP. Overall, Digital State Capacity is positively correlated with broader political and development indicators, such as civil society participation, multiparty elections, access to electricity, and urbanisation, but these correlations are generally weaker than its correlations with the closest state-capacity and digital-government measures (Appendix Tables~\ref{tab:app_pearson_convergent_per_capita_2008_2023}–\ref{tab:app_pearson_convergent_per_worker_2008_2023}). The V-Dem internet-control variables are especially useful discriminant comparators because they measure legal, political, technical, or practised control over the online information environment. DSC instead measures public-sector digital infrastructure, which may be used for service delivery, monitoring, enforcement, censorship, or other state objectives.

\paragraph{OLS Correlations}
We next estimate controlled associations between DSC and existing measures of state capacity and digital government using the following specification.
Because population, GDP, and internet use are themselves correlated with DSC, these specifications should be interpreted as controlled descriptive correlations rather than causal estimates.

\begin{equation}\label{eq:baseline}
    Y_{c,t} = \beta_0 + \beta_1 DSC_{c,t} + X_{c,t}'\gamma + \lambda_{r,t} + \varepsilon_{c,t}
\end{equation}

where $Y_{c,t}$ is an existing measure of state capacity and $X_{c,t}$ is a vector of country-year controls with coefficient vector $\gamma$. The broad validation specifications control for log population, log GDP per capita, and population internet access. Specifications that additionally control for democracy identify this separately in their table notes. $\lambda_{r,t}$ is region (r) times year (t) fixed effects capturing shocks that affect the entire region in that year.\footnote{Regions are continents: Americas, Europe, Asia and Oceania, and Africa. Because the historical series contains limited within-country variation, country fixed effects would absorb much of the cross-country level variation that the measure is designed to capture. We therefore use region-by-year fixed effects to absorb common regional shocks while preserving cross-country level comparisons. Figure~\ref{fig:2019_break_ip_trend} shows the limited year-to-year variation in the historical government IP count and the 2019 change in data source and construction.} Because region-by-year fixed effects absorb common year shocks, we do not include separate year fixed effects in the specification. Standard errors are two-way clustered by country and region-year unless otherwise indicated.

Table~\ref{tab:percapita_capacity_cal} reports controlled associations between per-capita DSC and existing measures of state capacity. Per-capita DSC is positively associated with the U.N. E-Government Index, the World Bank Statistical Performance Index, and the Hanson--Sigman State Capacity Index. The postal delivery-time and returned-letter measures from \textcite{chong2014letter} are not statistically distinguishable from zero. Because the comparator measures cover different periods, this table provides extended convergent-validity evidence rather than a common-sample comparison. Table~\ref{tab:perworker_capacity_cal} reports the corresponding per-public-sector-worker specification, which measures the ICT capital intensity of the public-sector workforce. Per-worker DSC is positively associated with the U.N. E-Government Index, the World Bank Statistical Performance Index, and the Hanson--Sigman State Capacity Index. The postal delivery-time and returned-letter coefficients are not statistically distinguishable from zero. The similarity to the per-capita results indicates that the validation pattern is similar under the public-sector capital-intensity interpretation. The subindicator tables report the same pattern in more detail. Tables~\ref{tab:percapita_stat_sub} and~\ref{tab:perworker_stat_sub} show that the Statistical Performance composite index and Data Sources are positive and statistically significant. The coefficient on Data Use is positive but not statistically significant after controls and region-by-year fixed effects. Data Services and Data Products are not significant. This pattern provides partial convergent-validity evidence: the association is clearest for overall statistical performance and the underlying sources available to the statistical system, rather than uniformly across all selected pillars. Table~\ref{tab:percapita_egov_sub} examines the components of the U.N. E-Government measure. Per-capita DSC is associated with a better E-Government Rank, where a lower numerical rank denotes better performance. The coefficients for the overall E-Government Index, E-Participation, Telecommunications Infrastructure, and Online Services have the expected positive signs but are not statistically distinguishable from zero. The overall-index specification uses a smaller sample because it additionally controls for tax revenue as a share of GDP. Relative to the broader specification in Table~\ref{tab:percapita_capacity_cal}, these results show that the association with E-Government outcomes is sensitive to the control set and estimation sample. Table~\ref{tab:perworker_egov_sub} reports the corresponding per-worker normalisation. The per-worker results follow the same pattern: DSC is associated with a better E-Government Rank, while the coefficients for the overall index, E-Participation, Telecommunications Infrastructure, and Online Services are not statistically distinguishable from zero.

Appendix Tables~\ref{tab:app_bad_controls_percapita_capacity_cal} and~\ref{tab:app_bad_controls_perworker_capacity_cal} report validation specifications without log GDP per capita, log population, or population internet access. They retain region-by-year fixed effects. The positive associations with the U.N. E-Government Index, the World Bank Statistical Performance Index, and the Hanson-Sigman State Capacity Index remain under both normalisations, while the postal measures do not show a consistent validation pattern. Because the appendix and main tables use different estimation samples, the comparison does not isolate the contribution of any single control. The controlled specifications remain the primary validation estimates.

\subsection{Subnational, Level, and Function Variation}
\label{sec:subnational_level_function}

The validation exercises above treat DSC primarily as a country-year measure. A central contribution of the dataset, however, is that the same entity-level IP records can be aggregated within countries by subnational geography, administrative level, and government function. We therefore use these classifications to describe how observed digital state capacity is distributed across the state apparatus. These patterns are descriptive: they show where government IPs recorded in DB-IP are concentrated, not whether those resources are effectively used. These subnational comparisons are not adjusted for cross-country differences in fiscal decentralisation or formal responsibility for service provision. Countries that assign more responsibilities to regional or local governments may naturally display different vertical allocations of government IPs.

\begin{figure}[ht]
    \centering
    \caption{Admin-1 Log Government IPs per Capita by Administrative Level, 2020}
    \includegraphics[width=\linewidth]{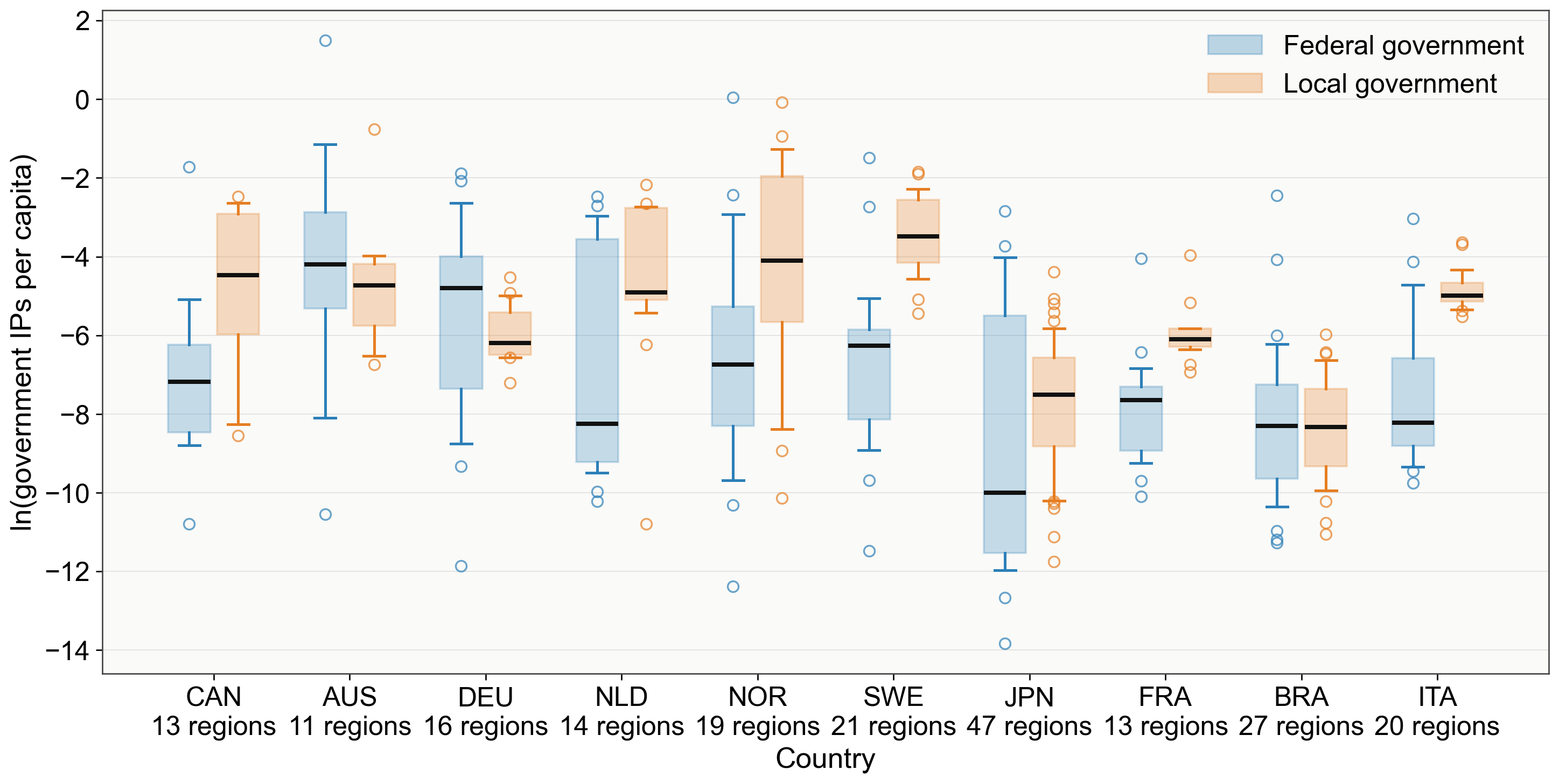}
    \fignote{The figure reports the distribution across Admin-1 regions of the natural logarithm of federal- and local-government IPs per capita. The sample comprises the ten democracies with the largest total government IP counts in 2020 among countries meeting the regional-coverage requirements. Eligibility requires more than five population-covered Admin-1 regions and at least five regions with positive government IPs at each administrative level. Boxes span the 25th to 75th percentiles, black lines indicate medians, whiskers indicate the 10th and 90th percentiles, and markers denote observations outside the whiskers. Regime categories use the last observed Polity V classification, carried forward to 2020.}
    \label{fig:levels_democracy_logpop}
\end{figure}

Figure~\ref{fig:levels_democracy_logpop} illustrates the vertical allocation of government IPs per capita across federal and local entities for the ten democracies with the largest total government IP counts in 2020 among countries meeting the regional-coverage requirements. The figure shows substantial within-country dispersion across Admin-1 regions and clear differences in the balance between federal and local capacity. The appendix reports related federal-only and local-only distributions, together with non-democracy and broader caterpillar diagnostics for countries with sufficient Admin-1 coverage (Appendix~\ref{app:additional_level_function_descriptives}). We treat those extended figures as diagnostic evidence because level classifications and subnational geolocation are more sensitive to country-specific administrative structure and DB-IP geolocation coverage.

\begin{figure}[ht]
    \centering
    \caption{Country-Level Log Government IPs per Capita by Government Function and Regime Type, 2020}
    \includegraphics[width=\linewidth]{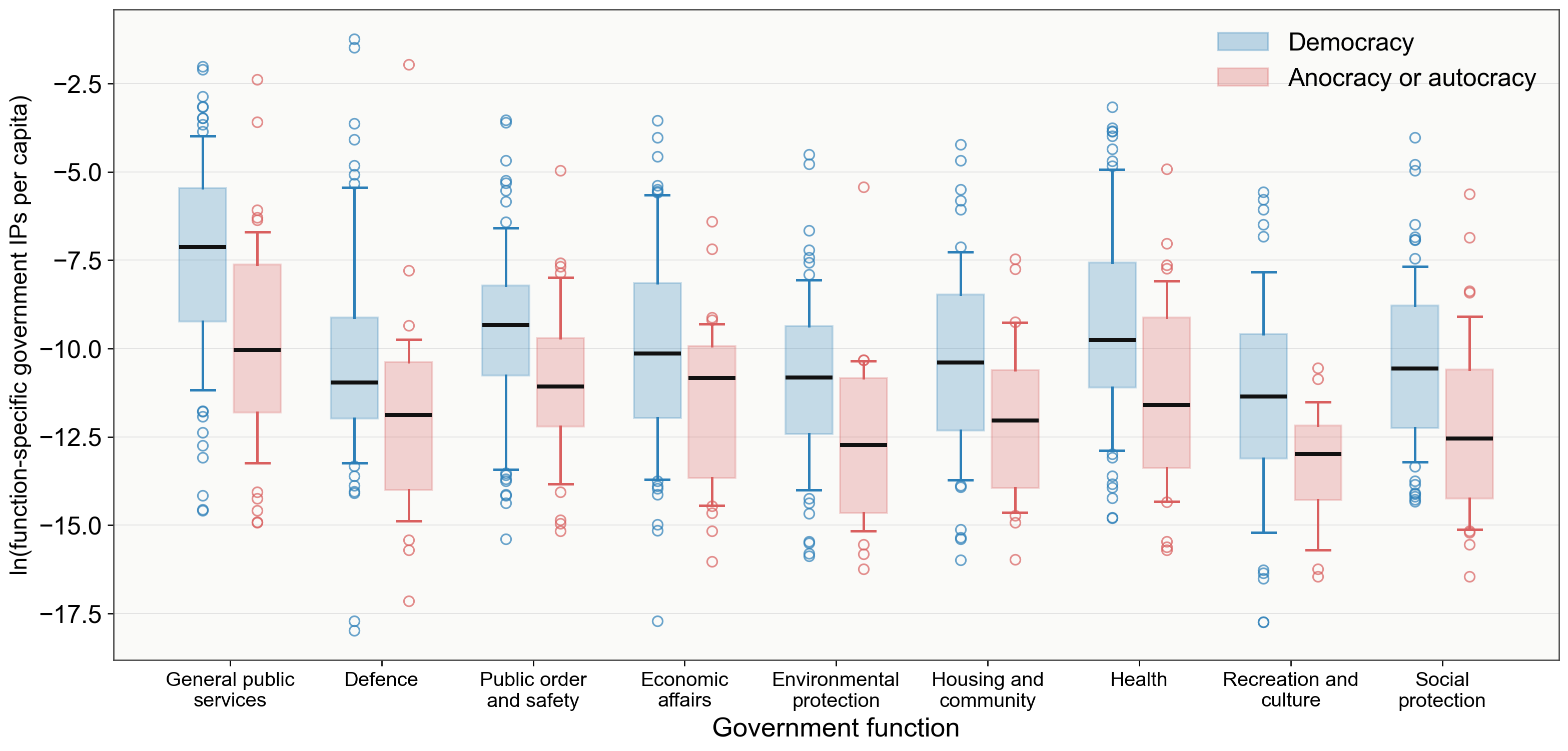}
    \fignote{The unit of observation is a country-function pair. Function-specific government IPs are summed across Admin-1 regions and divided by the corresponding aggregate population; the vertical axis reports the natural logarithm of this ratio. Country-function pairs with zero government IPs are omitted because the logarithm is undefined. Boxes span the 25th to 75th percentiles, black lines indicate medians, whiskers indicate the 10th and 90th percentiles, and markers denote observations outside the whiskers. Blue denotes democracies and red denotes anocracies or autocracies. Regime categories use the last observed Polity V classification, carried forward to 2020.}
    \label{fig:cofog_ip_distribution_dem_country}
\end{figure}

We next classify government entities by COFOG function. Figure~\ref{fig:cofog_ip_distribution_dem_country} compares government IPs per capita across major COFOG functions for democracies and non-democracies, using country-function pairs as the unit of observation. The figure shows that DSC is not evenly distributed across functions: general public services, public order and safety, health, environmental protection, housing, and social protection tend to have higher digital capacity in democracies, while defence is closer across regime types. This functional heterogeneity motivates the function-specific outcome tables in Section~\ref{sec:level_funct_outcomes}. Appendix~\ref{app:additional_level_function_descriptives} reports the corresponding top-country subnational distributions by function, separately for per-capita and raw-count transformations.

\section{National Level Outcomes}
\label{sec:nat_outcomes}

We conduct regressions in two stages. First, we further validate the data by showing correlations with other related measures of state capacity. Second, we estimate controlled associations between Digital State Capacity and selected outcomes for which public-sector information-processing capacity is theoretically relevant. \textcite{LeeZhang2017StateCapacity} test their measure of informational state capacity on tax revenue, mortality, literacy, and educational enrolment. \textcite{hanson2021leviathan} regress their aggregate measure of state capacity on infant mortality, life expectancy, the number of roads, water availability, and the number of hospitals as measures of public-service delivery. Following this measurement-validation approach, we regress these country-year outcomes on our measures of Digital State Capacity. We consider country-level measures for vaccinations, infant mortality, corruption, taxation, and social insurance coverage as indicators of public-service delivery. \textcite{odone2021use} highlight the importance of digital technologies in vaccination programmes, while \textcite{khelfaoui2022information} emphasise the role of ICT in reducing infant mortality. Moreover, \textcite{andersen2009government} suggest that e-government can be an effective strategy to reduce corruption, and \textcite{dzansi2020technology} show that providing technology to tax collectors can improve tax collection, particularly at the local level in developing countries, by allowing collectors to learn more about otherwise hard-to-observe household characteristics.

We use several indicators from the World Bank. Specifically, we examine vaccination rates for hepatitis, measles, and diphtheria, tax revenue as a percentage of GDP, and infant and neonatal mortality rates. We also include bribery incidence, measured as the percentage of firms reporting at least one bribery request in public transactions involving utilities, permits, licences, or taxes. Furthermore, we examine the percentage of the population participating in social insurance programmes, including contributory pensions and health insurance \parencite{WorldBankASPIRE}. Finally, we incorporate the Corruption Perceptions Index from Transparency International \parencite{TransparencyInternational2023CPI}, for which higher values correspond to lower perceived corruption.

Tables~\ref{tab:dsc_pc_general_baseline} and~\ref{tab:dsc_pw_general_baseline} report the baseline conditional associations. Under both normalisations, higher DSC is associated with higher CPI scores (lower perceived corruption), lower bribery incidence, and higher tax revenue as a share of GDP; each association is statistically significant at the 5 percent level or better. Social insurance coverage is not statistically distinguishable from zero. A doubling of DSC per capita is associated with a 0.93-point increase in CPI, a 0.68 percentage-point reduction in bribery incidence, and a 0.36 percentage-point increase in the tax share. The corresponding per-worker estimates are a 0.88-point increase in CPI, a 0.71-percentage-point reduction in bribery incidence, and a 0.34-percentage-point increase in the tax share. The results are therefore similar across the two normalisations.\footnote{A one-standard-deviation difference in log DSC corresponds to approximately 19 times as many government IPs per resident or 15 times as many per public-sector worker. The corresponding CPI differences are 4.0 and 3.4 points.}

Next, we consider health outcomes that are common across countries. Table~\ref{tab:dsc_pc_health_baseline} reports the baseline regressions for the per-capita measure of Digital State Capacity with general health outcomes. Table~\ref{tab:dsc_pw_health_baseline} reports the per-worker measure. For interpretation, vaccination coefficients are expressed as percentage-point changes, while the log mortality coefficients are converted to percentage changes. In the per-capita specification, a one-standard-deviation increase is associated with a 2.7 percentage-point increase in measles coverage (Column 1), a 2.9 percentage-point increase in diphtheria coverage (Column 2), a 2.8 percentage-point increase in hepatitis B coverage (Column 3), a 13.0 percent decrease in neonatal mortality (Column 4), and a 14.8 percent decrease in infant mortality (Column 5). In the per-worker specification, a one-standard-deviation increase is associated with increases of 2.3, 2.6, and 2.5 percentage points in measles, diphtheria, and hepatitis B coverage, respectively, and decreases of 10.8 and 12.5 percent in neonatal and infant mortality. All coefficients are statistically significant at the 5 percent level or better. As a robustness check, we control for existing measures of state capacity to test whether our per capita and per-worker measures retain associations with the outcomes after conditioning on existing measures. These specifications provide an incremental-validation exercise. They ask whether DSC retains an association with theoretically relevant outcomes after conditioning on existing measures of similar constructs, such as the State Capacity Index, the E-Government Index, and the Statistical Capacity Index. The goal is not to show that DSC outperforms these measures, but to assess whether it contains non-redundant variation. Table~\ref{tab:dsc_pc_robust} reports the per-capita results for CPI and diphtheria vaccination, while Table~\ref{tab:dsc_pw_robust} reports the per-worker results. Columns 1 and 5 report the respective baseline specifications. Under both normalisations, the CPI coefficient remains significant after controlling for the E-Government Index (Column 3) or Statistical Performance Index (Column 4), but not after controlling for the Hanson--Sigman State Capacity Index (Column 2). The diphtheria coefficient remains significant with the Hanson--Sigman (Column 6) or E-Government control (Column 7), but not with the Statistical Performance Index (Column 8). Because the comparator measures cover different periods, these changes reflect both additional conditioning and changes in the estimation sample.

Table~\ref{tab:dsc_pc_vacc_robust} and Table~\ref{tab:dsc_pw_vacc_robust} show the per capita and per-worker robustness checks for hepatitis and measles vaccination rates. Columns 1 and 2 show the baseline specifications. Columns 3 and 4 control for the State Capacity Index. Columns 5 and 6 control for the E-Government Index. Columns 7 and 8 control for the Statistical Performance Index. For both specifications, the measles vaccination coefficient is robust to the inclusion of the State Capacity Index (Column 4), remaining significant at the 5 percent level. With the E-Government Index included (Column 6), the measles coefficient remains significant at the 5 percent level in the per-capita specification and at the 10 percent level in the per-worker specification, but it is not significant with the Statistical Performance Index (Column 8). The hepatitis coefficient is not significant with the State Capacity Index (Column 3), is significant at the 10 percent level with the E-Government Index (Column 5), and is not significant with the Statistical Performance Index (Column 7). The measles results therefore follow a similar pattern to diphtheria, while the hepatitis results are weaker. Table~\ref{tab:dsc_pc_mort_robust} and Table~\ref{tab:dsc_pw_mort_robust} show the per capita and per-worker robustness checks for infant and neonatal mortality. Columns 1 and 2 show the baseline specifications. Columns 3 and 4 control for the State Capacity Index. Columns 5 and 6 control for the E-Government Index. Columns 7 and 8 control for the Statistical Performance Index. In the per-worker specification, the DSC coefficient remains statistically significant when controlling for the State Capacity Index (Columns 3--4) and for infant mortality when controlling for the E-Government Index (Column 5), but it is not significant for neonatal mortality with the E-Government Index (Column 6) or once the Statistical Performance Index is included (Columns 7--8). In the per-capita specification, the DSC coefficient remains significant when controlling for the State Capacity Index (Columns 3--4) and the E-Government Index (Columns 5--6), although the neonatal mortality coefficient in Column 6 is significant only at the 10 percent level. It is not significant when the Statistical Performance Index is included (Columns 7--8).

Appendix Tables~\ref{tab:app_bad_controls_pc_general_baseline}--\ref{tab:app_bad_controls_pw_health_baseline} report outcome associations without country-year controls and with separate region and year fixed effects. The directions of the CPI, tax, vaccination, and mortality associations are similar to the main estimates, while bribery is less stable across normalisations and the social-insurance estimates differ from the controlled specifications. The coefficients are generally larger when the country-year controls are omitted. This difference should not be interpreted as an estimate of omitted-variable bias because the fixed effects, data coverage, and estimation samples also differ. The controlled specifications in the main tables remain the basis for interpreting magnitudes.

\subsection{Heterogeneity Analysis}
\label{sec:hetero_variables}
The capacity/use distinction articulated in Section~\ref{sec:dsc} has direct empirical implications. If DSC measures capacity rather than welfare-improving use, the same level of digital capacity may produce different outcomes in different institutional environments, depending on whether digital infrastructure is deployed for service delivery or for information control. The freedom-of-press and freedom-of-the-net interactions reported for corruption below speak to this. We first use level-specific decompositions to ask whether the national associations are more closely related to federal or local government IPs. We then turn to heterogeneity by institutional environment, region, and income group. Specifically, we examine interactions between our Digital State Capacity measures and region, level of democracy, and wealth, and include our baseline controls (log population, log GDP per capita, percent of households with internet, and the level of democracy). In addition, we explore descriptive heterogeneity related to freedom of information in terms of traditional press freedom and freedom on the internet. Since some countries do not have a state level of government, we do not include the state level in this analysis.

Our interaction terms for country wealth are based on historical World Bank data \parencite{worldbank2025}, which categorise countries into four income tiers: low income (L), lower middle income (LM), upper middle income (UM), and high income (H). The classification details are provided by \textcite{worldbank_classification}, based on GNI per capita in the respective financial year. For internet freedom, instead of the Freedom on the Net measure \parencite{house2022freedom}, which includes a limited number of countries, we use the V-Dem measures of internet filtering, internet censorship, and social media filtering, creating a binary indicator equal to 1 for country-years below the sample median, indicating a more restrictive information environment \parencite{VDemV13, Vdemcodebook2023}. For traditional media freedom, we use the Freedom of the Press measure \parencite{freedom2009freedom}, assigning an analogous binary indicator for lower press freedom. For regions, we use continents: Europe, the Americas, Asia and Oceania, and Africa.

\textcite{odone2021use} emphasise the role of digital technologies in vaccination programmes, including digitising programme structures, institutional websites for information diffusion, digital medical record infrastructure, and state institutions' ability to collaborate with relevant parties.\footnote{We do not address misinformation and hesitancy here.} Effective programmes are tailored to the target population's specific needs and characteristics, suggesting that local informational capacity is important. For infant mortality, \textcite{khelfaoui2022information} explain that ICT's effect on reducing infant mortality is strongest in low-income countries, facilitated by inter-organisational communication and relevant ICT skills. Citizens' internet access is also important. This implies that DSC and citizen internet access are complementary: government digital infrastructure can expand administrative capacity, but many welfare-improving uses require citizens or intermediaries to be reachable through digital channels. Similar to vaccinations \parencite{odone2021use} and taxation \parencite{dzansi2020technology, Balan2022TaxComplianceDRC}, we hypothesise that local level informational capacity is important. For both health outcomes, we expect associations to strengthen as development progresses from low to middle-income settings. Recent government efforts in more developed countries, for example, focus on digitising health ministries and implementing national digital health systems \parencite{odone2021use}. Such transformations may not yet have occurred in developing countries, where basic infrastructure such as a national health portal may exist but remain insufficient on its own \parencite{odone2021use}.

\begin{sloppypar}
\textcite{andersen2009government} highlight multiple channels through which ICT and e-government can affect corruption. Many simple tasks, such as accessing land records, can reduce opportunities for corruption by streamlining processes through digitisation and reducing face-to-face interaction. However, if corruption is embedded in institutions, digitisation could increase corruption. \textcite{bertot2010using} identify several anti-corruption channels: proactive dissemination of rules, decisions, actions, and expenditures; information release on request; public meetings; whistleblower leaks; and more open processes for taxes, land records, licences, and related transactions. The effectiveness in reducing corruption depends on digital access and literacy among the general population, as well as the level of trust citizens have in the government \parencite[e.g.][]{guriev20213g}.
\end{sloppypar}

Tables~\ref{tab:loca_dsc_pc_general} and~\ref{tab:fed_dsc_pc_general} report the level-specific decompositions for the general outcomes, while Tables~\ref{tab:loca_dsc_pc_health} and~\ref{tab:fed_dsc_pc_health} report the corresponding health-outcome specifications. For federal DSC, higher DSC is associated with higher CPI scores, lower bribery incidence, higher tax revenue, and higher social insurance coverage; all four coefficients are statistically significant at the 5 percent level. For local DSC, the CPI coefficient is positive and significant at the 10 percent level, the tax-revenue coefficient is positive and significant at the 1 percent level, and the social-insurance coefficient is negative and significant at the 10 percent level; the bribery coefficient is not statistically distinguishable from zero. For health outcomes, none of the federal DSC coefficients is statistically significant. For local DSC, the measles coefficient is positive and significant at the 10 percent level, and the neonatal- and infant-mortality coefficients are negative and significant at the 5 percent level; the diphtheria and hepatitis coefficients are not statistically distinguishable from zero. These patterns are best read as descriptive level decompositions of where government IPs recorded in DB-IP are located within the state, not as evidence that a single administrative level mechanically drives the national associations.

Table~\ref{tab:corruption_freedom} reports results for the total government IPs per capita measure of Digital State Capacity interacted with measures of media and internet freedom, where the binary variable equals one for a more restrictive information environment. Relative to countries with less restrictive information environments, countries with lower Freedom of the Press (Column 1), stronger Internet Filtering (Column 2), stronger Social Media Monitoring (Column 3), and stronger Internet Censorship (Column 4) have a significantly weaker association between Digital State Capacity and improved corruption perceptions. Table~\ref{tab:corruption_continent} reports continent splits; the association is significant only for the Americas and is insignificant for Europe, Africa, and Asia/Oceania. Table~\ref{tab:corruption_wealth} reports the results for the per-capita Digital State Capacity measure by country-year income group; the association is significant at the 5 percent level for upper-middle-income economies and at the 1 percent level for low-income economies. Tables~\ref{tab:infmort_wealth}, \ref{tab:dip_wealth},
\ref{tab:hep_wealth}, and~\ref{tab:measles_wealth} report the health outcome
results by income group. The positive vaccination coefficients indicate that
higher DSC is associated with greater vaccination coverage, consistent with the
role of digital records and coordination described by
\textcite{odone2021use}. The coefficients for all three vaccines are
statistically significant for upper-middle-income economies (Column 2) and the
pooled lower-middle- and low-income sample (Column 5), but none is significant
for lower-middle-income economies alone (Column 3). The infant mortality
coefficients are negative, consistent in direction with
\textcite{khelfaoui2022information}, but are statistically significant only for
high-income and upper-middle-income economies (Columns 1 and 2). Table~\ref{tab:vac_freedom} and Table~\ref{tab:mortality_freedom} report the results for vaccination rates and infant mortality. Overall, these results are consistent with DSC measuring capacity rather than welfare-improving use. The corruption association is attenuated in environments with stronger press and internet controls, consistent with the same digital capacity being deployed differently across institutional contexts.

\subsection{EU (NUTS 2) Level}

Subnational measures of government quality or state capacity are limited. \textcite{luna2017capturing} highlight the difficulties in capturing and validating subnational variation in state capacity. We follow the same validation approach as for the national level. Given limited data, we focus on the European Union, where two relevant sources are available: the European Quality of Government Index, conducted in 2010, 2013, 2017, 2021, and 2024, and the share of the population who have interacted with, or submitted a form to, government online \parencite{Charron2021EQI}. Table~\ref{tab:correlation_dsc_eu} reports pairwise correlations between our Digital State Capacity per-capita and per-worker measures and these subnational comparators. The correlations between Digital State Capacity and the aggregate Government Quality Index and its subindicators are relatively low, indicating that subnational digital government is not reducible to these survey-based measures of government quality. 
The correlation is stronger for interaction with government through the internet (46 percent for the per-capita measure and 41 percent for the per-worker measure). Correlations are also positive for the percentage of households using the internet (47 and 43 percent) and log GDP per capita (51 and 45 percent). Table~\ref{tab:dsc_gov_dig_qual_val_ols} reports the OLS regression results of log government IPs per capita and per worker on the Digital Government measure and the aggregate Government Quality Index, controlling for log GDP per capita, log population, and the percentage of households with broadband internet. The Digital Government measure is significant at the 10 percent level for both log government IPs per capita (Column 1) and log government IPs per public-sector worker (Column 2). The coefficient on the Government Quality Index is not statistically significant.

% NOTE: Paragraph commented out pending co-author review — it describes Admin-1 nightlight/OSM controls and the WBES summary table (tab:es_sum), but no analysis in the current draft uses them.
\begin{comment}
We control for the log of the 2020 population \parencite{CIESIN2016} and the log of nightlight intensity per capita at the yearly level from 2019 to 2023 (using 2020 population as the denominator). Nightlight intensity data, masked to remove background noise, is sourced from the VIIRS V2 dataset, with 2022 and 2023 data downloaded from the Earth Observation Group (Version 22) \parencite{EOG_NighttimeLights_2022, EOG_NighttimeLights_2023}. We calculate the average monthly radiance and aggregate the total counts to the Admin-1 level using \textcite{GADM2024} boundaries. We also use data from Daylight OpenStreetMap to identify government amenities and offices, following the approach of \textcite{ChangWang2021ReachOfState}. This variable is cross-sectional and was queried in 2023. These counts are similarly aggregated to the Admin-1 level using \textcite{GADM2024} boundaries. Table~\ref{tab:es_sum} presents summary statistics for all variables used in the analysis.
\end{comment}

\subsection{Level and Function of Government}
\label{sec:level_funct_outcomes}

%Section \ref{sec:level_funct_outcomes} details 

Table~\ref{tab:ghs_correlations} reports descriptive correlations between health-specific and total government IP measures and the Global Health Security indicators. These correlations provide context for the function-specific outcome tables. Tables~\ref{tab:neo_mort}–\ref{tab:inf_mort} show that health-specific IPs are more strongly associated with mortality outcomes than total government IPs. For vaccination coverage, Tables~\ref{tab:vacc_meas}–\ref{tab:vacc_dip} show the reverse pattern in the normalised columns: total government IPs per capita and per public-sector worker are associated with outcomes as strongly as, or more strongly than, the corresponding health-specific measures. The raw-count columns compare total government IP holdings with general-public-services IP holdings and provide a scale-sensitive descriptive check rather than the main health-specific comparison. Together, these patterns suggest vaccination coverage correlates with broader government digitisation beyond the health sector alone.

Table~\ref{tab:inf_mort_het} shows that health IPs per capita exhibit the strongest association with infant mortality (18.1 percent, column 7), followed closely by Public Order and Safety (17.5 percent, column 3), while Defence (column 2) and Recreation (column 8) show notably weaker correlations. Table~\ref{tab:neonat_het} also shows negative and statistically significant coefficients across government functions, with the largest associations for Health and Environmental Protection (13.8 percent, Columns 7 and 5), followed closely by General Public Services (13.7 percent, Column 1) and Public Order and Safety (12.7 percent, Column 3).

These patterns are consistent with child mortality outcomes being associated with digital capacity across multiple government sectors rather than digitisation of the healthcare sector alone. Health sector digitisation corresponds to direct clinical service provision such as prenatal health monitoring, birth records, and maternal health records. Public Order and Safety digital systems relate to emergency response infrastructure, including ambulance dispatch. Social protection digital systems correspond to targeted health and welfare programmes. Environmental protection relates to provision and monitoring of clean water access and sanitation systems. General Public Services digitisation corresponds to foundational administrative infrastructure, including civil registration, censuses and core statistics systems, inter-governmental communication and coordination platforms, and data systems that underpin other sectors' operations and information sharing. Defence and recreation show relatively smaller associations as they play less direct roles in healthcare provision and child mortality outcomes.

Table~\ref{tab:measles_vacc_het} demonstrates that vaccination coverage is positively associated with digital capacity across multiple government functions, with significant coefficients for Health (column 7), Housing (column 6), General Public Services (column 1), Public Order (column 3), and Social Protection (column 9). This pattern is further illustrated in Table~\ref{tab:dip_vacc_het}, which shows health IPs are associated with the largest coefficient (1.6 percent, column 7), while General Public Services (column 1), Recreation (column 8), Public Order (column 3), and Social Protection (column 9) all show significant positive associations. Tables~\ref{tab:cpi_het}, \ref{tab:cpi_pw_het}, and~\ref{tab:cpi_total_het} report the corresponding function-level decompositions for the Corruption Perceptions Index in per-capita, per-worker, and raw-count terms.

\section{Conclusion}
This paper introduces Digital State Capacity, a new measure of government ICT capacity based on the public network infrastructure held by government organisations. By identifying government-held public IPv4 address space, DSC captures an observable stock of usable digital infrastructure that can be disaggregated by country, subnational region, organisation, administrative level, and government function. The measure adds a digital-infrastructure dimension to the state-capacity literature, complementing existing fiscal, administrative, informational, and digital-government measures.

The validation exercises show that DSC is positively related to adjacent measures of e-government, statistical capacity, government effectiveness, and broader state capacity, while remaining conceptually distinct from them. The descriptive applications show that countries with larger government ICT infrastructure stocks tend to have governance and public-service outcomes in the expected directions, including lower measured corruption and higher vaccination coverage. These associations demonstrate the empirical relevance of the measure, but they are not causal estimates. The measure captures government ICT infrastructure, not the use to which that capacity is put. The corruption heterogeneity results are consistent with this distinction: the association between DSC and lower corruption is attenuated in environments with stronger information controls, consistent with similar digital capacity being deployed differently across institutional contexts.

The measure also has important limits. DSC captures government-held public IPv4 resources and the capacity implied by acquiring and maintaining them; it does not directly observe software quality, staff expertise, service uptake, private addressing, cloud infrastructure, or IPv6 deployment. The historical 2008–2018 component should be interpreted as a lower-bound estimate relative to the monthly DB-IP data from 2019 onward. Classification uncertainty is also part of the measurement problem, especially for universities, state-owned telecommunications firms, broadcasters, utilities, and other public-sector boundary cases. These limitations define the appropriate interpretation of DSC: it is a scalable measure of public-sector ICT capital stock and digital-infrastructure capacity that complements measures of digital-government performance.

\pagebreak

\printbibliography[heading=subbibliography]

\pagebreak
\section*{Figures}
\clearpage

\clearpage

\clearpage

\clearpage

\clearpage

\clearpage

\clearpage
\pagebreak 
\section*{Tables}

\begin{table}[ht]
\centering
\caption{Regional Coverage of Government IP Observations, 2020}
\label{tab:dsc_regional_coverage}
\resizebox{\linewidth}{!}{%
\begin{tabular}{lccc}
\toprule
Region & \shortstack{Countries and territories\\with government IP} & \shortstack{Admin-1 regions\\with government IP} & \shortstack{Admin-1 population\\coverage} \\
\midrule
Africa & 57/58 & 250/877 & 47.8\% \\
Americas & 47/50 & 522/769 & 98.5\% \\
Asia & 47/57 & 618/964 & 93.8\% \\
Europe & 47/51 & 585/829 & 97.6\% \\
Oceania & 19/21 & 66/204 & 83.9\% \\
\midrule
Total & 217/237 & 2,041/3,643 & 86.7\% \\
\bottomrule
\end{tabular}
}
        \tabnote{Entries report countries and territories, and Admin-1 regions, with positive government IP ownership in calendar-year 2020 out of the population-covered denominator in each region. The denominator is restricted to countries, territories, and GADM Admin-1 regions with usable 2020 population data from \textcite{CIESIN2016}. Admin-1 population coverage is population-weighted and uses the same denominator. These are government-positive coverage statistics, not broader any-DB-IP source-visible counts. Missing government IP observations should be interpreted as non-visibility in this source and classification pipeline, not necessarily as absence of government digital infrastructure.}
\end{table}

\clearpage

\begin{table}[ht]
\centering
\small
\caption{Pearson Correlations with Government IPs per Capita and Their Natural Logarithm, 2019–2023}
\resizebox{\linewidth}{!}{%
\begin{tabularx}{\linewidth}{>{\raggedright\arraybackslash}Xcccc}
\toprule
\textbf{Variable} & \textbf{Year} & \textbf{Gov IP/Capita} & \textbf{ln(Gov IP/Capita)} & \textbf{N} \\
\midrule
\multicolumn{5}{l}{\textbf{Statistical Capacity WB}} \\
Data Infrastructure & 2019--2023 & 0.28*** & 0.60*** & 877 \\
Data Products & 2019--2023 & 0.07** & -0.00 & 979 \\
Data Services & 2019--2023 & 0.23*** & 0.51*** & 861 \\
Data Sources & 2019--2023 & 0.25*** & 0.64*** & 857 \\
Data Use & 2019--2023 & 0.15*** & 0.19*** & 979 \\
Statistical Index & 2019--2023 & 0.26*** & 0.59*** & 857 \\
\addlinespace
\multicolumn{5}{l}{\textbf{UN E-Gov Indicators}} \\
E-Gov Index & 2020--2022 & 0.29*** & 0.73*** & 355 \\
E-Gov Rank & 2020--2022 & -0.31*** & -0.73*** & 355 \\
E-Participation Index & 2020--2022 & 0.29*** & 0.57*** & 355 \\
Human Capital Index & 2020--2022 & 0.25*** & 0.75*** & 355 \\
Online Service Index & 2020--2022 & 0.27*** & 0.59*** & 355 \\
Telecom Infra Index & 2020--2022 & 0.28*** & 0.69*** & 355 \\
\addlinespace
\multicolumn{5}{l}{\textbf{Worldwide Governance}} \\
Ctrl Corruption & 2019--2023 & 0.34*** & 0.68*** & 913 \\
Gov Effectiveness & 2019--2023 & 0.32*** & 0.69*** & 913 \\
Political Stability & 2019--2023 & 0.17*** & 0.62*** & 913 \\
Regulatory Quality & 2019--2023 & 0.33*** & 0.71*** & 913 \\
Rule of Law & 2019--2023 & 0.34*** & 0.71*** & 913 \\
Voice \& Accountability & 2019--2023 & 0.29*** & 0.64*** & 901 \\
\bottomrule
\end{tabularx}

}
\tabnote{This table reports pairwise Pearson correlations with related measures of state capacity, informational capacity, and e-government over the years shown. The Worldwide Governance Indicator rows use the World Bank 2024 WGI update, which reports estimates through 2023 \parencite{WorldBank_WGI_2024}. Natural logarithms are used, and the N column reports the common pairwise sample for the raw and logged correlations. Stars indicate rejection of the null hypothesis that the pairwise correlation is zero: $^{*}p<0.10$, $^{**}p<0.05$, and $^{***}p<0.01$.}
\label{tab:pearson_correlation_per_capita_2019}
\end{table}

\clearpage

\begin{table}[ht]
\centering
\small
\caption{Pearson Correlations with Government IPs per Public-Sector Worker and Their Natural Logarithm, 2019–2023}
\resizebox{0.95\linewidth}{!}{%
\begin{tabularx}{\linewidth}{>{\raggedright\arraybackslash}Xcccc}
\toprule
\textbf{Variable} & \textbf{Year} & \textbf{Gov IP/Worker} & \textbf{ln(Gov IP/Worker)} & \textbf{N} \\
\midrule
\multicolumn{5}{l}{\textbf{Statistical Capacity WB}} \\
Data Infrastructure & 2019--2023 & 0.27*** & 0.62*** & 834 \\
Data Products & 2019--2023 & 0.09*** & 0.17*** & 874 \\
Data Services & 2019--2023 & 0.20*** & 0.53*** & 828 \\
Data Sources & 2019--2023 & 0.24*** & 0.61*** & 828 \\
Data Use & 2019--2023 & 0.16*** & 0.34*** & 874 \\
Statistical Index & 2019--2023 & 0.25*** & 0.61*** & 828 \\
\addlinespace
\multicolumn{5}{l}{\textbf{UN E-Gov Indicators}} \\
E-Gov Index & 2020--2022 & 0.26*** & 0.68*** & 336 \\
E-Gov Rank & 2020--2022 & -0.27*** & -0.68*** & 336 \\
E-Participation Index & 2020--2022 & 0.26*** & 0.57*** & 336 \\
Human Capital Index & 2020--2022 & 0.23*** & 0.68*** & 336 \\
Online Service Index & 2020--2022 & 0.25*** & 0.59*** & 336 \\
Telecom Infra Index & 2020--2022 & 0.25*** & 0.64*** & 336 \\
\addlinespace
\multicolumn{5}{l}{\textbf{Worldwide Governance}} \\
Ctrl Corruption & 2019--2023 & 0.32*** & 0.63*** & 845 \\
Gov Effectiveness & 2019--2023 & 0.29*** & 0.64*** & 845 \\
Political Stability & 2019--2023 & 0.16*** & 0.58*** & 845 \\
Regulatory Quality & 2019--2023 & 0.30*** & 0.67*** & 845 \\
Rule of Law & 2019--2023 & 0.32*** & 0.66*** & 845 \\
Voice \& Accountability & 2019--2023 & 0.28*** & 0.60*** & 839 \\
\bottomrule
\end{tabularx}

}
\tabnote{This table reports pairwise Pearson correlations with related measures of state capacity, informational capacity, and e-government over the years shown. The Worldwide Governance Indicator rows use the World Bank 2024 WGI update, which reports estimates through 2023 \parencite{WorldBank_WGI_2024}. Natural logarithms are used, and the N column reports the common pairwise sample for the raw and logged correlations. Stars indicate rejection of the null hypothesis that the pairwise correlation is zero: $^{*}p<0.10$, $^{**}p<0.05$, and $^{***}p<0.01$. The per-worker sample is restricted to country-years with observed public-sector employment.}
\label{tab:pearson_correlation_per_worker_2019}
\end{table}

\clearpage

\begin{table}[htbp]
    \centering
    \scriptsize
    \setlength{\tabcolsep}{3pt}
    \renewcommand{\arraystretch}{0.9}
    \begin{adjustbox}{max width=\linewidth,max totalheight=0.78\textheight}
    \begin{tabular}{l 
                D{.}{.}{2} 
                D{.}{.}{2} 
                D{.}{.}{2} 
                D{.}{.}{2} 
                D{.}{.}{0} 
               }
    \toprule
                        &        mean&          sd&         min&         max&       count\\
\midrule
ln Total Gov IPs p.c.&       -8.05&        2.96&      -18.94&       -0.19&        2206\\
ln Fed. Gov IPs p.c.&       -8.49&        3.12&      -16.95&       -0.26&         821\\
ln State Gov IPs p.c.&       -9.95&        2.82&      -17.99&       -2.72&         539\\
ln Local Gov IPs p.c.&       -8.58&        3.18&      -16.59&       -1.97&         831\\
ln Total Gov IPs Per Worker&        1.01&        2.73&       -9.53&        8.59&        2005\\
ln Federal Gov IPs Per Worker&        0.78&        2.79&       -6.95&        8.53&         767\\
ln State Gov IPs Per Worker&       -0.81&        2.54&       -7.57&        5.93&         528\\
ln Local Gov IPs Per Worker&        0.55&        2.76&       -6.22&        6.77&         755\\
ln GDP p.c.&        9.26&        1.37&        5.46&       12.24&        2169\\
ln Pop.&       15.61&        2.28&        9.25&       21.09&        2206\\
Internet percent Pop.&        0.61&        0.26&        0.04&        1.00&        2053\\
E-Gov Index&        0.61&        0.19&        0.00&        0.98&         966\\
E-Participation Index&        0.50&        0.28&        0.00&        1.00&         966\\
Govt Online Reg Cap.&        0.53&        0.17&        0.00&        0.96&        1727\\
Govt Internet Shut Cap.&        0.51&        0.23&        0.00&        1.00&        1727\\
Govt Internet Filter Cap.&        0.54&        0.19&        0.04&        1.00&        1727\\
Govt Cyber Security&        0.53&        0.18&        0.03&        1.00&        1727\\
Online Service Index&        0.57&        0.23&        0.00&        1.00&         966\\
Human Capital Index&        0.76&        0.17&        0.00&        1.00&         966\\
Telecom Infra Index&        0.49&        0.25&        0.00&        1.00&         966\\
Avg Days&      170.83&      102.15&       16.20&      418.80&          87\\
ln(Avg Days)&        4.96&        0.64&        2.79&        6.04&          87\\
Returned&        0.74&        0.27&        0.00&        1.00&          87\\
Statistical Index (WB)&        0.68&        0.17&        0.22&        0.95&        1186\\
Data Use (WB)&        0.67&        0.26&        0.00&        1.00&        2206\\
Data Services (WB)&        0.68&        0.22&        0.00&        1.00&        1190\\
Data Products (WB)&        0.60&        0.19&        0.07&        0.94&        2206\\
Data Sources (WB)&        0.56&        0.19&        0.03&        0.94&        1246\\
Data Infrastructure (WB)&        0.61&        0.27&        0.05&        1.00&        1213\\
Vacc. Meas. Prcnt.&        0.89&        0.12&        0.18&        0.99&        2000\\
Vacc. Dip. Prcnt.&        0.89&        0.12&        0.19&        0.99&        2000\\
Vacc. Hep. Prcnt.&        0.88&        0.13&        0.20&        0.99&        1871\\
Corruption Perception Index&        0.47&        0.20&        0.08&        0.92&        1548\\
Bribery Incidence (prcnt)&        0.12&        0.12&        0.00&        0.70&         201\\
Tax Prcnt. GDP. p.c.&        0.17&        0.07&        0.00&        0.40&        1511\\
ln Neonatal Mortality&        1.86&        0.96&       -0.51&        3.93&        2021\\
ln Infant Mortality&        2.50&        1.01&        0.26&        5.68&        2021\\
Registration Time (Days)&       43.52&       53.24&        1.00&      391.00&         633\\
Registration Cost&        4.61&        4.06&        0.00&       27.82&         633\\
Social Insurance Coverage&        0.23&        0.18&        0.01&        0.60&         335\\
\bottomrule

    \end{tabular}
    \end{adjustbox}
    \caption{Country-Level Summary Statistics, 2008--2023}
    \label{tab:country_sum}
    \tabnote{This table reports variable-specific summary statistics for the extended 2008--2023 country-year panel. The row universe consists of country-years with positive observed government IPs. Government-level measures are available only for 2019--2023. Natural logarithms are used. Counts vary with the temporal and geographic coverage of each measure; per-worker measures additionally require positive public-sector employment.}
\end{table}

\clearpage
\begin{table}[ht]
\centering
\scriptsize
\caption{Pairwise Correlations with Federal Government IP Measures, 2019--2023}
\begin{tabularx}{\linewidth}{>{\raggedright\arraybackslash}X c c c c}
\toprule
\textbf{Variable}
  & \makecell{\textbf{Federal IPs}\\\textbf{per capita}}
  & \makecell{\textbf{ln(Federal IPs}\\\textbf{per capita)}}
  & \makecell{\textbf{Federal IPs}\\\textbf{per worker}}
  & \makecell{\textbf{ln(Federal IPs}\\\textbf{per worker)}} \\
\midrule
\multicolumn{5}{l}{\textbf{Statistical Capacity WB (2019--2023)}} \\
Data Infrastructure & \makecell{0.20***\\(872)} & \makecell{0.57***\\(763)} & \makecell{0.18***\\(830)} & \makecell{0.57***\\(738)} \\
Data Products & \makecell{0.08**\\(884)} & \makecell{0.20***\\(768)} & \makecell{0.06*\\(835)} & \makecell{0.28***\\(743)} \\
Data Services & \makecell{0.16***\\(856)} & \makecell{0.50***\\(752)} & \makecell{0.13***\\(824)} & \makecell{0.49***\\(732)} \\
Data Sources & \makecell{0.18***\\(852)} & \makecell{0.62***\\(748)} & \makecell{0.16***\\(824)} & \makecell{0.59***\\(732)} \\
Data Use & \makecell{0.12***\\(884)} & \makecell{0.29***\\(768)} & \makecell{0.12***\\(835)} & \makecell{0.37***\\(743)} \\
Statistical Index & \makecell{0.19***\\(852)} & \makecell{0.57***\\(748)} & \makecell{0.16***\\(824)} & \makecell{0.58***\\(732)} \\
\addlinespace
\multicolumn{5}{l}{\textbf{UN E-Gov Indicators (2020--2022)}} \\
E-Gov Index & \makecell{0.20***\\(355)} & \makecell{0.67***\\(309)} & \makecell{0.18***\\(336)} & \makecell{0.61***\\(299)} \\
E-Gov Rank & \makecell{-0.21***\\(355)} & \makecell{-0.66***\\(309)} & \makecell{-0.18***\\(336)} & \makecell{-0.61***\\(299)} \\
E-Participation Index & \makecell{0.22***\\(355)} & \makecell{0.56***\\(309)} & \makecell{0.19***\\(336)} & \makecell{0.54***\\(299)} \\
Human Capital Index & \makecell{0.17***\\(355)} & \makecell{0.68***\\(309)} & \makecell{0.15***\\(336)} & \makecell{0.60***\\(299)} \\
Online Service Index & \makecell{0.20***\\(355)} & \makecell{0.55***\\(309)} & \makecell{0.17***\\(336)} & \makecell{0.53***\\(299)} \\
Telecom Infra Index & \makecell{0.19***\\(355)} & \makecell{0.63***\\(309)} & \makecell{0.17***\\(336)} & \makecell{0.58***\\(299)} \\
\addlinespace
\multicolumn{5}{l}{\textbf{Worldwide Governance (2019--2023)}} \\
Control of Corruption & \makecell{0.23***\\(691)} & \makecell{0.65***\\(611)} & \makecell{0.21***\\(657)} & \makecell{0.58***\\(592)} \\
Government Effectiveness & \makecell{0.22***\\(691)} & \makecell{0.62***\\(611)} & \makecell{0.20***\\(657)} & \makecell{0.56***\\(592)} \\
Political Stability & \makecell{0.09**\\(691)} & \makecell{0.54***\\(611)} & \makecell{0.09**\\(657)} & \makecell{0.47***\\(592)} \\
Regulatory Quality & \makecell{0.23***\\(691)} & \makecell{0.62***\\(611)} & \makecell{0.21***\\(657)} & \makecell{0.57***\\(592)} \\
Rule of Law & \makecell{0.24***\\(691)} & \makecell{0.64***\\(611)} & \makecell{0.22***\\(657)} & \makecell{0.58***\\(592)} \\
Voice \& Accountability & \makecell{0.19***\\(691)} & \makecell{0.57***\\(611)} & \makecell{0.18***\\(657)} & \makecell{0.52***\\(592)} \\
\bottomrule
\end{tabularx}

\tabnote{This table shows pairwise correlations with related measures of state capacity, informational capacity, and e-government for 2019--2023 in the 193 countries covered by the U.N. E-Government Survey. Each cell reports a pairwise Pearson correlation, with the transformation-specific number of country-year observations in parentheses. Raw measures retain observed zeros, while logged measures require positive government IP values. Per-worker measures additionally require positive observed public-sector employment. Natural logarithms are used. Stars indicate rejection of the null hypothesis that the pairwise correlation is zero: $^{*}p<0.10$, $^{**}p<0.05$, and $^{***}p<0.01$.}
\label{tab:pearson_fed}
\end{table}

\begin{table}[ht]
\centering
\scriptsize
\caption{Pairwise Correlations with Local Government IP Measures, 2019--2023}
\begin{tabularx}{\linewidth}{>{\raggedright\arraybackslash}X c c c c}
\toprule
\textbf{Variable}
  & \makecell{\textbf{Local IPs}\\\textbf{per capita}}
  & \makecell{\textbf{ln(Local IPs}\\\textbf{per capita)}}
  & \makecell{\textbf{Local IPs}\\\textbf{per worker}}
  & \makecell{\textbf{ln(Local IPs}\\\textbf{per worker)}} \\
\midrule
\multicolumn{5}{l}{\textbf{Statistical Capacity WB (2019--2023)}} \\
Data Infrastructure & \makecell{0.26***\\(872)} & \makecell{0.63***\\(739)} & \makecell{0.35***\\(830)} & \makecell{0.64***\\(715)} \\
Data Products & \makecell{0.00\\(884)} & \makecell{0.15***\\(751)} & \makecell{0.22***\\(835)} & \makecell{0.29***\\(720)} \\
Data Services & \makecell{0.23***\\(856)} & \makecell{0.54***\\(731)} & \makecell{0.30***\\(824)} & \makecell{0.56***\\(711)} \\
Data Sources & \makecell{0.25***\\(852)} & \makecell{0.62***\\(731)} & \makecell{0.32***\\(824)} & \makecell{0.60***\\(711)} \\
Data Use & \makecell{0.10***\\(884)} & \makecell{0.31***\\(751)} & \makecell{0.24***\\(835)} & \makecell{0.40***\\(720)} \\
Statistical Index & \makecell{0.25***\\(852)} & \makecell{0.61***\\(731)} & \makecell{0.35***\\(824)} & \makecell{0.63***\\(711)} \\
\addlinespace
\multicolumn{5}{l}{\textbf{UN E-Gov Indicators (2020--2022)}} \\
E-Gov Index & \makecell{0.32***\\(355)} & \makecell{0.74***\\(301)} & \makecell{0.35***\\(336)} & \makecell{0.69***\\(288)} \\
E-Gov Rank & \makecell{-0.33***\\(355)} & \makecell{-0.74***\\(301)} & \makecell{-0.36***\\(336)} & \makecell{-0.68***\\(288)} \\
E-Participation Index & \makecell{0.23***\\(355)} & \makecell{0.59***\\(301)} & \makecell{0.31***\\(336)} & \makecell{0.59***\\(288)} \\
Human Capital Index & \makecell{0.29***\\(355)} & \makecell{0.74***\\(301)} & \makecell{0.32***\\(336)} & \makecell{0.67***\\(288)} \\
Online Service Index & \makecell{0.24***\\(355)} & \makecell{0.62***\\(301)} & \makecell{0.31***\\(336)} & \makecell{0.61***\\(288)} \\
Telecom Infra Index & \makecell{0.34***\\(355)} & \makecell{0.70***\\(301)} & \makecell{0.33***\\(336)} & \makecell{0.64***\\(288)} \\
\addlinespace
\multicolumn{5}{l}{\textbf{Worldwide Governance (2019--2023)}} \\
Control of Corruption & \makecell{0.42***\\(691)} & \makecell{0.67***\\(589)} & \makecell{0.44***\\(657)} & \makecell{0.62***\\(569)} \\
Government Effectiveness & \makecell{0.38***\\(691)} & \makecell{0.70***\\(589)} & \makecell{0.39***\\(657)} & \makecell{0.64***\\(569)} \\
Political Stability & \makecell{0.31***\\(691)} & \makecell{0.61***\\(589)} & \makecell{0.31***\\(657)} & \makecell{0.56***\\(569)} \\
Regulatory Quality & \makecell{0.39***\\(691)} & \makecell{0.70***\\(589)} & \makecell{0.40***\\(657)} & \makecell{0.65***\\(569)} \\
Rule of Law & \makecell{0.40***\\(691)} & \makecell{0.68***\\(589)} & \makecell{0.42***\\(657)} & \makecell{0.63***\\(569)} \\
Voice \& Accountability & \makecell{0.37***\\(691)} & \makecell{0.63***\\(589)} & \makecell{0.41***\\(657)} & \makecell{0.60***\\(569)} \\
\bottomrule
\end{tabularx}

\tabnote{This table shows pairwise correlations with related measures of state capacity, informational capacity, and e-government for 2019--2023 in the 193 countries covered by the U.N. E-Government Survey. Each cell reports a pairwise Pearson correlation, with the transformation-specific number of country-year observations in parentheses. Raw measures retain observed zeros, while logged measures require positive government IP values. Per-worker measures additionally require positive observed public-sector employment. Natural logarithms are used. Stars indicate rejection of the null hypothesis that the pairwise correlation is zero: $^{*}p<0.10$, $^{**}p<0.05$, and $^{***}p<0.01$.}
\label{tab:pearson_local}
\end{table}

\clearpage

\begin{table}[htbp]
    \centering
    \caption{Conditional Associations Between Digital State Capacity per Capita and Existing State-Capacity Measures}
    \label{tab:percapita_capacity_cal}
        \resizebox{\linewidth}{!}{{
\def\sym#1{\ifmmode^{#1}\else\(^{#1}\)\fi}
\begin{tabular}{l*{5}{c}}
\toprule
                    &\\  Dependent variables: \\ Adjacent capacity measures         &                     &                     &                     &                     \\
                    &\multicolumn{1}{c}{(1)}&\multicolumn{1}{c}{(2)}&\multicolumn{1}{c}{(3)}&\multicolumn{1}{c}{(4)}&\multicolumn{1}{c}{(5)}\\
                    &\multicolumn{1}{c}{U.N. E-Gov}&\multicolumn{1}{c}{World Bank SPI}&\multicolumn{1}{c}{ln Delivery Time}&\multicolumn{1}{c}{Returned Share}&\multicolumn{1}{c}{Hanson--Sigman Capacity}\\
\midrule
ln government IPs per capita&     0.006\sym{**}&     0.016\sym{***}&    -0.028&     0.002&     0.007\sym{**}\\
                    &    (0.002)&    (0.004)&    (0.029)&    (0.012)&    (0.003)\\
\addlinespace
ln GDP p.c.&     0.045\sym{***}&     0.012&     0.012&     0.048&     0.051\sym{***}\\
                    &    (0.009)&    (0.013)&    (0.053)&    (0.031)&    (0.016)\\
\addlinespace
ln Pop.&     0.022\sym{***}&     0.032\sym{***}&    -0.027&    -0.001&     0.018\sym{***}\\
                    &    (0.002)&    (0.004)&    (0.027)&    (0.013)&    (0.006)\\
\addlinespace
Internet percent Pop.&     0.376\sym{***}&     0.097&    -1.231\sym{*}&     0.166&     0.179\sym{**}\\
                    &    (0.046)&    (0.062)&    (0.487)&    (0.157)&    (0.079)\\
\midrule
Observations        &       940&      1132&        87&        87&       688\\
Dep. Var. Mean      &     0.614&     0.690&     4.957&     0.740&     0.639\\
Countries           &       173&       172&        87&        87&        94\\
Region-year clusters&        40&        40&         5&         5&        40\\
Adjusted \(R^2\)   &     0.860&     0.602&     0.344&     0.272&     0.762\\
Region-Year FE      &       Yes&       Yes&       Yes&       Yes&       Yes\\
Years               &2008-2022, biennial& 2016-2023&      2010&      2010& 2008-2015\\
\bottomrule
\multicolumn{6}{l}{\footnotesize Standard errors in parentheses}\\
\multicolumn{6}{l}{\footnotesize \sym{*} \(p<0.1\), \sym{**} \(p<0.05\), \sym{***} \(p<0.01\)}\\
\end{tabular}
}
}
    \tabnote{This table reports country-year conditional associations between the natural logarithm of government-held public IPv4 addresses per capita and existing measures of state capacity. All specifications control for the natural logarithms of GDP per capita and population, and for population internet access, and include region-by-year fixed effects. Column 1 reports the U.N. E-Government Index for biennial observations from 2008 to 2022. Column 2 reports the World Bank Statistical Performance Index for 2016--2023. Columns 3 and 4 report the natural logarithm of average delivery time and the share of letters returned in the 2010 postal experiment from \textcite{chong2014letter}. Column 5 reports the Hanson--Sigman State Capacity Index for 2008--2015. Standard errors are two-way clustered by country and region-year.}
\end{table}

\clearpage

\begin{table}[htbp]
    \centering
    \caption{OLS Regression: Correlation Between Government IPs per Public-Sector Worker and Existing State Capacity Measures}
    \label{tab:perworker_capacity_cal}
    \resizebox{\linewidth}{!}{{
\def\sym#1{\ifmmode^{#1}\else\(^{#1}\)\fi}
\begin{tabular}{l*{5}{c}}
\toprule
                    &\\  Dependent variables: \\ Adjacent capacity measures         &                     &                     &                     &                     \\
                    &\multicolumn{1}{c}{(1)}&\multicolumn{1}{c}{(2)}&\multicolumn{1}{c}{(3)}&\multicolumn{1}{c}{(4)}&\multicolumn{1}{c}{(5)}\\
                    &\multicolumn{1}{c}{U.N. E-Gov}&\multicolumn{1}{c}{World Bank SPI}&\multicolumn{1}{c}{Log Delivery Time}&\multicolumn{1}{c}{Returned Share}&\multicolumn{1}{c}{Hanson--Sigman Capacity}\\
\midrule
Log government IPs per public-sector worker&     0.005\sym{**}&     0.015\sym{***}&    -0.033&     0.003&     0.008\sym{***}\\
                    &    (0.002)&    (0.004)&    (0.033)&    (0.013)&    (0.003)\\
\addlinespace
Log GDP p.c.&     0.049\sym{***}&     0.018&     0.009&     0.047&     0.052\sym{***}\\
                    &    (0.009)&    (0.013)&    (0.053)&    (0.029)&    (0.016)\\
\addlinespace
Log Pop.&     0.020\sym{***}&     0.028\sym{***}&    -0.027&    -0.001&     0.018\sym{***}\\
                    &    (0.002)&    (0.005)&    (0.026)&    (0.013)&    (0.006)\\
\addlinespace
Internet percent Pop.&     0.367\sym{***}&     0.090&    -1.238\sym{*}&     0.165&     0.182\sym{**}\\
                    &    (0.047)&    (0.063)&    (0.481)&    (0.157)&    (0.079)\\
\midrule
Observations        &       902&      1103&        87&        87&       688\\
Dep. Var. Mean      &     0.615&     0.695&     4.957&     0.740&     0.639\\
Countries           &       164&       165&        87&        87&        94\\
Region-year clusters&        40&        40&         5&         5&        40\\
Adjusted \(R^2\)   &     0.863&     0.594&     0.348&     0.272&     0.764\\
Region-Year FE      &       Yes&       Yes&       Yes&       Yes&       Yes\\
Years               &2008-2022, biennial& 2016-2023&      2010&      2010& 2008-2015\\
\bottomrule
\multicolumn{6}{l}{\footnotesize Standard errors in parentheses}\\
\multicolumn{6}{l}{\footnotesize \sym{*} \(p<0.1\), \sym{**} \(p<0.05\), \sym{***} \(p<0.01\)}\\
\end{tabular}
}
}
    \tabnote{This table reports country-year conditional associations between log government-held public IPv4 addresses per public-sector worker and existing measures of state capacity. This normalisation measures public-sector ICT capital relative to the size of the public-sector workforce and is available only where public-sector employment is observed. All specifications control for log GDP per capita, log population, and population internet access, and include region-by-year fixed effects. Column 1 reports the U.N. E-Government Index for biennial observations from 2008 to 2022. Column 2 reports the World Bank Statistical Performance Index for 2016--2023. Columns 3 and 4 report log average delivery time and the share of letters returned in the 2010 postal experiment from \textcite{chong2014letter}. Column 5 reports the \textcite{hanson2021leviathan} State Capacity Index for 2008--2015. Standard errors are two-way clustered by country and region-year. The specifications provide controlled convergent-validity evidence and should not be interpreted as causal estimates.}
\end{table}

\clearpage
\begin{table}[htbp]
    \centering
    \caption{OLS Regression: Correlation Between Government IPs per Capita and World Bank Statistical Performance Indicator Measures}
        \resizebox{\linewidth}{!}{{
\def\sym#1{\ifmmode^{#1}\else\(^{#1}\)\fi}
\begin{tabular}{l*{5}{c}}
\toprule
                    &\\  Dependent variables: \\ World Bank SPI and selected pillars         &                     &                     &                     &                     \\
                    &\multicolumn{1}{c}{(1)}&\multicolumn{1}{c}{(2)}&\multicolumn{1}{c}{(3)}&\multicolumn{1}{c}{(4)}&\multicolumn{1}{c}{(5)}\\
                    &\multicolumn{1}{c}{Overall SPI}&\multicolumn{1}{c}{Data Use}&\multicolumn{1}{c}{Data Services}&\multicolumn{1}{c}{Data Products}&\multicolumn{1}{c}{Data Sources}\\
\midrule
Log government IPs per capita&     0.011\sym{**}&     0.005&     0.008&     0.002&     0.012\sym{**}\\
                    &    (0.004)&    (0.004)&    (0.006)&    (0.003)&    (0.005)\\
\addlinespace
Log GDP p.c.&     0.010&    -0.001&     0.024&    -0.021\sym{**}&     0.016\\
                    &    (0.012)&    (0.014)&    (0.018)&    (0.010)&    (0.015)\\
\addlinespace
Log Pop.&     0.027\sym{***}&     0.018\sym{***}&     0.035\sym{***}&     0.017\sym{***}&     0.024\sym{***}\\
                    &    (0.005)&    (0.006)&    (0.008)&    (0.004)&    (0.007)\\
\addlinespace
Internet percent Pop.&     0.136\sym{**}&     0.001&     0.154&    -0.005&     0.270\sym{***}\\
                    &    (0.064)&    (0.074)&    (0.091)&    (0.048)&    (0.076)\\
\addlinespace
Democracy (Polity)&     0.186\sym{***}&     0.148\sym{***}&     0.279\sym{***}&     0.134\sym{***}&     0.169\sym{***}\\
                    &    (0.034)&    (0.039)&    (0.049)&    (0.027)&    (0.036)\\
\midrule
Observations        &       997&      1676&       997&      1676&      1048\\
Dep. Var. Mean      &     0.705&     0.724&     0.712&     0.662&     0.581\\
Countries           &       147&       148&       147&       148&       147\\
Region-year clusters&        40&        80&        40&        80&        44\\
Adjusted \(R^2\)   &     0.657&     0.708&     0.492&     0.618&     0.656\\
Region-Year FE      &       Yes&       Yes&       Yes&       Yes&       Yes\\
Years               & 2016-2023& 2008-2023& 2016-2023& 2008-2023& 2015-2023\\
\bottomrule
\multicolumn{6}{l}{\footnotesize Standard errors in parentheses}\\
\multicolumn{6}{l}{\footnotesize \sym{*} \(p<0.1\), \sym{**} \(p<0.05\), \sym{***} \(p<0.01\)}\\
\end{tabular}
}
}
    \label{tab:percapita_stat_sub}
    \tabnote{This table reports country-year conditional associations between log government-held public IPv4 addresses per capita and the World Bank Statistical Performance Index and selected pillars, each normalised to range from zero to one. Column 1 reports the overall index for 2016--2023. Columns 2--5 report Data Use for 2008--2023, Data Services for 2016--2023, Data Products for 2008--2023, and Data Sources for 2015--2023. The Data Infrastructure pillar is not included in this table. All specifications control for log GDP per capita, log population, population internet access, and democracy, and include region-by-year fixed effects. Standard errors are two-way clustered by country and region-year. The specifications provide controlled convergent-validity evidence and should not be interpreted as causal estimates.}
\end{table}

\clearpage

\begin{table}[htbp]
    \centering
    \caption{OLS Regression: Correlation Between Government IPs per Public-Sector Worker and World Bank Statistical Performance Indicator Measures}
        \resizebox{\linewidth}{!}{{
\def\sym#1{\ifmmode^{#1}\else\(^{#1}\)\fi}
\begin{tabular}{l*{5}{c}}
\toprule
                    &\\  Dependent variables: \\ World Bank SPI and selected pillars         &                     &                     &                     &                     \\
                    &\multicolumn{1}{c}{(1)}&\multicolumn{1}{c}{(2)}&\multicolumn{1}{c}{(3)}&\multicolumn{1}{c}{(4)}&\multicolumn{1}{c}{(5)}\\
                    &\multicolumn{1}{c}{Overall SPI}&\multicolumn{1}{c}{Data Use}&\multicolumn{1}{c}{Data Services}&\multicolumn{1}{c}{Data Products}&\multicolumn{1}{c}{Data Sources}\\
\midrule
Log government IPs per public-sector worker&     0.011\sym{**}&     0.005&     0.008&     0.002&     0.012\sym{**}\\
                    &    (0.004)&    (0.004)&    (0.006)&    (0.003)&    (0.005)\\
\addlinespace
Log GDP p.c.&     0.012&     0.000&     0.024&    -0.021\sym{**}&     0.018\\
                    &    (0.012)&    (0.014)&    (0.017)&    (0.009)&    (0.014)\\
\addlinespace
Log Pop.&     0.027\sym{***}&     0.018\sym{***}&     0.035\sym{***}&     0.017\sym{***}&     0.024\sym{***}\\
                    &    (0.005)&    (0.006)&    (0.008)&    (0.004)&    (0.007)\\
\addlinespace
Internet percent Pop.&     0.149\sym{**}&     0.004&     0.171\sym{*}&    -0.003&     0.286\sym{***}\\
                    &    (0.064)&    (0.075)&    (0.090)&    (0.048)&    (0.076)\\
\addlinespace
Democracy (Polity)&     0.187\sym{***}&     0.148\sym{***}&     0.278\sym{***}&     0.133\sym{***}&     0.169\sym{***}\\
                    &    (0.034)&    (0.039)&    (0.049)&    (0.027)&    (0.036)\\
\midrule
Observations        &       995&      1674&       995&      1674&      1046\\
Dep. Var. Mean      &     0.705&     0.724&     0.712&     0.662&     0.581\\
Countries           &       147&       148&       147&       148&       147\\
Region-year clusters&        40&        80&        40&        80&        44\\
Adjusted \(R^2\)   &     0.657&     0.707&     0.494&     0.617&     0.658\\
Region-Year FE      &       Yes&       Yes&       Yes&       Yes&       Yes\\
Years               & 2016-2023& 2008-2023& 2016-2023& 2008-2023& 2015-2023\\
\bottomrule
\multicolumn{6}{l}{\footnotesize Standard errors in parentheses}\\
\multicolumn{6}{l}{\footnotesize \sym{*} \(p<0.1\), \sym{**} \(p<0.05\), \sym{***} \(p<0.01\)}\\
\end{tabular}
}
}
    \label{tab:perworker_stat_sub}
    \tabnote{This table reports country-year conditional associations between log government-held public IPv4 addresses per public-sector worker and the World Bank Statistical Performance Index and selected pillars, each normalised to range from zero to one. The per-worker normalisation measures public-sector ICT capital relative to the size of the public-sector workforce and is available only where public-sector employment is observed. Column 1 reports the overall index for 2016--2023. Columns 2--5 report Data Use for 2008--2023, Data Services for 2016--2023, Data Products for 2008--2023, and Data Sources for 2015--2023. The Data Infrastructure pillar is not included in this table. All specifications control for log GDP per capita, log population, population internet access, and democracy, and include region-by-year fixed effects. Standard errors are two-way clustered by country and region-year. The specifications provide controlled convergent-validity evidence.}
\end{table}

\clearpage

\begin{table}[htbp]
    \centering
    \caption{OLS Regression: Correlation Between Government IPs per Capita and U.N. E-Government Measures.}
        \resizebox{\linewidth}{!}{{
\def\sym#1{\ifmmode^{#1}\else\(^{#1}\)\fi}
\begin{tabular}{l*{5}{c}}
\toprule
                    &\\  Dependent variables: \\ U.N. E-Government measures         &                     &                     &                     &                     \\
                    &\multicolumn{1}{c}{(1)}&\multicolumn{1}{c}{(2)}&\multicolumn{1}{c}{(3)}&\multicolumn{1}{c}{(4)}&\multicolumn{1}{c}{(5)}\\
                    &\multicolumn{1}{c}{E-Gov Index}&\multicolumn{1}{c}{E-Gov Rank}&\multicolumn{1}{c}{E-Participation}&\multicolumn{1}{c}{Telecom Infrastructure}&\multicolumn{1}{c}{Online Services}\\
\midrule
Log government IPs per capita&     0.003&    -1.591\sym{**}&     0.007&     0.002&     0.006\\
                    &    (0.002)&    (0.594)&    (0.004)&    (0.003)&    (0.004)\\
\addlinespace
Log GDP p.c.&     0.051\sym{***}&   -13.099\sym{***}&     0.046\sym{***}&     0.063\sym{***}&     0.054\sym{***}\\
                    &    (0.009)&    (2.219)&    (0.015)&    (0.009)&    (0.013)\\
\addlinespace
Log Pop.&     0.020\sym{***}&    -4.566\sym{***}&     0.044\sym{***}&     0.009\sym{***}&     0.041\sym{***}\\
                    &    (0.003)&    (0.766)&    (0.007)&    (0.003)&    (0.005)\\
\addlinespace
Internet percent Pop.&     0.341\sym{***}&   -81.558\sym{***}&     0.352\sym{***}&     0.467\sym{***}&     0.348\sym{***}\\
                    &    (0.045)&   (12.663)&    (0.100)&    (0.043)&    (0.091)\\
\addlinespace
Democracy (Polity)&     0.009&   -13.166\sym{*}&     0.128\sym{***}&     0.082\sym{**}&     0.108\sym{***}\\
                    &    (0.028)&    (6.643)&    (0.039)&    (0.031)&    (0.039)\\
\midrule
Observations        &       641&       813&       813&       813&       813\\
Dep. Var. Mean      &     0.653&    70.461&     0.523&     0.496&     0.595\\
Countries           &       119&       146&       146&       146&       146\\
Region-year clusters&        40&        40&        40&        40&        40\\
Adjusted \(R^2\)   &     0.875&     0.871&     0.666&     0.903&     0.685\\
Region-Year FE      &       Yes&       Yes&       Yes&       Yes&       Yes\\
Years               &2008-2022, biennial&2008-2022, biennial&2008-2022, biennial&2008-2022, biennial&2008-2022, biennial\\
\bottomrule
\multicolumn{6}{l}{\footnotesize Standard errors in parentheses}\\
\multicolumn{6}{l}{\footnotesize \sym{*} \(p<0.1\), \sym{**} \(p<0.05\), \sym{***} \(p<0.01\)}\\
\end{tabular}
}
}
    \label{tab:percapita_egov_sub}
    \tabnote{This table reports country-year conditional associations between log government-held public IPv4 addresses per capita and U.N. E-Government measures for biennial observations from 2008 to 2022. Column 1 reports the E-Government Development Index. Column 2 reports E-Government Rank, for which a lower numerical value denotes better performance. Columns 3--5 report the E-Participation, Telecommunications Infrastructure, and Online Service indices. All specifications control for log GDP per capita, log population, population internet access, and democracy. Column 1 additionally controls for tax revenue as a share of GDP and therefore has a smaller estimation sample. All specifications include region-by-year fixed effects, and standard errors are two-way clustered by country and region-year. }
\end{table}

\clearpage

\begin{table}[htbp]
    \centering
    \caption{OLS Regression: Correlation Between Government IPs per Public-Sector Worker and E-Government Sub-Indicators}
        \resizebox{\linewidth}{!}{{
\def\sym#1{\ifmmode^{#1}\else\(^{#1}\)\fi}
\begin{tabular}{l*{5}{c}}
\toprule
                    &\\  Dependent variables: \\ U.N. E-Government measures         &                     &                     &                     &                     \\
                    &\multicolumn{1}{c}{(1)}&\multicolumn{1}{c}{(2)}&\multicolumn{1}{c}{(3)}&\multicolumn{1}{c}{(4)}&\multicolumn{1}{c}{(5)}\\
                    &\multicolumn{1}{c}{E-Gov Index}&\multicolumn{1}{c}{E-Gov Rank}&\multicolumn{1}{c}{E-Participation}&\multicolumn{1}{c}{Telecom Infrastructure}&\multicolumn{1}{c}{Online Services}\\
\midrule
Log government IPs per public-sector worker&     0.002&    -1.344\sym{**}&     0.006&     0.002&     0.005\\
                    &    (0.003)&    (0.582)&    (0.005)&    (0.003)&    (0.004)\\
\addlinespace
Log GDP p.c.&     0.053\sym{***}&   -13.682\sym{***}&     0.048\sym{***}&     0.064\sym{***}&     0.056\sym{***}\\
                    &    (0.009)&    (2.217)&    (0.015)&    (0.009)&    (0.013)\\
\addlinespace
Log Pop.&     0.020\sym{***}&    -4.450\sym{***}&     0.044\sym{***}&     0.008\sym{***}&     0.040\sym{***}\\
                    &    (0.003)&    (0.778)&    (0.007)&    (0.003)&    (0.005)\\
\addlinespace
Internet percent Pop.&     0.342\sym{***}&   -82.722\sym{***}&     0.359\sym{***}&     0.467\sym{***}&     0.350\sym{***}\\
                    &    (0.045)&   (12.691)&    (0.100)&    (0.042)&    (0.090)\\
\addlinespace
Democracy (Polity)&     0.010&   -13.341\sym{*}&     0.128\sym{***}&     0.082\sym{**}&     0.109\sym{***}\\
                    &    (0.028)&    (6.603)&    (0.039)&    (0.031)&    (0.039)\\
\midrule
Observations        &       640&       812&       812&       812&       812\\
Dep. Var. Mean      &     0.653&    70.491&     0.523&     0.495&     0.595\\
Countries           &       119&       146&       146&       146&       146\\
Region-year clusters&        40&        40&        40&        40&        40\\
Adjusted \(R^2\)   &     0.875&     0.870&     0.666&     0.903&     0.685\\
Region-Year FE      &       Yes&       Yes&       Yes&       Yes&       Yes\\
Years               &2008-2022, biennial&2008-2022, biennial&2008-2022, biennial&2008-2022, biennial&2008-2022, biennial\\
\bottomrule
\multicolumn{6}{l}{\footnotesize Standard errors in parentheses}\\
\multicolumn{6}{l}{\footnotesize \sym{*} \(p<0.1\), \sym{**} \(p<0.05\), \sym{***} \(p<0.01\)}\\
\end{tabular}
}
}
    \label{tab:perworker_egov_sub}
    \tabnote{This table reports country-year conditional associations between log government-held public IPv4 addresses per public-sector worker and U.N. E-Government measures for biennial observations from 2008 to 2022. The normalisation measures public-sector ICT capital relative to the size of the public-sector workforce. Column 1 reports the E-Government Development Index. Column 2 reports E-Government Rank, for which a lower numerical value denotes better performance. Columns 3--5 report the E-Participation, Telecommunications Infrastructure, and Online Service indices. All specifications control for log GDP per capita, log population, population internet access, and democracy. Column 1 additionally controls for tax revenue as a share of GDP and therefore has a smaller estimation sample. All specifications include region-by-year fixed effects, and standard errors are two-way clustered by country and region-year.}
\end{table}

\clearpage

\begin{table}[ht]
    \centering
    \caption{Baseline OLS Regression: Correlation Between Government IPs per Capita and General Government Outcomes}
    \resizebox{\linewidth}{!}{%
        {
\def\sym#1{\ifmmode^{#1}\else\(^{#1}\)\fi}
\begin{tabular}{l*{4}{c}}
\toprule
                    &\\  Dependent Variables: \\ General Outcomes         &                     &                     &                     \\
                    &\multicolumn{1}{c}{(1)}&\multicolumn{1}{c}{(2)}&\multicolumn{1}{c}{(3)}&\multicolumn{1}{c}{(4)}\\
                    &\multicolumn{1}{c}{CPI}&\multicolumn{1}{c}{Bribery Incidence}&\multicolumn{1}{c}{Tax/GDP}&\multicolumn{1}{c}{Soc. Ins. Cov.}\\
\midrule
Log government IPs per capita&     0.013\sym{***}&    -0.010\sym{**}&     0.005\sym{**}&     0.012\\
                    &    (0.005)&    (0.004)&    (0.002)&    (0.008)\\
\addlinespace
Log GDP p.c.&     0.116\sym{***}&    -0.070\sym{***}&    -0.005&    -0.025\\
                    &    (0.016)&    (0.018)&    (0.008)&    (0.038)\\
\addlinespace
Log Pop.&    -0.000&     0.010&    -0.007\sym{**}&    -0.001\\
                    &    (0.006)&    (0.007)&    (0.003)&    (0.012)\\
\addlinespace
Internet percent Pop.&    -0.016&     0.174\sym{***}&     0.028&     0.429\sym{**}\\
                    &    (0.084)&    (0.044)&    (0.035)&    (0.175)\\
\addlinespace
Democracy (Polity)&     0.095\sym{**}&    -0.024&     0.096\sym{***}&     0.032\\
                    &    (0.043)&    (0.063)&    (0.020)&    (0.058)\\
\midrule
Observations        &      1055&       169&      1309&       320\\
Dep. Var. Mean      &     0.496&     0.122&     0.171&     0.231\\
Countries           &       118&        94&       120&        65\\
Region-year clusters&        60&        23&        80&        47\\
Adjusted \(R^2\)   &     0.744&     0.432&     0.446&     0.631\\
Region-year FE      &       Yes&       Yes&       Yes&       Yes\\
Years               & 2012-2023&2008-2023, available survey years& 2008-2023& 2008-2022\\
\bottomrule
\multicolumn{5}{l}{\footnotesize Standard errors in parentheses}\\
\multicolumn{5}{l}{\footnotesize \sym{*} \(p<0.1\), \sym{**} \(p<0.05\), \sym{***} \(p<0.01\)}\\
\end{tabular}
}

    }
    \tabnote{This table reports country-year associations between log government-held public IPv4 addresses per capita and four government outcomes. Column 1 reports the Transparency International Corruption Perceptions Index divided by 100, with higher values denoting lower perceived public-sector corruption, for 2012--2023. Column 2 reports bribery incidence among firms for the available survey years from 2008 to 2023. Columns 3 and 4 report tax revenue as a share of GDP for 2008--2023 and social insurance coverage as a share of the population for 2008--2022. The outcomes in all four columns are expressed on a zero-to-one scale. All specifications control for log GDP per capita, log population, population internet access, and democracy; Column 1 additionally controls for tax revenue as a share of GDP. All specifications include region-by-year fixed effects. Standard errors are two-way clustered by country and region-year.}
    \label{tab:dsc_pc_general_baseline}
\end{table}

\clearpage

\begin{table}[ht]
    \centering
    \caption{Baseline OLS Regression: Correlation Between Government IPs per Public-Sector Worker and General Government Outcomes}
    \resizebox{\linewidth}{!}{%
        {
\def\sym#1{\ifmmode^{#1}\else\(^{#1}\)\fi}
\begin{tabular}{l*{4}{c}}
\toprule
                    &\\  Dependent Variables: \\ General Outcomes         &                     &                     &                     \\
                    &\multicolumn{1}{c}{(1)}&\multicolumn{1}{c}{(2)}&\multicolumn{1}{c}{(3)}&\multicolumn{1}{c}{(4)}\\
                    &\multicolumn{1}{c}{CPI}&\multicolumn{1}{c}{Bribery Incidence}&\multicolumn{1}{c}{Tax/GDP}&\multicolumn{1}{c}{Soc. Ins. Cov.}\\
\midrule
Log government IPs per public-sector worker&     0.013\sym{**}&    -0.010\sym{**}&     0.005\sym{**}&     0.009\\
                    &    (0.005)&    (0.004)&    (0.002)&    (0.008)\\
\addlinespace
Log GDP p.c.&     0.119\sym{***}&    -0.072\sym{***}&    -0.004&    -0.023\\
                    &    (0.015)&    (0.017)&    (0.008)&    (0.038)\\
\addlinespace
Log Pop.&    -0.001&     0.010&    -0.007\sym{**}&    -0.003\\
                    &    (0.006)&    (0.007)&    (0.003)&    (0.012)\\
\addlinespace
Internet percent Pop.&    -0.005&     0.167\sym{***}&     0.033&     0.452\sym{**}\\
                    &    (0.084)&    (0.045)&    (0.035)&    (0.175)\\
\addlinespace
Democracy (Polity)&     0.094\sym{**}&    -0.021&     0.096\sym{***}&     0.035\\
                    &    (0.043)&    (0.064)&    (0.020)&    (0.059)\\
\midrule
Observations        &      1053&       169&      1307&       320\\
Dep. Var. Mean      &     0.496&     0.122&     0.171&     0.231\\
Countries           &       118&        94&       120&        65\\
Region-year clusters&        60&        23&        80&        47\\
Adjusted \(R^2\)   &     0.743&     0.433&     0.443&     0.626\\
Region-year FE      &       Yes&       Yes&       Yes&       Yes\\
Years               & 2012-2023&2008-2023, available survey years& 2008-2023& 2008-2022\\
\bottomrule
\multicolumn{5}{l}{\footnotesize Standard errors in parentheses}\\
\multicolumn{5}{l}{\footnotesize \sym{*} \(p<0.1\), \sym{**} \(p<0.05\), \sym{***} \(p<0.01\)}\\
\end{tabular}
}

    }
    \tabnote{This table reports country-year associations between log government-held public IPv4 addresses per public-sector worker and four government outcomes. This normalisation measures public-sector ICT capital relative to the size of the public-sector workforce and restricts the estimation sample to country-years with observed public-sector employment. Column 1 reports the Transparency International Corruption Perceptions Index divided by 100, with higher values denoting lower perceived public-sector corruption, for 2012--2023. Column 2 reports bribery incidence among firms for the available survey years from 2008 to 2023. Columns 3 and 4 report tax revenue as a share of GDP for 2008--2023 and social insurance coverage as a share of the population for 2008--2022. The outcomes in all four columns are expressed on a zero-to-one scale. All specifications control for log GDP per capita, log population, population internet access, and democracy; Column 1 additionally controls for tax revenue as a share of GDP. All specifications include region-by-year fixed effects. Standard errors are two-way clustered by country and region-year.}
    \label{tab:dsc_pw_general_baseline}
\end{table}

\clearpage

\begin{table}[ht]
    \centering
    \caption{Baseline OLS Regression: Correlation Between Government IPs per Capita and General Health Outcomes}
    \resizebox{\linewidth}{!}{%
        {
\def\sym#1{\ifmmode^{#1}\else\(^{#1}\)\fi}
\begin{tabular}{l*{5}{c}}
\toprule
                    &\\  Dependent Variables: \\ Health Outcomes         &                     &                     &                     &                     \\
                    &\multicolumn{1}{c}{(1)}&\multicolumn{1}{c}{(2)}&\multicolumn{1}{c}{(3)}&\multicolumn{1}{c}{(4)}&\multicolumn{1}{c}{(5)}\\
                    &\multicolumn{1}{c}{Vacc. Meas.}&\multicolumn{1}{c}{Vacc. Dip.}&\multicolumn{1}{c}{Vacc. Hep.}&\multicolumn{1}{c}{Neonatal Mort.}&\multicolumn{1}{c}{Infant Mort.}\\
\midrule
Log government IPs per capita&     0.009\sym{**}&     0.010\sym{***}&     0.009\sym{**}&    -0.047\sym{**}&    -0.054\sym{***}\\
                    &    (0.003)&    (0.003)&    (0.004)&    (0.019)&    (0.018)\\
\addlinespace
Log GDP p.c.&     0.018&     0.021\sym{*}&     0.021&    -0.295\sym{***}&    -0.285\sym{***}\\
                    &    (0.013)&    (0.012)&    (0.014)&    (0.060)&    (0.052)\\
\addlinespace
Log Pop.&     0.001&    -0.006&    -0.011\sym{*}&    -0.001&    -0.000\\
                    &    (0.005)&    (0.004)&    (0.006)&    (0.024)&    (0.021)\\
\addlinespace
Internet percent Pop.&     0.032&     0.022&    -0.036&    -0.502\sym{*}&    -0.638\sym{**}\\
                    &    (0.062)&    (0.059)&    (0.080)&    (0.300)&    (0.253)\\
\addlinespace
Democracy (Polity)&    -0.020&     0.009&    -0.019&     0.189&     0.047\\
                    &    (0.028)&    (0.031)&    (0.035)&    (0.244)&    (0.183)\\
\midrule
Observations        &      1676&      1676&      1563&      1676&      1676\\
Dep. Var. Mean      &     0.890&     0.896&     0.876&     1.820&     2.463\\
Countries           &       148&       148&       145&       148&       148\\
Region-year clusters&        80&        80&        80&        80&        80\\
Adjusted \(R^2\)   &     0.319&     0.310&     0.150&     0.807&     0.873\\
Region-year FE      &       Yes&       Yes&       Yes&       Yes&       Yes\\
Years               & 2008-2023& 2008-2023& 2008-2023& 2008-2023& 2008-2023\\
\bottomrule
\multicolumn{6}{l}{\footnotesize Standard errors in parentheses}\\
\multicolumn{6}{l}{\footnotesize \sym{*} \(p<0.1\), \sym{**} \(p<0.05\), \sym{***} \(p<0.01\)}\\
\end{tabular}
}

    }
    \tabnote{This table reports country-year associations between log government-held public IPv4 addresses per capita and health outcomes for 2008--2023. Columns 1--3 report measles, diphtheria, and hepatitis B vaccination coverage, respectively, expressed as shares from zero to one. Columns 4 and 5 report the natural logarithms of the neonatal and infant mortality rates per 1,000 live births. All specifications control for log GDP per capita, log population, population internet access, and democracy, and include region-by-year fixed effects. Standard errors are two-way clustered by country and region-year.}
    \label{tab:dsc_pc_health_baseline}
\end{table}

\clearpage

\begin{table}[ht]
    \centering
    \caption{Baseline OLS Regression: Correlation Between Government IPs per Public-Sector Worker and General Health Outcomes}
    \resizebox{\linewidth}{!}{%
        {
\def\sym#1{\ifmmode^{#1}\else\(^{#1}\)\fi}
\begin{tabular}{l*{5}{c}}
\toprule
                    &\\  Dependent Variables: \\ Health Outcomes         &                     &                     &                     &                     \\
                    &\multicolumn{1}{c}{(1)}&\multicolumn{1}{c}{(2)}&\multicolumn{1}{c}{(3)}&\multicolumn{1}{c}{(4)}&\multicolumn{1}{c}{(5)}\\
                    &\multicolumn{1}{c}{Vacc. Meas.}&\multicolumn{1}{c}{Vacc. Dip.}&\multicolumn{1}{c}{Vacc. Hep.}&\multicolumn{1}{c}{Neonatal Mort.}&\multicolumn{1}{c}{Infant Mort.}\\
\midrule
Log government IPs per public-sector worker&     0.009\sym{**}&     0.010\sym{***}&     0.009\sym{**}&    -0.042\sym{**}&    -0.049\sym{***}\\
                    &    (0.003)&    (0.003)&    (0.004)&    (0.020)&    (0.018)\\
\addlinespace
Log GDP p.c.&     0.020&     0.022\sym{*}&     0.022&    -0.307\sym{***}&    -0.299\sym{***}\\
                    &    (0.012)&    (0.012)&    (0.014)&    (0.061)&    (0.052)\\
\addlinespace
Log Pop.&     0.000&    -0.006&    -0.011\sym{*}&     0.001&     0.002\\
                    &    (0.005)&    (0.004)&    (0.006)&    (0.024)&    (0.021)\\
\addlinespace
Internet percent Pop.&     0.040&     0.031&    -0.026&    -0.549\sym{*}&    -0.688\sym{***}\\
                    &    (0.062)&    (0.059)&    (0.081)&    (0.297)&    (0.252)\\
\addlinespace
Democracy (Polity)&    -0.021&     0.008&    -0.021&     0.189&     0.048\\
                    &    (0.028)&    (0.031)&    (0.035)&    (0.247)&    (0.188)\\
\midrule
Observations        &      1674&      1674&      1561&      1674&      1674\\
Dep. Var. Mean      &     0.890&     0.896&     0.876&     1.820&     2.463\\
Countries           &       148&       148&       145&       148&       148\\
Region-year clusters&        80&        80&        80&        80&        80\\
Adjusted \(R^2\)   &     0.317&     0.309&     0.149&     0.805&     0.871\\
Region-year FE      &       Yes&       Yes&       Yes&       Yes&       Yes\\
Years               & 2008-2023& 2008-2023& 2008-2023& 2008-2023& 2008-2023\\
\bottomrule
\multicolumn{6}{l}{\footnotesize Standard errors in parentheses}\\
\multicolumn{6}{l}{\footnotesize \sym{*} \(p<0.1\), \sym{**} \(p<0.05\), \sym{***} \(p<0.01\)}\\
\end{tabular}
}

    }
    \tabnote{This table reports country-year associations between log government-held public IPv4 addresses per public-sector worker and health outcomes for 2008--2023. This normalisation measures public-sector ICT capital relative to the size of the public-sector workforce and restricts the estimation sample to country-years with observed public-sector employment. Columns 1--3 report measles, diphtheria, and hepatitis B vaccination coverage, respectively, expressed as shares from zero to one. Columns 4 and 5 report the natural logarithms of the neonatal and infant mortality rates per 1,000 live births. All specifications control for log GDP per capita, log population, population internet access, and democracy, and include region-by-year fixed effects. Standard errors are two-way clustered by country and region-year.}
    \label{tab:dsc_pw_health_baseline}
\end{table}

\clearpage

%%% Working Point

\subsection{Country-Level Outcome Robustness Checks}

\begin{table}[ht]
    \centering
    \caption{Robustness OLS Regression: Correlation Between Government IPs per Capita and General Outcomes Controlling for State Capacity Measures}
    \resizebox{\linewidth}{!}{%
        {
\def\sym#1{\ifmmode^{#1}\else\(^{#1}\)\fi}
\begin{tabular}{l*{8}{c}}
\toprule
                    &\\  Dependent Variables: \\ General Outcomes         &                     &                     &                     &                     &                     &                     &                     \\
                    &\multicolumn{1}{c}{(1)}&\multicolumn{1}{c}{(2)}&\multicolumn{1}{c}{(3)}&\multicolumn{1}{c}{(4)}&\multicolumn{1}{c}{(5)}&\multicolumn{1}{c}{(6)}&\multicolumn{1}{c}{(7)}&\multicolumn{1}{c}{(8)}\\
                    &\multicolumn{1}{c}{CPI}&\multicolumn{1}{c}{CPI}&\multicolumn{1}{c}{CPI}&\multicolumn{1}{c}{CPI}&\multicolumn{1}{c}{Diph. Vacc.}&\multicolumn{1}{c}{Diph. Vacc.}&\multicolumn{1}{c}{Diph. Vacc.}&\multicolumn{1}{c}{Diph. Vacc.}\\
\midrule
Log government IPs per capita&     0.013\sym{***}&     0.006&     0.012\sym{**}&     0.012\sym{**}&     0.010\sym{***}&     0.008\sym{**}&     0.007\sym{**}&     0.005\\
                    &    (0.005)&    (0.004)&    (0.005)&    (0.005)&    (0.003)&    (0.003)&    (0.003)&    (0.004)\\
\addlinespace
Log GDP p.c.&     0.116\sym{***}&     0.013&     0.097\sym{***}&     0.117\sym{***}&     0.021\sym{*}&     0.005&     0.005&     0.014\\
                    &    (0.016)&    (0.019)&    (0.017)&    (0.014)&    (0.012)&    (0.019)&    (0.013)&    (0.012)\\
\addlinespace
Log Pop.&    -0.000&    -0.019\sym{***}&    -0.006&    -0.008&    -0.006&    -0.011\sym{*}&    -0.014\sym{***}&    -0.013\sym{**}\\
                    &    (0.007)&    (0.007)&    (0.007)&    (0.007)&    (0.004)&    (0.006)&    (0.005)&    (0.006)\\
\addlinespace
Internet percent Pop.&    -0.016&     0.056&    -0.106&    -0.120&     0.022&    -0.082&    -0.136\sym{**}&     0.041\\
                    &    (0.086)&    (0.087)&    (0.101)&    (0.073)&    (0.061)&    (0.067)&    (0.059)&    (0.073)\\
\addlinespace
Democracy (Polity)&     0.095\sym{**}&     0.036&     0.106\sym{**}&     0.030&     0.009&    -0.060&    -0.015&    -0.026\\
                    &    (0.046)&    (0.052)&    (0.047)&    (0.048)&    (0.033)&    (0.039)&    (0.032)&    (0.035)\\
\addlinespace
State Capacity&&     1.251\sym{***}&&&&     0.325\sym{**}&&\\
                    &&    (0.146)&&&&    (0.123)&&\\
\addlinespace
E-Gov Index&&&     0.328\sym{**}&&&&     0.393\sym{***}&\\
                    &&&    (0.148)&&&&    (0.092)&\\
\addlinespace
Statistical Index (WB)&&&&     0.297\sym{***}&&&&     0.327\sym{***}\\
                    &&&&    (0.103)&&&&    (0.083)\\
\midrule
Observations        &      1055&       288&       515&       761&      1676&       663&       813&       997\\
Dep. Var. Mean      &     0.496&     0.516&     0.500&     0.487&     0.896&     0.919&     0.898&     0.881\\
Countries           &       118&        77&       117&       118&       148&        91&       146&       147\\
Country clusters&       118&        77&       117&       118&       148&        91&       146&       147\\
Adjusted \(R^2\)   &     0.744&     0.866&     0.753&     0.766&     0.310&     0.339&     0.360&     0.357\\
Region-year FE      &       Yes&       Yes&       Yes&       Yes&       Yes&       Yes&       Yes&       Yes\\
Years               & 2012-2023& 2012-2015&2012-2022, biennial& 2016-2023& 2008-2023& 2008-2015&2008-2022, biennial& 2016-2023\\
\bottomrule
\multicolumn{9}{l}{\footnotesize Standard errors in parentheses}\\
\multicolumn{9}{l}{\footnotesize \sym{*} \(p<0.1\), \sym{**} \(p<0.05\), \sym{***} \(p<0.01\)}\\
\end{tabular}
}

    }
    \tabnote{This table reports country-year associations between log government-held public IPv4 addresses per capita and the Corruption Perceptions Index in Columns 1--4 and diphtheria vaccination coverage in Columns 5--8. The Corruption Perceptions Index is divided by 100, with higher values denoting lower perceived public-sector corruption; vaccination coverage is expressed as a share from zero to one. Columns 1 and 5 report the baseline specifications. Columns 2 and 6 add the normalised Hanson--Sigman State Capacity Index and use its 2008--2015 coverage, with the CPI sample beginning in 2012. Columns 3 and 7 add the U.N. E-Government Index for biennial observations through 2022. Columns 4 and 8 add the World Bank Statistical Performance Index for 2016--2023. All specifications control for log GDP per capita, log population, population internet access, and democracy; Columns 1--4 additionally control for tax revenue as a share of GDP. All specifications include region-by-year fixed effects. Standard errors are clustered by country.}
    \label{tab:dsc_pc_robust}
\end{table}

\clearpage

\begin{table}[ht]
    \centering
    \caption{Robustness OLS Regression: Correlation Between Government IPs per Public-Sector Worker and General Outcomes Controlling for State Capacity Measures}
    \resizebox{\linewidth}{!}{%
        {
\def\sym#1{\ifmmode^{#1}\else\(^{#1}\)\fi}
\begin{tabular}{l*{8}{c}}
\toprule
                    &\\  Dependent Variables: \\ General Outcomes         &                     &                     &                     &                     &                     &                     &                     \\
                    &\multicolumn{1}{c}{(1)}&\multicolumn{1}{c}{(2)}&\multicolumn{1}{c}{(3)}&\multicolumn{1}{c}{(4)}&\multicolumn{1}{c}{(5)}&\multicolumn{1}{c}{(6)}&\multicolumn{1}{c}{(7)}&\multicolumn{1}{c}{(8)}\\
                    &\multicolumn{1}{c}{CPI}&\multicolumn{1}{c}{CPI}&\multicolumn{1}{c}{CPI}&\multicolumn{1}{c}{CPI}&\multicolumn{1}{c}{Diph. Vacc.}&\multicolumn{1}{c}{Diph. Vacc.}&\multicolumn{1}{c}{Diph. Vacc.}&\multicolumn{1}{c}{Diph. Vacc.}\\
\midrule
Log government IPs per public-sector worker&     0.013\sym{**}&     0.006&     0.011\sym{**}&     0.011\sym{**}&     0.010\sym{***}&     0.008\sym{***}&     0.007\sym{**}&     0.005\\
                    &    (0.005)&    (0.005)&    (0.005)&    (0.005)&    (0.003)&    (0.003)&    (0.003)&    (0.004)\\
\addlinespace
Log GDP p.c.&     0.119\sym{***}&     0.014&     0.099\sym{***}&     0.120\sym{***}&     0.022\sym{*}&     0.007&     0.005&     0.014\\
                    &    (0.015)&    (0.016)&    (0.017)&    (0.014)&    (0.012)&    (0.018)&    (0.014)&    (0.013)\\
\addlinespace
Log Pop.&    -0.001&    -0.019\sym{***}&    -0.007&    -0.008&    -0.006&    -0.011\sym{*}&    -0.014\sym{***}&    -0.013\sym{**}\\
                    &    (0.006)&    (0.006)&    (0.006)&    (0.006)&    (0.004)&    (0.006)&    (0.005)&    (0.006)\\
\addlinespace
Internet percent Pop.&    -0.005&     0.060&    -0.099&    -0.111&     0.031&    -0.077&    -0.130\sym{**}&     0.047\\
                    &    (0.084)&    (0.095)&    (0.102)&    (0.072)&    (0.059)&    (0.059)&    (0.056)&    (0.069)\\
\addlinespace
Democracy (Polity)&     0.094\sym{**}&     0.034&     0.104\sym{**}&     0.029&     0.008&    -0.062\sym{*}&    -0.018&    -0.026\\
                    &    (0.043)&    (0.052)&    (0.042)&    (0.045)&    (0.031)&    (0.036)&    (0.029)&    (0.032)\\
\addlinespace
State Capacity&&     1.251\sym{***}&&&&     0.323\sym{**}&&\\
                    &&    (0.130)&&&&    (0.139)&&\\
\addlinespace
E-Gov Index&&&     0.336\sym{**}&&&&     0.400\sym{***}&\\
                    &&&    (0.138)&&&&    (0.102)&\\
\addlinespace
Statistical Index (WB)&&&&     0.300\sym{***}&&&&     0.327\sym{***}\\
                    &&&&    (0.109)&&&&    (0.091)\\
\midrule
Observations        &      1053&       288&       514&       759&      1674&       663&       812&       995\\
Dep. Var. Mean      &     0.496&     0.516&     0.500&     0.487&     0.896&     0.919&     0.898&     0.881\\
Countries           &       118&        77&       117&       118&       148&        91&       146&       147\\
Region-year clusters&        60&        20&        30&        40&        80&        40&        40&        40\\
Adjusted \(R^2\)   &     0.743&     0.866&     0.753&     0.764&     0.309&     0.340&     0.361&     0.357\\
Region-year FE      &       Yes&       Yes&       Yes&       Yes&       Yes&       Yes&       Yes&       Yes\\
Years               & 2012-2023& 2012-2015&2012-2022, biennial& 2016-2023& 2008-2023& 2008-2015&2008-2022, biennial& 2016-2023\\
\bottomrule
\multicolumn{9}{l}{\footnotesize Standard errors in parentheses}\\
\multicolumn{9}{l}{\footnotesize \sym{*} \(p<0.1\), \sym{**} \(p<0.05\), \sym{***} \(p<0.01\)}\\
\end{tabular}
}

    }
    \tabnote{This table reports country-year associations between log government-held public IPv4 addresses per public-sector worker and the Corruption Perceptions Index in Columns 1--4 and diphtheria vaccination coverage in Columns 5--8. This normalisation measures public-sector ICT capital relative to the size of the public-sector workforce and restricts the estimation sample to country-years with observed public-sector employment. The Corruption Perceptions Index is divided by 100, with higher values denoting lower perceived public-sector corruption; vaccination coverage is expressed as a share from zero to one. Columns 1 and 5 report the baseline specifications. Columns 2 and 6 add the normalised Hanson--Sigman State Capacity Index and use its 2008--2015 coverage, with the CPI sample beginning in 2012. Columns 3 and 7 add the U.N. E-Government Index for biennial observations through 2022. Columns 4 and 8 add the World Bank Statistical Performance Index for 2016--2023. All specifications control for log GDP per capita, log population, population internet access, and democracy; Columns 1--4 additionally control for tax revenue as a share of GDP. All specifications include region-by-year fixed effects. Standard errors are two-way clustered by country and region-year.}
    \label{tab:dsc_pw_robust}
\end{table}

\clearpage

\begin{table}[ht]
    \centering
    \caption{Robustness OLS Regression: Correlation Between Government IPs per Capita and Vaccinations Controlling for State Capacity Measures}
    \resizebox{0.85\linewidth}{!}{%
        {
\def\sym#1{\ifmmode^{#1}\else\(^{#1}\)\fi}
\begin{tabular}{l*{8}{c}}
\toprule
                    &\\  Dependent Variables: \\ Vaccination Outcomes         &                     &                     &                     &                     &                     &                     &                     \\
                    &\multicolumn{1}{c}{(1)}&\multicolumn{1}{c}{(2)}&\multicolumn{1}{c}{(3)}&\multicolumn{1}{c}{(4)}&\multicolumn{1}{c}{(5)}&\multicolumn{1}{c}{(6)}&\multicolumn{1}{c}{(7)}&\multicolumn{1}{c}{(8)}\\
                    &\multicolumn{1}{c}{Vacc. Hep.}&\multicolumn{1}{c}{Vacc. Meas.}&\multicolumn{1}{c}{Vacc. Hep.}&\multicolumn{1}{c}{Vacc. Meas.}&\multicolumn{1}{c}{Vacc. Hep.}&\multicolumn{1}{c}{Vacc. Meas.}&\multicolumn{1}{c}{Vacc. Hep.}&\multicolumn{1}{c}{Vacc. Meas.}\\
\midrule
Log government IPs per capita&     0.009\sym{**}&     0.009\sym{**}&     0.007&     0.006\sym{**}&     0.007\sym{*}&     0.007\sym{**}&     0.005&     0.005\\
                    &    (0.004)&    (0.003)&    (0.006)&    (0.003)&    (0.004)&    (0.003)&    (0.004)&    (0.005)\\
\addlinespace
Log GDP p.c.&     0.021&     0.018&     0.020&    -0.000&     0.004&     0.005&     0.010&     0.014\\
                    &    (0.014)&    (0.013)&    (0.029)&    (0.015)&    (0.017)&    (0.017)&    (0.015)&    (0.015)\\
\addlinespace
Log Pop.&    -0.011\sym{*}&     0.001&    -0.024\sym{***}&    -0.008&    -0.019\sym{***}&    -0.007&    -0.013\sym{*}&    -0.004\\
                    &    (0.006)&    (0.005)&    (0.009)&    (0.005)&    (0.006)&    (0.006)&    (0.007)&    (0.007)\\
\addlinespace
Internet percent Pop.&    -0.036&     0.032&    -0.233\sym{*}&    -0.042&    -0.199\sym{**}&    -0.132\sym{**}&     0.036&     0.049\\
                    &    (0.080)&    (0.062)&    (0.116)&    (0.053)&    (0.078)&    (0.056)&    (0.075)&    (0.075)\\
\addlinespace
Democracy (Polity)&    -0.019&    -0.020&    -0.088\sym{*}&    -0.086\sym{***}&    -0.045&    -0.045\sym{*}&    -0.038&    -0.051\\
                    &    (0.035)&    (0.028)&    (0.050)&    (0.030)&    (0.033)&    (0.023)&    (0.035)&    (0.035)\\
\addlinespace
State Capacity&&&     0.227&     0.255\sym{*}&&&&\\
                    &&&    (0.217)&    (0.132)&&&&\\
\addlinespace
E-Gov Index&&&&&     0.404\sym{***}&     0.387\sym{***}&&\\
                    &&&&&    (0.114)&    (0.112)&&\\
\addlinespace
Statistical Index (WB)&&&&&&&     0.328\sym{***}&     0.328\sym{***}\\
                    &&&&&&&    (0.099)&    (0.097)\\
\midrule
Observations        &      1563&      1676&       589&       663&       753&       813&       958&       997\\
Dep. Var. Mean      &     0.876&     0.890&     0.889&     0.918&     0.877&     0.892&     0.868&     0.871\\
Countries           &       145&       148&        83&        91&       143&       146&       144&       147\\
Region-year clusters&        80&        80&        40&        40&        40&        40&        40&        40\\
Adjusted \(R^2\)   &     0.150&     0.319&     0.134&     0.369&     0.189&     0.368&     0.260&     0.348\\
Region-year FE      &       Yes&       Yes&       Yes&       Yes&       Yes&       Yes&       Yes&       Yes\\
Years               & 2008-2023& 2008-2023& 2008-2015& 2008-2015&2008-2022, biennial&2008-2022, biennial& 2016-2023& 2016-2023\\
\bottomrule
\multicolumn{9}{l}{\footnotesize Standard errors in parentheses}\\
\multicolumn{9}{l}{\footnotesize \sym{*} \(p<0.1\), \sym{**} \(p<0.05\), \sym{***} \(p<0.01\)}\\
\end{tabular}
}

    }
    \tabnote{This table reports country-year associations between log government-held public IPv4 addresses per capita and hepatitis B and measles vaccination coverage, each expressed as a share from zero to one. Columns 1 and 2 report the baseline specifications for 2008--2023. Columns 3 and 4 add the normalised Hanson--Sigman State Capacity Index for 2008--2015. Columns 5 and 6 add the U.N. E-Government Index for biennial observations from 2008 to 2022. Columns 7 and 8 add the World Bank Statistical Performance Index for 2016--2023. All specifications control for log GDP per capita, log population, population internet access, and democracy, and include region-by-year fixed effects. Standard errors are two-way clustered by country and region-year.}
    \label{tab:dsc_pc_vacc_robust}
\end{table}

\clearpage

\begin{table}[ht]
    \centering
    \caption{Robustness OLS Regression: Correlation Between Government IPs per Public-Sector Worker and Vaccinations Controlling for State Capacity Measures}
    \resizebox{\linewidth}{!}{%
        {
\def\sym#1{\ifmmode^{#1}\else\(^{#1}\)\fi}
\begin{tabular}{l*{8}{c}}
\toprule
                    &\\  Dependent Variables: \\ Vaccination Outcomes         &                     &                     &                     &                     &                     &                     &                     \\
                    &\multicolumn{1}{c}{(1)}&\multicolumn{1}{c}{(2)}&\multicolumn{1}{c}{(3)}&\multicolumn{1}{c}{(4)}&\multicolumn{1}{c}{(5)}&\multicolumn{1}{c}{(6)}&\multicolumn{1}{c}{(7)}&\multicolumn{1}{c}{(8)}\\
                    &\multicolumn{1}{c}{Vacc. Hep.}&\multicolumn{1}{c}{Vacc. Meas.}&\multicolumn{1}{c}{Vacc. Hep.}&\multicolumn{1}{c}{Vacc. Meas.}&\multicolumn{1}{c}{Vacc. Hep.}&\multicolumn{1}{c}{Vacc. Meas.}&\multicolumn{1}{c}{Vacc. Hep.}&\multicolumn{1}{c}{Vacc. Meas.}\\
\midrule
Log government IPs per public-sector worker&     0.009\sym{**}&     0.009\sym{**}&     0.008&     0.006\sym{**}&     0.007\sym{*}&     0.006\sym{*}&     0.005&     0.005\\
                    &    (0.004)&    (0.003)&    (0.005)&    (0.003)&    (0.004)&    (0.003)&    (0.004)&    (0.005)\\
\addlinespace
Log GDP p.c.&     0.022&     0.020&     0.021&     0.001&     0.003&     0.005&     0.010&     0.016\\
                    &    (0.014)&    (0.012)&    (0.029)&    (0.016)&    (0.016)&    (0.016)&    (0.014)&    (0.015)\\
\addlinespace
Log Pop.&    -0.011\sym{*}&     0.000&    -0.023\sym{***}&    -0.008&    -0.019\sym{***}&    -0.008&    -0.013\sym{*}&    -0.004\\
                    &    (0.006)&    (0.005)&    (0.008)&    (0.005)&    (0.006)&    (0.006)&    (0.007)&    (0.007)\\
\addlinespace
Internet percent Pop.&    -0.026&     0.040&    -0.230\sym{*}&    -0.038&    -0.192\sym{**}&    -0.126\sym{**}&     0.045&     0.054\\
                    &    (0.081)&    (0.062)&    (0.117)&    (0.053)&    (0.079)&    (0.056)&    (0.074)&    (0.074)\\
\addlinespace
Democracy (Polity)&    -0.021&    -0.021&    -0.092\sym{*}&    -0.088\sym{***}&    -0.049&    -0.047\sym{*}&    -0.038&    -0.050\\
                    &    (0.035)&    (0.028)&    (0.049)&    (0.029)&    (0.032)&    (0.023)&    (0.035)&    (0.035)\\
\addlinespace
State Capacity&&&     0.221&     0.254\sym{*}&&&&\\
                    &&&    (0.218)&    (0.132)&&&&\\
\addlinespace
E-Gov Index&&&&&     0.413\sym{***}&     0.395\sym{***}&&\\
                    &&&&&    (0.112)&    (0.112)&&\\
\addlinespace
Statistical Index (WB)&&&&&&&     0.327\sym{***}&     0.330\sym{***}\\
                    &&&&&&&    (0.099)&    (0.097)\\
\midrule
Observations        &      1561&      1674&       589&       663&       752&       812&       956&       995\\
Dep. Var. Mean      &     0.876&     0.890&     0.889&     0.918&     0.878&     0.892&     0.868&     0.871\\
Countries           &       145&       148&        83&        91&       143&       146&       144&       147\\
Region-year clusters&        80&        80&        40&        40&        40&        40&        40&        40\\
Adjusted \(R^2\)   &     0.149&     0.317&     0.137&     0.369&     0.191&     0.368&     0.259&     0.347\\
Region-year FE      &       Yes&       Yes&       Yes&       Yes&       Yes&       Yes&       Yes&       Yes\\
Years               & 2008-2023& 2008-2023& 2008-2015& 2008-2015&2008-2022, biennial&2008-2022, biennial& 2016-2023& 2016-2023\\
\bottomrule
\multicolumn{9}{l}{\footnotesize Standard errors in parentheses}\\
\multicolumn{9}{l}{\footnotesize \sym{*} \(p<0.1\), \sym{**} \(p<0.05\), \sym{***} \(p<0.01\)}\\
\end{tabular}
}

    }
    \tabnote{This table reports country-year associations between log government-held public IPv4 addresses per public-sector worker and hepatitis B and measles vaccination coverage, each expressed as a share from zero to one. This normalisation measures public-sector ICT capital relative to the size of the public-sector workforce and restricts the estimation sample to country-years with observed public-sector employment. Columns 1 and 2 report the baseline specifications for 2008--2023. Columns 3 and 4 add the normalised Hanson--Sigman State Capacity Index for 2008--2015. Columns 5 and 6 add the U.N. E-Government Index for biennial observations from 2008 to 2022. Columns 7 and 8 add the World Bank Statistical Performance Index for 2016--2023. All specifications control for log GDP per capita, log population, population internet access, and democracy, and include region-by-year fixed effects. Standard errors are two-way clustered by country and region-year.}
    \label{tab:dsc_pw_vacc_robust}
\end{table}

\clearpage

\begin{table}[ht]
    \centering
    \caption{Robustness OLS Regression: Correlation Between Government IPs per Capita and Infant Mortality Controlling for State Capacity Measures}
    \resizebox{\linewidth}{!}{%
        {
\def\sym#1{\ifmmode^{#1}\else\(^{#1}\)\fi}
\begin{tabular}{l*{8}{c}}
\toprule
                    &\\  Dependent Variables: \\ Mortality Outcomes         &                     &                     &                     &                     &                     &                     &                     \\
                    &\multicolumn{1}{c}{(1)}&\multicolumn{1}{c}{(2)}&\multicolumn{1}{c}{(3)}&\multicolumn{1}{c}{(4)}&\multicolumn{1}{c}{(5)}&\multicolumn{1}{c}{(6)}&\multicolumn{1}{c}{(7)}&\multicolumn{1}{c}{(8)}\\
                    &\multicolumn{1}{c}{Infant Mort.}&\multicolumn{1}{c}{Neonatal Mort.}&\multicolumn{1}{c}{Infant Mort.}&\multicolumn{1}{c}{Neonatal Mort.}&\multicolumn{1}{c}{Infant Mort.}&\multicolumn{1}{c}{Neonatal Mort.}&\multicolumn{1}{c}{Infant Mort.}&\multicolumn{1}{c}{Neonatal Mort.}\\
\midrule
Log government IPs per capita&    -0.054\sym{***}&    -0.047\sym{**}&    -0.065\sym{***}&    -0.062\sym{***}&    -0.043\sym{**}&    -0.035\sym{*}&    -0.018&    -0.010\\
                    &    (0.018)&    (0.019)&    (0.018)&    (0.021)&    (0.019)&    (0.020)&    (0.018)&    (0.017)\\
\addlinespace
Log GDP p.c.&    -0.285\sym{***}&    -0.295\sym{***}&    -0.211\sym{***}&    -0.190\sym{**}&    -0.192\sym{***}&    -0.193\sym{***}&    -0.267\sym{***}&    -0.289\sym{***}\\
                    &    (0.052)&    (0.060)&    (0.066)&    (0.075)&    (0.045)&    (0.056)&    (0.058)&    (0.070)\\
\addlinespace
Log Pop.&    -0.000&    -0.001&     0.012&    -0.003&     0.040\sym{*}&     0.042\sym{*}&     0.036&     0.043\\
                    &    (0.021)&    (0.024)&    (0.024)&    (0.029)&    (0.020)&    (0.024)&    (0.023)&    (0.027)\\
\addlinespace
Internet percent Pop.&    -0.638\sym{**}&    -0.502\sym{*}&    -0.248&    -0.435&     0.100&     0.245&    -0.690\sym{**}&    -0.389\\
                    &    (0.253)&    (0.300)&    (0.382)&    (0.493)&    (0.307)&    (0.329)&    (0.279)&    (0.310)\\
\addlinespace
Democracy (Polity)&     0.047&     0.189&     0.335&     0.512&     0.190&     0.342&     0.188&     0.317\\
                    &    (0.183)&    (0.244)&    (0.239)&    (0.326)&    (0.184)&    (0.248)&    (0.159)&    (0.218)\\
\addlinespace
State Capacity&&&    -1.298\sym{**}&    -1.171\sym{*}&&&&\\
                    &&&    (0.517)&    (0.627)&&&&\\
\addlinespace
E-Gov Index&&&&&    -2.076\sym{***}&    -2.196\sym{***}&&\\
                    &&&&&    (0.362)&    (0.443)&&\\
\addlinespace
Statistical Index (WB)&&&&&&&    -1.495\sym{***}&    -1.522\sym{***}\\
                    &&&&&&&    (0.359)&    (0.405)\\
\midrule
Observations        &      1676&      1676&       663&       663&       813&       813&       997&       997\\
Dep. Var. Mean      &     2.463&     1.820&     2.338&     1.705&     2.442&     1.801&     2.558&     1.909\\
Countries           &       148&       148&        91&        91&       146&       146&       147&       147\\
Region-year clusters&        80&        80&        40&        40&        40&        40&        40&        40\\
Adjusted \(R^2\)   &     0.873&     0.807&     0.873&     0.794&     0.890&     0.828&     0.894&     0.836\\
Region-year FE      &       Yes&       Yes&       Yes&       Yes&       Yes&       Yes&       Yes&       Yes\\
Years               & 2008-2023& 2008-2023& 2008-2015& 2008-2015&2008-2022, biennial&2008-2022, biennial& 2016-2023& 2016-2023\\
\bottomrule
\multicolumn{9}{l}{\footnotesize Standard errors in parentheses}\\
\multicolumn{9}{l}{\footnotesize \sym{*} \(p<0.1\), \sym{**} \(p<0.05\), \sym{***} \(p<0.01\)}\\
\end{tabular}
}

    }
    \tabnote{This table reports country-year associations between log government-held public IPv4 addresses per capita and the natural logarithms of the infant and neonatal mortality rates per 1,000 live births. Columns 1 and 2 report the baseline specifications for 2008--2023. Columns 3 and 4 add the normalised Hanson--Sigman State Capacity Index for 2008--2015. Columns 5 and 6 add the U.N. E-Government Index for biennial observations from 2008 to 2022. Columns 7 and 8 add the World Bank Statistical Performance Index for 2016--2023. All specifications control for log GDP per capita, log population, population internet access, and democracy, and include region-by-year fixed effects. Standard errors are two-way clustered by country and region-year.}
    \label{tab:dsc_pc_mort_robust}
\end{table}

\clearpage

\begin{table}[ht]
    \centering
    \caption{Robustness OLS Regression: Correlation Between Government IPs per Public-Sector Worker and Infant Mortality Controlling for State Capacity Measures}
    \resizebox{\linewidth}{!}{%
        {
\def\sym#1{\ifmmode^{#1}\else\(^{#1}\)\fi}
\begin{tabular}{l*{8}{c}}
\toprule
                    &\\  Dependent Variables: \\ Mortality Outcomes         &                     &                     &                     &                     &                     &                     &                     \\
                    &\multicolumn{1}{c}{(1)}&\multicolumn{1}{c}{(2)}&\multicolumn{1}{c}{(3)}&\multicolumn{1}{c}{(4)}&\multicolumn{1}{c}{(5)}&\multicolumn{1}{c}{(6)}&\multicolumn{1}{c}{(7)}&\multicolumn{1}{c}{(8)}\\
                    &\multicolumn{1}{c}{Infant Mort.}&\multicolumn{1}{c}{Neonatal Mort.}&\multicolumn{1}{c}{Infant Mort.}&\multicolumn{1}{c}{Neonatal Mort.}&\multicolumn{1}{c}{Infant Mort.}&\multicolumn{1}{c}{Neonatal Mort.}&\multicolumn{1}{c}{Infant Mort.}&\multicolumn{1}{c}{Neonatal Mort.}\\
\midrule
Log government IPs per public-sector worker&    -0.049\sym{***}&    -0.042\sym{**}&    -0.062\sym{***}&    -0.058\sym{**}&    -0.040\sym{**}&    -0.032&    -0.013&    -0.005\\
                    &    (0.018)&    (0.020)&    (0.018)&    (0.022)&    (0.018)&    (0.020)&    (0.018)&    (0.017)\\
\addlinespace
Log GDP p.c.&    -0.299\sym{***}&    -0.307\sym{***}&    -0.225\sym{***}&    -0.204\sym{***}&    -0.199\sym{***}&    -0.198\sym{***}&    -0.272\sym{***}&    -0.293\sym{***}\\
                    &    (0.052)&    (0.061)&    (0.065)&    (0.075)&    (0.047)&    (0.058)&    (0.057)&    (0.070)\\
\addlinespace
Log Pop.&     0.002&     0.001&     0.015&     0.000&     0.042\sym{**}&     0.044\sym{*}&     0.037&     0.044\\
                    &    (0.021)&    (0.024)&    (0.024)&    (0.029)&    (0.020)&    (0.024)&    (0.023)&    (0.027)\\
\addlinespace
Internet percent Pop.&    -0.688\sym{***}&    -0.549\sym{*}&    -0.290&    -0.476&     0.073&     0.221&    -0.713\sym{**}&    -0.406\\
                    &    (0.252)&    (0.297)&    (0.384)&    (0.493)&    (0.305)&    (0.325)&    (0.280)&    (0.309)\\
\addlinespace
Democracy (Polity)&     0.048&     0.189&     0.353&     0.528&     0.197&     0.348&     0.185&     0.315\\
                    &    (0.188)&    (0.247)&    (0.249)&    (0.336)&    (0.188)&    (0.252)&    (0.158)&    (0.217)\\
\addlinespace
State Capacity&&&    -1.312\sym{**}&    -1.187\sym{*}&&&&\\
                    &&&    (0.523)&    (0.634)&&&&\\
\addlinespace
E-Gov Index&&&&&    -2.123\sym{***}&    -2.239\sym{***}&&\\
                    &&&&&    (0.357)&    (0.439)&&\\
\addlinespace
Statistical Index (WB)&&&&&&&    -1.514\sym{***}&    -1.541\sym{***}\\
                    &&&&&&&    (0.357)&    (0.403)\\
\midrule
Observations        &      1674&      1674&       663&       663&       812&       812&       995&       995\\
Dep. Var. Mean      &     2.463&     1.820&     2.338&     1.705&     2.442&     1.801&     2.559&     1.910\\
Countries           &       148&       148&        91&        91&       146&       146&       147&       147\\
Region-year clusters&        80&        80&        40&        40&        40&        40&        40&        40\\
Adjusted \(R^2\)   &     0.871&     0.805&     0.871&     0.792&     0.889&     0.827&     0.893&     0.836\\
Region-year FE      &       Yes&       Yes&       Yes&       Yes&       Yes&       Yes&       Yes&       Yes\\
Years               & 2008-2023& 2008-2023& 2008-2015& 2008-2015&2008-2022, biennial&2008-2022, biennial& 2016-2023& 2016-2023\\
\bottomrule
\multicolumn{9}{l}{\footnotesize Standard errors in parentheses}\\
\multicolumn{9}{l}{\footnotesize \sym{*} \(p<0.1\), \sym{**} \(p<0.05\), \sym{***} \(p<0.01\)}\\
\end{tabular}
}

    }
    \tabnote{This table reports country-year associations between log government-held public IPv4 addresses per public-sector worker and the natural logarithms of the infant and neonatal mortality rates per 1,000 live births. This normalisation measures public-sector ICT capital relative to the size of the public-sector workforce and restricts the estimation sample to country-years with observed public-sector employment. Columns 1 and 2 report the baseline specifications for 2008--2023. Columns 3 and 4 add the normalised Hanson--Sigman State Capacity Index for 2008--2015. Columns 5 and 6 add the U.N. E-Government Index for biennial observations from 2008 to 2022. Columns 7 and 8 add the World Bank Statistical Performance Index for 2016--2023. All specifications control for log GDP per capita, log population, population internet access, and democracy, and include region-by-year fixed effects. Standard errors are two-way clustered by country and region-year.}
    \label{tab:dsc_pw_mort_robust}
\end{table}

\clearpage

\begin{table}[ht]
    \centering
    \caption{OLS Regression: Correlation Between Local-Level Government IPs per Capita and Health Outcomes}
    \resizebox{\linewidth}{!}{%
        {
\def\sym#1{\ifmmode^{#1}\else\(^{#1}\)\fi}
\begin{tabular}{l*{5}{c}}
\toprule
 & \multicolumn{1}{c}{(1)} & \multicolumn{1}{c}{(2)} & \multicolumn{1}{c}{(3)} & \multicolumn{1}{c}{(4)} & \multicolumn{1}{c}{(5)}\\
 & \multicolumn{1}{c}{Vacc. Meas.} & \multicolumn{1}{c}{Vacc. Dip.} & \multicolumn{1}{c}{Vacc. Hep.} & \multicolumn{1}{c}{Neonatal Mort.} & \multicolumn{1}{c}{Infant Mort.}\\
\midrule
Log Local Gov IPs p.c. & 0.009\sym{*} & 0.006 & 0.006 & -0.044\sym{**} & -0.040\sym{**}\\
 & (0.005) & (0.005) & (0.005) & (0.018) & (0.020)\\
\addlinespace
Log GDP p.c. & 0.026 & 0.025\sym{*} & 0.025 & -0.280\sym{***} & -0.271\sym{***}\\
 & (0.016) & (0.014) & (0.015) & (0.071) & (0.062)\\
\addlinespace
Log Pop. & 0.006 & -0.004 & -0.002 & 0.018 & 0.010\\
 & (0.008) & (0.005) & (0.006) & (0.027) & (0.025)\\
\addlinespace
Internet percent Pop. & 0.041 & 0.026 & 0.021 & -0.370 & -0.729\sym{**}\\
 & (0.090) & (0.085) & (0.089) & (0.326) & (0.317)\\
\addlinespace
Democracy (Polity) & 0.017 & 0.041 & 0.031 & 0.092 & -0.014\\
 & (0.045) & (0.043) & (0.044) & (0.223) & (0.185)\\
\midrule
Observations & 643 & 643 & 626 & 643 & 643\\
Dep. Var. Mean & 0.867 & 0.878 & 0.867 & 1.907 & 2.558\\
Clusters & 133 & 133 & 130 & 133 & 133\\
Adjusted \(R^2\) & 0.229 & 0.236 & 0.152 & 0.822 & 0.874\\
Region-year FE & Yes & Yes & Yes & Yes & Yes\\
Years & 2019--2023 & 2019--2023 & 2019--2023 & 2019--2023 & 2019--2023\\
\bottomrule
\multicolumn{6}{l}{\footnotesize Standard errors in parentheses}\\
\multicolumn{6}{l}{\footnotesize \sym{*} \(p<0.1\), \sym{**} \(p<0.05\), \sym{***} \(p<0.01\)}\\
\end{tabular}
}

    }
    \tabnote{This table presents baseline regressions for log local government IPs per capita with vaccination rates, neonatal mortality, and infant mortality. Standard errors are clustered at the country level because there are insufficient clusters with multi-way clustering at the country and region-year level. Region-year fixed effects are included.}
    \label{tab:loca_dsc_pc_health}

\end{table}

\clearpage

\begin{table}[ht]
    \centering
    \caption{OLS Regression: Correlation Between Federal-Level Government IPs per Capita and General Outcomes}

    \resizebox{\linewidth}{!}{%
        {
\def\sym#1{\ifmmode^{#1}\else\(^{#1}\)\fi}
\begin{tabular}{l*{4}{c}}
\toprule
 & \multicolumn{1}{c}{(1)} & \multicolumn{1}{c}{(2)} & \multicolumn{1}{c}{(3)} & \multicolumn{1}{c}{(4)}\\
 & \multicolumn{1}{c}{CPI} & \multicolumn{1}{c}{Bribery Incidence} & \multicolumn{1}{c}{Tax/GDP} & \multicolumn{1}{c}{Soc. Ins. Cov.}\\
\midrule
Log Fed. Gov IPs p.c. & 0.010\sym{**} & -0.013\sym{**} & 0.007\sym{**} & 0.029\sym{**}\\
 & (0.004) & (0.006) & (0.003) & (0.012)\\
\addlinespace
Log GDP p.c. & 0.129\sym{***} & -0.072\sym{***} & -0.011 & -0.009\\
 & (0.014) & (0.022) & (0.008) & (0.046)\\
\addlinespace
Log Pop. & -0.003 & 0.007 & -0.009\sym{**} & -0.027\sym{*}\\
 & (0.007) & (0.009) & (0.004) & (0.016)\\
\addlinespace
Internet percent Pop. & -0.116 & 0.208\sym{**} & 0.068 & 0.243\\
 & (0.077) & (0.103) & (0.044) & (0.216)\\
\addlinespace
Democracy (Polity) & 0.076\sym{*} & -0.013 & 0.088\sym{***} & 0.063\\
 & (0.045) & (0.055) & (0.021) & (0.070)\\
\midrule
Observations & 483 & 84 & 483 & 90\\
Dep. Var. Mean & 0.484 & 0.104 & 0.169 & 0.210\\
Clusters & 110 & 72 & 110 & 43\\
Adjusted \(R^2\) & 0.764 & 0.417 & 0.420 & 0.591\\
Region-year FE & Yes & Yes & Yes & Yes\\
Years & 2019--2023 & 2019--2023 & 2019--2023 & 2019--2022\\
\bottomrule
\multicolumn{5}{l}{\footnotesize Standard errors in parentheses}\\
\multicolumn{5}{l}{\footnotesize \sym{*} \(p<0.1\), \sym{**} \(p<0.05\), \sym{***} \(p<0.01\)}\\
\end{tabular}
}

    }
    \tabnote{This table presents baseline regressions for log federal government IPs per capita with corruption, bribery, tax revenue as a percent of GDP, and social insurance coverage. Standard errors are clustered at the country level because there are insufficient clusters with multi-way clustering at the country and region-year level. Region-year fixed effects are included.}
    \label{tab:fed_dsc_pc_general}

\end{table}

\clearpage

\begin{table}[ht]
    \centering
    \caption{OLS Regression: Correlation Between Local-Level Government IPs per Capita and General Outcomes}

    \resizebox{\linewidth}{!}{%
        {
\def\sym#1{\ifmmode^{#1}\else\(^{#1}\)\fi}
\begin{tabular}{l*{4}{c}}
\toprule
 & \multicolumn{1}{c}{(1)} & \multicolumn{1}{c}{(2)} & \multicolumn{1}{c}{(3)} & \multicolumn{1}{c}{(4)}\\
 & \multicolumn{1}{c}{CPI} & \multicolumn{1}{c}{Bribery Incidence} & \multicolumn{1}{c}{Tax/GDP} & \multicolumn{1}{c}{Soc. Ins. Cov.}\\
\midrule
Log Local Gov IPs p.c. & 0.010\sym{*} & -0.010 & 0.007\sym{***} & -0.011\sym{*}\\
 & (0.006) & (0.007) & (0.002) & (0.006)\\
\addlinespace
Log GDP p.c. & 0.125\sym{***} & -0.082\sym{***} & -0.014\sym{*} & -0.037\\
 & (0.015) & (0.021) & (0.008) & (0.041)\\
\addlinespace
Log Pop. & -0.003 & 0.009 & -0.009\sym{***} & -0.025\\
 & (0.008) & (0.009) & (0.003) & (0.019)\\
\addlinespace
Internet percent Pop. & -0.106 & 0.264\sym{***} & 0.074\sym{*} & 0.632\sym{***}\\
 & (0.079) & (0.091) & (0.041) & (0.232)\\
\addlinespace
Democracy (Polity) & 0.085 & -0.002 & 0.100\sym{***} & 0.108\\
 & (0.053) & (0.054) & (0.021) & (0.081)\\
\midrule
Observations & 471 & 79 & 471 & 85\\
Dep. Var. Mean & 0.493 & 0.098 & 0.172 & 0.216\\
Clusters & 106 & 66 & 106 & 41\\
Adjusted \(R^2\) & 0.723 & 0.458 & 0.457 & 0.581\\
Region-year FE & Yes & Yes & Yes & Yes\\
Years & 2019--2023 & 2019--2023 & 2019--2023 & 2019--2022\\
\bottomrule
\multicolumn{5}{l}{\footnotesize Standard errors in parentheses}\\
\multicolumn{5}{l}{\footnotesize \sym{*} \(p<0.1\), \sym{**} \(p<0.05\), \sym{***} \(p<0.01\)}\\
\end{tabular}
}

    }
    \tabnote{This table presents baseline regressions for log local government IPs per capita with corruption, bribery, tax revenue as a percent of GDP, and social insurance coverage. Standard errors are clustered at the country level because there are insufficient clusters with multi-way clustering at the country and region-year level. Region-year fixed effects are included.}
    \label{tab:loca_dsc_pc_general}
\end{table}

\clearpage

\begin{table}[ht]
    \centering
    \caption{OLS Regression: Correlation Between Federal-Level Government IPs per Capita and Health Outcomes}

    \resizebox{\linewidth}{!}{%
        {
\def\sym#1{\ifmmode^{#1}\else\(^{#1}\)\fi}
\begin{tabular}{l*{5}{c}}
\toprule
 & \multicolumn{1}{c}{(1)} & \multicolumn{1}{c}{(2)} & \multicolumn{1}{c}{(3)} & \multicolumn{1}{c}{(4)} & \multicolumn{1}{c}{(5)}\\
 & \multicolumn{1}{c}{Vacc. Meas.} & \multicolumn{1}{c}{Vacc. Dip.} & \multicolumn{1}{c}{Vacc. Hep.} & \multicolumn{1}{c}{Neonatal Mort.} & \multicolumn{1}{c}{Infant Mort.}\\
\midrule
Log Fed. Gov IPs p.c. & 0.003 & 0.001 & -0.000 & -0.008 & -0.006\\
 & (0.004) & (0.004) & (0.004) & (0.017) & (0.014)\\
\addlinespace
Log GDP p.c. & 0.030\sym{*} & 0.029\sym{**} & 0.029\sym{*} & -0.327\sym{***} & -0.308\sym{***}\\
 & (0.016) & (0.013) & (0.015) & (0.069) & (0.056)\\
\addlinespace
Log Pop. & 0.005 & -0.004 & -0.002 & 0.013 & 0.007\\
 & (0.008) & (0.005) & (0.006) & (0.029) & (0.025)\\
\addlinespace
Internet percent Pop. & 0.095 & 0.083 & 0.085 & -0.485 & -0.879\sym{***}\\
 & (0.090) & (0.086) & (0.089) & (0.312) & (0.285)\\
\addlinespace
Democracy (Polity) & 0.030 & 0.050 & 0.043 & 0.089 & -0.020\\
 & (0.039) & (0.041) & (0.042) & (0.213) & (0.163)\\
\midrule
Observations & 658 & 658 & 641 & 658 & 658\\
Dep. Var. Mean & 0.865 & 0.874 & 0.863 & 1.942 & 2.597\\
Clusters & 138 & 138 & 135 & 138 & 138\\
Adjusted \(R^2\) & 0.265 & 0.284 & 0.198 & 0.826 & 0.882\\
Region-year FE & Yes & Yes & Yes & Yes & Yes\\
Years & 2019--2023 & 2019--2023 & 2019--2023 & 2019--2023 & 2019--2023\\
\bottomrule
\multicolumn{6}{l}{\footnotesize Standard errors in parentheses}\\
\multicolumn{6}{l}{\footnotesize \sym{*} \(p<0.1\), \sym{**} \(p<0.05\), \sym{***} \(p<0.01\)}\\
\end{tabular}
}

    }
    \tabnote{This table presents baseline regressions for log federal government IPs per capita with vaccination rates, neonatal mortality, and infant mortality. Standard errors are clustered at the country level because there are insufficient clusters with multi-way clustering at the country and region-year level. Region-year fixed effects are included.}
        \label{tab:fed_dsc_pc_health}

\end{table}

\clearpage

\subsection{Country-Level Outcome Heterogeneity Analysis}

\begin{table}[ht]
    \centering
    \caption{Corruption Heterogeneity by Internet and Press Freedom}
    \resizebox{\linewidth}{!}{%
        {
\def\sym#1{\ifmmode^{#1}\else\(^{#1}\)\fi}
\begin{tabular}{l*{4}{c}}
\toprule
&\multicolumn{1}{c}{(1)}&\multicolumn{1}{c}{(2)}&\multicolumn{1}{c}{(3)}&\multicolumn{1}{c}{(4)}\\
&\multicolumn{1}{c}{CPI}&\multicolumn{1}{c}{CPI}&\multicolumn{1}{c}{CPI}&\multicolumn{1}{c}{CPI}\\
\midrule
Natural log of government IPs per capita&0.020\sym{***}&0.019\sym{***}&0.021\sym{***}&0.018\sym{***}\\
&(0.005)&(0.005)&(0.005)&(0.005)\\
\addlinespace
DSC $\times$ lower press freedom&-0.015\sym{**}&&&\\
&(0.006)&&&\\
\addlinespace
DSC $\times$ internet filtering&&-0.015\sym{***}&&\\
&&(0.005)&&\\
\addlinespace
DSC $\times$ social-media monitoring&&&-0.015\sym{**}&\\
&&&(0.006)&\\
\addlinespace
DSC $\times$ internet censorship&&&&-0.012\sym{**}\\
&&&&(0.005)\\
\addlinespace
Log GDP p.c.&0.080\sym{***}&0.088\sym{***}&0.089\sym{***}&0.097\sym{***}\\
&(0.016)&(0.014)&(0.014)&(0.014)\\
\addlinespace
Log Pop.&-0.005&-0.005&-0.004&-0.003\\
&(0.005)&(0.005)&(0.005)&(0.006)\\
\addlinespace
Internet percent Pop.&0.058&0.052&0.053&0.037\\
&(0.065)&(0.065)&(0.064)&(0.065)\\
\addlinespace
Democracy (Polity)&0.095\sym{***}&0.106\sym{***}&0.115\sym{***}&0.133\sym{***}\\
&(0.033)&(0.032)&(0.033)&(0.034)\\
\midrule
Observations&1545&1545&1545&1545\\
Dep. Var. Mean&0.452&0.452&0.452&0.452\\
Countries&151&151&151&151\\
Region-year clusters&60&60&60&60\\
Adjusted \(R^2\)&0.741&0.738&0.737&0.728\\
Region-year FE&Yes&Yes&Yes&Yes\\
Baseline controls&Yes&Yes&Yes&Yes\\
\bottomrule
\multicolumn{5}{l}{\footnotesize Standard errors in parentheses}\\
\multicolumn{5}{l}{\footnotesize \sym{*} \(p<0.10\), \sym{**} \(p<0.05\), \sym{***} \(p<0.01\)}\\
\end{tabular}
}

    }
    \tabnote{This table presents a heterogeneity analysis based on the level of press and internet freedom in a country. The analysis uses interaction terms for countries below the median level of freedom, where the dummy variable equals 1 for countries with press censorship or monitoring. All four interaction coefficients are negative; the interaction with internet filtering is significant at the 1 percent level, and the interactions with lower press freedom, social media monitoring, and internet censorship are significant at the 5 percent level.}
     \label{tab:corruption_freedom}
\end{table}

\clearpage

\begin{table}[ht]
    \centering
    \caption{Corruption Heterogeneity by Continent}
    \resizebox{\linewidth}{!}{%
        {
\def\sym#1{\ifmmode^{#1}\else\(^{#1}\)\fi}
\begin{tabular}{l*{4}{c}}
\toprule
Dependent variable: CPI&\multicolumn{1}{c}{(1)}&\multicolumn{1}{c}{(2)}&\multicolumn{1}{c}{(3)}&\multicolumn{1}{c}{(4)}\\
\midrule
Natural log of government IPs per capita $\times$ Africa&0.007&&&\\
&(0.005)&&&\\
\addlinespace
Natural log of government IPs per capita $\times$ Asia/Oceania&&0.009&&\\
&&(0.007)&&\\
\addlinespace
Natural log of government IPs per capita $\times$ Americas&&&0.021\sym{***}&\\
&&&(0.007)&\\
\addlinespace
Natural log of government IPs per capita $\times$ Europe&&&&0.001\\
&&&&(0.008)\\
\midrule
Observations&329&487&298&431\\
Dep. Var. Mean&0.341&0.427&0.404&0.596\\
Countries&43&46&26&36\\
Country clusters&43&46&26&36\\
Adjusted \(R^2\)&0.365&0.708&0.695&0.840\\
Region-year FE&Yes&Yes&Yes&Yes\\
Baseline controls&Yes&Yes&Yes&Yes\\
\bottomrule
\multicolumn{5}{l}{\footnotesize Standard errors in parentheses}\\
\multicolumn{5}{l}{\footnotesize \sym{*} \(p<0.10\), \sym{**} \(p<0.05\), \sym{***} \(p<0.01\)}\\
\end{tabular}
}

    }
    \tabnote{This table shows heterogeneity analysis by continent using interaction terms. The interactions capture whether the effect varies by region, with the dummy equal to 1 for countries within each respective continent. Only Column 3, which interacts with the Americas, shows a significant effect. }
    \label{tab:corruption_continent}
\end{table}

\clearpage
\begin{table}[ht]
    \caption{Corruption Heterogeneity by Level of Wealth}
    \label{tab:corruption_wealth}
    \centering
    
    \resizebox{\linewidth}{!}{%
        {
\def\sym#1{\ifmmode^{#1}\else\(^{#1}\)\fi}
\begin{tabular}{l*{4}{c}}
\toprule
Dependent variable: CPI&\multicolumn{1}{c}{(1)}&\multicolumn{1}{c}{(2)}&\multicolumn{1}{c}{(3)}&\multicolumn{1}{c}{(4)}\\
\midrule
Natural log of government IPs per capita $\times$ high income&0.002&&&\\
&(0.009)&&&\\
\addlinespace
Natural log of government IPs per capita $\times$ upper-middle income&&0.017\sym{**}&&\\
&&(0.007)&&\\
\addlinespace
Natural log of government IPs per capita $\times$ lower-middle income&&&0.002&\\
&&&(0.003)&\\
\addlinespace
Natural log of government IPs per capita $\times$ low income&&&&0.022\sym{***}\\
&&&&(0.008)\\
\midrule
Observations&517&456&382&165\\
Dep. Var. Mean&0.659&0.382&0.334&0.295\\
Countries&48&57&54&30\\
Country clusters&48&57&54&30\\
Adjusted \(R^2\)&0.704&0.262&0.381&0.448\\
Region-year FE&Yes&Yes&Yes&Yes\\
Baseline controls&Yes&Yes&Yes&Yes\\
\bottomrule
\multicolumn{5}{l}{\footnotesize Standard errors in parentheses}\\
\multicolumn{5}{l}{\footnotesize \sym{*} \(p<0.10\), \sym{**} \(p<0.05\), \sym{***} \(p<0.01\)}\\
\end{tabular}
}

    }
    \tabnote{This table presents heterogeneity analysis by World Bank income group. Each column restricts the sample to the listed income group, with DSC measured as log government IPs per capita. All specifications include the baseline controls and region-year fixed effects shown in the table; coefficients are descriptive associations.}
\end{table}

\clearpage
\begin{landscape}
    \begin{table}[ht]
        \centering
        \caption{Infant Mortality Heterogeneity by Level of Wealth}
        \label{tab:infmort_wealth}
        \begin{adjustbox}{max width=\linewidth,max totalheight=0.76\textheight}
            {
\def\sym#1{\ifmmode^{#1}\else\(^{#1}\)\fi}
\begin{tabular}{l*{5}{c}}
\toprule
Dependent variable: infant mortality&\multicolumn{1}{c}{(1)}&\multicolumn{1}{c}{(2)}&\multicolumn{1}{c}{(3)}&\multicolumn{1}{c}{(4)}&\multicolumn{1}{c}{(5)}\\
\midrule
Natural log of government IPs per capita $\times$ high income&-0.054\sym{**}&&&&\\
&(0.021)&&&&\\
\addlinespace
Natural log of government IPs per capita $\times$ upper-middle income&&-0.067\sym{**}&&&\\
&&(0.033)&&&\\
\addlinespace
Natural log of government IPs per capita $\times$ lower-middle income&&&-0.008&&\\
&&&(0.020)&&\\
\addlinespace
Natural log of government IPs per capita $\times$ low income&&&&-0.039&\\
&&&&(0.023)&\\
\addlinespace
Natural log of government IPs per capita $\times$ lower-income pool&&&&&-0.022\\
&&&&&(0.017)\\
\midrule
Observations&833&694&660&272&952\\
Dep. Var. Mean&1.565&2.583&3.302&4.283&3.592\\
Countries&48&64&77&39&96\\
Country clusters&48&64&77&39&96\\
Adjusted \(R^2\)&0.706&0.622&0.580&0.548&0.709\\
Region-year FE&Yes&Yes&Yes&Yes&Yes\\
Baseline controls&Yes&Yes&Yes&Yes&Yes\\
\bottomrule
\multicolumn{6}{l}{\footnotesize Standard errors in parentheses}\\
\multicolumn{6}{l}{\footnotesize \sym{*} \(p<0.10\), \sym{**} \(p<0.05\), \sym{***} \(p<0.01\)}\\
\end{tabular}
}

        \end{adjustbox}
        \tabnote{This table presents heterogeneity analysis by World Bank income group. Each column restricts the sample to the listed income group, with DSC measured as log government IPs per capita. The final column pools lower-middle- and low-income countries because the low-income cell is small. All specifications include the baseline controls and region-year fixed effects shown in the table; coefficients are descriptive associations.}
    \end{table}
\end{landscape}
\clearpage

\begin{landscape}
    \begin{table}[ht]
        \centering
        \caption{Diphtheria Vaccination Heterogeneity by Level of Wealth}
        \label{tab:dip_wealth}
        \begin{adjustbox}{max width=\linewidth,max totalheight=0.76\textheight}
            {
\def\sym#1{\ifmmode^{#1}\else\(^{#1}\)\fi}
\begin{tabular}{l*{5}{c}}
\toprule
Dependent variable: diphtheria vaccination&\multicolumn{1}{c}{(1)}&\multicolumn{1}{c}{(2)}&\multicolumn{1}{c}{(3)}&\multicolumn{1}{c}{(4)}&\multicolumn{1}{c}{(5)}\\
\midrule
Natural log of government IPs per capita $\times$ high income&0.003&&&&\\
&(0.002)&&&&\\
\addlinespace
Natural log of government IPs per capita $\times$ upper-middle income&&0.011\sym{**}&&&\\
&&(0.005)&&&\\
\addlinespace
Natural log of government IPs per capita $\times$ lower-middle income&&&0.006&&\\
&&&(0.005)&&\\
\addlinespace
Natural log of government IPs per capita $\times$ low income&&&&0.023\sym{**}&\\
&&&&(0.011)&\\
\addlinespace
Natural log of government IPs per capita $\times$ lower-income pool&&&&&0.014\sym{***}\\
&&&&&(0.005)\\
\midrule
Observations&833&694&660&272&952\\
Dep. Var. Mean&0.952&0.902&0.870&0.764&0.837\\
Countries&48&64&77&39&96\\
Country clusters&48&64&77&39&96\\
Adjusted \(R^2\)&0.141&0.202&0.219&0.308&0.300\\
Region-year FE&Yes&Yes&Yes&Yes&Yes\\
Baseline controls&Yes&Yes&Yes&Yes&Yes\\
\bottomrule
\multicolumn{6}{l}{\footnotesize Standard errors in parentheses}\\
\multicolumn{6}{l}{\footnotesize \sym{*} \(p<0.10\), \sym{**} \(p<0.05\), \sym{***} \(p<0.01\)}\\
\end{tabular}
}

        \end{adjustbox}
        \tabnote{This table presents heterogeneity analysis by World Bank income group. Each column restricts the sample to the listed income group, with DSC measured as log government IPs per capita. The final column pools lower-middle- and low-income countries because the low-income cell is small. All specifications include the baseline controls and region-year fixed effects shown in the table; coefficients are descriptive associations.}
    \end{table}
\end{landscape}
\clearpage

\begin{landscape}
    \begin{table}[ht]
        \centering
        \caption{Hepatitis Vaccination Heterogeneity by Level of Wealth}
        \label{tab:hep_wealth}
        \begin{adjustbox}{max width=\linewidth,max totalheight=0.76\textheight}
            {
\def\sym#1{\ifmmode^{#1}\else\(^{#1}\)\fi}
\begin{tabular}{l*{5}{c}}
\toprule
Dependent variable: hepatitis B vaccination&\multicolumn{1}{c}{(1)}&\multicolumn{1}{c}{(2)}&\multicolumn{1}{c}{(3)}&\multicolumn{1}{c}{(4)}&\multicolumn{1}{c}{(5)}\\
\midrule
Natural log of government IPs per capita $\times$ high income&0.001&&&&\\
&(0.008)&&&&\\
\addlinespace
Natural log of government IPs per capita $\times$ upper-middle income&&0.012\sym{**}&&&\\
&&(0.005)&&&\\
\addlinespace
Natural log of government IPs per capita $\times$ lower-middle income&&&0.007&&\\
&&&(0.005)&&\\
\addlinespace
Natural log of government IPs per capita $\times$ low income&&&&0.016&\\
&&&&(0.013)&\\
\addlinespace
Natural log of government IPs per capita $\times$ lower-income pool&&&&&0.012\sym{**}\\
&&&&&(0.006)\\
\midrule
Observations&663&675&644&245&912\\
Dep. Var. Mean&0.909&0.889&0.858&0.759&0.827\\
Countries&45&63&76&38&95\\
Country clusters&45&63&76&38&95\\
Adjusted \(R^2\)&0.212&0.133&0.175&0.247&0.230\\
Region-year FE&Yes&Yes&Yes&Yes&Yes\\
Baseline controls&Yes&Yes&Yes&Yes&Yes\\
\bottomrule
\multicolumn{6}{l}{\footnotesize Standard errors in parentheses}\\
\multicolumn{6}{l}{\footnotesize \sym{*} \(p<0.10\), \sym{**} \(p<0.05\), \sym{***} \(p<0.01\)}\\
\end{tabular}
}

        \end{adjustbox}
        \tabnote{This table presents heterogeneity analysis by World Bank income group. Each column restricts the sample to the listed income group, with DSC measured as log government IPs per capita. The final column pools lower-middle- and low-income countries because the low-income cell is small. All specifications include the baseline controls and region-year fixed effects shown in the table; coefficients are descriptive associations.}
    \end{table}
\end{landscape}
\clearpage

\begin{landscape}
    \begin{table}[ht]
        \centering
        \caption{Measles Vaccination Heterogeneity by Level of Wealth}
        \label{tab:measles_wealth}
        \begin{adjustbox}{max width=\linewidth,max totalheight=0.76\textheight}
            {
\def\sym#1{\ifmmode^{#1}\else\(^{#1}\)\fi}
\begin{tabular}{l*{5}{c}}
\toprule
Dependent variable: measles vaccination&\multicolumn{1}{c}{(1)}&\multicolumn{1}{c}{(2)}&\multicolumn{1}{c}{(3)}&\multicolumn{1}{c}{(4)}&\multicolumn{1}{c}{(5)}\\
\midrule
Natural log of government IPs per capita $\times$ high income&0.003&&&&\\
&(0.003)&&&&\\
\addlinespace
Natural log of government IPs per capita $\times$ upper-middle income&&0.009\sym{**}&&&\\
&&(0.004)&&&\\
\addlinespace
Natural log of government IPs per capita $\times$ lower-middle income&&&0.007&&\\
&&&(0.005)&&\\
\addlinespace
Natural log of government IPs per capita $\times$ low income&&&&0.022\sym{*}&\\
&&&&(0.012)&\\
\addlinespace
Natural log of government IPs per capita $\times$ lower-income pool&&&&&0.012\sym{**}\\
&&&&&(0.005)\\
\midrule
Observations&833&694&660&272&952\\
Dep. Var. Mean&0.940&0.896&0.872&0.741&0.833\\
Countries&48&64&77&39&96\\
Country clusters&48&64&77&39&96\\
Adjusted \(R^2\)&0.191&0.225&0.252&0.256&0.331\\
Region-year FE&Yes&Yes&Yes&Yes&Yes\\
Baseline controls&Yes&Yes&Yes&Yes&Yes\\
\bottomrule
\multicolumn{6}{l}{\footnotesize Standard errors in parentheses}\\
\multicolumn{6}{l}{\footnotesize \sym{*} \(p<0.10\), \sym{**} \(p<0.05\), \sym{***} \(p<0.01\)}\\
\end{tabular}
}

        \end{adjustbox}
        \tabnote{This table presents heterogeneity analysis by World Bank income group. Each column restricts the sample to the listed income group, with DSC measured as log government IPs per capita. The final column pools lower-middle- and low-income countries because the low-income cell is small. All specifications include the baseline controls and region-year fixed effects shown in the table; coefficients are descriptive associations.}
    \end{table}
\end{landscape}
\clearpage

\clearpage

\clearpage
\begin{landscape}
    \begin{table}[ht]
        \centering
        \caption{Vaccination Heterogeneity by Internet and Press Freedom}
        \resizebox{\linewidth}{!}{%
            {
\def\sym#1{\ifmmode^{#1}\else\(^{#1}\)\fi}
\begin{tabular}{l*{12}{c}}
\toprule
&\multicolumn{1}{c}{(1)}&\multicolumn{1}{c}{(2)}&\multicolumn{1}{c}{(3)}&\multicolumn{1}{c}{(4)}&\multicolumn{1}{c}{(5)}&\multicolumn{1}{c}{(6)}&\multicolumn{1}{c}{(7)}&\multicolumn{1}{c}{(8)}&\multicolumn{1}{c}{(9)}&\multicolumn{1}{c}{(10)}&\multicolumn{1}{c}{(11)}&\multicolumn{1}{c}{(12)}\\
&\multicolumn{1}{c}{Vacc. Dip.}&\multicolumn{1}{c}{Vacc. Meas.}&\multicolumn{1}{c}{Vacc. Hep.}&\multicolumn{1}{c}{Vacc. Dip.}&\multicolumn{1}{c}{Vacc. Meas.}&\multicolumn{1}{c}{Vacc. Hep.}&\multicolumn{1}{c}{Vacc. Dip.}&\multicolumn{1}{c}{Vacc. Meas.}&\multicolumn{1}{c}{Vacc. Hep.}&\multicolumn{1}{c}{Vacc. Dip.}&\multicolumn{1}{c}{Vacc. Meas.}&\multicolumn{1}{c}{Vacc. Hep.}\\
\midrule
Natural log of government IPs per public-sector worker&0.004&0.003&-0.002&0.003&0.002&-0.003&0.003&0.003&-0.008&0.003&0.003&-0.008\\
&(0.003)&(0.004)&(0.007)&(0.002)&(0.003)&(0.006)&(0.002)&(0.003)&(0.007)&(0.002)&(0.003)&(0.007)\\
\addlinespace
DSC $\times$ lower press freedom&0.002&0.001&0.010&&&&&&&&&\\
&(0.005)&(0.005)&(0.008)&&&&&&&&&\\
\addlinespace
DSC $\times$ internet filtering&&&&0.006&0.006&0.012&&&&&&\\
&&&&(0.004)&(0.005)&(0.007)&&&&&&\\
\addlinespace
DSC $\times$ social-media monitoring&&&&&&&0.005&0.004&0.018\sym{**}&&&\\
&&&&&&&(0.003)&(0.003)&(0.009)&&&\\
\addlinespace
DSC $\times$ internet censorship&&&&&&&&&&0.005&0.003&0.018\sym{**}\\
&&&&&&&&&&(0.004)&(0.004)&(0.009)\\
\midrule
Observations&1074&1074&1011&1254&1254&1142&1254&1254&1142&1254&1254&1142\\
Dep. Var. Mean&0.908&0.899&0.884&0.910&0.902&0.880&0.910&0.902&0.880&0.910&0.902&0.880\\
Countries&122&122&118&124&124&119&124&124&119&124&124&119\\
Region-year clusters&77&77&77&90&90&90&90&90&90&90&90&90\\
Adjusted \(R^2\)&0.221&0.167&0.038&0.230&0.162&0.066&0.234&0.167&0.087&0.230&0.156&0.086\\
Region-year FE&Yes&Yes&Yes&Yes&Yes&Yes&Yes&Yes&Yes&Yes&Yes&Yes\\
Baseline controls&Yes&Yes&Yes&Yes&Yes&Yes&Yes&Yes&Yes&Yes&Yes&Yes\\
\bottomrule
\multicolumn{13}{l}{\footnotesize Standard errors in parentheses}\\
\multicolumn{13}{l}{\footnotesize \sym{*} \(p<0.10\), \sym{**} \(p<0.05\), \sym{***} \(p<0.01\)}\\
\end{tabular}
}

        }
        \tabnote{This table presents heterogeneity by press and internet freedom. The interaction terms equal 1 for countries below the median level of freedom, indicating monitoring, censorship, or lower press freedom. Only the hepatitis vaccination interactions with social media monitoring (Column 9) and internet censorship (Column 12) are statistically significant, both at the 5 percent level; the remaining interaction coefficients are not statistically significant.}
        \label{tab:vac_freedom}
    \end{table}
\end{landscape}

\clearpage
\begin{landscape}
    \begin{table}[ht]
        \centering
        \caption{Infant Mortality Heterogeneity by Internet and Press Freedom}
        \resizebox{\linewidth}{!}{%
            {
\def\sym#1{\ifmmode^{#1}\else\(^{#1}\)\fi}
\begin{tabular}{l*{8}{c}}
\toprule
&\multicolumn{1}{c}{(1)}&\multicolumn{1}{c}{(2)}&\multicolumn{1}{c}{(3)}&\multicolumn{1}{c}{(4)}&\multicolumn{1}{c}{(5)}&\multicolumn{1}{c}{(6)}&\multicolumn{1}{c}{(7)}&\multicolumn{1}{c}{(8)}\\
&\multicolumn{1}{c}{Inf. Mort.}&\multicolumn{1}{c}{Neo. Mort.}&\multicolumn{1}{c}{Inf. Mort.}&\multicolumn{1}{c}{Neo. Mort.}&\multicolumn{1}{c}{Inf. Mort.}&\multicolumn{1}{c}{Neo. Mort.}&\multicolumn{1}{c}{Inf. Mort.}&\multicolumn{1}{c}{Neo. Mort.}\\
\midrule
Natural log of government IPs per public-sector worker&-0.026&-0.012&-0.010&-0.009&-0.016&-0.014&-0.021&-0.016\\
&(0.023)&(0.030)&(0.018)&(0.021)&(0.018)&(0.020)&(0.018)&(0.019)\\
\addlinespace
DSC $\times$ lower press freedom&0.010&-0.019&&&&&&\\
&(0.031)&(0.032)&&&&&&\\
\addlinespace
DSC $\times$ internet filtering&&&-0.021&-0.034&&&&\\
&&&(0.030)&(0.029)&&&&\\
\addlinespace
DSC $\times$ social-media monitoring&&&&&-0.014&-0.027&&\\
&&&&&(0.019)&(0.018)&&\\
\addlinespace
DSC $\times$ internet censorship&&&&&&&-0.001&-0.019\\
&&&&&&&(0.024)&(0.021)\\
\midrule
Observations&1074&1074&1254&1254&1254&1254&1254&1254\\
Dep. Var. Mean&2.565&1.944&2.553&1.930&2.553&1.930&2.553&1.930\\
Countries&122&122&124&124&124&124&124&124\\
Region-year clusters&77&77&90&90&90&90&90&90\\
Adjusted \(R^2\)&0.822&0.763&0.819&0.764&0.824&0.775&0.818&0.765\\
Region-year FE&Yes&Yes&Yes&Yes&Yes&Yes&Yes&Yes\\
Baseline controls&Yes&Yes&Yes&Yes&Yes&Yes&Yes&Yes\\
\bottomrule
\multicolumn{9}{l}{\footnotesize Standard errors in parentheses}\\
\multicolumn{9}{l}{\footnotesize \sym{*} \(p<0.10\), \sym{**} \(p<0.05\), \sym{***} \(p<0.01\)}\\
\end{tabular}
}

        }
        \tabnote{This table presents heterogeneity by press and internet freedom. The interaction terms equal 1 for countries below the median level of freedom, indicating monitoring, censorship, or lower press freedom. None of the Digital State Capacity or interaction coefficients is statistically significant at conventional levels.}
        \label{tab:mortality_freedom}
    \end{table}
\end{landscape}
\clearpage

\clearpage
\begin{table}[ht]
    \centering
    \caption{NUTS2 Correlations Between Digital State Capacity and Subnational Digital-Government Measures}
    \label{tab:correlation_dsc_eu}
    \begin{adjustbox}{max width=\linewidth}
        \begin{threeparttable}
            \begin{tabular}{lcc|cc}
                \toprule
                & \multicolumn{2}{c|}{\textbf{Log Gov IP per capita}} & \multicolumn{2}{c}{\textbf{Log Gov IP per worker}} \\
                \cmidrule(lr){2-3} \cmidrule(lr){4-5}
                \textbf{Variable} & \textbf{Correlation} & \textbf{Observations} & \textbf{Correlation} & \textbf{Observations} \\
                \midrule
                \textbf{Internet Gov.\ Interact.}     & 0.46\textsuperscript{***} & 372 & 0.41\textsuperscript{***} & 372 \\
                \textbf{Government Quality Index.}    & 0.26\textsuperscript{***} & 624 & 0.18\textsuperscript{***} & 629 \\
                \textbf{Control Corruption.}          & 0.26\textsuperscript{***} & 624 & 0.19\textsuperscript{***} & 629 \\
                \textbf{Corruption Experience}        & 0.16\textsuperscript{***} & 491 & 0.13\textsuperscript{**}  & 491 \\
                \textbf{Corruption Perception}        & 0.30\textsuperscript{***} & 491 & 0.24\textsuperscript{***} & 491 \\
                \textbf{Pub Adm.\ Qual.}              & 0.25\textsuperscript{***} & 624 & 0.18\textsuperscript{***} & 629 \\
                \textbf{Impartial Pub.\ Admn.}        & 0.23\textsuperscript{***} & 624 & 0.15\textsuperscript{***} & 629 \\
                \textbf{Log GDP per capita}           & 0.51\textsuperscript{***} & 635 & 0.45\textsuperscript{***} & 634 \\
                \textbf{Log Population}               & 0.01                     & 635 & 0.03                     & 634 \\
                \textbf{Household Internet prcnt}     & 0.47\textsuperscript{***} & 536 & 0.43\textsuperscript{***} & 536 \\
                \textbf{Household Broadband prcnt}    & 0.37\textsuperscript{***} & 344 & 0.33\textsuperscript{***} & 344 \\
                \bottomrule
            \end{tabular}
            \begin{tablenotes}
                \footnotesize
                \item \textit{Note:} \textsuperscript{*} \( p < 0.10 \), \textsuperscript{**} \( p < 0.05 \), \textsuperscript{***} \( p < 0.01 \).
            \end{tablenotes}
        \end{threeparttable}
    \end{adjustbox}
\end{table}

\clearpage

\subsection{EU Subnational Analysis}

\begin{table}[htbp]
    \centering
    \caption{NUTS2 Validation Regressions: Digital State Capacity, Digital Government, and Government Quality}
    \resizebox{\textwidth}{!}{%
        {
\def\sym#1{\ifmmode^{#1}\else\(^{#1}\)\fi}
\begin{tabular}{l*{4}{c}}
\toprule
 & \multicolumn{1}{c}{(1)} & \multicolumn{1}{c}{(2)} & \multicolumn{1}{c}{(3)} & \multicolumn{1}{c}{(4)}\\
 & \multicolumn{1}{c}{Digital Gov} & \multicolumn{1}{c}{Digital Gov} & \multicolumn{1}{c}{Gov Quality} & \multicolumn{1}{c}{Gov Quality}\\
\midrule
Log Gov IP per capita & 0.004\sym{*} &  & -0.005 & \\
 & (0.002) &  & (0.003) & \\
\addlinespace
Log Gov IP per worker &  & 0.003\sym{*} &  & -0.005\\
 &  & (0.002) &  & (0.003)\\
\addlinespace
Log GDP. p.c. & 0.062\sym{***} & 0.063\sym{***} & 0.056 & 0.057\\
 & (0.018) & (0.018) & (0.037) & (0.037)\\
\addlinespace
Log Pop. & 0.009\sym{**} & 0.008\sym{**} & -0.015\sym{**} & -0.015\sym{**}\\
 & (0.004) & (0.004) & (0.007) & (0.007)\\
\addlinespace
Household Broadband prcnt & 0.150 & 0.149 & 0.129 & 0.126\\
 & (0.125) & (0.125) & (0.096) & (0.096)\\
\midrule
Observations & 364 & 364 & 356 & 356\\
Dep. Var. Mean & 0.357 & 0.357 & 0.505 & 0.505\\
Clusters & 65 & 65 & 57 & 57\\
Adjusted \(R^2\) & 0.933 & 0.933 & 0.884 & 0.884\\
Country FE & Yes & Yes & Yes & Yes \\
Year FE & Yes & Yes & Yes & Yes \\
Years & 2013--2024 & 2013--2024 & 2013--2024 & 2013--2024 \\
\bottomrule
\multicolumn{5}{l}{\footnotesize Standard errors in parentheses}\\
\multicolumn{5}{l}{\footnotesize \sym{*} \(p<0.1\), \sym{**} \(p<0.05\), \sym{***} \(p<0.01\)}\\
\end{tabular}
}

    }
    \label{tab:dsc_gov_dig_qual_val_ols}
    \tabnote{This table reports NUTS 2 validation regressions for the Digital Government measure and the aggregate Government Quality Index. All specifications control for log GDP per capita, log population, and household broadband access, and include country and year fixed effects. Standard errors are clustered by country-year. The sample spans 2013--2024, with 364 observations in Columns 1 and 2 and 356 observations in Columns 3 and 4.}
\end{table}

\clearpage

% NOTE: WBES summary-stats table commented out pending co-author review — the Enterprise Survey data are not analysed in the current draft (see commented paragraph in the EU NUTS-2 subsection).
\begin{comment}
\begin{table}[htbp]
    \centering
    \begin{tabular}{l
                D{.}{.}{2}
                D{.}{.}{2}
                D{.}{.}{2}
                D{.}{.}{2}
                D{.}{.}{0}
               }
    \hline
    \input{tables/main/enterprise_survey/wbes_panel_summary_statistics}
    \end{tabular}
    \caption{ Summary statistics of key variables }
    \label{tab:es_sum}
\end{table}
\end{comment}

\clearpage

\subsection{Additional Country-Level Outcome Heterogeneity}

\clearpage
\begin{table}[htbp]
\centering
\caption{Pairwise Correlations: Corruption Perceptions, Health, and GHS Indicators with Government IP Measures (2019)}
\label{tab:ghs_correlations}
\begin{adjustbox}{max width=\linewidth,max totalheight=0.78\textheight}
\begin{threeparttable}
\footnotesize
\renewcommand{\arraystretch}{1.08}
\setlength{\tabcolsep}{3pt}
%% Panel A
\textbf{Panel A: Health IPs (COFOG 707)}
\vspace{0.5em}
\begin{tabular}{p{5.0cm}cccccc}
\toprule
\textbf{Indicator} &
\multicolumn{2}{c}{\makecell{Health IPs\\Per Capita}} &
\multicolumn{2}{c}{\makecell{Health IPs\\Per 1k Public Emp}} &
\multicolumn{2}{c}{\makecell{Health IPs\\(Log Total)}} \\
\cmidrule(lr){2-3} \cmidrule(lr){4-5} \cmidrule(lr){6-7}
 & $r$ & $N$ & $r$ & $N$ & $r$ & $N$ \\
\midrule
Corruption Perceptions & {\footnotesize0.78$\textsuperscript{***}$} & {\scriptsize(124)} & {\footnotesize0.72$\textsuperscript{***}$} & {\scriptsize(67)} & {\footnotesize0.60$\textsuperscript{***}$} & {\scriptsize(126)} \\
Global Health Security Index & {\footnotesize0.73$\textsuperscript{***}$} & {\scriptsize(122)} & {\footnotesize0.74$\textsuperscript{***}$} & {\scriptsize(67)} & {\footnotesize0.79$\textsuperscript{***}$} & {\scriptsize(123)} \\
Prevention of Pathogen Emergence & {\footnotesize0.71$\textsuperscript{***}$} & {\scriptsize(122)} & {\footnotesize0.71$\textsuperscript{***}$} & {\scriptsize(67)} & {\footnotesize0.77$\textsuperscript{***}$} & {\scriptsize(123)} \\
Biosecurity & {\footnotesize0.56$\textsuperscript{***}$} & {\scriptsize(122)} & {\footnotesize0.51$\textsuperscript{***}$} & {\scriptsize(67)} & {\footnotesize0.63$\textsuperscript{***}$} & {\scriptsize(123)} \\
Immunization & {\footnotesize0.36$\textsuperscript{***}$} & {\scriptsize(122)} & {\footnotesize0.33$\textsuperscript{***}$} & {\scriptsize(67)} & {\footnotesize0.36$\textsuperscript{***}$} & {\scriptsize(123)} \\
Early Detection Reporting & {\footnotesize0.49$\textsuperscript{***}$} & {\scriptsize(122)} & {\footnotesize0.53$\textsuperscript{***}$} & {\scriptsize(67)} & {\footnotesize0.63$\textsuperscript{***}$} & {\scriptsize(123)} \\
Laboratory Systems & {\footnotesize0.35$\textsuperscript{***}$} & {\scriptsize(122)} & {\footnotesize0.37$\textsuperscript{***}$} & {\scriptsize(67)} & {\footnotesize0.51$\textsuperscript{***}$} & {\scriptsize(123)} \\
Surveillance Data Accessibility & {\footnotesize0.55$\textsuperscript{***}$} & {\scriptsize(122)} & {\footnotesize0.56$\textsuperscript{***}$} & {\scriptsize(67)} & {\footnotesize0.59$\textsuperscript{***}$} & {\scriptsize(123)} \\
Rapid Response Mitigation & {\footnotesize0.62$\textsuperscript{***}$} & {\scriptsize(122)} & {\footnotesize0.65$\textsuperscript{***}$} & {\scriptsize(67)} & {\footnotesize0.67$\textsuperscript{***}$} & {\scriptsize(123)} \\
Risk Communication & {\footnotesize0.39$\textsuperscript{***}$} & {\scriptsize(122)} & {\footnotesize0.34$\textsuperscript{***}$} & {\scriptsize(67)} & {\footnotesize0.45$\textsuperscript{***}$} & {\scriptsize(123)} \\
Health Sector Robustness & {\footnotesize0.64$\textsuperscript{***}$} & {\scriptsize(122)} & {\footnotesize0.60$\textsuperscript{***}$} & {\scriptsize(67)} & {\footnotesize0.70$\textsuperscript{***}$} & {\scriptsize(123)} \\
Healthcare Worker Communication & {\footnotesize0.30$\textsuperscript{***}$} & {\scriptsize(122)} & {\footnotesize0.36$\textsuperscript{***}$} & {\scriptsize(67)} & {\footnotesize0.38$\textsuperscript{***}$} & {\scriptsize(123)} \\
\makecell[l]{National Capacity\\Commitments} & {\footnotesize0.55$\textsuperscript{***}$} & {\scriptsize(122)} & {\footnotesize0.54$\textsuperscript{***}$} & {\scriptsize(67)} & {\footnotesize0.63$\textsuperscript{***}$} & {\scriptsize(123)} \\
\bottomrule
\end{tabular}
\vspace{1em}

\textbf{Panel B: Total Government IPs}
\vspace{0.5em}
\begin{tabular}{p{5.0cm}cccccc}
\toprule
\textbf{Indicator} &
\multicolumn{2}{c}{\makecell{Total Gov IPs\\Per Capita}} &
\multicolumn{2}{c}{\makecell{Total Gov IPs\\Per 1k Public Emp}} &
\multicolumn{2}{c}{\makecell{Total Gov IPs\\(Log Total)}} \\
\cmidrule(lr){2-3} \cmidrule(lr){4-5} \cmidrule(lr){6-7}
 & $r$ & $N$ & $r$ & $N$ & $r$ & $N$ \\
\midrule
Corruption Perceptions & {\footnotesize0.71$\textsuperscript{***}$} & {\scriptsize(163)} & {\footnotesize0.59$\textsuperscript{***}$} & {\scriptsize(88)} & {\footnotesize0.54$\textsuperscript{***}$} & {\scriptsize(167)} \\
Global Health Security Index & {\footnotesize0.64$\textsuperscript{***}$} & {\scriptsize(169)} & {\footnotesize0.62$\textsuperscript{***}$} & {\scriptsize(88)} & {\footnotesize0.76$\textsuperscript{***}$} & {\scriptsize(173)} \\
Prevention of Pathogen Emergence & {\footnotesize0.60$\textsuperscript{***}$} & {\scriptsize(169)} & {\footnotesize0.59$\textsuperscript{***}$} & {\scriptsize(88)} & {\footnotesize0.74$\textsuperscript{***}$} & {\scriptsize(173)} \\
Biosecurity & {\footnotesize0.53$\textsuperscript{***}$} & {\scriptsize(169)} & {\footnotesize0.47$\textsuperscript{***}$} & {\scriptsize(88)} & {\footnotesize0.63$\textsuperscript{***}$} & {\scriptsize(173)} \\
Immunization & {\footnotesize0.25$\textsuperscript{***}$} & {\scriptsize(169)} & {\footnotesize0.32$\textsuperscript{***}$} & {\scriptsize(88)} & {\footnotesize0.36$\textsuperscript{***}$} & {\scriptsize(173)} \\
Early Detection Reporting & {\footnotesize0.41$\textsuperscript{***}$} & {\scriptsize(169)} & {\footnotesize0.40$\textsuperscript{***}$} & {\scriptsize(88)} & {\footnotesize0.64$\textsuperscript{***}$} & {\scriptsize(173)} \\
Laboratory Systems & {\footnotesize0.28$\textsuperscript{***}$} & {\scriptsize(169)} & {\footnotesize0.23$\textsuperscript{**}$} & {\scriptsize(88)} & {\footnotesize0.55$\textsuperscript{***}$} & {\scriptsize(173)} \\
Surveillance Data Accessibility & {\footnotesize0.54$\textsuperscript{***}$} & {\scriptsize(169)} & {\footnotesize0.54$\textsuperscript{***}$} & {\scriptsize(88)} & {\footnotesize0.65$\textsuperscript{***}$} & {\scriptsize(173)} \\
Rapid Response Mitigation & {\footnotesize0.54$\textsuperscript{***}$} & {\scriptsize(169)} & {\footnotesize0.52$\textsuperscript{***}$} & {\scriptsize(88)} & {\footnotesize0.66$\textsuperscript{***}$} & {\scriptsize(173)} \\
Risk Communication & {\footnotesize0.35$\textsuperscript{***}$} & {\scriptsize(169)} & {\footnotesize0.46$\textsuperscript{***}$} & {\scriptsize(88)} & {\footnotesize0.50$\textsuperscript{***}$} & {\scriptsize(173)} \\
Health Sector Robustness & {\footnotesize0.60$\textsuperscript{***}$} & {\scriptsize(169)} & {\footnotesize0.61$\textsuperscript{***}$} & {\scriptsize(88)} & {\footnotesize0.72$\textsuperscript{***}$} & {\scriptsize(173)} \\
Healthcare Worker Communication & {\footnotesize0.25$\textsuperscript{***}$} & {\scriptsize(169)} & {\footnotesize0.34$\textsuperscript{***}$} & {\scriptsize(88)} & {\footnotesize0.37$\textsuperscript{***}$} & {\scriptsize(173)} \\
\makecell[l]{National Capacity\\Commitments} & {\footnotesize0.40$\textsuperscript{***}$} & {\scriptsize(169)} & {\footnotesize0.40$\textsuperscript{***}$} & {\scriptsize(88)} & {\footnotesize0.57$\textsuperscript{***}$} & {\scriptsize(173)} \\
\bottomrule
\end{tabular}
\vspace{0.4em}
\begin{tablenotes}
\footnotesize
\item \emph{Note}: Entries are Pearson correlations of log-transformed government IP metrics and corruption perceptions (CPI), health, and Global Health Security indicators for 2019. $N$ is sample size. $r$ is the correlation coefficient. $^{***}p<0.01$, $^{**}p<0.05$, $^{*}p<0.10$. Measures per 1k public employees and per capita as shown. Gov IPs = all government sub-categories; Health IPs = COFOG 707. "Log Total" is the log of total government IP counts in the category. Corruption Perceptions = CPI score (Transparency International).
\end{tablenotes}
\end{threeparttable}
\end{adjustbox}
\end{table}

\clearpage
\begin{sidewaystable}[htbp]
\centering
\caption{Neonatal Mortality (2019--2020)}
  \label{tab:neo_mort}
  \resizebox{0.92\textheight}{!}{{
\def\sym#1{\ifmmode^{#1}\else\(^{#1}\)\fi}
\begin{tabular}{l*{6}{c}}
\toprule
                    &\multicolumn{1}{c}{(1)}&\multicolumn{1}{c}{(2)}&\multicolumn{1}{c}{(3)}&\multicolumn{1}{c}{(4)}&\multicolumn{1}{c}{(5)}&\multicolumn{1}{c}{(6)}\\
                    &\multicolumn{1}{c}{} &\multicolumn{1}{c}{} &\multicolumn{1}{c}{} &\multicolumn{1}{c}{} &\multicolumn{1}{c}{} &\multicolumn{1}{c}{} \\
\midrule
Log Total Gov IPs p.c.&      -0.129\sym{***}&                     &                     &                     &                     &                     \\
                    &     (0.037)         &                     &                     &                     &                     &                     \\
\addlinespace
Log Health IPs p.c. &                     &      -0.171\sym{***}&                     &                     &                     &                     \\
                    &                     &     (0.031)         &                     &                     &                     &                     \\
\addlinespace
Log Total Gov IPs Per Worker&                     &                     &      -0.111\sym{**} &                     &                     &                     \\
                    &                     &                     &     (0.040)         &                     &                     &                     \\
\end{tabular}
}
}
\end{sidewaystable}

\clearpage

\begin{sidewaystable}[htbp]
\centering
\caption{Infant Mortality (2019--2020)}
\label{tab:inf_mort}
\resizebox{0.92\textheight}{!}{{
\def\sym#1{\ifmmode^{#1}\else\(^{#1}\)\fi}
\begin{tabular}{l*{6}{c}}
\toprule
                    &\multicolumn{1}{c}{(1)}&\multicolumn{1}{c}{(2)}&\multicolumn{1}{c}{(3)}&\multicolumn{1}{c}{(4)}&\multicolumn{1}{c}{(5)}&\multicolumn{1}{c}{(6)}\\
                    &\multicolumn{1}{c}{} &\multicolumn{1}{c}{} &\multicolumn{1}{c}{} &\multicolumn{1}{c}{} &\multicolumn{1}{c}{} &\multicolumn{1}{c}{} \\
\midrule
Log Total Gov IPs p.c.&      -0.146\sym{***}&                     &                     &                     &                     &                     \\
                    &     (0.032)         &                     &                     &                     &                     &                     \\
\addlinespace
Log Health IPs p.c. &                     &      -0.181\sym{***}&                     &                     &                     &                     \\
                    &                     &     (0.025)         &                     &                     &                     &                     \\
\addlinespace
Log Total Gov IPs Per Worker&                     &                     &      -0.126\sym{***}&                     &                     &                     \\
                    &                     &                     &     (0.038)         &                     &                     &                     \\
\end{tabular}
}
}
\end{sidewaystable}

  \clearpage
  \begin{sidewaystable}[htbp]
  \centering
  \caption{Measles Vaccination (2019--2020)}
  \label{tab:vacc_meas}
  \resizebox{0.92\textheight}{!}{{
\def\sym#1{\ifmmode^{#1}\else\(^{#1}\)\fi}
\begin{tabular}{l*{6}{c}}
\toprule
                    &\multicolumn{1}{c}{(1)}&\multicolumn{1}{c}{(2)}&\multicolumn{1}{c}{(3)}&\multicolumn{1}{c}{(4)}&\multicolumn{1}{c}{(5)}&\multicolumn{1}{c}{(6)}\\
                    &\multicolumn{1}{c}{} &\multicolumn{1}{c}{} &\multicolumn{1}{c}{} &\multicolumn{1}{c}{} &\multicolumn{1}{c}{} &\multicolumn{1}{c}{} \\
\midrule
Log Total Gov IPs p.c.&       0.020\sym{***}&                     &                     &                     &                     &                     \\
                    &     (0.004)         &                     &                     &                     &                     &                     \\
\addlinespace
Log Health IPs p.c. &                     &       0.019\sym{***}&                     &                     &                     &                     \\
                    &                     &     (0.006)         &                     &                     &                     &                     \\
\addlinespace
Log Total Gov IPs Per Worker&                     &                     &       0.019\sym{***}&                     &                     &                     \\
                    &                     &                     &     (0.004)         &                     &                     &                     \\
\end{tabular}
}
}
  \end{sidewaystable}

\clearpage

  \begin{sidewaystable}[htbp]
  \centering
  \caption{Hepatitis B Vaccination (2019--2020)}
  \label{tab:vacc_hep}
  \resizebox{0.92\textheight}{!}{{
\def\sym#1{\ifmmode^{#1}\else\(^{#1}\)\fi}
\begin{tabular}{l*{6}{c}}
\toprule
                    &\multicolumn{1}{c}{(1)}&\multicolumn{1}{c}{(2)}&\multicolumn{1}{c}{(3)}&\multicolumn{1}{c}{(4)}&\multicolumn{1}{c}{(5)}&\multicolumn{1}{c}{(6)}\\
                    &\multicolumn{1}{c}{} &\multicolumn{1}{c}{} &\multicolumn{1}{c}{} &\multicolumn{1}{c}{} &\multicolumn{1}{c}{} &\multicolumn{1}{c}{} \\
\midrule
Log Total Gov IPs p.c.&       0.018\sym{***}&                     &                     &                     &                     &                     \\
                    &     (0.005)         &                     &                     &                     &                     &                     \\
\addlinespace
Log Health IPs p.c. &                     &       0.014\sym{*}  &                     &                     &                     &                     \\
                    &                     &     (0.007)         &                     &                     &                     &                     \\
\addlinespace
Log Total Gov IPs Per Worker&                     &                     &       0.017\sym{***}&                     &                     &                     \\
                    &                     &                     &     (0.005)         &                     &                     &                     \\
\end{tabular}
}
}
  \end{sidewaystable}

  \clearpage
\begin{sidewaystable}[htbp]
\centering
  \caption{Diphtheria Vaccination (2019--2020)}
  \label{tab:vacc_dip}
  \resizebox{0.92\textheight}{!}{{
\def\sym#1{\ifmmode^{#1}\else\(^{#1}\)\fi}
\begin{tabular}{l*{6}{c}}
\toprule
                    &\multicolumn{1}{c}{(1)}&\multicolumn{1}{c}{(2)}&\multicolumn{1}{c}{(3)}&\multicolumn{1}{c}{(4)}&\multicolumn{1}{c}{(5)}&\multicolumn{1}{c}{(6)}\\
                    &\multicolumn{1}{c}{} &\multicolumn{1}{c}{} &\multicolumn{1}{c}{} &\multicolumn{1}{c}{} &\multicolumn{1}{c}{} &\multicolumn{1}{c}{} \\
\midrule
Log Total Gov IPs p.c.&       0.018\sym{***}&                     &                     &                     &                     &                     \\
                    &     (0.004)         &                     &                     &                     &                     &                     \\
\addlinespace
Log Health IPs p.c. &                     &       0.016\sym{**} &                     &                     &                     &                     \\
                    &                     &     (0.006)         &                     &                     &                     &                     \\
\addlinespace
Log Total Gov IPs Per Worker&                     &                     &       0.018\sym{***}&                     &                     &                     \\
                    &                     &                     &     (0.005)         &                     &                     &                     \\
\end{tabular}
}
}
\end{sidewaystable}

\clearpage
\begin{sidewaystable}[htbp]
\centering
\caption{Infant Mortality (2019--2020) Function of Government}
\label{tab:inf_mort_het}
\resizebox{0.92\textheight}{!}{{
\def\sym#1{\ifmmode^{#1}\else\(^{#1}\)\fi}
\begin{tabular}{l*{9}{c}}
\toprule
                    &\multicolumn{1}{c}{(1)}&\multicolumn{1}{c}{(2)}&\multicolumn{1}{c}{(3)}&\multicolumn{1}{c}{(4)}&\multicolumn{1}{c}{(5)}&\multicolumn{1}{c}{(6)}&\multicolumn{1}{c}{(7)}&\multicolumn{1}{c}{(8)}&\multicolumn{1}{c}{(9)}\\
                    &\multicolumn{1}{c}{} &\multicolumn{1}{c}{} &\multicolumn{1}{c}{} &\multicolumn{1}{c}{} &\multicolumn{1}{c}{} &\multicolumn{1}{c}{} &\multicolumn{1}{c}{} &\multicolumn{1}{c}{} &\multicolumn{1}{c}{} \\
\midrule
Log General Public Services IPs p.c.&      -0.143\sym{***}&                     &                     &                     &                     &                     &                     &                     &                     \\
                    &     (0.029)         &                     &                     &                     &                     &                     &                     &                     &                     \\
\addlinespace
Log Defense IPs p.c.&                     &      -0.092\sym{**} &                     &                     &                     &                     &                     &                     &                     \\
                    &                     &     (0.037)         &                     &                     &                     &                     &                     &                     &                     \\
\addlinespace
Log Public Order and Safety IPs p.c.&                     &                     &      -0.175\sym{***}&                     &                     &                     &                     &                     &                     \\
                    &                     &                     &     (0.031)         &                     &                     &                     &                     &                     &                     \\
\end{tabular}
}
}
\end{sidewaystable}

\clearpage
\begin{sidewaystable}[htbp]
\centering
\caption{Measles Vaccinations (2019--2020) Function of Government}
\label{tab:measles_vacc_het}
\resizebox{0.92\textheight}{!}{{
\def\sym#1{\ifmmode^{#1}\else\(^{#1}\)\fi}
\begin{tabular}{l*{9}{c}}
\toprule
                    &\multicolumn{1}{c}{(1)}&\multicolumn{1}{c}{(2)}&\multicolumn{1}{c}{(3)}&\multicolumn{1}{c}{(4)}&\multicolumn{1}{c}{(5)}&\multicolumn{1}{c}{(6)}&\multicolumn{1}{c}{(7)}&\multicolumn{1}{c}{(8)}&\multicolumn{1}{c}{(9)}\\
                    &\multicolumn{1}{c}{} &\multicolumn{1}{c}{} &\multicolumn{1}{c}{} &\multicolumn{1}{c}{} &\multicolumn{1}{c}{} &\multicolumn{1}{c}{} &\multicolumn{1}{c}{} &\multicolumn{1}{c}{} &\multicolumn{1}{c}{} \\
\midrule
Log General Public Services IPs p.c.&       0.014\sym{***}&                     &                     &                     &                     &                     &                     &                     &                     \\
                    &     (0.003)         &                     &                     &                     &                     &                     &                     &                     &                     \\
\addlinespace
Log Defense IPs p.c.&                     &       0.004         &                     &                     &                     &                     &                     &                     &                     \\
                    &                     &     (0.005)         &                     &                     &                     &                     &                     &                     &                     \\
\addlinespace
Log Public Order and Safety IPs p.c.&                     &                     &       0.011\sym{**} &                     &                     &                     &                     &                     &                     \\
                    &                     &                     &     (0.004)         &                     &                     &                     &                     &                     &                     \\
\end{tabular}
}
}
\end{sidewaystable}

\clearpage
\begin{sidewaystable}[htbp]
\centering
\caption{Diphtheria Vaccinations (2019--2020) Function of Government}
\label{tab:dip_vacc_het}
\resizebox{0.92\textheight}{!}{{
\def\sym#1{\ifmmode^{#1}\else\(^{#1}\)\fi}
\begin{tabular}{l*{9}{c}}
\toprule
                    &\multicolumn{1}{c}{(1)}&\multicolumn{1}{c}{(2)}&\multicolumn{1}{c}{(3)}&\multicolumn{1}{c}{(4)}&\multicolumn{1}{c}{(5)}&\multicolumn{1}{c}{(6)}&\multicolumn{1}{c}{(7)}&\multicolumn{1}{c}{(8)}&\multicolumn{1}{c}{(9)}\\
                    &\multicolumn{1}{c}{} &\multicolumn{1}{c}{} &\multicolumn{1}{c}{} &\multicolumn{1}{c}{} &\multicolumn{1}{c}{} &\multicolumn{1}{c}{} &\multicolumn{1}{c}{} &\multicolumn{1}{c}{} &\multicolumn{1}{c}{} \\
\midrule
Log General Public Services IPs p.c.&       0.011\sym{***}&                     &                     &                     &                     &                     &                     &                     &                     \\
                    &     (0.003)         &                     &                     &                     &                     &                     &                     &                     &                     \\
\addlinespace
Log Defense IPs p.c.&                     &       0.006         &                     &                     &                     &                     &                     &                     &                     \\
                    &                     &     (0.003)         &                     &                     &                     &                     &                     &                     &                     \\
\addlinespace
Log Public Order and Safety IPs p.c.&                     &                     &       0.008\sym{**} &                     &                     &                     &                     &                     &                     \\
                    &                     &                     &     (0.003)         &                     &                     &                     &                     &                     &                     \\
\end{tabular}
}
}
\end{sidewaystable}

\clearpage
\begin{landscape}
\begin{table}[htbp]
\centering
\caption{Neonatal Mortality (2019--2023) Function of Government}
\label{tab:neonat_het}
\begin{adjustbox}{max width=\linewidth,max totalheight=0.78\textheight}
{
\def\sym#1{\ifmmode^{#1}\else\(^{#1}\)\fi}
\begin{tabular}{l*{9}{c}}
\toprule
                    &\multicolumn{1}{c}{(1)}         &\multicolumn{1}{c}{(2)}         &\multicolumn{1}{c}{(3)}         &\multicolumn{1}{c}{(4)}         &\multicolumn{1}{c}{(5)}         &\multicolumn{1}{c}{(6)}         &\multicolumn{1}{c}{(7)}         &\multicolumn{1}{c}{(8)}         &\multicolumn{1}{c}{(9)}         \\
\midrule
Log General Public Services IPs p.c.&      -0.137\sym{***}&                     &                     &                     &                     &                     &                     &                     &                     \\
                    &     (0.025)         &                     &                     &                     &                     &                     &                     &                     &                     \\
\addlinespace
Log Defense IPs p.c.&                     &      -0.077\sym{***}&                     &                     &                     &                     &                     &                     &                     \\
                    &                     &     (0.028)         &                     &                     &                     &                     &                     &                     &                     \\
\addlinespace
Log Public Order and Safety IPs p.c.&                     &                     &      -0.127\sym{***}&                     &                     &                     &                     &                     &                     \\
                    &                     &                     &     (0.028)         &                     &                     &                     &                     &                     &                     \\
\addlinespace
Log Economic Affairs IPs p.c.&                     &                     &                     &      -0.111\sym{***}&                     &                     &                     &                     &                     \\
\end{tabular}
}

\end{adjustbox}
\end{table}
\end{landscape}

\clearpage
\begin{landscape}
\begin{table}[htbp]
\centering
\caption{CPI and Function of Government, per Capita}
\label{tab:cpi_het}
\begin{adjustbox}{max width=\linewidth,max totalheight=0.78\textheight}
{
\def\sym#1{\ifmmode^{#1}\else\(^{#1}\)\fi}
\begin{tabular}{l*{9}{c}}
\toprule
                    &\multicolumn{1}{c}{(1)}         &\multicolumn{1}{c}{(2)}         &\multicolumn{1}{c}{(3)}         &\multicolumn{1}{c}{(4)}         &\multicolumn{1}{c}{(5)}         &\multicolumn{1}{c}{(6)}         &\multicolumn{1}{c}{(7)}         &\multicolumn{1}{c}{(8)}         &\multicolumn{1}{c}{(9)}         \\
\midrule
Log General Public Services IPs p.c.&       0.036\sym{***}&                     &                     &                     &                     &                     &                     &                     &                     \\
                    &     (0.005)         &                     &                     &                     &                     &                     &                     &                     &                     \\
\addlinespace
Log Defense IPs p.c.&                     &       0.031\sym{***}&                     &                     &                     &                     &                     &                     &                     \\
                    &                     &     (0.006)         &                     &                     &                     &                     &                     &                     &                     \\
\addlinespace
Log Public Order and Safety IPs p.c.&                     &                     &       0.039\sym{***}&                     &                     &                     &                     &                     &                     \\
                    &                     &                     &     (0.006)         &                     &                     &                     &                     &                     &                     \\
\addlinespace
Log Economic Affairs IPs p.c.&                     &                     &                     &       0.035\sym{***}&                     &                     &                     &                     &                     \\
\end{tabular}
}

\end{adjustbox}
\end{table}
\end{landscape}

\clearpage
\begin{landscape}
\begin{table}[htbp]
\centering
\caption{CPI and Function of Government, per Public-Sector Worker}
\label{tab:cpi_pw_het}
\begin{adjustbox}{max width=\linewidth,max totalheight=0.78\textheight}
{
\def\sym#1{\ifmmode^{#1}\else\(^{#1}\)\fi}
\begin{tabular}{l*{9}{c}}
\toprule
                    &\multicolumn{1}{c}{(1)}         &\multicolumn{1}{c}{(2)}         &\multicolumn{1}{c}{(3)}         &\multicolumn{1}{c}{(4)}         &\multicolumn{1}{c}{(5)}         &\multicolumn{1}{c}{(6)}         &\multicolumn{1}{c}{(7)}         &\multicolumn{1}{c}{(8)}         &\multicolumn{1}{c}{(9)}         \\
\midrule
Log General Public Services IPs Per Worker&       0.031\sym{***}&                     &                     &                     &                     &                     &                     &                     &                     \\
                    &     (0.006)         &                     &                     &                     &                     &                     &                     &                     &                     \\
\addlinespace
Log Defense IPs Per Worker&                     &       0.027\sym{***}&                     &                     &                     &                     &                     &                     &                     \\
                    &                     &     (0.006)         &                     &                     &                     &                     &                     &                     &                     \\
\addlinespace
Log Public Order and Safety IPs Per Worker&                     &                     &       0.031\sym{***}&                     &                     &                     &                     &                     &                     \\
                    &                     &                     &     (0.008)         &                     &                     &                     &                     &                     &                     \\
\end{tabular}
}

\end{adjustbox}
\end{table}
\end{landscape}
\clearpage
\begin{landscape}
\begin{table}[htbp]
\centering
\caption{CPI and Function of Government, Raw Counts}
\label{tab:cpi_total_het}
\begin{adjustbox}{max width=\linewidth,max totalheight=0.78\textheight}
{
\def\sym#1{\ifmmode^{#1}\else\(^{#1}\)\fi}
\begin{tabular}{l*{9}{c}}
\toprule
                    &\multicolumn{1}{c}{(1)}         &\multicolumn{1}{c}{(2)}         &\multicolumn{1}{c}{(3)}         &\multicolumn{1}{c}{(4)}         &\multicolumn{1}{c}{(5)}         &\multicolumn{1}{c}{(6)}         &\multicolumn{1}{c}{(7)}         &\multicolumn{1}{c}{(8)}         &\multicolumn{1}{c}{(9)}         \\
\midrule
Log General Public Services IPs&       0.016\sym{***}&                     &                     &                     &                     &                     &                     &                     &                     \\
                    &     (0.005)         &                     &                     &                     &                     &                     &                     &                     &                     \\
\addlinespace
Log Defense IPs     &                     &       0.022\sym{***}&                     &                     &                     &                     &                     &                     &                     \\
                    &                     &     (0.004)         &                     &                     &                     &                     &                     &                     &                     \\
\addlinespace
Log Public Order and Safety IPs&                     &                     &       0.017\sym{***}&                     &                     &                     &                     &                     &                     \\
                    &                     &                     &     (0.005)         &                     &                     &                     &                     &                     &                     \\
\addlinespace
Log Economic Affairs IPs&                     &                     &                     &       0.024\sym{***}&                     &                     &                     &                     &                     \\
\end{tabular}
}

\end{adjustbox}
\end{table}
\end{landscape}

\appendix

\begin{center}
    \begin{Large}
ONLINE APPENDIX    
\end{Large}
\end{center}

% Reset table numbering to include appendix letter (A.1, B.1, etc.)
\counterwithin{table}{section}
\counterwithin{figure}{section}
\renewcommand{\thetable}{\Alph{section}.\arabic{table}}
\renewcommand{\thefigure}{\Alph{section}.\arabic{figure}}
  
\section{Large Language Model Prompts}
We take a hybrid approach to categorising government entities. First, we predict likely government entities using the PaLM 2 large language model \parencite{Chowdhery2023PALM, Anil2023PaLM2}. We cross-validate this with the 70-billion-parameter Llama 3 model \textcite{dubey2024llama} by examining the level of agreement. To recover historical allocation dates, we match owner-name and IP-range pairs to RDAP queries, which are discussed in more detail in Appendix~\ref{app:historical}. We match these records to the IP ranges that we have categorised in the DB-IP database starting from 2019. We confirm matches in several steps:

\subsection{Classification Prompt: Llama 3 and PaLM 2}
\label{app:prompt1}
The intuition is to use the simplest and most generalisable means of categorisation, starting with keyword definitions in the context of the organisation and then using prior knowledge of the organisation or similar organisations. Given the risk of hallucination, we develop a methodology that combines several large knowledge bases to provide additional organisational context for each classification, using LLMs for semantic comparison of the initial output with knowledge-base facts. We use the prompt below (Appendix~\ref{app:mainprompt}) for the first stage of the categorisation. The structure of the prompt aims to be as generalisable as possible while providing information and suggestions about the reasoning for a specific suggested category, which can be tested as true or false to identify the hallucination rate. The columns that help with validation are keyword definitions for n-grams (one or more words) that the model used to determine the suggested category. We use these for additional validation. The Boolean columns, such as type and short word, help filter out more ambiguous categories such as people and acronyms, which require a higher level of evidence to accept the category.

The inputs are: (i) the DB-IP owner name in its original language with no text preprocessing (PaLM and Llama both perform well with messy tokens such as misread Unicode characters and missing spaces); and (ii) the ISO2 country code from DB-IP for country-specific context of the entity. This prompt was run using PaLM 2 between August and October 2023. This initial query ran for 80 days with 6 concurrent API keys. PaLM 2 builds on PaLM, whose documented pretraining mixture includes filtered webpages, books, Wikipedia, news articles, source code, and social media conversations \parencite[p.~6]{Chowdhery2023PALM}. The PaLM 2 report describes a multilingual mixture that also includes code, mathematics, conversational data, and parallel multilingual data \parencite{Anil2023PaLM2}. These training sources include Wikipedia, although the PaLM 2 report does not report the share of its training tokens drawn from Wikipedia. The multilingual training makes the model suitable for use on our dataset without requiring pre-translation, ensuring the original name of the organisation is retained.

\subsubsection{Main Prompt}
\label{app:mainprompt}
\paragraph{Task:} \textbf{Analyze the given entity to extract and define its keywords, considering the country context via the ISO3 code and the relevance of international organizations.}

\vspace{1mm}

\paragraph{Input:}
\begin{itemize}
    \item \textbf{Entity}: \{\texttt{[Entity Name]}\}
    \item \textbf{ISO3 Code}: \{\texttt{[ISO3 Code]}\}
\end{itemize}

\vspace{1mm}

\paragraph{Instructions:}
\begin{enumerate}
    \item \textbf{Understand the overarching theme of the entity} to guide subsequent steps. Consider the overarching theme and use the ISO3 code for contextual nuance.
    \item \textbf{Dissect the entity into its individual terms} without compounding words. Examples include breaking ``BrandName ProductType XYZ'' into separate terms \texttt{'Brandname'}, \texttt{'Product'}, and \texttt{'Category'}. Dissect compounds into unigrams, e.g., ``securityagency'' to \texttt{'security'} and \texttt{'agency'}, but retain terms like ``New York''.
    \item \textbf{Identify any term that is an acronym or abbreviation} and provide the full form.
    \item \textbf{Initially define each term} using the ISO3 code for context.
    \item \textbf{Start with unigrams, then ngrams} in defining terms.
    \item \textbf{Address unclear terms} and list potential alternative interpretations.
    \item \textbf{After holistic understanding, refine each term's definition}. Understand the overarching theme of the entity to guide term definition and categorization. \textbf{Term Priority}: Start with unigrams, then ngrams.
    \item \textbf{Organize findings into a structured JSON output}.
    \item \textbf{Ensure accuracy} of terms, acronyms, and categories.
\end{enumerate}

\vspace{1mm}

\paragraph{Output Format (in JSON structure):}

\begin{quote}
\ttfamily
\scriptsize
\sloppy
{
  "category": "[Relevant categories for the entity]",\\
  "type": "Entity's classification",\\
  "source": "Keywords",\\
  "is\_gov": "Boolean indicating if the entity is a public service, governmental or state-run organization",\\
  "is\_international": "Boolean indicating if the entity is an international organization",\\
  "definitions": [\\
    {\\
      "term": "Keyword or 'Uncertain'/'No definition' If two equally likely definitions give both.",\\
      "definition": "Contextual definition",\\
      "short\_word": "Boolean indicating if the term is an acronym/abbreviation",\\
      "knowledge\_base\_category": "Primary category reflecting the entity's context, e.g., 'Organizational Functions', 'Technical Terms', etc.",\\
      "secondary\_knowledge\_base\_category": "Secondary category (if applicable)"\\
    }
  ]
}
\end{quote}

\subsubsection{Validation Prompt}
\label{app:prompt2}
Classify the following entities into their appropriate IMF COFOG categories and extract key keywords with contextual definitions. For each entity, the input includes the \textbf{entity ID}, \textbf{name}, \textbf{snippet} (short google description), \textbf{Predicted category}, and \textbf{iso}. 

\textbf{Note:} The Predicted category is a prediction and might be incorrect; it should be considered but not heavily weighted in your classification. Additionally, verify whether the \textbf{Title} accurately represents the entity. If it does not, prioritize the entity name over the snippet in your analysis.

\paragraph{Part 1: IMF COFOG Classification}
\begin{enumerate}[label=\arabic*.]
    \item \textbf{Title Accuracy Verification:} Determine if the 'Title' accurately represents the entity.
    \begin{itemize}
        \item \textbf{Accurate:} If the 'Title' clearly and correctly identifies the entity.
        \item \textbf{Inaccurate:} If the 'Title' is misleading or does not accurately describe the entity.
        \item \textbf{Flagging:} Add a flag in the output indicating the accuracy of the 'Title'.
    \end{itemize}
    \item \textbf{Classification:} Assign the entity to an IMF COFOG category using the 'COFOG Description' and 'COFOG Definitions'.
    \item \textbf{Entity Type:} Determine if the entity is one of the following:
    \begin{itemize}
        \item \textbf{Government Entity:} Any organization part of a governmental structure (e.g., federal, state, local government).
        \item \textbf{Geographic Government Entity (GeoGov):} Geographic entities such as prefectures, cities, provinces, counties, or municipalities.
        \item \textbf{Telecom Entity:} Organizations providing telecommunications services such as telephone, radio, television, and internet.
    \end{itemize}
    \item \textbf{Level of Government:} If identified as a government or GeoGov entity, specify the level of government (e.g., federal, state, local).
\end{enumerate}

\paragraph{Part 2: Keyword Extraction and Definition}
\begin{enumerate}[label=\arabic*.]
    \item Extract relevant keywords from the entity name and snippet.
    \item Define each keyword using the iso code for context. Expand acronyms or abbreviations.
    \item Handle ambiguities by listing alternative interpretations for unclear terms.
    \item Structure the keywords and definitions in the output.
\end{enumerate}

\clearpage

\paragraph{Output Format}
Provide the results in the following JSON format:
\begin{quote}
\ttfamily
\scriptsize
\sloppy
[
  {
    "entity\_id": "entity\_id",\\
    "entity": "entity\_name",\\
    "translated\_name": "translated\_name",\\
    "title\_accurate": true/false,\\
    "category": "COFOG\_category",\\
    "secondary\_category": "Secondary\_COFOG\_category",\\
    "reasoning": "reasoning\_for\_classification",\\
    "is\_government": true/false,\\
    "is\_geogov": true/false,\\
    "level\_of\_government": "Level\_of\_Government",\\
    "level\_of\_government\_reasoning": "Reasoning\_for\_Level\_of\_Government",\\
    "is\_telecom": true/false,\\
    "keywords\_analysis": {\\
      "keywords": [\\
        {
          "term": "Keyword",\\
          "definition": "Contextual definition",\\
          "is\_acronym": true/false,\\
          "category": "Primary category",\\
          "secondary\_category": "Secondary category if applicable"\\
        }
      ]
    }
  }
]
\end{quote}

\clearpage
\textbf{Additional Instructions:}
\begin{itemize}
    \item Ensure clarity and accuracy in all reasoning steps.
    \item Follow the JSON format strictly for consistency.
    \item Prioritize both the 'entity' name and snippet based on the title's accuracy.
\end{itemize}

\subsection{Example Entities Level of Government}
\label{app:gov_examples_level}
\subsubsection*{Federal}
\begin{itemize}
    \item DSU NASNI
    \item Federal Network Serv
    \item Ditco Navy Recruiting Evst
    \item Ditco Navy Reserve RCC SW
    \item DeptHomelandSecurity (Department of Homeland Security, USA)
    \item General Services Administration (USA)
    \item US GOVT FEMA (Federal Emergency Management Agency)
    \item US GOVT NAVY
    \item FEMA (Federal Emergency Management Agency, USA)
    \item US Attorney Office (USA)
\end{itemize}
\subsubsection*{State}
\begin{itemize}
    \item NNEPRA
    \item Aopc-crt of Common Pleas (USA)
    \item New York State Senate (USA)
    \item State of Colorado DOT (Department of Transportation, Colorado, USA)
    \item Wisconsin Department of Justice (USA)
    \item State of MN (Minnesota, USA)
    \item ST OF TEXAS COMPTROLLERS (State of Texas Comptroller)
    \item Tribunal Regional Eleitoral do Goias (Regional Electoral Tribunal of Goiás, Brazil)
    \item Secretaria de Estado do Planejamento (SEPOG RO; State Secretariat of Planning, Rondônia, Brazil)
    \item Fundação Carlos Chagas Filho de Amparo à Pesquisa (Carlos Chagas Filho Foundation for Research Support, Brazil)
\end{itemize}

\subsubsection*{Local}
\begin{itemize}
    \item Fayette County Housing Authority (USA)
    \item Jackson County DSS (Department of Social Services, USA)
    \item City of SAN Jose California (USA)
    \item LAS VEGAS CITY (USA)
    \item City of LAS Vegas (USA)
    \item Comune di Roma (City of Rome, Italy)
    \item Comune di Lucca (City of Lucca, Italy)
    \item Comune di Fara in Sabina (City of Fara in Sabina, Italy)
    \item Comune di Caltavuturo (City of Caltavuturo, Italy)
    \item Comune di Pescara (City of Pescara, Italy)
\end{itemize}

\subsubsection*{International}
\begin{itemize}
    \item Programa de las Naciones Unidas para el Desarrollo (United Nations Development Programme)
    \item UN Icty (United Nations International Criminal Tribunal for the former Yugoslavia)
    \item International Maritime Organization (IMO)
    \item European Fisheries Control Agency
    \item European Space Agency (ESA)
    \item United Nations (UN)
    \item Frontex bv (European Border and Coast Guard Agency)
    \item NATO Support Procurement Agency (NSPA)
    \item Inter-american Defense Board
    \item Commonwealth War Graves Commission (UK)
\end{itemize}

\subsubsection*{Other Administrative}
\begin{itemize}
    \item Vander Meyden DA
    \item Conseil Général (General Council, France)
    \item Serco LTD
    \item Blueberry River First Nations (Canada)
    \item Quanzhou Frontier Troop (China)
    \item Lbsec SCE Departemental Incendie ET Secours (Fire Department and Rescue, France)
    \item The GEO GROUP (USA)
    \item Réserve Indienne de Odanak (Odanak Indian Reserve, Canada)
    \item Grand Council of the Crees (Canada)
    \item Fides
\end{itemize}

\subsubsection*{Unknown}
\begin{itemize}
    \item Partitodemocratico (Democratic Party, Italy)
    \item \makecell[l]{Amministrazione Provinciale di Viterbo\\(Provincial Administration of Viterbo, Italy)}
    \item AUTOSTRADE PER L'ITALIA SPA (Italian Highway Authority)
    \item Istituto di Tutela ed Assistenza Lavoratori (Institute for Protection and Assistance to Workers, Italy)
    \item Società per Imprese Pubbliche (Company for Public Enterprises, Italy)
    \item Ente Nazionale di Assistenza Sociale (National Institute for Social Assistance, Italy)
    \item CONS. NAZ. INGEGNERI D.L.LGT 23/11944 N.382 ART. 2 (National Council of Engineers, Italy)
    \item D.G. per la Coop Cultural Scientific e Tecnic Min D (Directorate General for Cultural, Scientific, and Technical Cooperation, Italy)
    \item National Medical Center (USA)
    \item Ministry of Strategy and Finance (South Korea)
\end{itemize}

\subsection{Example Entities Function of Government}

\section{Additional Data Processing}

This appendix documents the data-processing steps used to move from raw public IPv4 ranges to the analysis measures. Appendix Table~\ref{tab:dsc_process_overview} provides a guide to the full pipeline and points readers to the appendix sections documenting each step.

\clearpage
\begin{landscape}

\begingroup
\small
\setlength{\tabcolsep}{5pt}
\setlength{\LTleft}{0pt}
\setlength{\LTright}{0pt}
\setlength{\LTcapwidth}{\linewidth}
\renewcommand{\arraystretch}{1.25}

\refstepcounter{table}
\label{tab:dsc_process_overview}
\begin{center}
\textbf{\tablename~\thetable: Guide to the Digital State Capacity Data Construction Pipeline}
\end{center}
\vspace{0.5em}

\begin{longtable}{@{}
>{\raggedright\arraybackslash}p{0.18\linewidth}
>{\raggedright\arraybackslash}p{0.55\linewidth}
>{\raggedright\arraybackslash}p{0.22\linewidth}
@{}}

\toprule
\textbf{Stage} & \textbf{Operation} & \textbf{Output / documentation} \\
\midrule
\endfirsthead

\multicolumn{3}{c}%
{{\tablename\ \thetable{} continued from previous page}} \\
\toprule
\textbf{Stage} & \textbf{Operation} & \textbf{Output / documentation} \\
\midrule
\endhead

\midrule
\multicolumn{3}{r}{\footnotesize Continued on next page} \\
\endfoot

\bottomrule
\multicolumn{3}{@{}p{0.95\linewidth}@{}}{\footnotesize
\textit{Note:} This guide table summarises the construction of the Digital State Capacity measure and points readers to the appendix sections documenting each step. DB-IP monthly files provide the post-2019 ownership and geolocation source; RDAP and Regional Internet Registry records provide the source records for the lower-coverage historical extension.} \\
\endlastfoot

\multicolumn{3}{@{}l}{\textbf{A. Source data and enrichment}} \\
\midrule

Public IPv4 ranges
& Use DB-IP monthly files to extract the start address, end address, and number of addresses in each public IPv4 range.
& Raw public IPv4 resources; primary data source. \\

Owner names and geography
& Use DB-IP fields to recover the registered organisation name, country, and subnational location associated with each public IPv4 range.
& Owner-name, country, and Admin-1 fields; main analysis geography. \\

Search enrichment
& For candidate organisations, retrieve Google search information through Serper, including the title, URL, snippet, and sitelinks where available.
& Additional organisational context; Appendix~\ref{app:serper}. \\

\addlinespace[0.6em]
\multicolumn{3}{@{}l}{\textbf{B. Entity classification}} \\
\midrule

Government status
& Classify each owner-name–country pair as government or non-government using the DB-IP owner name, country context, and search-enriched organisational information.
& Government entity indicator; Appendices~\ref{app:prompt1} and~\ref{app:prompt2}. \\

Level of government
& For retained government entities, classify the level of government represented by the organisation.
& Federal, state, local, and other government labels; Appendix~\ref{app:gov_examples_level}. \\

Function of government
& Assign each government entity to the closest function of government using COFOG categories.
& COFOG function labels; Appendices~\ref{app:prompt2} and~\ref{app:data_cleaning}. \\

Exclusions
& Remove education entities and other categories excluded from the main measure because they cannot be classified consistently across countries.
& Final analysis sample; Appendix~\ref{app:data_cleaning}. \\

\addlinespace[0.6em]
\multicolumn{3}{@{}l}{\textbf{C. Historical extension}} \\
\midrule

Allocation-date matching
& Match owner-name and IP-range pairs to RDAP and Regional Internet Registry records in order to recover historical allocation dates.
& Historical allocation records; Appendix~\ref{app:historical}. \\

Historical coverage checks
& Compare historical matches with the post-2019 DB-IP data to assess the coverage and consistency of the historical extension.
& Historical validity checks; Appendix~\ref{app:historical_validation}. \\

\addlinespace[0.6em]
\multicolumn{3}{@{}l}{\textbf{D. Digital State Capacity measures}} \\
\midrule

Raw DSC counts
& Aggregate government IPv4 resources by country, year, subnational region, level of government, and function, depending on the analysis.
& Country-year, regional, level, and function counts; main data product. \\

Normalised DSC measures
& Divide raw DSC counts by population or public-sector employment to construct comparable intensity measures.
& Per capita and per-worker measures; Section~\ref{sec:val}. \\

\end{longtable}

\endgroup
\end{landscape}
\clearpage

\subsection{Google Serper Query}
\label{app:serper}

After the initial government/non-government classification described in Appendix~\ref{app:prompt1}, we query candidate government organisations using Serper \textcite{serper_dev_2023}. Serper provides access to Google search results, allowing us to collect additional information about each organisation. We use the organisation name and country as query inputs and retain the top organic result.

For our classification task, the key fields are the matched title or organisation name, the URL, and the search-result snippet. The snippet often provides a short description of the organisation, such as whether it is a ministry, department, municipality, public authority, firm, school, university, network provider, or other type of entity. Where available, we also retain sitelinks and knowledge-panel information \parencite{google_knowledgepanel_2023}. Less than 30 percent of entities have a knowledge panel, so the organic search result is the main search-enrichment field used consistently across entities.

Figure~\ref{fig:detailed-info} shows an example for the Department of Transport and Planning. The search result identifies the organisation, links to the official website, and provides a snippet describing the department's transport, planning, land, precinct, and policy functions. This information is used as additional context in the final classification step.

\begin{figure}[ht]
    \centering
    \footnotesize
    \caption{Example of Google Information for Government Entity}
    \label{fig:detailed-info}
    \begin{itemize}
        \item \textbf{Knowledge Graph: The Department of Transport and Planning}
        \begin{itemize}
            \item \textbf{Type}: Government Department
            \item \textbf{Image URL}: \url{https://encrypted-tbn0.gstatic.com/images?q=tbn:ANd9GcRyBOv75SMkmDThaaPtUcRg1ekX-ROzOeU&s=0}
            \item \textbf{Attributes}:
            \begin{itemize}
                \item \textbf{Subsidiaries}: VicRoads and Public Transport Victoria
                \item \textbf{Department Executive}: Paul Younis, Secretary
                \item \textbf{Founded}: January 1, 2019
                \item \textbf{Annual Budget}: \$9.1 billion (FY 19–20)
                \item \textbf{Preceding Department}: Department of Economic Development, Jobs, Transport and Resources
            \end{itemize}
        \end{itemize}
        
        \item \textbf{Organic Search Results}
        \begin{itemize}
            \item \textbf{Result 1: Department of Transport and Planning | vic.gov.au}
            \begin{itemize}
                \item \textbf{Link}: \url{https://www.vic.gov.au/department-transport-and-planning}
                \item \textbf{Snippet}: Victoria's Department of Transport and Planning (DTP) brings together key transport, planning, land, precinct and policy functions within a ...
                \item \textbf{Date}: August 30, 2024
                \item \textbf{Sitelinks}:
                \begin{itemize}
                    \item \textbf{Contact Us}: \url{https://www.vic.gov.au/contact-department-transport-and-planning}
                    \item \textbf{Our Strategic Plan 2024-2028}: \url{https://www.vic.gov.au/department-of-transport-and-planning-strategic-plan}
                    \item \textbf{Document Archive}: \url{https://www.vic.gov.au/department-transport-and-planning-document-archive}
                \end{itemize}
                \item \textbf{Position}: 1
            \end{itemize}
        \end{itemize}
    \end{itemize}
\end{figure}

\subsection{Final Classification and COFOG Labels}
\label{app:final_classification}

We then combine the DB-IP owner name, country, initial model predictions, and Serper-enriched information in a final LLM classification step. This step determines whether the entity is a government organisation, whether it is a telecommunications or network provider, its level of government, and its closest function of government. Government function is classified using COFOG categories.

These labels are used in the cleaning rules below. The education exclusion relies on the COFOG function label. The telecommunications exclusion relies on the entity-type label. The international-organisation exclusion relies on the level-of-government classification. The final classification therefore produces the labels used to construct the analysis sample.

\subsection{Data Cleaning}
\label{app:data_cleaning}

After the final classification step, we apply a small set of cleaning rules to construct the analysis sample. We exclude education and telecommunications entities from the main DSC measure because, without additional evidence, it is difficult to determine consistently whether they are government organisations. This is a measurement decision rather than a claim that these sectors are outside the public sector. Schools, universities, network providers, utilities, and communications entities vary substantially across countries and may be public agencies, state-owned enterprises, regulated firms, private providers, public-private entities, or mixed institutions. These distinctions are often not visible from the registered owner name and search snippets alone. Including these entities would therefore introduce a large source of cross-country classification error where direct state ownership or administrative control cannot be verified consistently.

We retain unambiguous government network and communications entities, such as defence networks, public communications agencies, and government utilities, where the entity can be directly linked to the state. Appendix~\ref{sec:govisp} lists the government network and telecommunications entities retained in the analysis sample. We also drop records where a government organisation is associated with IP ranges geolocated outside the country of the registering government, because the analysis relies on geographic attribution. Finally, we exclude international organisations that cannot be attributed to a single national government or to the country corresponding to their geolocation.

After these cleaning rules, the final 2019–2024 monthly DB-IP sample contains approximately 150,000 country-entity records classified as government. The separate historical sample contains approximately 30,000 matched government entities as of December 2018. The first count is a set of country-entity records in the monthly data; the second is the smaller set of entities that can be matched to historical allocation records.

\clearpage

\section{Historical Extension and Supplementary Validation}

\subsection{Historical Allocation Information}
\label{app:historical}
Our historical extension queries the Registration Data Access Protocol (RDAP) to obtain organisational information and registration dates for IP address ranges \parencite{icannoct2021}. In practice, we run these RDAP lookups against APNIC infrastructure that federates historical registration data across all five Regional Internet Registries.\footnote{RDAP provides standardised access to registration data from Regional Internet Registries, with APNIC serving as a unified endpoint for national registries in Indonesia, Japan, Korea, and Taiwan \parencite{apnic2024rdap, apnicblog2020}.}

The query first identifies the relevant organisation by searching the ``entities'' section of the JSON data returned by the RDAP service. We look for entities with the role of ``registrant,'' which typically identifies the organisation that registered the IP address. After identifying the organisation, we search the ``events'' section for an event labelled ``registration,'' which indicates when the IP address was first registered by the organisation.\footnote{RFC 7485 documents that the five RIRs expose organisation, contact, and registration-date fields under different labels and with incomplete coverage across registries \parencite{zhou2015rfc7485}.} We then link the extracted registration date to the organisation identified in DB-IP, assuming that the registration event corresponds to the organisation's first association with the IP address.

For DB-IP ranges that match an RDAP registrant record, we check whether the range is observed in DB-IP on or after the queried registration date. If it is, the observation moves to the next matching stage. Because the query process uses longest-prefix lookup,\footnote{RFC 9224 \parencite{rfc9224validation} explains that IP address lookup uses the longest match, returning the largest IP CIDR block or range. RDAP queries return the largest IP CIDR block contained withing the queried address range, and we query by IP address rather than entity to minimise matching uncertainty \parencite{rfc9224validation}. Our implementation processes two queries per request, with cloud-based scaling available for higher throughput \parencite{microsoft2021rdap}. Querying by entity is technically possible but has more unknowns than querying by IP address.} we then check whether the DB-IP range associated with the matched owner is a superset of, or equivalent to, the range returned in the RDAP query. If the RDAP range is larger than the DB-IP range, we drop the match.

\subsection{Historical Government IP Address Validation}
\label{app:historical_validation}
\begin{figure}[htbp]
    \centering
    \includegraphics[width=\linewidth,height=0.78\textheight,keepaspectratio]{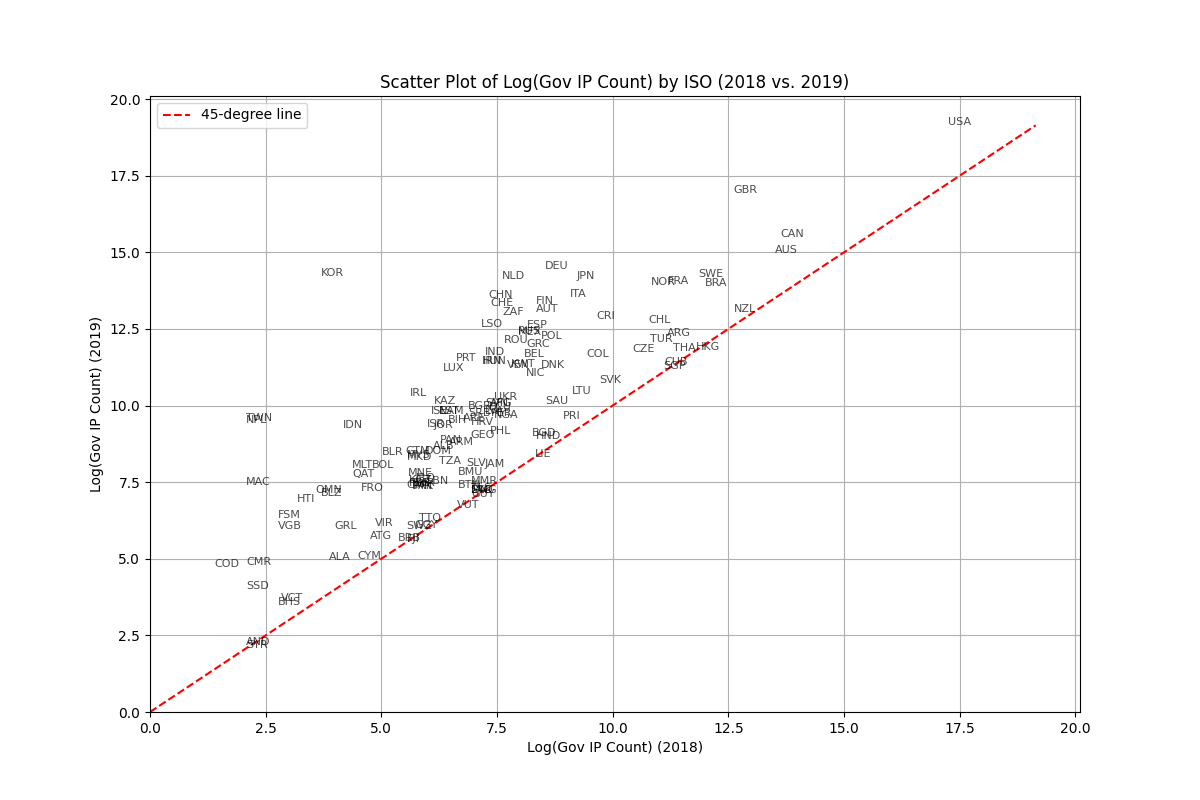}
    \caption{Country-level comparison of log total government IP counts between the last year covered by the historical extension (2018) and the first year of monthly DB-IP data (2019).}
    \label{fig:govip_2018_2019}
\end{figure}

\begin{figure}[htbp]
    \centering
    \includegraphics[width=\linewidth,height=0.78\textheight,keepaspectratio]{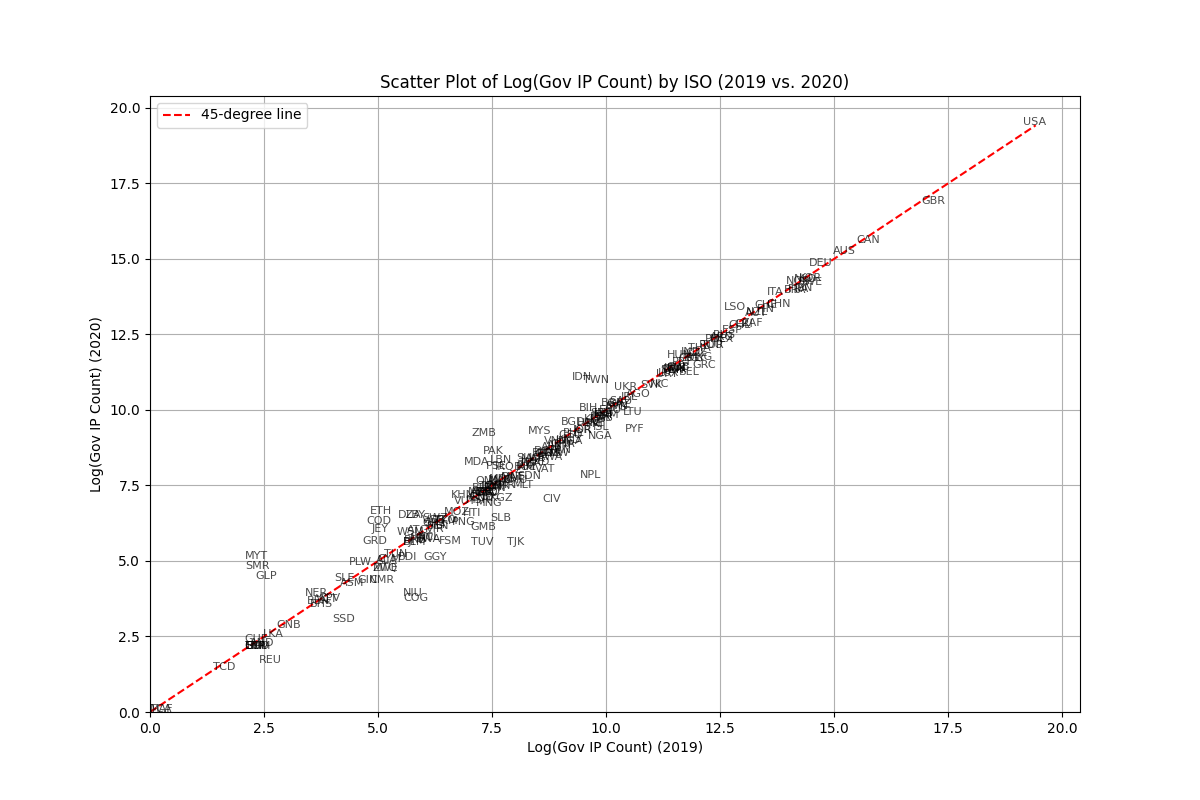}
    \caption{Country-level comparison of log total government IP counts in the monthly DB-IP data, 2019 and 2020.}
    \label{fig:govip_2019_2020}
\end{figure}

Our digital state capacity data for 2019–2023 are constructed from text classifications of owner names reported in the monthly DB-IP data. The historical extension omits unmatched entities and therefore provides a lower-bound estimate of government IPs before 2019. Figure~\ref{fig:govip_2018_2019} shows a scatter plot of 2018 and 2019 values at the country level, while Figure~\ref{fig:govip_2019_2020} shows the comparison between 2019 and 2020.
The Pearson correlation between reconstructed 2018 log counts and DB-IP-based 2019 log counts is 0.81. Although the lower-bound historical measure omits many observations, this correlation indicates that it recovers substantial cross-country variation in levels. It does not show that the two components recover identical levels or within-country changes. Between the DB-IP-based 2019 and 2020 log counts, the correlation is 0.983. Given data limitations, we do not observe 72 of 205 countries and territories until 2019 or later. 

 \clearpage
 
\clearpage

\clearpage
\begin{figure}
    \centering
    \caption{Illustration of the 2019 change in data source and construction for log government IPs per capita in selected countries.}
    \includegraphics[width=\linewidth]{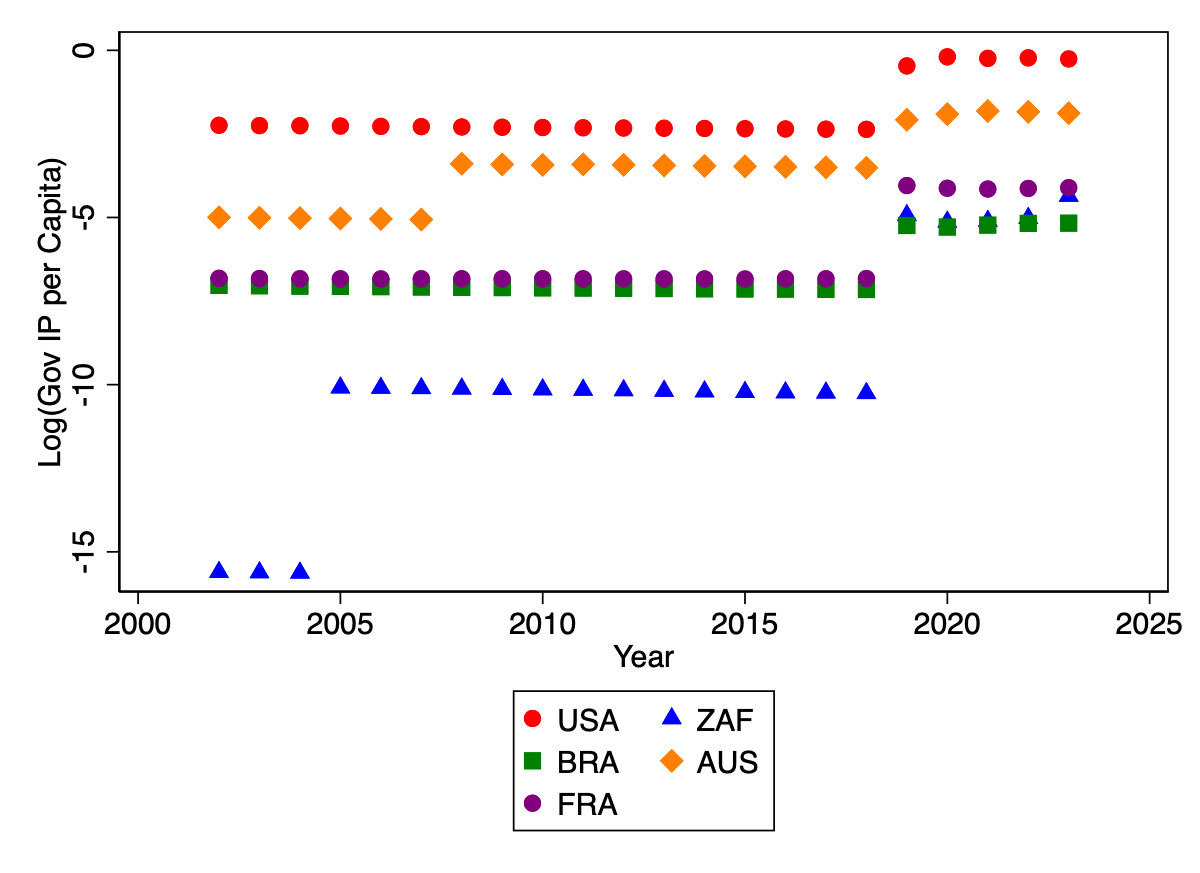}
    \fignote{The figure compares values before and after the transition from the historical allocation data to the monthly DB-IP data. The visible 2019 level shift reflects the change in source coverage and construction and should not be interpreted as within-country growth in digital state capacity.}
    \label{fig:2019_break_ip_trend}
\end{figure}

\clearpage
\subsection{Extended-Sample Validation Tables}
\label{app:extended_validation_tables}

Tables~\ref{tab:app_pearson_correlation_per_capita_2008_2023}–\ref{tab:app_pearson_convergent_per_worker_2008_2023} report validation correlations using the 2008–2023 extended sample. These tables combine the lower-coverage historical data through 2018 with monthly DB-IP observations from 2019 onward and should therefore be interpreted as supplementary extended-sample evidence. The main text reports the preferred 2019–2023 tables based only on monthly DB-IP observations.

\begin{table}[ht]
\centering
\small
\caption{Extended-Sample Pearson Correlations with Government IPs per Capita, 2008–2023}
\begin{adjustbox}{max width=\linewidth,max totalheight=0.78\textheight}

\begin{tabularx}{\linewidth}{>{\raggedright\arraybackslash}Xcccc}
\toprule
\textbf{Variable} & \textbf{Year} & \textbf{Gov IP/Capita} & \textbf{Log(Gov IP/Capita)} & \textbf{N} \\
\midrule
\multicolumn{5}{l}{\textbf{Brambor et al.}} \\
Info Capacity & 2008--2015 & -0.21*** & -0.02 & 401 \\
\addlinespace
\multicolumn{5}{l}{\textbf{Doing Business}} \\
Registration Cost & 2008--2015 & -0.15*** & -0.26*** & 640 \\
Registration Procedures & 2008--2015 & -0.29*** & -0.25*** & 640 \\
Registration Time (Days) & 2008--2015 & -0.16*** & -0.12*** & 640 \\
\addlinespace
\multicolumn{5}{l}{\textbf{Hanson Sigman}} \\
State Capacity & 2008--2015 & 0.29*** & 0.47*** & 695 \\
\addlinespace
\multicolumn{5}{l}{\textbf{Letter Grading}} \\
Avg Days & 2010 & -0.26** & -0.35*** & 87 \\
Log(Avg Days) & 2010 & -0.45*** & -0.43*** & 87 \\
Returned & 2010 & 0.19* & 0.31*** & 87 \\
\addlinespace
\multicolumn{5}{l}{\textbf{Statistical Capacity WB}} \\
Data Infrastructure & 2016--2023 & 0.24*** & 0.52*** & 1194 \\
Data Products & 2008--2023 & 0.07*** & -0.01 & 2161 \\
Data Services & 2016--2023 & 0.19*** & 0.44*** & 1173 \\
Data Sources & 2015--2023 & 0.22*** & 0.54*** & 1230 \\
Data Use & 2008--2023 & 0.13*** & 0.17*** & 2161 \\
Statistical Index & 2016--2023 & 0.23*** & 0.52*** & 1170 \\
\addlinespace
\multicolumn{5}{l}{\textbf{UN E-Gov Indicators}} \\
E-Gov Index & 2008--2022 & 0.25*** & 0.55*** & 963 \\
E-Gov Rank & 2008--2022 & -0.19*** & -0.43*** & 963 \\
E-Participation Index & 2008--2022 & 0.20*** & 0.36*** & 963 \\
Human Capital Index & 2008--2022 & 0.16*** & 0.43*** & 963 \\
Online Service Index & 2008--2022 & 0.20*** & 0.39*** & 963 \\
Telecom Infra Index & 2008--2022 & 0.25*** & 0.57*** & 963 \\
\addlinespace
\multicolumn{5}{l}{\textbf{VDEM}} \\
Impartial Pub. Admn. & 2008--2015 & 0.30*** & 0.52*** & 731 \\
State Auth. Terr. & 2008--2015 & 0.12*** & 0.33*** & 731 \\
State Fiscal Cap. & 2008--2015 & 0.23*** & 0.29*** & 731 \\
\addlinespace
\multicolumn{5}{l}{\textbf{Worldwide Governance}} \\
Ctrl Corruption & 2008--2023 & 0.25*** & 0.55*** & 2040 \\
Gov Effectiveness & 2008--2023 & 0.23*** & 0.51*** & 2040 \\
Political Stability & 2008--2023 & 0.14*** & 0.53*** & 2040 \\
Regulatory Quality & 2008--2023 & 0.24*** & 0.52*** & 2040 \\
Rule of Law & 2008--2023 & 0.25*** & 0.55*** & 2040 \\
Voice \& Accountability & 2008--2023 & 0.21*** & 0.49*** & 2040 \\
\bottomrule
\end{tabularx}

\end{adjustbox}
\tabnote{This table shows pairwise correlations with related measures of state capacity, informational capacity, and e-government for 2008–2023.}
\label{tab:app_pearson_correlation_per_capita_2008_2023}
\end{table}

\clearpage

\begin{table}[ht]
\centering
\small
\caption{Extended-Sample Pearson Correlations with Government IPs per Public-Sector Worker, 2008–2023}
\begin{adjustbox}{max width=\linewidth,max totalheight=0.78\textheight}

\begin{tabularx}{\linewidth}{>{\raggedright\arraybackslash}Xcccc}
\toprule
\textbf{Variable} & \textbf{Year} & \textbf{Gov IP/Worker} & \textbf{Log(Gov IP/Worker)} & \textbf{N} \\
\midrule
\multicolumn{5}{l}{\textbf{Brambor et al.}} \\
Info Capacity & 2008--2015 & -0.22*** & -0.09* & 362 \\
\addlinespace
\multicolumn{5}{l}{\textbf{Doing Business}} \\
Registration Cost & 2008--2015 & -0.16*** & -0.26*** & 593 \\
Registration Procedures & 2008--2015 & -0.27*** & -0.21*** & 593 \\
Registration Time (Days) & 2008--2015 & -0.16*** & -0.11*** & 593 \\
\addlinespace
\multicolumn{5}{l}{\textbf{Hanson Sigman}} \\
State Capacity & 2008--2015 & 0.29*** & 0.44*** & 656 \\
\addlinespace
\multicolumn{5}{l}{\textbf{Letter Grading}} \\
Avg Days & 2010 & -0.25** & -0.33*** & 82 \\
Log(Avg Days) & 2010 & -0.44*** & -0.41*** & 82 \\
Returned & 2010 & 0.18 & 0.28** & 82 \\
\addlinespace
\multicolumn{5}{l}{\textbf{Statistical Capacity WB}} \\
Data Infrastructure & 2016--2023 & 0.19*** & 0.51*** & 1113 \\
Data Products & 2008--2023 & 0.10*** & 0.22*** & 1867 \\
Data Services & 2016--2023 & 0.12*** & 0.46*** & 1103 \\
Data Sources & 2015--2023 & 0.17*** & 0.50*** & 1157 \\
Data Use & 2008--2023 & 0.14*** & 0.33*** & 1867 \\
Statistical Index & 2016--2023 & 0.17*** & 0.53*** & 1103 \\
\addlinespace
\multicolumn{5}{l}{\textbf{UN E-Gov Indicators}} \\
E-Gov Index & 2008--2022 & 0.17*** & 0.49*** & 888 \\
E-Gov Rank & 2008--2022 & -0.11*** & -0.35*** & 888 \\
E-Participation Index & 2008--2022 & 0.16*** & 0.38*** & 888 \\
Human Capital Index & 2008--2022 & 0.09*** & 0.35*** & 888 \\
Online Service Index & 2008--2022 & 0.14*** & 0.39*** & 888 \\
Telecom Infra Index & 2008--2022 & 0.18*** & 0.52*** & 888 \\
\addlinespace
\multicolumn{5}{l}{\textbf{VDEM}} \\
Impartial Pub. Admn. & 2008--2015 & 0.29*** & 0.50*** & 692 \\
State Auth. Terr. & 2008--2015 & 0.11*** & 0.30*** & 692 \\
State Fiscal Cap. & 2008--2015 & 0.24*** & 0.28*** & 692 \\
\addlinespace
\multicolumn{5}{l}{\textbf{Worldwide Governance}} \\
Ctrl Corruption & 2008--2023 & 0.18*** & 0.46*** & 1832 \\
Gov Effectiveness & 2008--2023 & 0.15*** & 0.42*** & 1832 \\
Political Stability & 2008--2023 & 0.09*** & 0.48*** & 1832 \\
Regulatory Quality & 2008--2023 & 0.17*** & 0.44*** & 1832 \\
Rule of Law & 2008--2023 & 0.18*** & 0.47*** & 1832 \\
Voice \& Accountability & 2008--2023 & 0.16*** & 0.44*** & 1832 \\
\bottomrule
\end{tabularx}

\end{adjustbox}
\tabnote{This table shows pairwise correlations with related measures of state capacity, informational capacity, and e-government for 2008–2023. Due to limited data coverage, there are fewer observations for the per-worker transformations than for the per-capita measures.}
\label{tab:app_pearson_correlation_per_worker_2008_2023}
\end{table}

\clearpage

\begin{table}[ht]
\centering
\small
\caption{Extended-Sample Pearson Correlations with Government IPs per Capita, 2008–2023}
\resizebox{0.95\linewidth}{!}{%

\begin{tabular}{lcccc}
\toprule
\textbf{Variable} & \textbf{Year Range} & \textbf{Gov IP/Capita} & \textbf{Log(Gov IP/Capita)} & \textbf{N} \\
\midrule
\multicolumn{5}{l}{\textbf{Polity}} \\
Polity V Score & 2008--2023 & 0.10*** & 0.29*** & 1683 \\
\addlinespace
\multicolumn{5}{l}{\textbf{VDEM}} \\
Civ. Society Partic. & 2008--2022 & 0.17*** & 0.25*** & 1720 \\
Female Civ. Society Partic. & 2008--2022 & 0.12*** & 0.33*** & 1720 \\
Govt Cyber Security & 2008--2022 & 0.14*** & 0.26*** & 1720 \\
Govt Internet Filter Cap. & 2008--2022 & 0.03 & -0.09*** & 1720 \\
Govt Internet Shut Cap. & 2008--2022 & -0.05* & -0.22*** & 1720 \\
Govt Online Reg Cap. & 2008--2022 & 0.09*** & 0.11*** & 1720 \\
Multi Party Elect. & 2008--2022 & 0.10*** & 0.19*** & 1651 \\
Pres. Elec. Consective & 2008--2022 & 0.15*** & 0.08*** & 1720 \\
\addlinespace
\multicolumn{5}{l}{\textbf{World Bank}} \\
Access to Electricity (\%) & 2019--2020 & 0.11** & 0.58*** & 380 \\
Exports Growth (\%) & 2019--2020 & -0.02 & -0.02 & 303 \\
Exports of Goods/Services (\% GDP) & 2019--2020 & -0.01 & 0.34*** & 331 \\
Inflation Rate (\%) & 2019--2020 & -0.04 & -0.16*** & 319 \\
Internet Users (\%) & 2008--2020 & 0.20*** & 0.53*** & 1486 \\
Log GDP per Capita & 2008--2023 & 0.23*** & 0.53*** & 2161 \\
Log Population & 2008--2023 & 0.04** & -0.30*** & 2161 \\
Oil Rents (\% GDP) & 2019--2020 & -0.06 & -0.11** & 356 \\
Tariff Rate (Simple Mean) & 2008--2020 & -0.08*** & -0.23*** & 1272 \\
Urban Population (\%) & 2008--2023 & 0.09*** & 0.28*** & 2159 \\
\bottomrule
\end{tabular}

}
\tabnote{This table reports pairwise correlations with selected related and unrelated measures for 2008–2023.}
\label{tab:app_pearson_convergent_per_capita_2008_2023}
\end{table}

\clearpage

\begin{table}[ht]
\centering
\small
\caption{Extended-Sample Pearson Correlations with Government IPs per Public-Sector Worker, 2008–2023}
\resizebox{0.95\linewidth}{!}{%

\begin{tabular}{lcccc}
\toprule
\textbf{Variable} & \textbf{Year Range} & \textbf{Gov IP/Worker} & \textbf{Log(Gov IP/Worker)} & \textbf{N} \\
\midrule
\multicolumn{5}{l}{\textbf{Polity}} \\
Polity V Score & 2008--2023 & 0.09*** & 0.30*** & 1640 \\
\addlinespace
\multicolumn{5}{l}{\textbf{VDEM}} \\
Civ. Society Partic. & 2008--2022 & 0.15*** & 0.27*** & 1621 \\
Female Civ. Society Partic. & 2008--2022 & 0.09*** & 0.32*** & 1621 \\
Govt Cyber Security & 2008--2022 & 0.10*** & 0.19*** & 1621 \\
Govt Internet Filter Cap. & 2008--2022 & 0.00 & -0.10*** & 1621 \\
Govt Internet Shut Cap. & 2008--2022 & -0.04* & -0.21*** & 1621 \\
Govt Online Reg Cap. & 2008--2022 & 0.07*** & 0.06** & 1621 \\
Multi Party Elect. & 2008--2022 & 0.09*** & 0.19*** & 1559 \\
Pres. Elec. Consective & 2008--2022 & 0.13*** & 0.12*** & 1621 \\
\addlinespace
\multicolumn{5}{l}{\textbf{World Bank}} \\
Access to Electricity (\%) & 2019--2020 & 0.04 & 0.51*** & 331 \\
Exports Growth (\%) & 2019--2020 & -0.02 & -0.00 & 277 \\
Exports of Goods/Services (\% GDP) & 2019--2020 & -0.01 & 0.32*** & 294 \\
Inflation Rate (\%) & 2019--2020 & -0.03 & -0.17*** & 296 \\
Internet Users (\%) & 2008--2020 & 0.14*** & 0.45*** & 1310 \\
Log GDP per Capita & 2008--2023 & 0.15*** & 0.40*** & 1848 \\
Log Population & 2008--2023 & 0.07*** & -0.21*** & 1848 \\
Oil Rents (\% GDP) & 2019--2020 & -0.07 & -0.13** & 319 \\
Tariff Rate (Simple Mean) & 2008--2020 & -0.07** & -0.26*** & 1161 \\
Urban Population (\%) & 2008--2023 & 0.08*** & 0.25*** & 1867 \\
\bottomrule
\end{tabular}

}
\tabnote{This table reports pairwise correlations with selected related and unrelated measures for 2008–2023.}
\label{tab:app_pearson_convergent_per_worker_2008_2023}
\end{table}

\clearpage

\subsection{Country-Universe Sensitivity}

\begin{table}[ht]
\centering
\small
\caption{Country-Universe Sensitivity of Federal and Local DSC Validation Correlations, 2019--2023}
\begin{adjustbox}{max width=\linewidth,max totalheight=0.78\textheight}
\begin{tabular}{lcccc}
\toprule
 & \multicolumn{2}{c}{ln(IPs per capita)} & \multicolumn{2}{c}{ln(IPs per public-sector worker)} \\
\cmidrule(lr){2-3}\cmidrule(lr){4-5}
SPI measure & All areas & UN members & All areas & UN members \\
\midrule
\multicolumn{5}{l}{\textbf{Federal government}} \\
Statistical Index & \makecell{0.57***\\(753)} & \makecell{0.57***\\(748)} & \makecell{0.58***\\(736)} & \makecell{0.58***\\(732)} \\
Data Use & \makecell{0.22***\\(821)} & \makecell{0.29***\\(768)} & \makecell{0.34***\\(767)} & \makecell{0.37***\\(743)} \\
Data Services & \makecell{0.50***\\(757)} & \makecell{0.50***\\(752)} & \makecell{0.49***\\(736)} & \makecell{0.49***\\(732)} \\
Data Products & \makecell{0.09***\\(821)} & \makecell{0.20***\\(768)} & \makecell{0.22***\\(767)} & \makecell{0.28***\\(743)} \\
Data Sources & \makecell{0.62***\\(753)} & \makecell{0.62***\\(748)} & \makecell{0.59***\\(736)} & \makecell{0.59***\\(732)} \\
Data Infrastructure & \makecell{0.57***\\(768)} & \makecell{0.57***\\(763)} & \makecell{0.57***\\(742)} & \makecell{0.57***\\(738)} \\
\addlinespace
\multicolumn{5}{l}{\textbf{Local government}} \\
Statistical Index & \makecell{0.61***\\(736)} & \makecell{0.61***\\(731)} & \makecell{0.63***\\(715)} & \makecell{0.63***\\(711)} \\
Data Use & \makecell{0.12***\\(831)} & \makecell{0.31***\\(751)} & \makecell{0.28***\\(755)} & \makecell{0.40***\\(720)} \\
Data Services & \makecell{0.54***\\(736)} & \makecell{0.54***\\(731)} & \makecell{0.56***\\(715)} & \makecell{0.56***\\(711)} \\
Data Products & \makecell{-0.10***\\(831)} & \makecell{0.15***\\(751)} & \makecell{0.08**\\(755)} & \makecell{0.29***\\(720)} \\
Data Sources & \makecell{0.62***\\(736)} & \makecell{0.62***\\(731)} & \makecell{0.60***\\(715)} & \makecell{0.60***\\(711)} \\
Data Infrastructure & \makecell{0.63***\\(744)} & \makecell{0.63***\\(739)} & \makecell{0.64***\\(719)} & \makecell{0.64***\\(715)} \\
\bottomrule
\end{tabular}

\end{adjustbox}
\tabnote{This table reports pairwise Pearson correlations between logged federal or local government IP intensity and the World Bank Statistical Performance Index and its five pillars. The all-area sample includes every source-visible ISO area with the required variables. The UN-member sample is restricted to the 193 countries appearing in the U.N. E-Government Survey. Per-worker measures also require positive public-sector employment. The number of country-year observations is reported in parentheses. Stars indicate rejection of the null hypothesis that the pairwise correlation is zero: $^{*}p<0.10$, $^{**}p<0.05$, and $^{***}p<0.01$.}
\label{tab:app_spi_country_universe_sensitivity}
\end{table}

\clearpage
\section{Classification Validation and Reproducibility}
\label{app:classification_validation}

\subsection{Two-Model Random-Sample Benchmark Audit}
LLM classification is a key step in the construction of DSC. Variables generated from unstructured data can introduce measurement error into empirical analysis \parencite{battaglia2024inference}; aggregated LLM predictions can rival human crowd accuracy while still containing systematic biases \parencite{schoenegger2024wisdom}; and LLMs can align with expert-coded sources while deviating systematically in particular cases \parencite{weidmann2025large}. The purpose of this appendix is to document benchmark agreement, specific disagreement patterns, and the extent to which disagreements are concentrated in countries or other aggregations that could affect aggregate scores. We report three complementary exercises: a two-model random-sample audit, a three-model benchmark using search context, and a comparison with locally run open-weight models. Model consensus is treated as a benchmark audit rather than as ground truth.

\begin{figure}
    \centering
    \includegraphics[width=\linewidth]{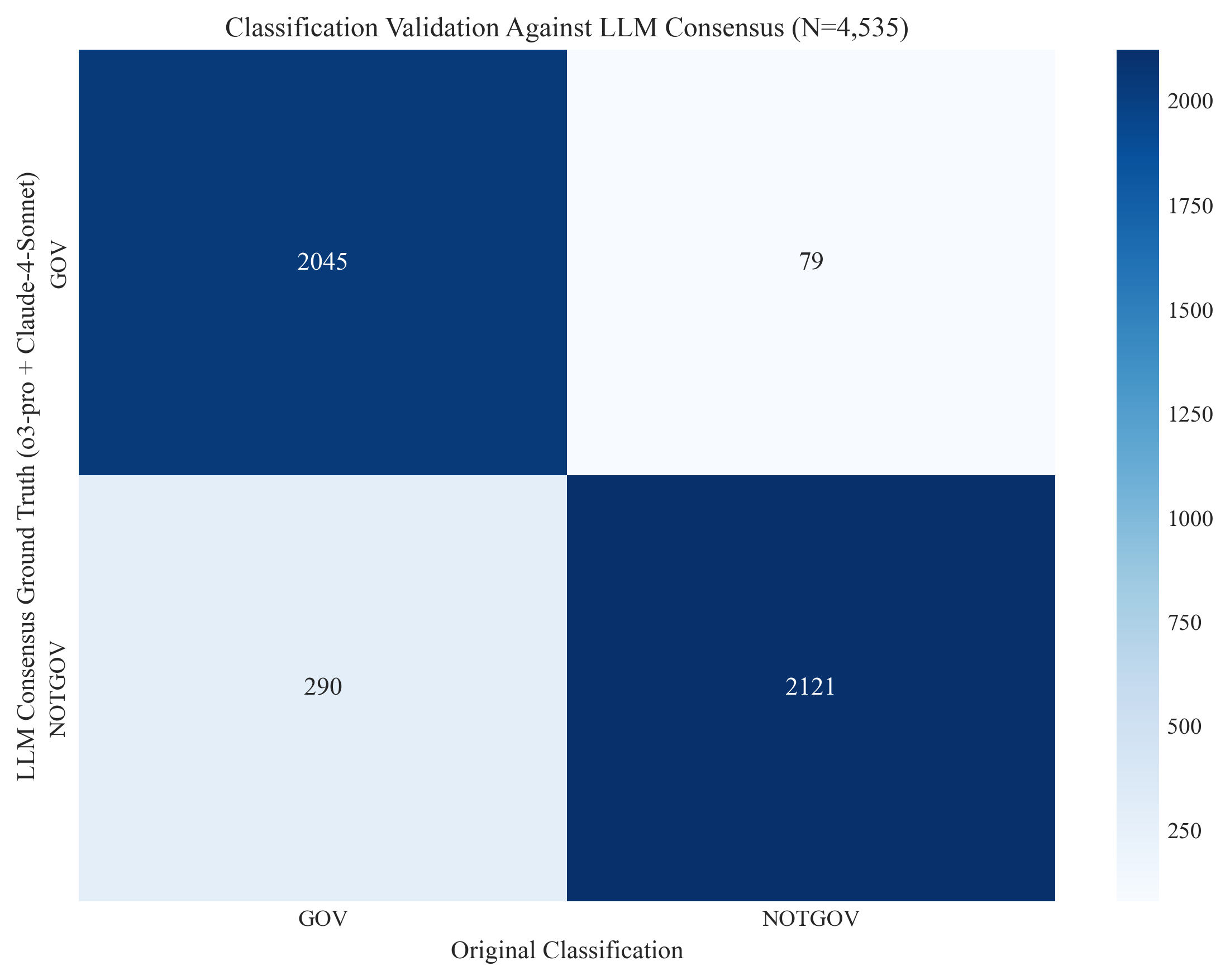}
    \caption{Confusion matrices comparing original labels with API consensus classifications based on OpenAI o3-pro and Claude Sonnet 4 model agreement. Full sample (n=4,535): 91.9\% agreement, $\kappa=0.838$. When the two models disagree, the output assigned higher confidence is used.}
    \label{fig:llm_validation1}
\end{figure}

\begin{figure}
    \centering
    \includegraphics[width=\linewidth]{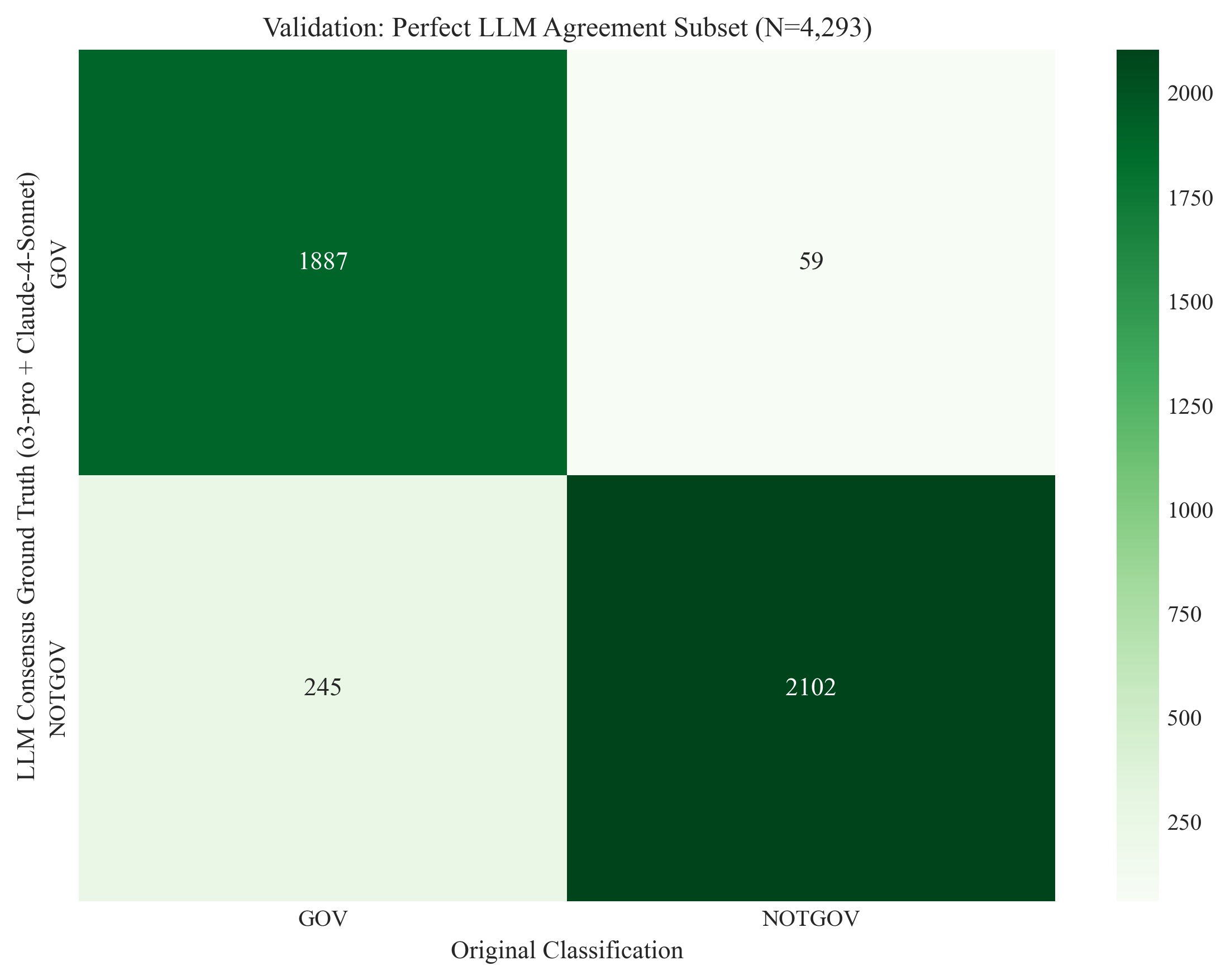}
    \caption{Confusion matrices comparing original labels with API consensus classifications based on OpenAI o3-pro and Claude Sonnet 4 model agreement. Model-agreement cases (n=4,293): 92.9\% agreement, $\kappa=0.858$.}
    \label{fig:llm_validation2}
\end{figure}

 \clearpage
As a first step, we characterise the main disagreement patterns in the classification pipeline, focusing on country-level patterns and common entity types associated with false positives and false negatives. We randomly sample 5,000 entities, with 2,500 predicted as government and 2,500 predicted as non-government, and run a simple classification prompt using OpenAI o3-pro and Claude Sonnet 4 as benchmark models. Both models returned results for 4,535 queries. Table~\ref{tab:api_consensus} shows the agreement between OpenAI o3-pro and Claude Sonnet 4. Table~\ref{tab:api_all} reports benchmark agreement between the labels used in the analysis and the two-model comparison, with 91.9 percent agreement overall. Figure~\ref{fig:llm_validation1} shows the confusion matrix using all model predictions; when OpenAI and Claude disagree, the more confident prediction is chosen. Figure~\ref{fig:llm_validation2} shows the confusion matrix for cases where the models agree. Here and below, false positives and false negatives are defined relative to the benchmark consensus.

Table~\ref{tab:fp_fn_descriptive} provides descriptive statistics on false positives and false negatives. Table~\ref{tab:false_positives} highlights common keyword patterns among false positives, and Table~\ref{tab:false_negative} highlights common keyword patterns among false negatives.

\begin{figure}
    \centering
    \includegraphics[width=\linewidth,height=0.78\textheight,keepaspectratio]{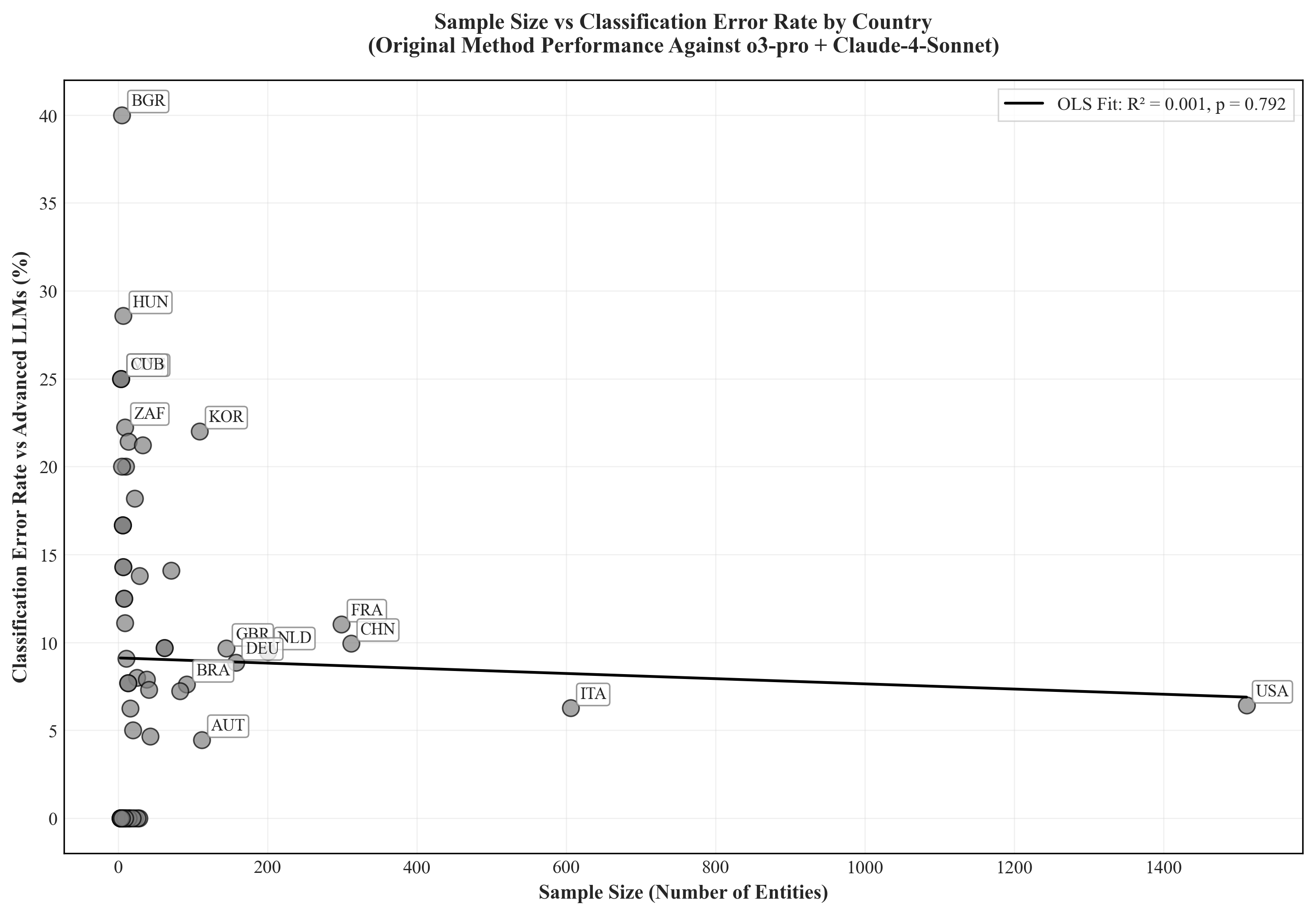}
    \caption{Relationship between sample size and benchmark disagreement rate by country. The OLS regression line is shown with $R^2$ and significance values. Outlier countries are labelled for identification of patterns requiring further investigation.}
    \label{fig:llm_validation3}
\end{figure}

\begin{figure}
    \centering
    \includegraphics[width=\linewidth,height=0.78\textheight,keepaspectratio]{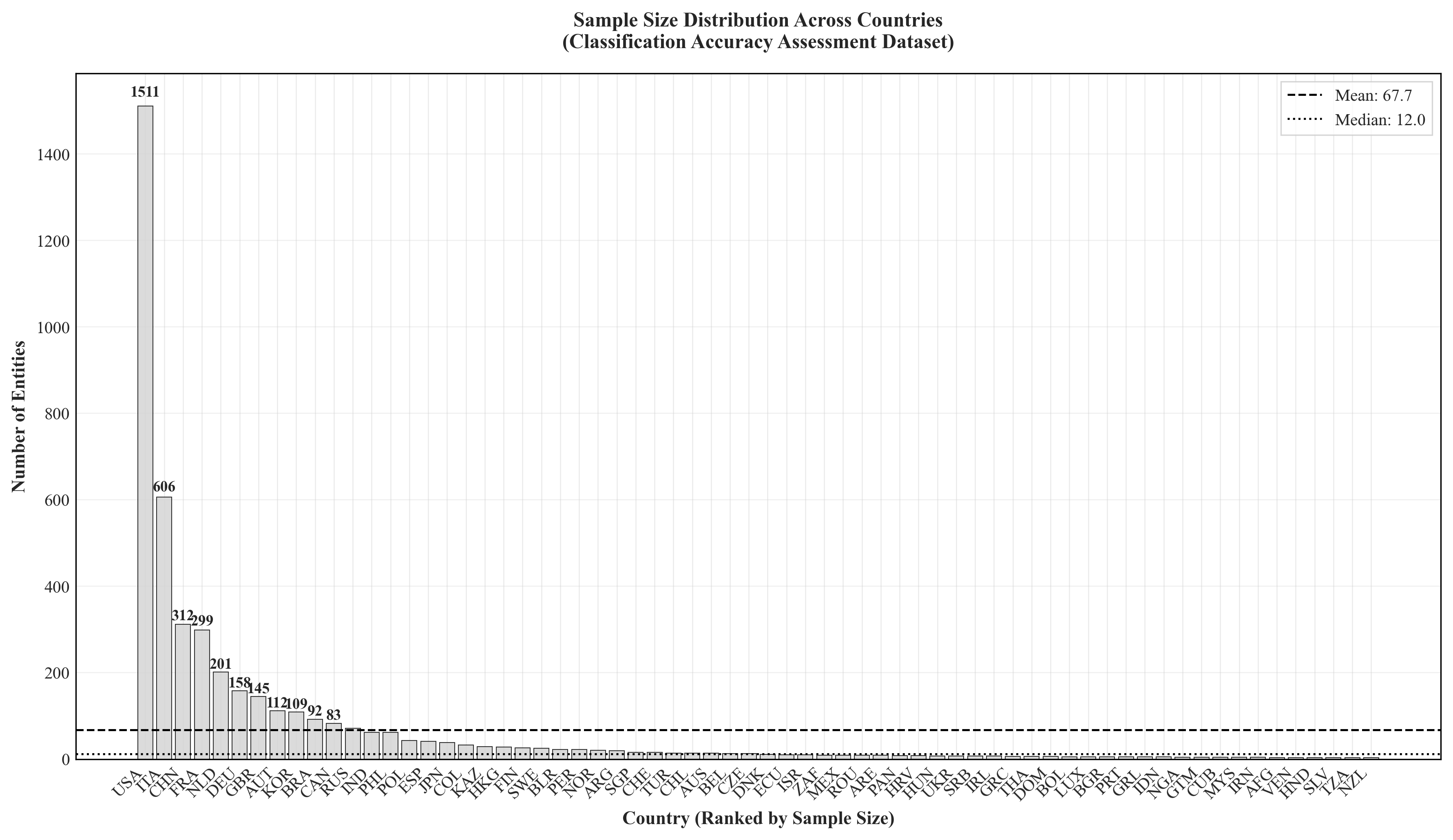}
    \caption{Sample size distribution across countries in the classification-agreement assessment. Mean and median sample sizes are indicated.}
    \label{fig:llm_validation4}
\end{figure}

\begin{figure}
    \centering
    \includegraphics[width=\linewidth,height=0.78\textheight,keepaspectratio]{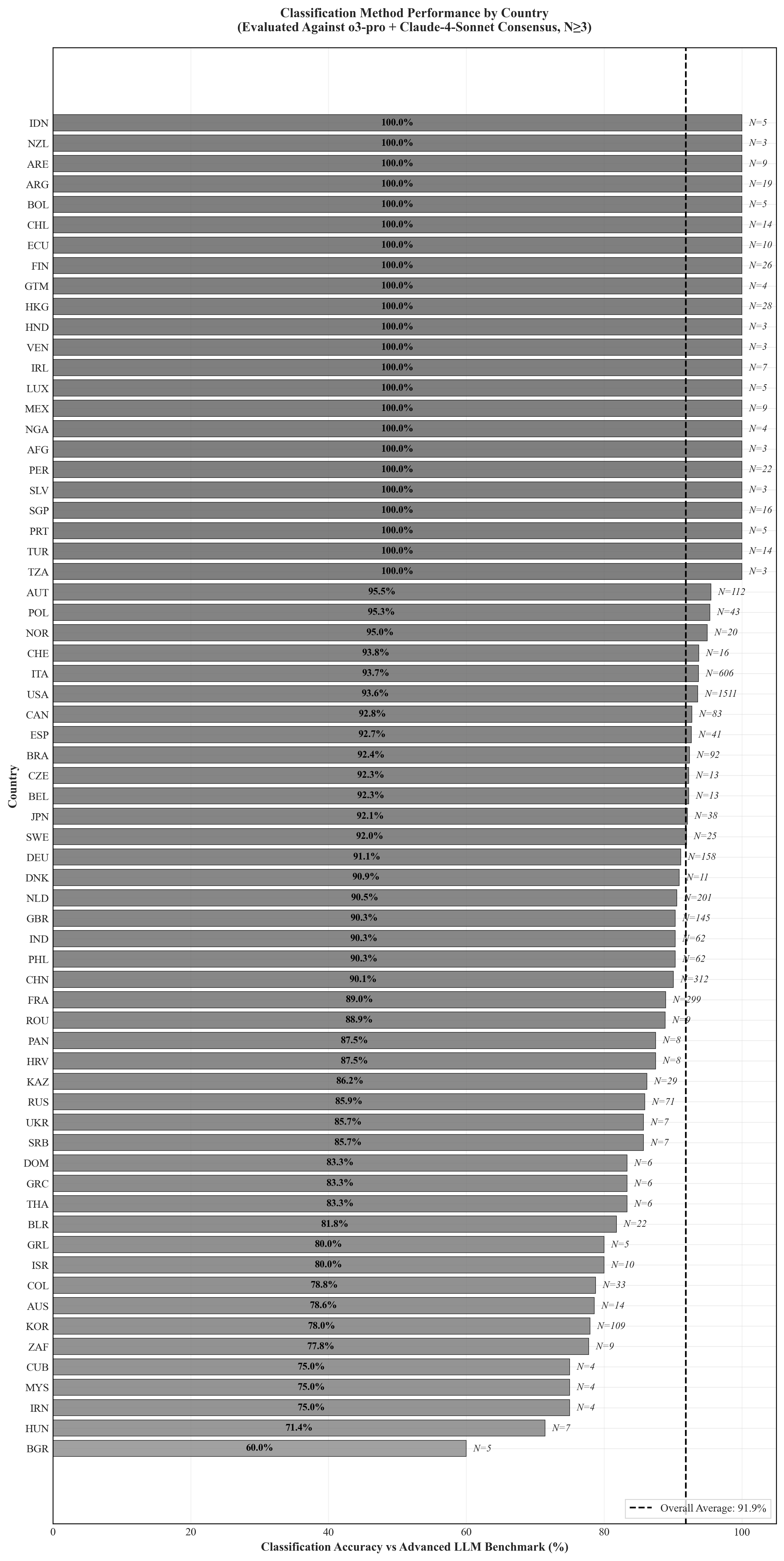}
    \caption{Classification agreement by country with the OpenAI o3-pro and Claude Sonnet 4 benchmark. Countries with $\geq$3 entities are shown. Overall agreement is 91.9\%. Sample sizes (N) are indicated.}
    \label{fig:llm_validation5}
\end{figure}

\begin{figure}
    \centering
    \includegraphics[width=\linewidth,height=0.78\textheight,keepaspectratio]{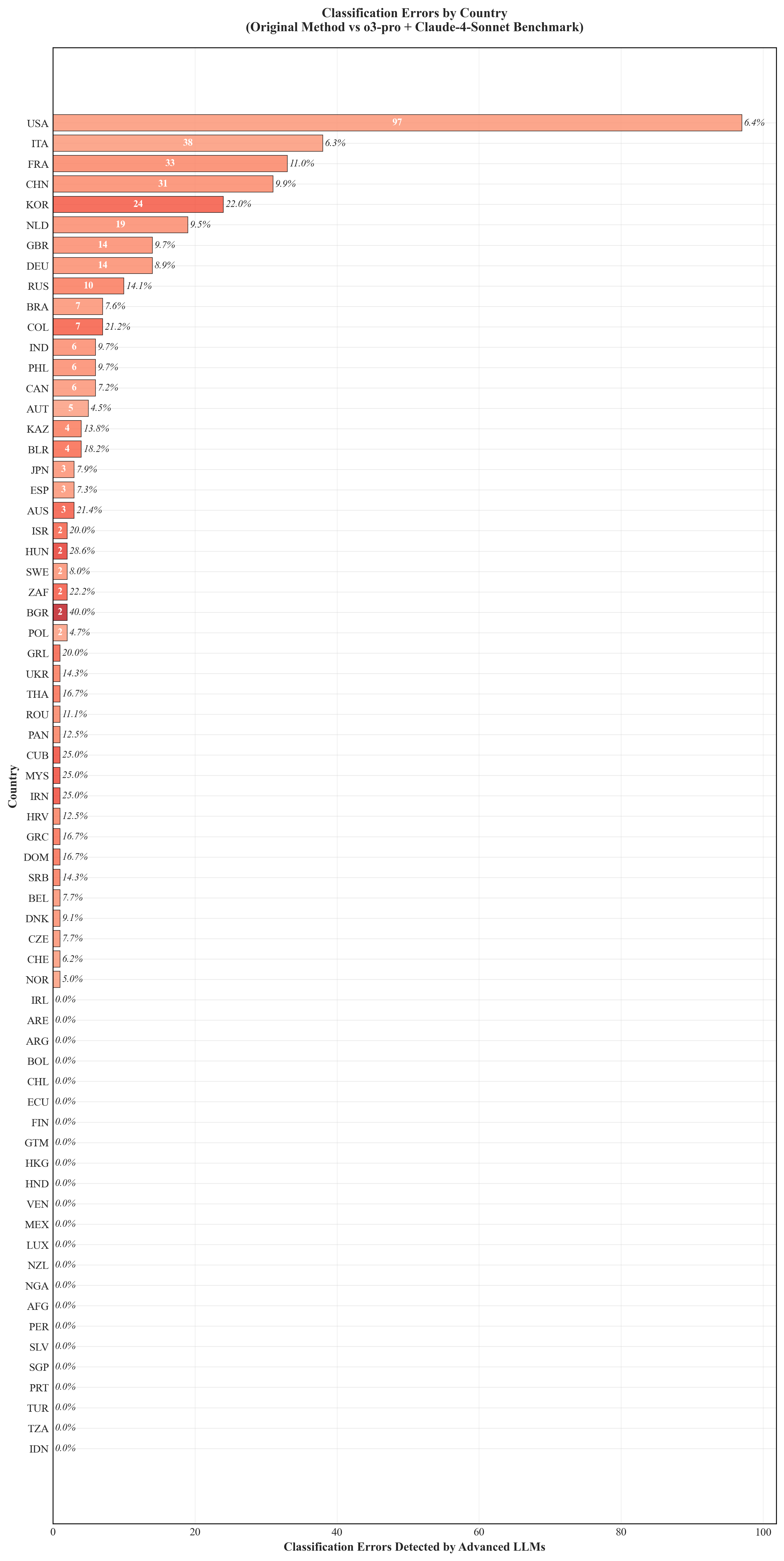}
    \caption{Classification disagreements with the OpenAI o3-pro and Claude Sonnet 4 benchmark, sorted by absolute disagreement count. Disagreement rates are shown as percentages.}
    \label{fig:llm_validation6}
\end{figure}

Figures~\ref{fig:llm_validation3} through~\ref{fig:llm_validation6} provide a detailed overview of country-level classification performance. Figure~\ref{fig:llm_validation3} illustrates the relationship between sample size and benchmark disagreement, with key outliers labelled for further scrutiny. Figure~\ref{fig:llm_validation4} displays the distribution of sample sizes across countries, highlighting central tendencies. Figure~\ref{fig:llm_validation5} presents agreement by country with the OpenAI o3-pro and Claude Sonnet 4 benchmark. Finally, Figure~\ref{fig:llm_validation6} summarises the distribution of classification disagreements by country, sorted by total disagreement count and accompanied by disagreement-rate percentages.
\clearpage

\begin{table}[h]
\centering
\caption{API Consensus Statistics (Rows Analysed: 4,535, Filtered Out: 465)}
\label{tab:api_consensus}
\begin{tabular}{lcc}
\hline
\textbf{Model Comparison} & \textbf{Count} & \textbf{Percentage} \\
\hline
Model-agreement cases & 4,293 & 94.7\% \\
Model-disagreement cases & 242 & 5.3\% \\
\hline
\textbf{Total} & \textbf{4,535} & \textbf{100.0\%} \\
\hline
\end{tabular}
\end{table}

\begin{table}[h]
\centering
\caption{Original Label Agreement with API Consensus}
\label{tab:api_all}
\begin{tabular}{lc}
\hline
\textbf{Metric} & \textbf{Value} \\
\hline
Agreement & 91.9\% \\
Correct Classifications & 4,166/4,535 \\
Original Rows & 5,000 \\
Rows Analysed & 4,535 \\
Rows Filtered Out & 465 \\
\hline
\end{tabular}
\end{table}

\clearpage
\begin{table}[h]
\centering
\caption{Classification Disagreement Analysis by Type}
\label{tab:fp_fn_descriptive}

\begin{tabular}{lcc}
\hline
\textbf{Metric} & \textbf{False Positives} & \textbf{False Negatives} \\
\hline
Total Count & 328 & 90 \\
\% of All Discrepancies & 78.5\% & 21.5\% \\
Missing Descriptions & 0 (0.0\%) & 90 (100.0\%) \\
Advanced Models Agree & 245/328 (74.7\%) & 59/90 (65.6\%) \\
Mean o3-pro Confidence & 0.811 & 0.737 \\
Mean Claude Sonnet 4 Confidence & 0.729 & 0.742 \\
\hline
\end{tabular}
\end{table}

\clearpage
\begin{table}[h]
\centering
\caption{Keywords in False Positive Classifications (Baseline Labels Government; Benchmark Labels Non-Government)}
\label{tab:false_positives}
\begin{tabularx}{\linewidth}{>{\raggedright\arraybackslash}X>{\centering\arraybackslash}p{2.4cm}>{\centering\arraybackslash}p{3.2cm}}
\hline
\textbf{Term} & \textbf{Frequency} & \makecell{\textbf{\% of}\\\textbf{False Positives}} \\
\hline
SERVICES & 7 & 2.1\% \\
PATRONATO & 6 & 1.8\% \\
COUNTY & 6 & 1.8\% \\
GMBH & 6 & 1.8\% \\
FIRE & 5 & 1.5\% \\
STATE & 5 & 1.5\% \\
SUDOGWONGANGNAMBONBUJANG & 5 & 1.5\% \\
ACLI & 4 & 1.2\% \\
\hline
\textbf{Total False Positives} & \textbf{328} & \textbf{78.5\%} \\
\hline
\end{tabularx}
\end{table}

\begin{table}[h]
\centering
\caption{Keywords in False Negative Classifications (Baseline Labels Non-Government; Benchmark Labels Government)}
\label{tab:false_negative}
\begin{tabularx}{\linewidth}{>{\raggedright\arraybackslash}X>{\centering\arraybackslash}p{2.4cm}>{\centering\arraybackslash}p{3.2cm}}
\hline
\textbf{Term} & \textbf{Frequency} & \makecell{\textbf{\% of}\\\textbf{False Negatives}} \\
\hline
SCHOOL & 14 & 15.6\% \\
COUNTY & 6 & 6.7\% \\
PRIMARY & 5 & 5.6\% \\
AND & 5 & 5.6\% \\
THE & 3 & 3.3\% \\
COMPANY & 3 & 3.3\% \\
ENGINEERING & 2 & 2.2\% \\
STATE & 2 & 2.2\% \\
\hline
\textbf{Total False Negatives} & \textbf{90} & \textbf{21.5\%} \\
\hline
\end{tabularx}
\end{table}

\clearpage

\clearpage

\clearpage

\clearpage

\clearpage

\clearpage

\subsection{Three-Model Random-Sample Benchmark with Search Context}
\begin{figure}
    \centering
    \includegraphics[width=\linewidth,height=0.78\textheight,keepaspectratio]{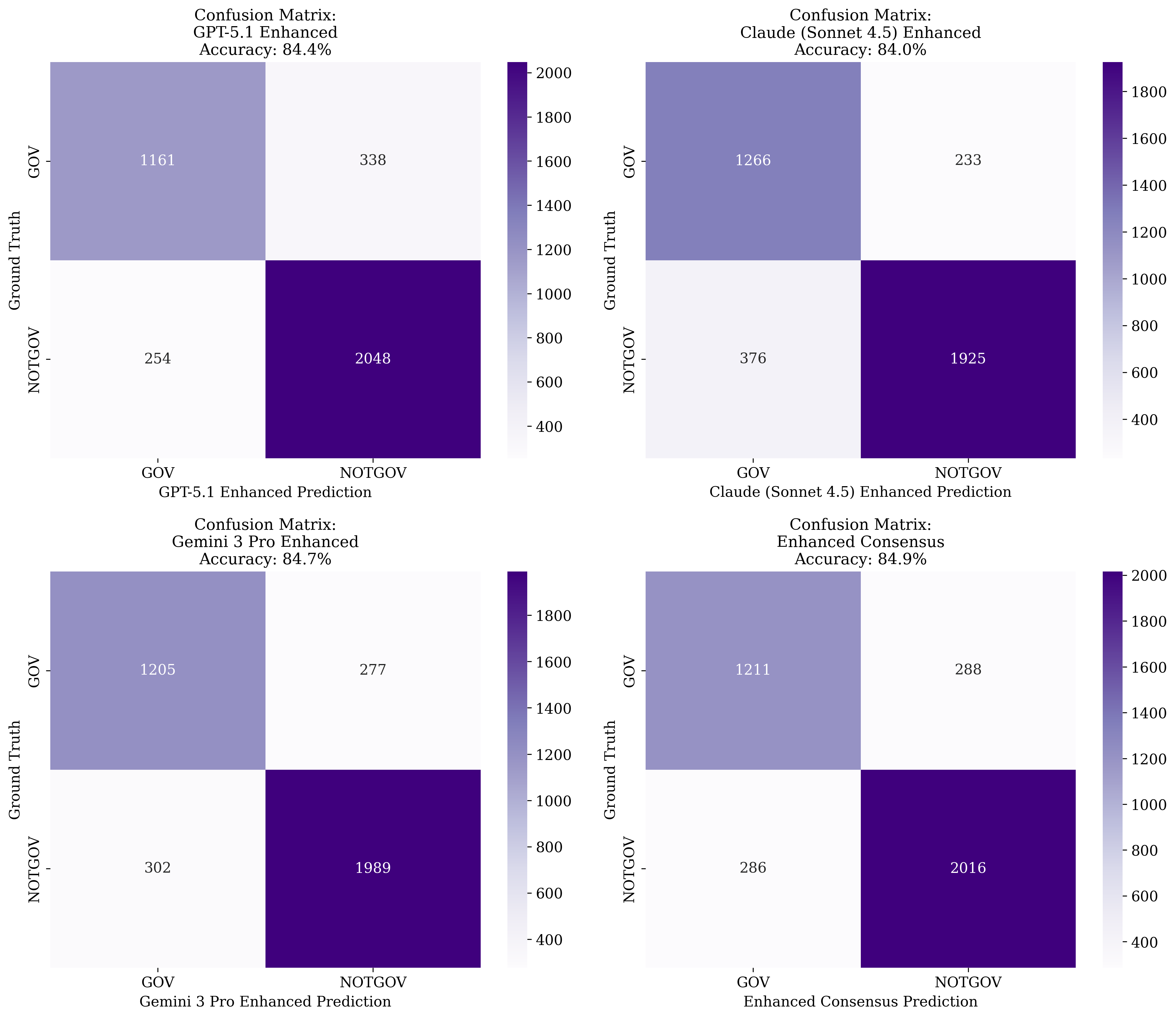}
    \caption{Agreement between baseline classifications and GPT-5.1, Claude Sonnet 4.5, Gemini 3 Pro, and their two-of-three consensus. Classification inputs include contextual information obtained through Serper search.}
    \label{fig:llm_validation_serper_heatmap}
\end{figure}

\begin{figure}
    \centering
    \includegraphics[width=\linewidth,height=0.78\textheight,keepaspectratio]{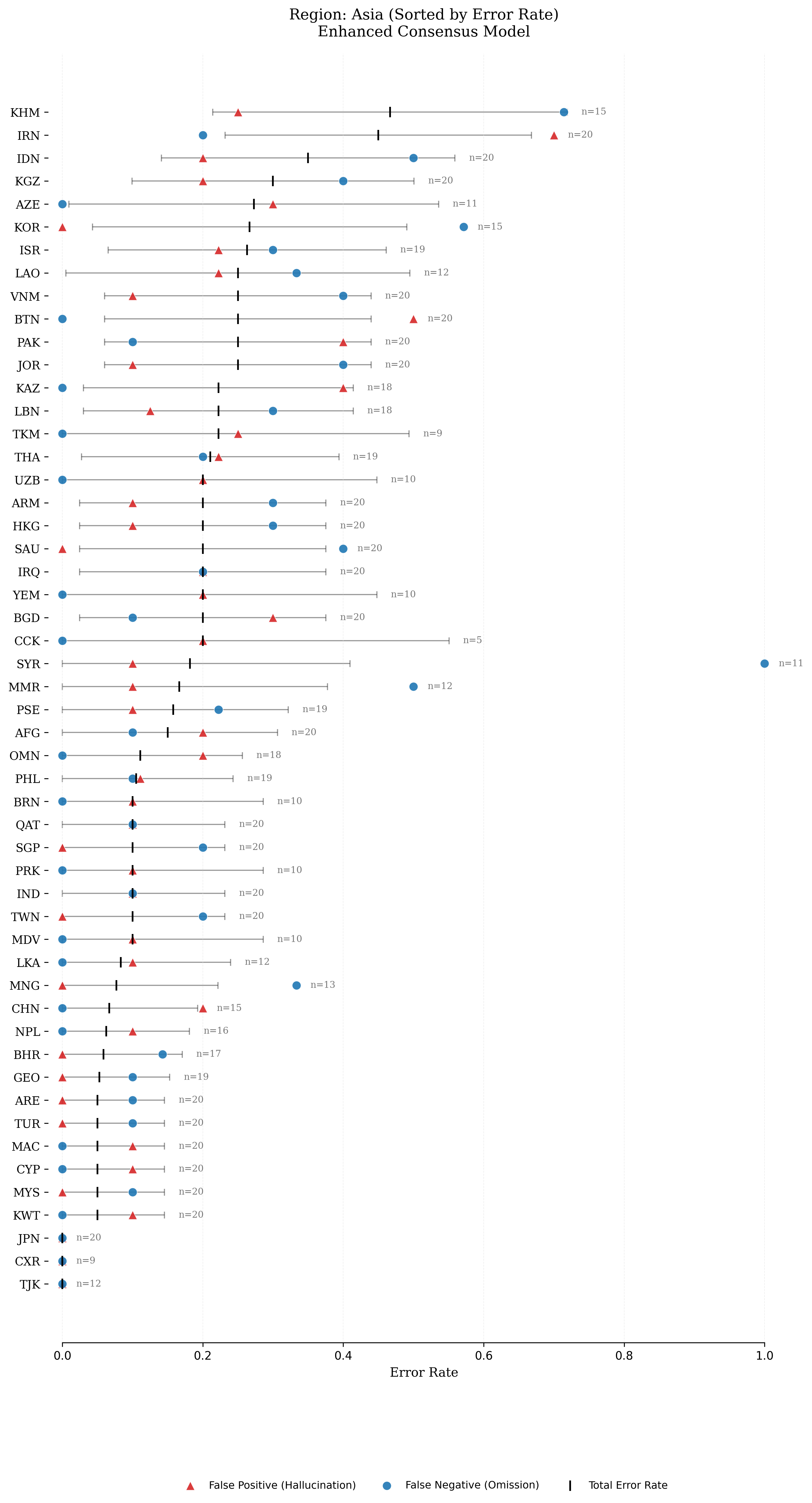}
    \caption{Classification disagreements with the three-model benchmark by type for countries in Asia.}
    \label{fig:llm_error_Asia}
\end{figure}

\begin{figure}
    \centering
    \includegraphics[width=\linewidth,height=0.78\textheight,keepaspectratio]{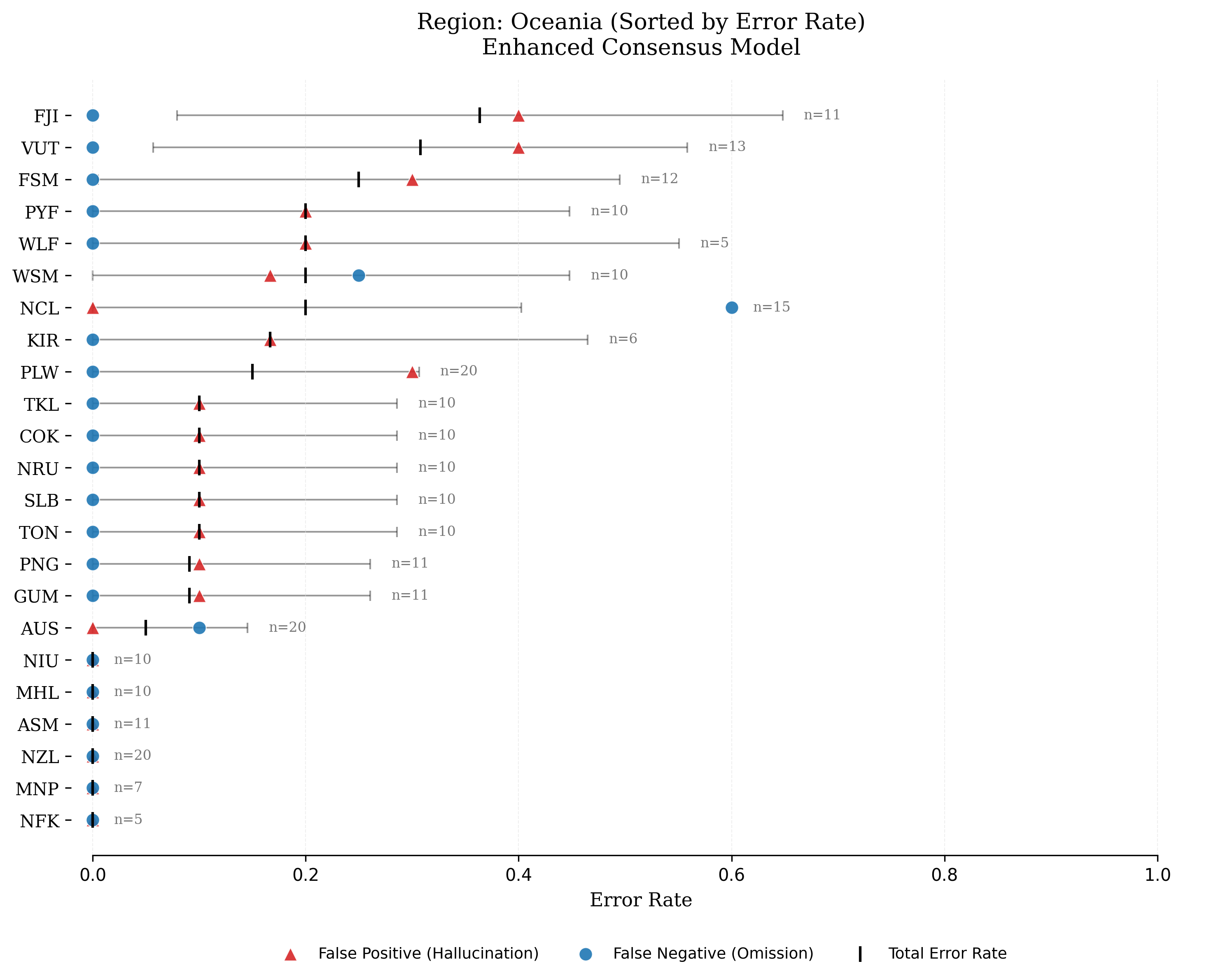}
    \caption{Classification disagreements with the three-model benchmark by type for countries in Oceania.}
    \label{fig:llm_error_Oceania}
\end{figure}

\begin{figure}
    \centering
    \includegraphics[width=\linewidth,height=0.78\textheight,keepaspectratio]{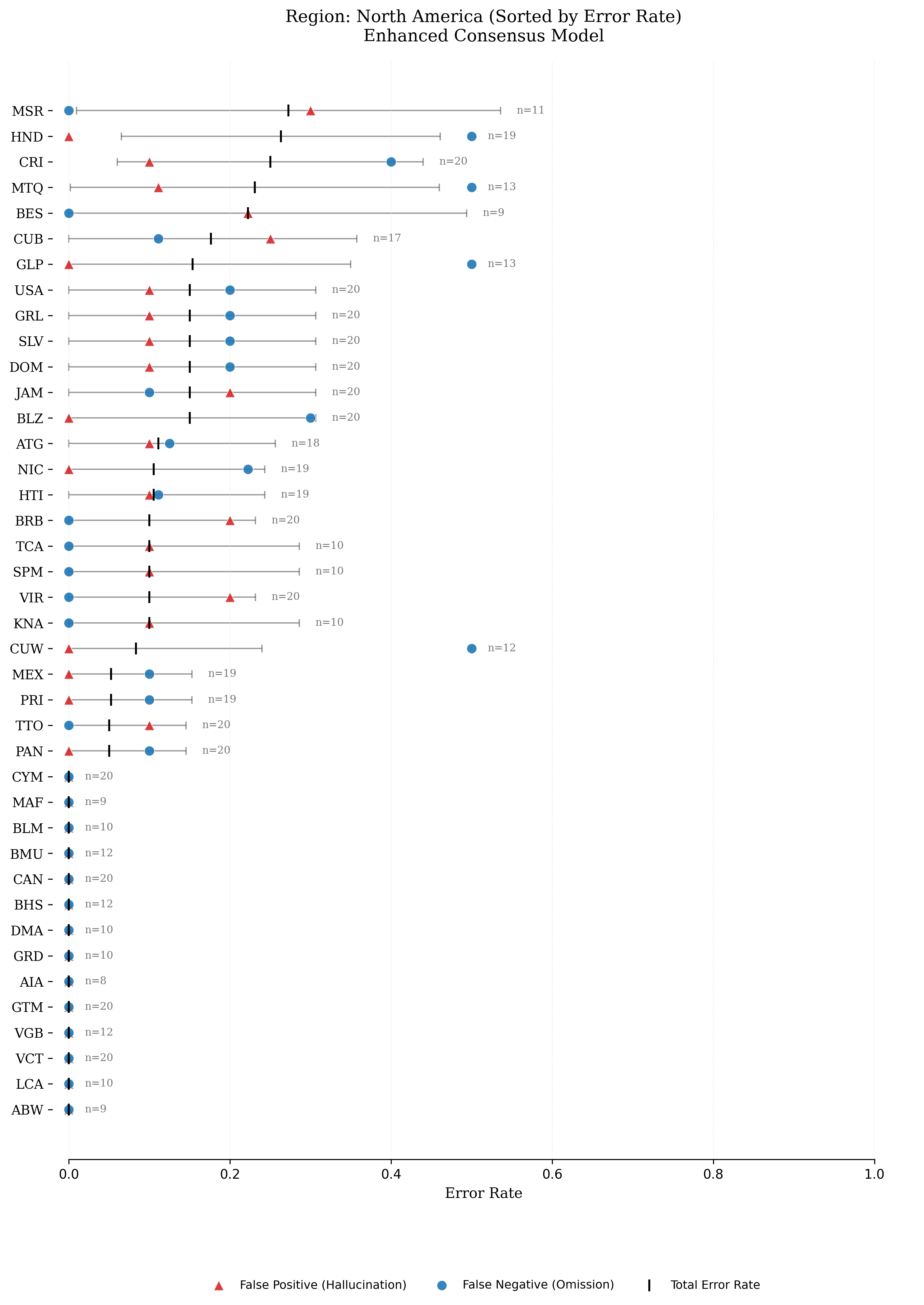}
    \caption{Classification disagreements with the three-model benchmark by type for countries in North America.}
    \label{fig:llm_error_NA}
\end{figure}

\begin{figure}
    \centering
    \includegraphics[width=\linewidth,height=0.78\textheight,keepaspectratio]{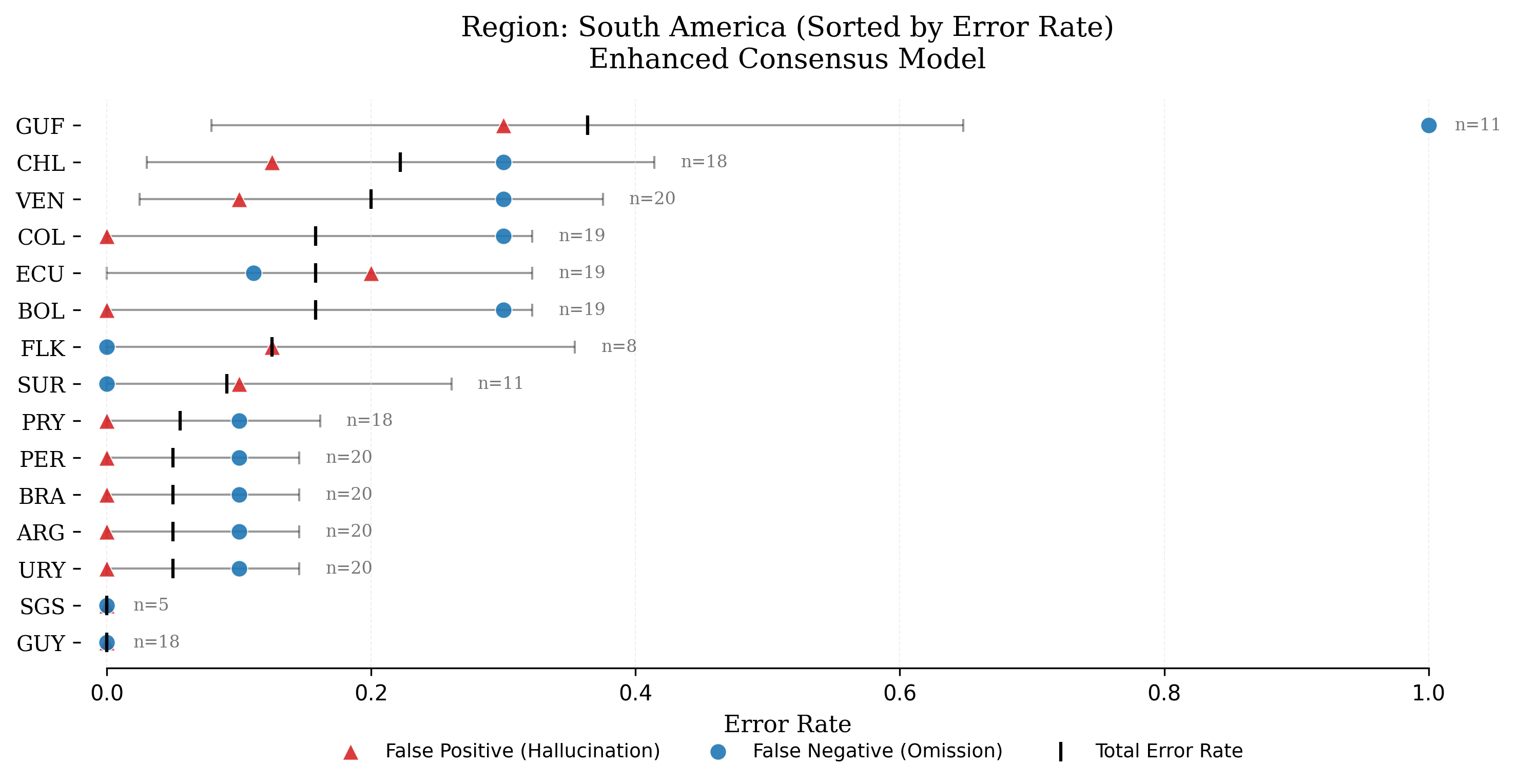}
    \caption{Classification disagreements with the three-model benchmark by type for countries in South America.}
    \label{fig:llm_error_SA}
\end{figure}

\begin{figure}
    \centering
    \includegraphics[width=\linewidth,height=0.78\textheight,keepaspectratio]{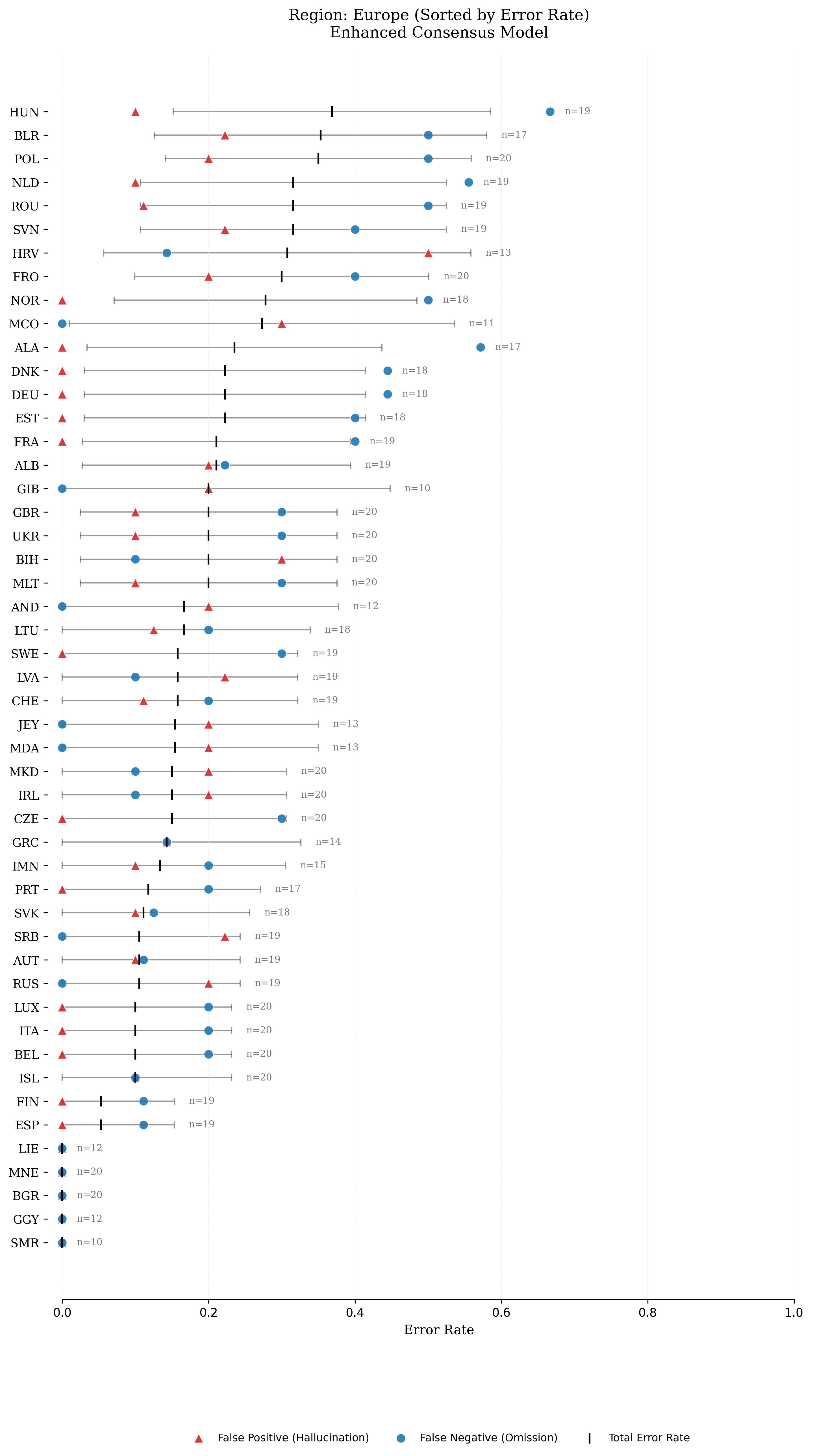}
    \caption{Classification disagreements with the three-model benchmark by type for countries in Europe.}
    \label{fig:llm_error_EU}
\end{figure}

\begin{figure}
    \centering
    \includegraphics[width=\linewidth,height=0.78\textheight,keepaspectratio]{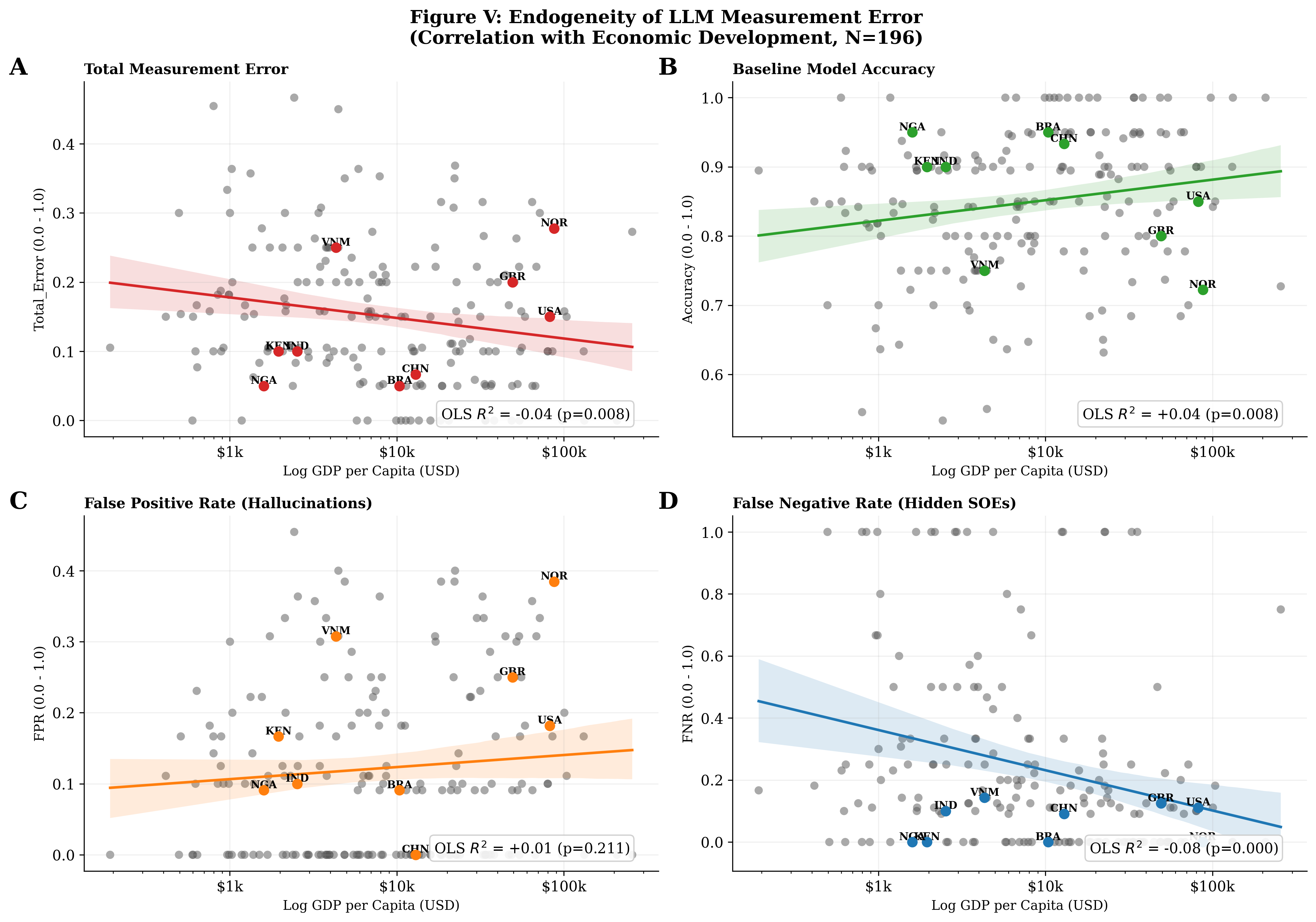}
    \caption{Country-level classification disagreement with the consensus of GPT-5.1, Claude Sonnet 4.5, and Gemini 3 Pro, and its association with GDP per capita in 2020. Countries with fewer than five government entities or no GDP per capita for 2020 are excluded.}
    \label{fig:llm_error_gdp}
\end{figure}

\begin{figure}
    \centering
    \includegraphics[width=\linewidth,height=0.78\textheight,keepaspectratio]{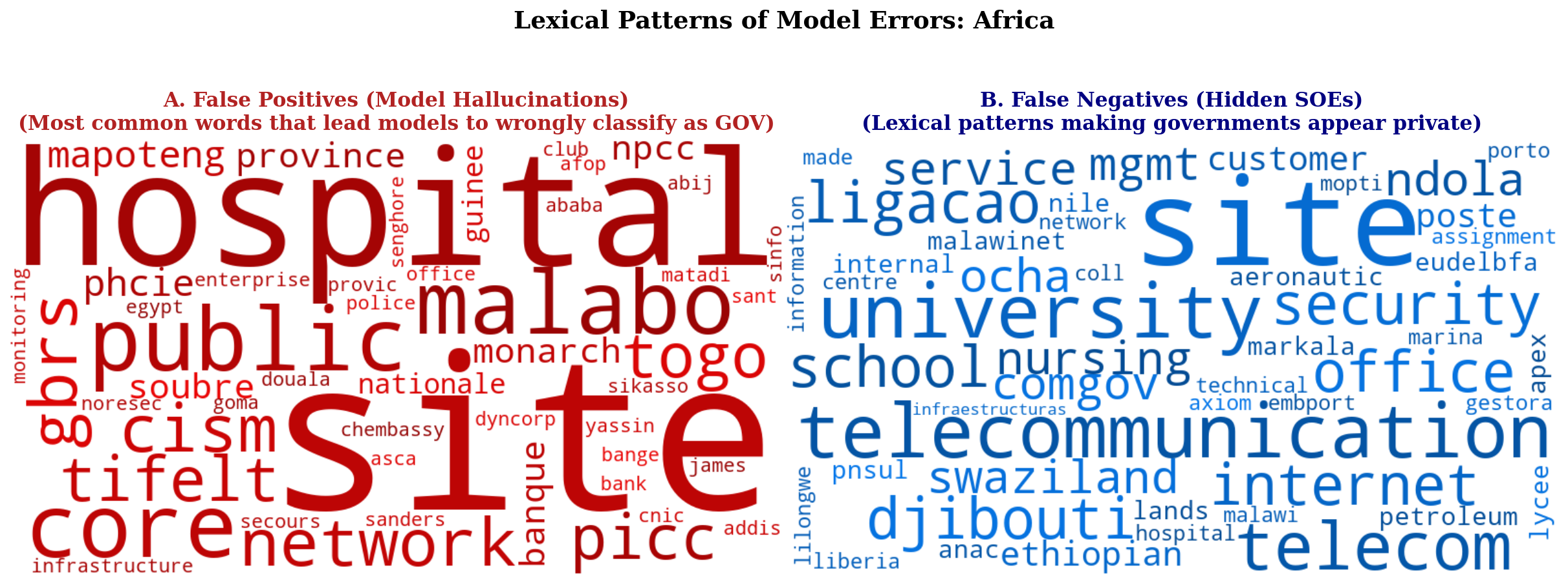}
    \caption{Lexical patterns of classification disagreement in Africa.}
    \label{fig:lexical_africa}
\end{figure}

\begin{figure}
    \centering
    \includegraphics[width=\linewidth,height=0.78\textheight,keepaspectratio]{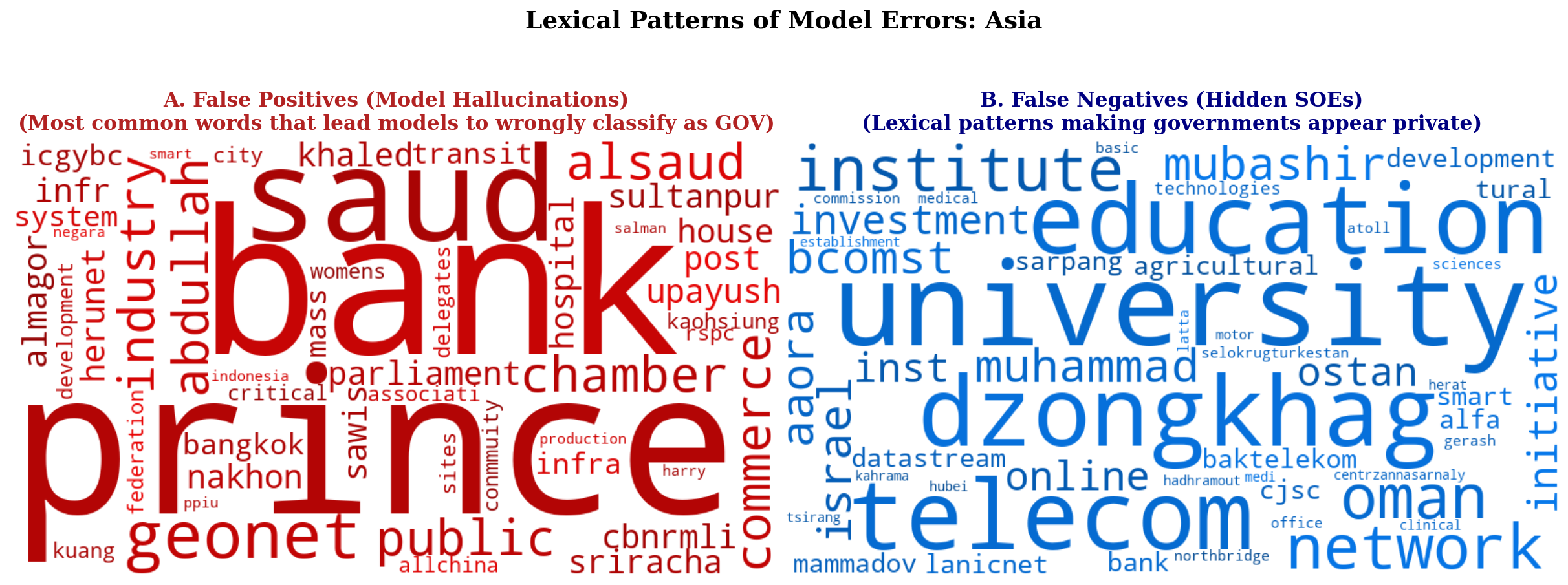}
    \caption{Lexical patterns of classification disagreement in Asia.}
    \label{fig:lexical_asia}
\end{figure}

\begin{figure}
    \centering
    \includegraphics[width=\linewidth,height=0.78\textheight,keepaspectratio]{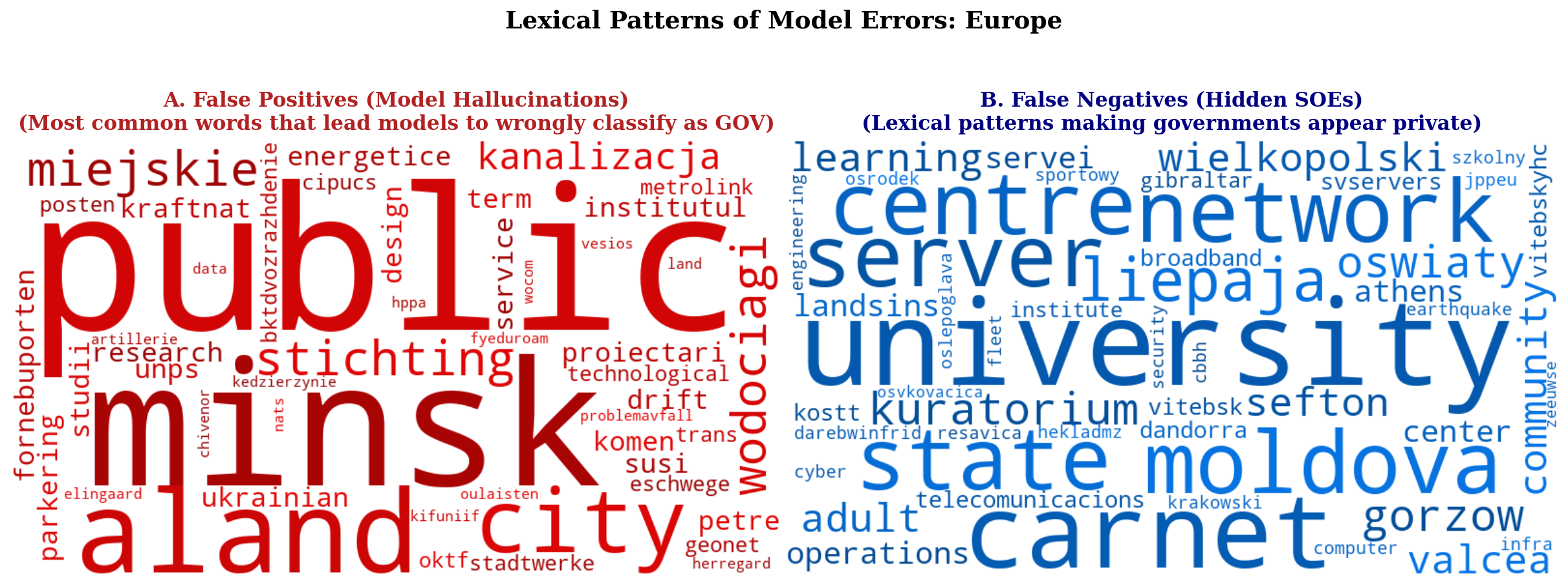}
    \caption{Lexical patterns of classification disagreement in Europe.}
    \label{fig:lexical_europe}
\end{figure}

\begin{figure}
    \centering
    \includegraphics[width=\linewidth,height=0.78\textheight,keepaspectratio]{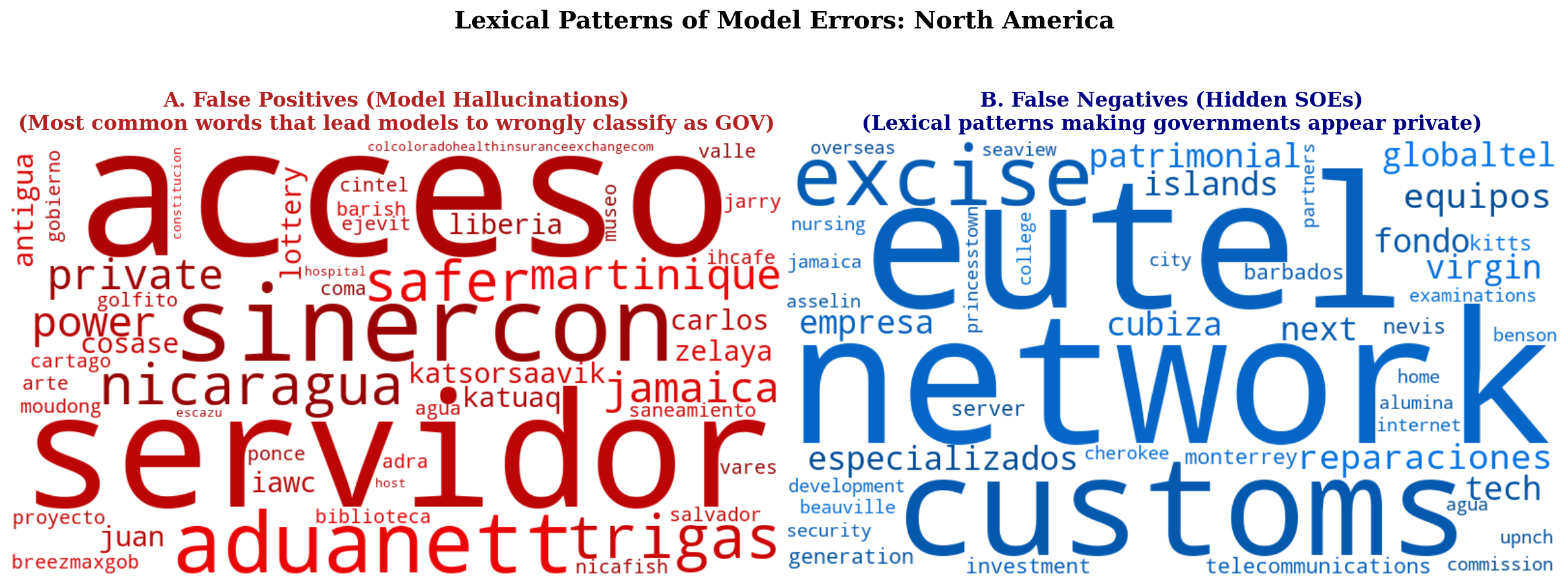}
    \caption{Lexical patterns of classification disagreement in North America.}
    \label{fig:lexical_north_america}
\end{figure}

\begin{figure}
    \centering
    \includegraphics[width=\linewidth,height=0.78\textheight,keepaspectratio]{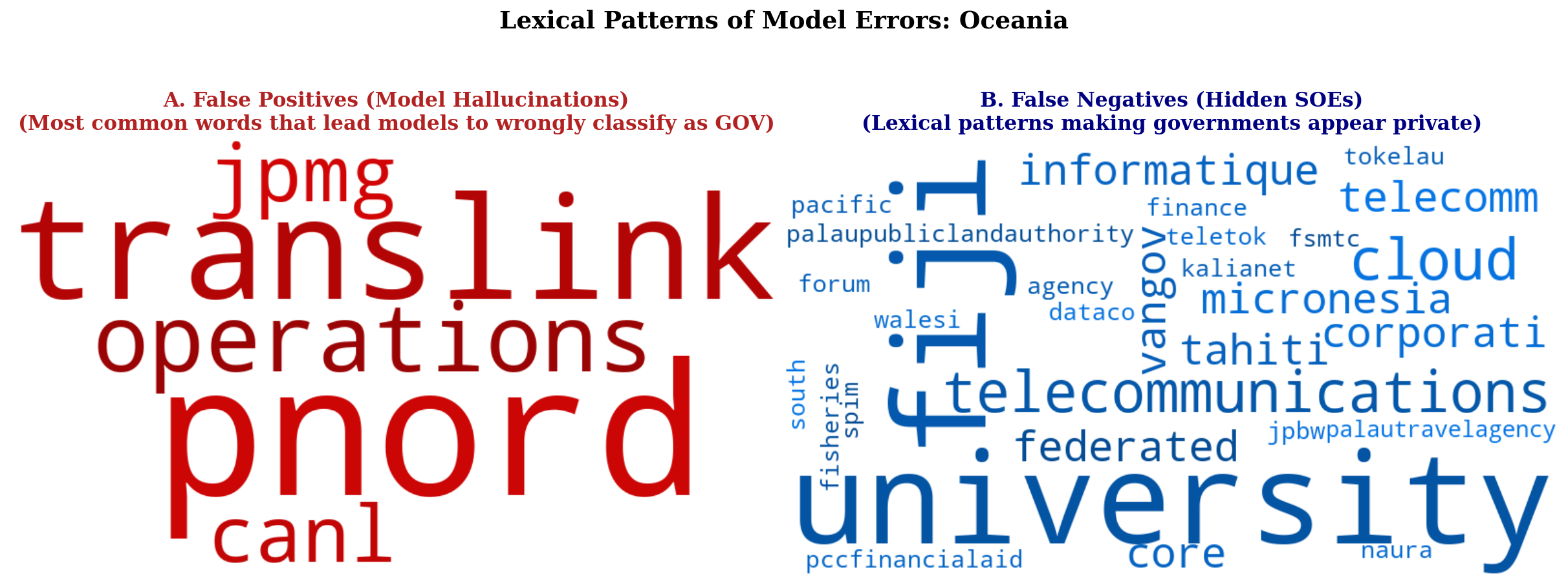}
    \caption{Lexical patterns of classification disagreement in Oceania.}
    \label{fig:lexical_oceania}
\end{figure}

\begin{figure}
    \centering
    \includegraphics[width=\linewidth,height=0.78\textheight,keepaspectratio]{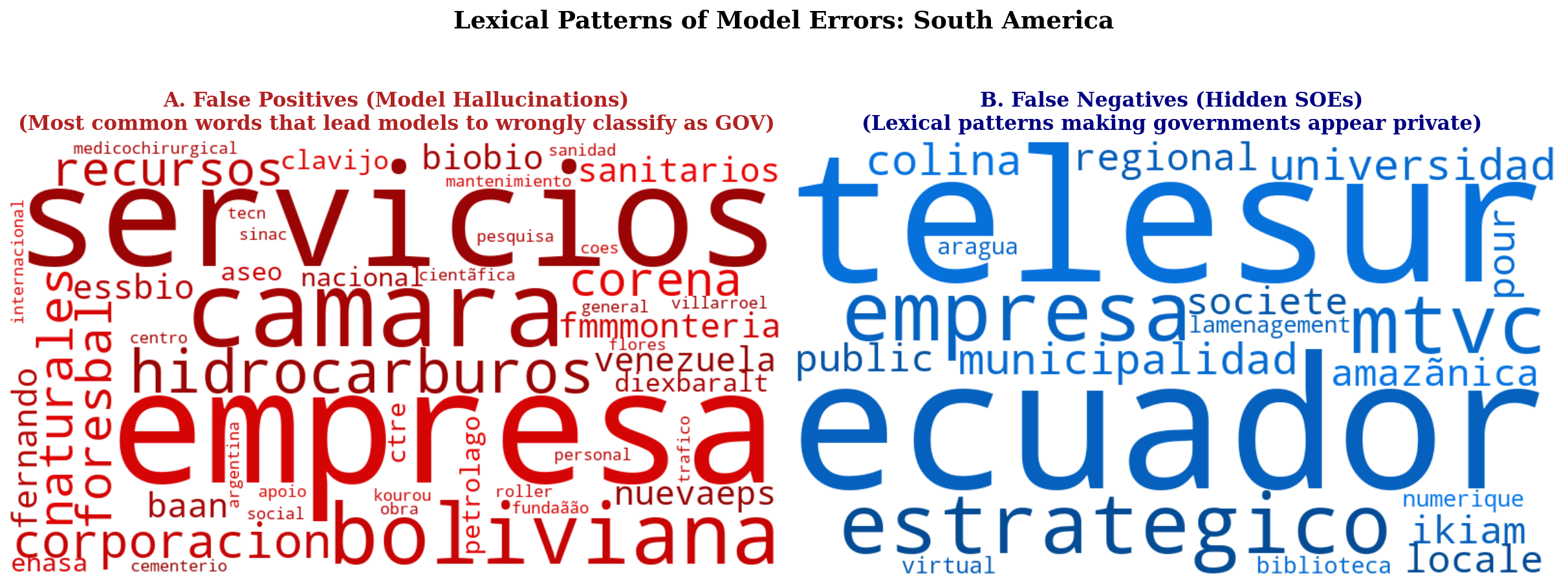}
    \caption{Lexical patterns of classification disagreement in South America.}
    \label{fig:lexical_south_america}
\end{figure}

Figure~\ref{fig:llm_validation_serper_heatmap} shows a heatmap of our baseline predictions compared with GPT-5.1, Claude Sonnet 4.5, and Gemini 3 Pro in the November 2025 validation run, as well as their majority consensus. Figure~\ref{fig:llm_error_gdp} examines whether benchmark disagreement is systematically related to economic development by correlating baseline classification metrics with GDP per capita across 196 countries. Panel A shows that total disagreement, measured by the difference between baseline and benchmark classifications, declines significantly with GDP per capita ($R^2 = 0.04$, $p=0.008$), while Panel B shows the inverse relationship for baseline agreement ($R^2 = 0.04$, $p=0.008$). Decomposing these disagreements reveals an asymmetry in their sources: false positive rates exhibit no significant correlation with development (Panel C), whereas false negative rates decline significantly with GDP per capita (Panel D, $R^2 = 0.08$, $p<0.001$). This pattern is consistent with greater benchmark disagreement over state-owned or state-linked entities in lower-income countries. Regional decompositions (Figures~\ref{fig:llm_error_Asia}–\ref{fig:llm_error_EU}) show corresponding patterns across geographic contexts, with particularly pronounced false negative rates in Asian and African economies.\footnote{The 25 highest-disagreement jurisdictions (disagreement rates 30–47\%) exhibit substantial heterogeneity in both development levels and disagreement types. High false negative rates appear in Liberia (FNR=1.00), Cambodia (FNR=0.71), and French Guiana (FNR=1.00), where the baseline model identifies fewer state-owned enterprises than the benchmark models. Conversely, high false positive rates appear in Iran (FPR=0.70), Croatia (FPR=0.50), and Fiji (FPR=0.40). Several transition economies (Hungary, Poland, Romania, Slovenia) exhibit elevated disagreement rates.} Figures~\ref{fig:lexical_africa}–\ref{fig:lexical_south_america} present word clouds of classification disagreements by region, revealing systematic patterns in where the classifications differ. The false negative word clouds (Panel B) are dominated by terms such as ``university,'' ``college,'' ``telecom,'' ``telecommunications,'' and ``broadcasting,'' indicating that many disagreements involve state-owned or state-linked educational and communications institutions. These cases are edge cases rather than ordinary classification noise: they define whether DSC measures core administrative government or a broader public-sector digital infrastructure stock. Our baseline measure is therefore best interpreted as a measure of core government ICT infrastructure, with some categories requiring separate sensitivity checks. Table~\ref{tab:lexical_drivers} shows examples of entities for each region with these key terms.

\clearpage

\clearpage
\begin{figure}
    \centering
    \includegraphics[width=\linewidth,height=0.78\textheight,keepaspectratio]{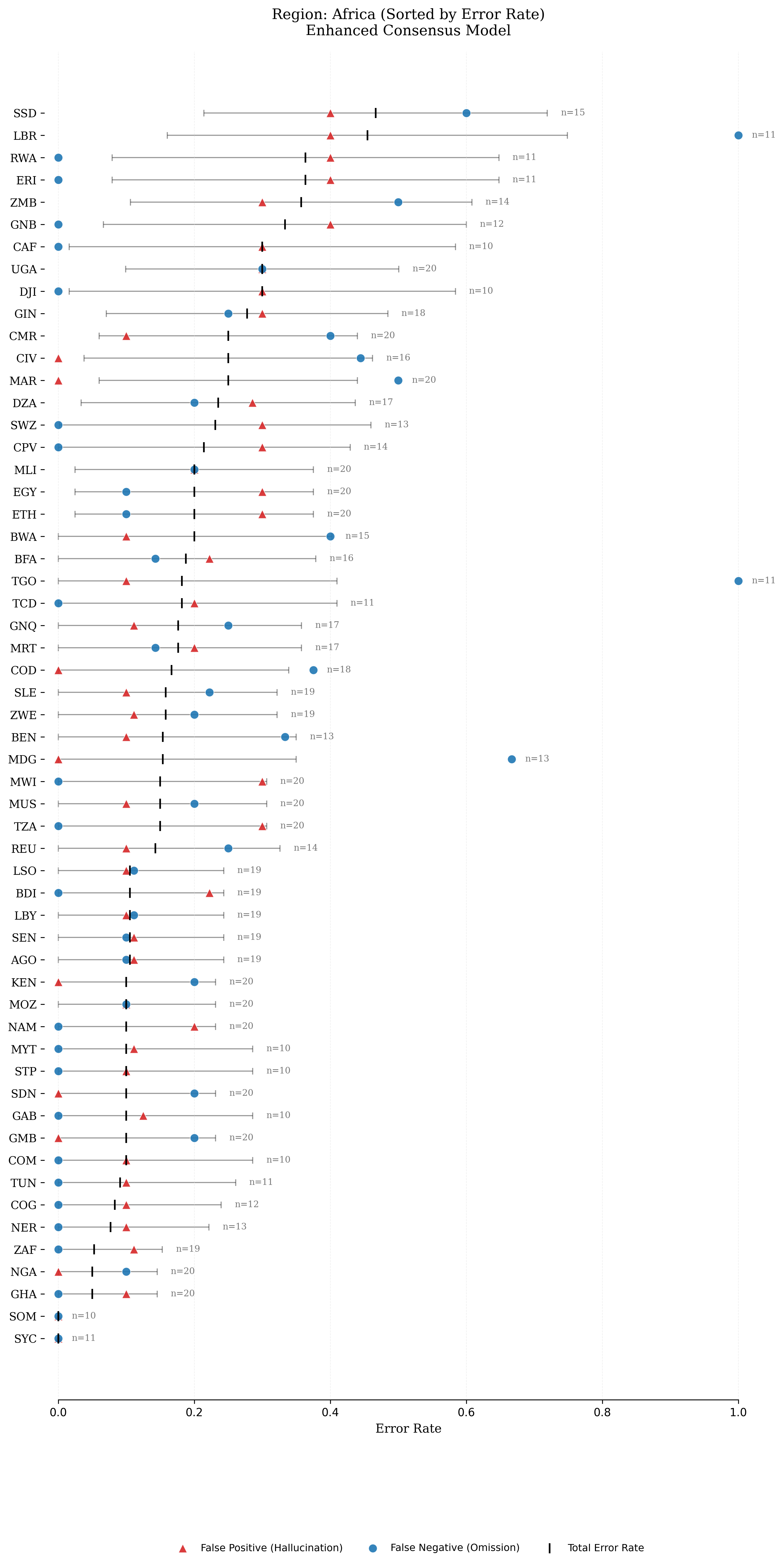}
    \caption{Classification disagreements with the three-model benchmark by type for countries in Africa.}
    \label{fig:llm_error_Africa}
\end{figure}

\clearpage

\clearpage
\clearpage

\clearpage

\clearpage

\clearpage
\clearpage
\clearpage
\clearpage
\clearpage

\clearpage
\begin{center}
\footnotesize
\begin{longtable}{>{\raggedright\arraybackslash}p{0.95\textwidth}}
\caption{\textbf{Lexical Patterns in Classification Disagreements.} This table lists frequent terms associated with disagreements between the baseline classification and the three-model benchmark in each region, along with representative examples. False positives are cases classified as government by the baseline and as non-government by the benchmark; false negatives are cases classified as non-government by the baseline and as government by the benchmark.}
\label{tab:lexical_drivers}\\
\hline \hline
\textbf{Panel Afr: Africa} \\ \hline
\textit{Terms in False Positives:} ``Hospital'', ``Public'', ``Core'', ``Malabo'', ``Site'', ``Cism'' \\
\textit{Examples:} Mapoteng Hospital (LSO); Sant James 2 Hospital (LBY); Enterprise public services (MDG); Public Health Institute (SDN); Core Malabo (GNQ); National Core Infrastructure (ZMB); Core Malabo (GNQ); Bange Malabo (GNQ) \\ [0.5em]
\textit{Terms in False Negatives:} ``Site'', ``University'', ``Telecom'', ``Djibouti'', ``Ligacao'', ``School'' \\
\textit{Examples:} Ocha Site21 (CAF); Ocha Site13 (ERI); Busitema University (UGA); Houari Boumedienes University of Science \& Technol (DZA); Sidi-Bel-Abbes University (DZA); Busitema University (UGA); Djibouti Telecom S.a. (DJI); Liberia Telecommunication Corporation (LBR) \\
\hline
\textbf{Panel Asi: Asia} \\ \hline
\textit{Terms in False Positives:} ``Prince'', ``Bank'', ``Saud'', ``Geonet'', ``Chamber'', ``Commerce'' \\
\textit{Examples:} Prince ABDULLAH ALSAUD (SAU); Prince Khaled AL Saud (SAU); Prince Salman Al Saud (SAU); Bank Negara Indonesia (IDN); Trade Bank of iraq Jame3a (IRQ); State Bank of India (QAT); Prince ABDULLAH ALSAUD (SAU); Prince Khaled AL Saud (SAU) \\ [0.5em]
\textit{Terms in False Negatives:} ``University'', ``Telecom'', ``Dzongkhag'', ``Education'', ``Institute'', ``Oman'' \\
\textit{Examples:} Gerash University of Medical Sciences (IRN); Hadhramout University (YEM); SVU Syrian Virtual University (SYR); Alfa Telecom CJSC (KGZ); Bhutan Telecom Ltd (BTN); Ttelecom (TKM); For Sarpang Dzongkhag LAN Services (BTN); For Tsirang Dzongkhag LAN Services (BTN) \\
\hline
\textbf{Panel Eur: Europe} \\ \hline
\textit{Terms in False Positives:} ``Public'', ``Minsk'', ``City'', ``Stichting'', ``Miejskie'', ``Wodociagi'' \\
\textit{Examples:} BGA Public (HUN); BM Oktf Public (HUN); Public Services (MKD); Bktdvozrazhdenie Minsk (BLR); Belnotarialpalata Minsk (BLR); Centrekspertiz Minsk (BLR); AB City Service (LTU); ASO Paide City2 (EST) \\ [0.5em]
\textit{Terms in False Negatives:} ``University'', ``Carnet'', ``Network'', ``State'', ``Moldova'', ``Centre'' \\
\textit{Examples:} Network Operations Center, University of Athens (GRC); State University of Moldova (MDA); Computer Centre of State University of Moldova (MDA); Carnet Oslepoglava (HRV); Carnet Osvkovacica (HRV); Carnet Pskrapanj (HRV); ISJ Valcea Network (ROU); Network Operations Center, University of Athens (GRC) \\
\hline
\textbf{Panel Nor: North America} \\ \hline
\textit{Terms in False Positives:} ``Acceso'', ``Servidor'', ``Sinercon'', ``Aduanett'', ``Trigas'', ``Nicaragua'' \\
\textit{Examples:} CGR Acceso (CUB); Servidor Acceso Golfito (CRI); Servidor Acceso Cartago 900 (CRI); Servidor Acceso Golfito (CRI); Servidor Acceso Cartago 900 (CRI); Servidor Acceso Escazu 900 (CRI); Sinercon (DOM); Aduanett DEI (HND) \\ [0.5em]
\textit{Terms in False Negatives:} ``Network'', ``Eutel'', ``Customs'', ``Excise'', ``Reparaciones'', ``Equipos'' \\
\textit{Examples:} Virgin Islands Next Generation Network (VIR); server network GW (MSR); Eutel N.V. (BES); Eutel N.V (BES); Customs and Excise Department (BRB); Customs and Excise Department (BRB); UEB Reparaciones y Equipos Especializados Cubiza (CUB); UEB Reparaciones y Equipos Especializados Cubiza (CUB) \\
\hline
\textbf{Panel Sou: South America} \\ \hline
\textit{Terms in False Positives:} ``Empresa'', ``Servicios'', ``Camara'', ``Boliviana'', ``Hidrocarburos'', ``Corporacion'' \\
\textit{Examples:} Empresa De Servicios Sanitarios Del Bio-bio Essbio (CHL); Empresa Nacional de Servicios de Aseo S.A. (Enasa) (CHL); Empresa De Servicios Sanitarios Del Bio-bio Essbio (CHL); Empresa Nacional de Servicios de Aseo S.A. (Enasa) (CHL); Camara Boliviana De Hidrocarburos (BOL); Camara Boliviana De Hidrocarburos (BOL); Camara Boliviana De Hidrocarburos (BOL); Corporacion Recursos Naturales Corena (ECU) \\ [0.5em]
\textit{Terms in False Negatives:} ``Telesur'', ``Ecuador'', ``Estrategico'', ``Empresa'', ``Mtvc'', ``Municipalidad'' \\
\textit{Examples:} Telesur (SUR); Ecuador Estrategico Empresa (ECU); Ecuador Estrategico Empresa (ECU); Ecuador Estrategico Empresa (ECU); FR Mtvc (GUF); I. Municipalidad de Colina (CHL) \\
\hline
\textbf{Panel Oce: Oceania} \\ \hline
\textit{Terms in False Positives:} ``Pnord'', ``Translink'', ``Operations'', ``Jpmg'', ``Canl'' \\
\textit{Examples:} Pnord (NCL); Translink Operations Pty Ltd (AUS); Translink Operations Pty Ltd (AUS); PNL WS CLR Jpmg (WSM); Canl ASL Nc11 (NCL) \\ [0.5em]
\textit{Terms in False Negatives:} ``Fiji'', ``University'', ``Telecommunications'', ``Cloud'', ``Vangov'', ``Core'' \\
\textit{Examples:} Fiji National University (FJI); Fiji Cloud (FJI); Fiji National University (FJI); The University of the South Pacific (FJI); Telecommunications Tokelau Corporation - Teletok (TKL); FSM Telecommunications Corporation (FSM); Fiji Cloud (FJI); Naura Cloud (NRU) \\
\hline
\end{longtable}
\end{center}

\clearpage

\subsection{Locally Run Open-Weight Model Reproducibility Check}

The local open-weight comparison is diagnostic rather than part of the main DSC construction. It asks whether a future researcher could reproduce the broad government/non-government classification task with locally runnable models and the same prompt structure, without relying on proprietary APIs or task-specific fine-tuning. As shown in Table~\ref{tab:baseline_models}, agreement with the three-model benchmark ranges from 75.2 percent for DeepSeek R1 Distill 14B to 82.8 percent for Qwen 3 14B, with a five-model majority-vote ensemble achieving 82.7 percent agreement. These results suggest that locally run open-weight models can provide a useful reproducibility benchmark, but the main validation evidence remains the two random-sample audits described above.

\begin{table}[htbp]
  \centering
  \caption{Agreement of Locally Run Open-Weight Models with the Three-Model Consensus}
  \label{tab:baseline_models}
  \begin{tabular}{lccc}
  \hline\hline
  Model & Parameters & Agreement & Government-Class F1 \\
  \hline
  GPT-OSS & 20B & 82.7 & 79.0 \\
  Mistral Small 3.1 & 24B & 79.1 & 77.2 \\
  Qwen 3 & 14B & 82.8 & 80.8 \\
  DeepSeek R1 Distill & 14B & 75.2 & 81.2 \\
  Gemma 3 & 12B & 76.6 & 70.5 \\
  \hline
  Majority Vote (3/5) & -- & 82.7 & -- \\
  \hline\hline
  \end{tabular}
  \begin{tablenotes}
  \small
  \item \textit{Note:} Agreement of locally run, off-the-shelf open-weight models with the two-of-three consensus from GPT-5.1, Claude Sonnet 4.5, and Gemini 3 Pro. The evaluation sample contains 500 entities, balanced between government and non-government according to the original labels. Models receive the entity name, country, and available snippet context. Metrics are calculated over parseable predictions, and government-class F1 treats government as the positive class. No model received task-specific fine-tuning.
  \end{tablenotes}
  \end{table}

\clearpage
\section{IP Details}

\subsection{Example IP Owner Information in DB-IP}
 \begin{table}[h!]
      \centering
      \begin{tabular}{@{}ll@{}}
          \toprule
          \textbf{Field}        & \textbf{Value}                  \\ \midrule
          ipStart               & 198.51.100.0                    \\
          ipEnd                 & 198.51.100.255                  \\
          continentCode         & NA                              \\
          continentName         & North America                   \\
          countryCode           & US                              \\
          countryName           & United States                   \\
          isEuMember            & False                           \\
          currencyCode          & USD                             \\
          currencyName          & Dollar                          \\
          phonePrefix           & 1                               \\
          languages             & en-US                           \\
          stateProvCode         & DC                              \\
          stateProv             & District of Columbia            \\
          district              & Northwest                       \\
          city                  & Washington                      \\
          geonameId             & 4140963                         \\
          zipCode               & 20500                           \\
          gmtOffset             & -5                              \\
          timeZone              & America/New\_York               \\
          latitude              & 38.8977                         \\
          longitude             & -77.0365                        \\
          weatherCode           & USDC0001                        \\
          asNumber              & 98765                           \\
          asName                & EXAMPLE-GOV-AS                  \\
          isp                   & Example Federal Network         \\
          usageType             & Government                      \\
          organization          & Example Executive Office        \\
          isCrawler             & False                           \\
          isProxy               & False                           \\
          threatLevel           & Low                             \\ \bottomrule
      \end{tabular}
      \caption{Example DB-IP Information for White House IP Address}
      \label{tab:ip-details}
  \end{table}

\clearpage

\subsection{Unambiguous Government ISP/Telecom Entities}
\label{sec:govisp}
\begin{enumerate}
    \item DITCO (Defense Information Technology Contracting Organization)
    \item DoD Network Information Center (Department of Defense Network Information Center)
    \item DOI/ESN (Department of the Interior/Enterprise Services Network)
    \item Federal Network Serv (Government Federal Network Services)
    \item USDA Telephone Invoices (USDA related network)
    \item Florida Information Resource Network (Florida state-run network)
    \item ESnet (Energy Sciences Network, U.S. government-funded research network)
    \item State of Tennesse-nettn (State of Tennessee Network)
    \item Connecticut Education Network (Connecticut state-run education network)
    \item California Public Utilities Commission (Government body overseeing utilities in California)
    \item Federal Communications Commission (U.S. government regulatory agency)
    \item CALNET2 Dept of General Services (California State Government network service)
    \item CALNET State of California CalTrans (California state network service for transportation)
    \item CALNET County of San Diego (San Diego County state network)
    \item CALNET County of Sacramento (Sacramento County state network)
    \item CALNET City of Madera (City of Madera California state network)
    \item CALNET City of Long Beach Data Center (Long Beach California state network service)
    \item GARR CSD BA (GARR, Italy government-run research and academic network)
    \item GARR Italian Research and Academic Network (Italian government-run research network)
    \item GIP RENATER (Government-funded RENATER network)
    \item RENATER Services (French government academic and research network)
    \item Government Service Network (A general government service network)
    \item Government Communications Network (General government communications network)
    \item Ente Publico De Radiotelevision De Les Illes Balea (Public broadcaster in Spain, Balearic Islands)
    \item Ente Publico Radio Television Canaria S.A. (Public broadcaster in Spain, Canary Islands)
    \item Instituto Cubano de Radio y Television ICRT (Cuban government-run radio and TV)
    \item Federal Radio Serv (U.S. Federal radio network services)
    \item State Tele-Radio Company (Russian state-run media organisation)
    \item TVRI (Indonesia's state-run television network)
    \item Government of Canada (Canadian government communications network)
    \item Ministry of ICT (Government ministry for information and communications technology)
    \item National Institute of Information and Communication (Government body for information and communication)
    \item Department of Transport and Communications (Various countries' government departments)
    \item Korean Central Government Communications Network (Government communications network in South Korea)
    \item Public Internet Access (Government-run public internet services)
\end{enumerate}

\section{Regional Internet Registries (RIR) Policies}
\label{app:RIRpol}
\subsection{LACNIC}

\subsubsection{Important Information for Members}
LACNIC offers important guidance for its members \cite{lacnicImportantInfo}.

\subsubsection{IPv4 Allocation Policy (Approved 06/08/2002)}
LACNIC’s IPv4 allocation policy was approved on 06/08/2002 \cite{lacnicIPv4Policy2002}.

\subsubsection{Policy Document: Section 3.2.5}
According to LACNIC’s official policy \cite{lacnicPolicyC3}:
\begin{quotation}
``Internet Registries shall use the group of IP addresses they have been allocated in an efficient manner. To this end, IRs shall document the justification for each IP address reallocation. 
... The documentation LACNIC may require includes: Engineering plans, subnetting and aggregation plan, 
description of network topology, description of network routing plans, etc.''
\end{quotation}

\subsection{AFRINIC}

\subsubsection{Registrant Member Contract (2005)}
AFRINIC created its registrant member contract in 2005 \cite{afrinicContract2005}. 
Key clauses include:
\begin{itemize}
  \item Section 2(b): ``Provide accurate and complete information...''
  \item Section 2(d): ``Any changes must be notified promptly...''
  \item Section 2(g): ``Provide and ensure accurate contact information in AFRINIC databases...''
\end{itemize}

\subsubsection{Membership Requirement}
To request IP resources, AFRINIC membership is mandatory \cite{afrinicMembership}.

\subsection{ARIN}

\subsubsection{Registration Services Agreement (RSA, 1997)}
ARIN’s RSA imposes several responsibilities \cite{arinRSA1997}:
\begin{itemize}
  \item Section 2(c): ``Holder must... promptly provide ARIN with complete and accurate information...''
  \item Section 3(a): ``Holder will promptly provide ARIN with accurate documentation...''
  \item Section 14(a): ``Holder may not assign or transfer... this Agreement... without ARIN’s prior written permission.''
\end{itemize}

\subsubsection{Resource Request Requirements}
ARIN does not require membership to request Internet number resources. Requests require an ARIN Online account linked to an authorised contact for a valid organisation identifier, and applicants must satisfy the relevant eligibility and documentation requirements \parencite{arinMembership, ARINResourceRequests}.

\subsection{RIPE NCC}

\subsubsection{Membership Requirement}
As stated by the RIPE NCC, ``You need to become a member to directly request Internet number resources from us.'' 
\cite{ripeMembership}

\subsubsection{RIPE-185 (26 October 1998)}
RIPE-185 \cite{ripe185} covers:
\begin{itemize}
  \item \textbf{3.2.1.1. Overview of Organisation}: Understanding organisational structure is critical.
  \item \textbf{3.2.1.2. Current Assignment Space Usage}: Past usage must be documented.
  \item \textbf{3.2.1.4. Addressing Plan}: A detailed addressing plan is required.
  \item \textbf{3.2.1.5. Database Information}: Registration details (organisation, persons) are needed.
  \item \textbf{3.2.2. Verification Information}: Verification of the requester’s network planning can be required.
\end{itemize}

\subsection{Completeness of Data}
\paragraph{Historical Allocation}
\textcite{apnic-rex} aggregates delegated-statistics files and transfer logs from the five RIRs to describe historical resource distributions \parencite{apnic-transfer-logs}. These sources do not provide a complete transaction history. APNIC documents that some transfer dates are inferred from changes between historical files when exact dates are unavailable \parencite{apnic-rex-infers}. The NRO governance framework records due-diligence and audit procedures across the RIRs, but these procedures do not eliminate error \parencite{nro-governance}. RIPE's registry data-quality assessment identifies particular limitations for legacy resources \parencite{ripe-data-quality}. AFRINIC's audit of its WHOIS database found that millions of IPv4 addresses had been attributed to organisations without justification, after which AFRINIC introduced remedial controls and additional verification procedures \parencite{arin-afrinic-audit, afrinic-ceo-statement}. The RIR statistics exchange format records a date and status for each allocation or assignment, but it is a snapshot of current registry status rather than a complete transaction history \parencite{apnic-record-format}. We therefore do not treat historical registration dates or registrant histories as error-free.

\section{Raw Government IP Statistics}
\label{app:raw_ip}
\begin{figure}
    \centering
    \includegraphics[width=\linewidth]{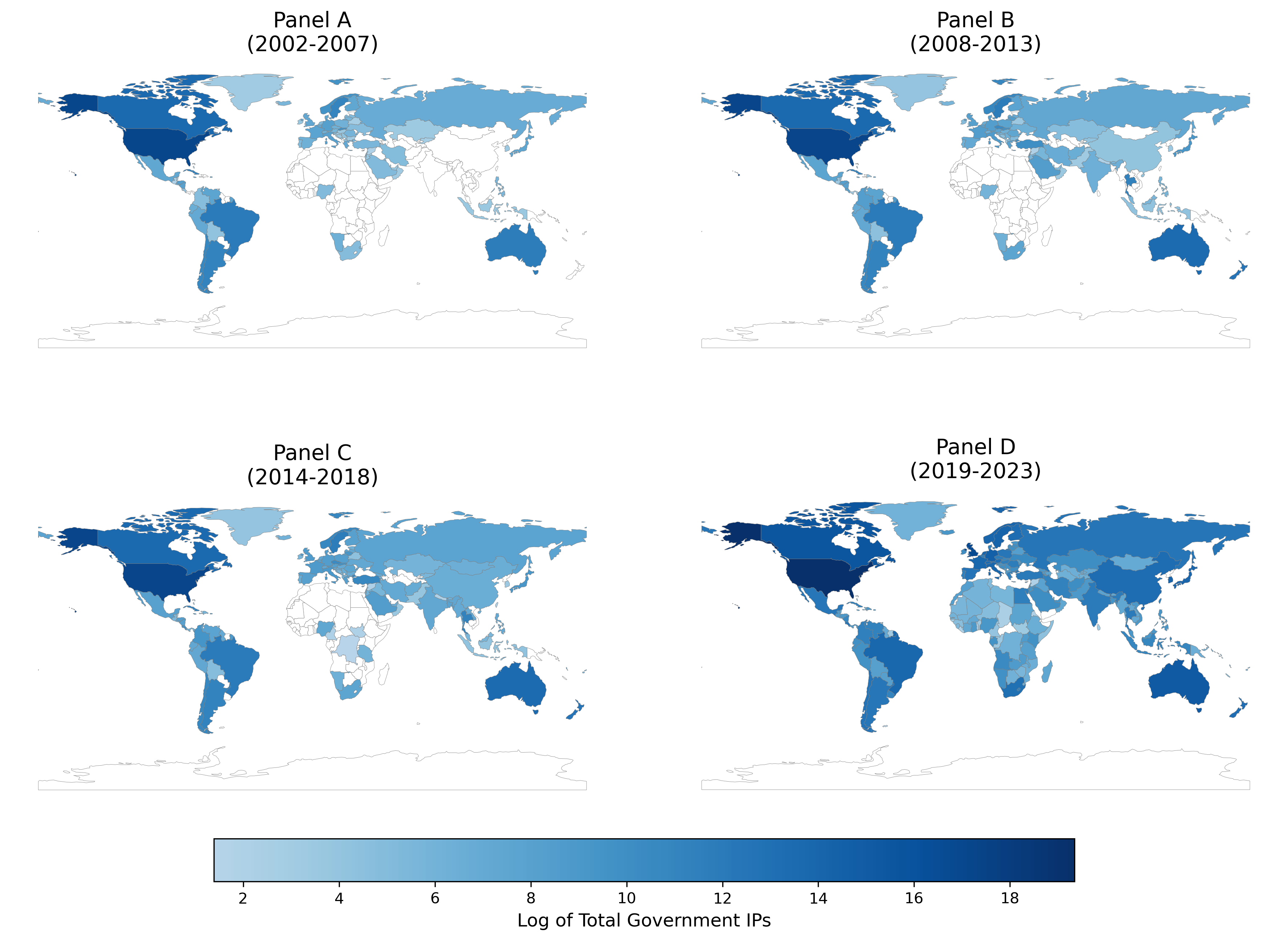}
    \caption{Map of Total Government IPs at the Country Level by Period}
    \label{fig:raw_map}
    \fignote{Country panels show country-period means of the annual log of raw government IP counts, without population or public-employment normalisation. A country is counted in a panel if it has positive raw government IP holdings in at least one country-year in that period: 93, 113, 132, and 219 countries in Panels A–D. This raw-count denominator is broader than the normalised national panels and the population-covered subnational sample.}
\end{figure}

\clearpage

\begin{figure}[htbp]
    \centering
    \includegraphics[width=\linewidth,height=0.76\textheight,keepaspectratio]{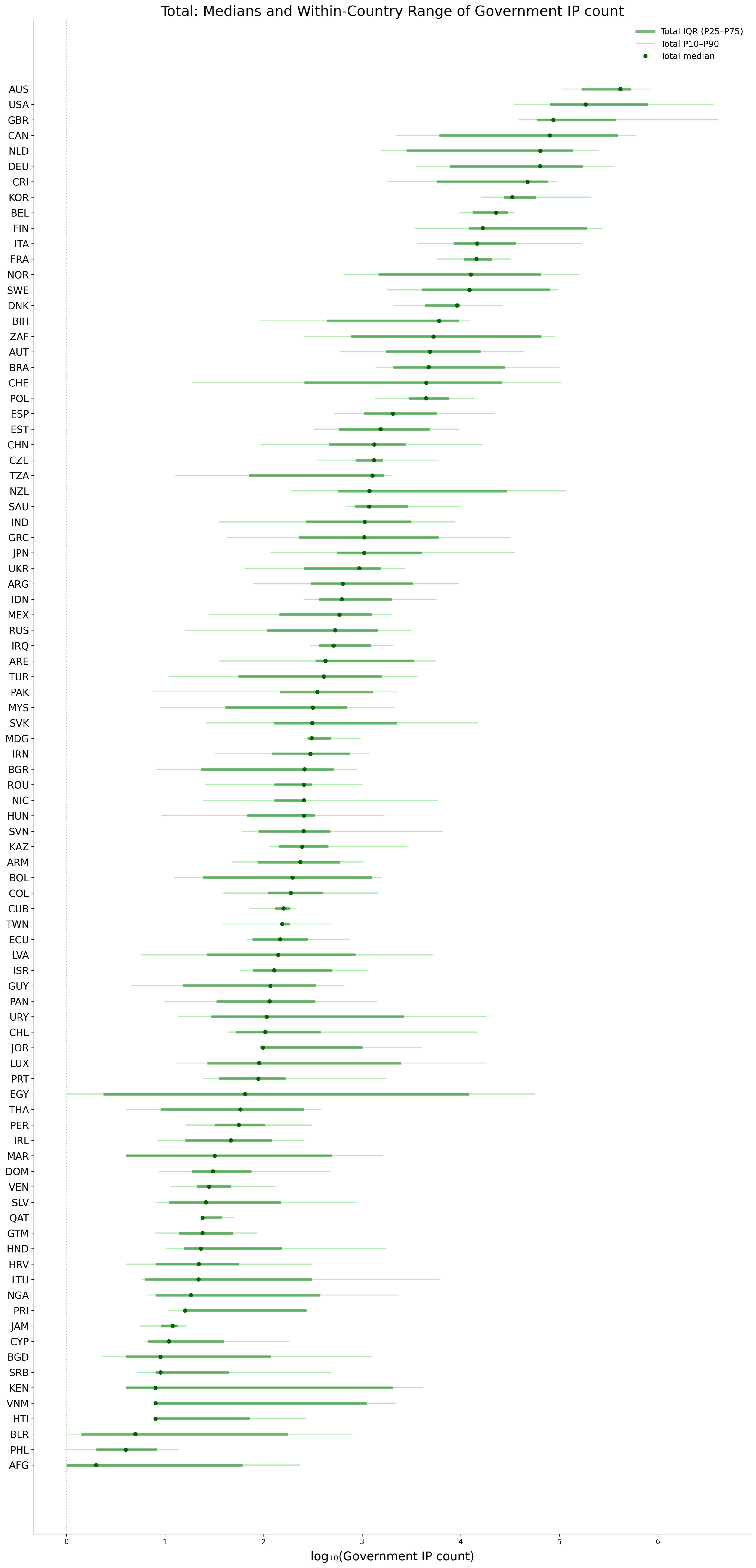}
    \caption{Within-country distribution of total government IP addresses (not population-adjusted). Each row shows a country's median (dots), interquartile range (thick bars), and 10th–90th percentile range (thin lines) of $\log_{10}(\text{Gov IP Count})$ across Admin-1 regions. Countries are sorted by ascending median. Coverage: 163 countries with data; 90 countries with $\geq$3 regions plotted; 1,314 Admin-1 regions.}
    \label{fig:caterpillar_total}
\end{figure}

\clearpage

\begin{table}[ht]
\centering
\small
\caption{Pearson Correlations with Government IP Counts and Log Government IP Counts, 2008--2023}
\begin{adjustbox}{max width=\linewidth,max totalheight=0.78\textheight}

\begin{tabular}{lcccc}
\toprule
\textbf{Variable} & \textbf{Year Range} & \textbf{Gov IP Count} & \textbf{Log(Gov IP Count)} & \textbf{N} \\
\midrule
\multicolumn{5}{l}{\textbf{Polity}} \\
Polity V Score & 2008--2023 & 0.02 & 0.24*** & 1703 \\
\addlinespace
\multicolumn{5}{l}{\textbf{VDEM}} \\
Civ. Society Partic. & 2008--2022 & 0.09*** & 0.21*** & 1733 \\
Female Civ. Society Partic. & 2008--2022 & 0.03 & 0.26*** & 1733 \\
Govt Cyber Security & 2008--2022 & 0.09*** & 0.38*** & 1733 \\
Govt Internet Filter Cap. & 2008--2022 & 0.04* & 0.09*** & 1733 \\
Govt Internet Shut Cap. & 2008--2022 & -0.02 & -0.10*** & 1733 \\
Govt Online Reg Cap. & 2008--2022 & 0.06** & 0.26*** & 1733 \\
Multi Party Elect. & 2008--2022 & 0.05** & 0.18*** & 1656 \\
Pres. Elec. Consective & 2008--2022 & 0.24*** & 0.17*** & 1733 \\
\addlinespace
\multicolumn{5}{l}{\textbf{World Bank}} \\
Access to Electricity (\%) & 2019--2020 & 0.05 & 0.39*** & 390 \\
Exports Growth (\%) & 2019--2020 & -0.01 & -0.04 & 303 \\
Exports of Goods/Services (\% GDP) & 2019--2020 & -0.08 & 0.08 & 331 \\
Inflation Rate (\%) & 2019--2020 & -0.01 & -0.12** & 321 \\
Internet Users (\%) & 2008--2020 & 0.07*** & 0.38*** & 1488 \\
Log GDP per Capita & 2008--2023 & 0.09*** & 0.28*** & 2150 \\
Log Population & 2008--2023 & 0.11*** & 0.42*** & 2150 \\
Oil Rents (\% GDP) & 2019--2020 & -0.03 & -0.05 & 356 \\
Tariff Rate (Simple Mean) & 2008--2020 & -0.02 & -0.26*** & 1267 \\
Urban Population (\%) & 2008--2023 & 0.05** & 0.27*** & 2202 \\
\bottomrule
\end{tabular}

\end{adjustbox}
\tabnote{This table reports pairwise correlations with raw government IP counts and their natural logarithm. Neither measure is normalised by population or public-sector employment.}
\label{tab:08raw_govip1}
\end{table}

\begin{table}[ht]
\centering
\small
\caption{Pearson Correlations with Gov IP Count and Log(Gov IP Count), 2008–2023}
\begin{adjustbox}{max width=\linewidth,max totalheight=0.78\textheight}

\begin{tabular}{lcccc}
\toprule
\textbf{Variable} & \textbf{Year} & \textbf{Gov IP Count} & \textbf{Log(Gov IP Count)} & \textbf{N} \\
\midrule
\multicolumn{5}{l}{\textbf{Brambor et al.}} \\
Info Capacity & 2008--2015 & -0.13** & -0.04 & 402 \\
\addlinespace
\multicolumn{5}{l}{\textbf{Hanson Sigman}} \\
State Capacity & 2008--2015 & 0.13*** & 0.46*** & 688 \\
\addlinespace
\multicolumn{5}{l}{\textbf{Letter Grading}} \\
Avg Days & 2010 & -0.17 & -0.26** & 86 \\
Returned & 2010 & 0.11 & 0.22** & 86 \\
\addlinespace
\multicolumn{5}{l}{\textbf{Statistical Capacity WB}} \\
Data Infrastructure & 2016--2023 & 0.11*** & 0.59*** & 1213 \\
Data Products & 2008--2023 & 0.05** & 0.40*** & 2206 \\
Data Services & 2016--2023 & 0.09*** & 0.56*** & 1190 \\
Data Sources & 2015--2023 & 0.11*** & 0.59*** & 1246 \\
Data Use & 2008--2023 & 0.07*** & 0.45*** & 2206 \\
Statistical Index & 2016--2023 & 0.11*** & 0.64*** & 1186 \\
\addlinespace
\multicolumn{5}{l}{\textbf{UN E-Gov Indicators}} \\
E-Gov Index & 2008--2022 & 0.11*** & 0.55*** & 966 \\
E-Gov Rank & 2008--2022 & -0.09*** & -0.43*** & 966 \\
E-Participation Index & 2008--2022 & 0.11*** & 0.50*** & 966 \\
Human Capital Index & 2008--2022 & 0.06** & 0.36*** & 966 \\
Online Service Index & 2008--2022 & 0.11*** & 0.54*** & 966 \\
Telecom Infra Index & 2008--2022 & 0.10*** & 0.47*** & 966 \\
\addlinespace
\multicolumn{5}{l}{\textbf{VDEM}} \\
Impartial Pub. Admn. & 2008--2015 & 0.13*** & 0.42*** & 724 \\
State Auth. Terr. & 2008--2015 & 0.00 & 0.12*** & 724 \\
State Fiscal Cap. & 2008--2015 & 0.14*** & 0.39*** & 724 \\
\addlinespace
\multicolumn{5}{l}{\textbf{Worldwide Governance}} \\
Ctrl Corruption & 2008--2023 & 0.07*** & 0.30*** & 2075 \\
Gov Effectiveness & 2008--2023 & 0.08*** & 0.33*** & 2075 \\
Political Stability & 2008--2023 & 0.01 & 0.14*** & 2075 \\
Regulatory Quality & 2008--2023 & 0.08*** & 0.34*** & 2075 \\
Rule of Law & 2008--2023 & 0.09*** & 0.31*** & 2075 \\
Voice \& Accountability & 2008--2023 & 0.06*** & 0.25*** & 2075 \\
\bottomrule
\end{tabular}

\end{adjustbox}
\tabnote{This table shows pairwise correlations with related measures of state capacity, informational capacity, and e-government for 2008–2023.}
\label{tab:19raw_govip1}
\end{table}

\clearpage
\begin{table}[ht]
\centering
\small
\caption{Pearson Correlations with Gov IP Count and Log(Gov IP Count), 2019–2023}
\begin{adjustbox}{max width=\linewidth,max totalheight=0.78\textheight}

\begin{tabular}{lcccc}
\toprule
\textbf{Variable} & \textbf{Year Range} & \textbf{Gov IP Count} & \textbf{Log(Gov IP Count)} & \textbf{N} \\
\midrule
\multicolumn{5}{l}{\textbf{Polity}} \\
Polity V Score & 2019--2023 & 0.02 & 0.31*** & 746 \\
\addlinespace
\multicolumn{5}{l}{\textbf{VDEM}} \\
Civ. Society Partic. & 2019--2022 & 0.12*** & 0.26*** & 661 \\
Female Civ. Society Partic. & 2019--2022 & 0.05 & 0.34*** & 661 \\
Govt Cyber Security & 2019--2022 & 0.11*** & 0.56*** & 661 \\
Govt Internet Filter Cap. & 2019--2022 & 0.04 & 0.10** & 661 \\
Govt Internet Shut Cap. & 2019--2022 & -0.05 & -0.23*** & 661 \\
Govt Online Reg Cap. & 2019--2022 & 0.08* & 0.38*** & 661 \\
Multi Party Elect. & 2019--2022 & 0.08** & 0.26*** & 628 \\
Pres. Elec. Consective & 2019--2022 & 0.30*** & 0.20*** & 661 \\
\addlinespace
\multicolumn{5}{l}{\textbf{World Bank}} \\
Access to Electricity (\%) & 2019--2020 & 0.05 & 0.39*** & 390 \\
Exports Growth (\%) & 2019--2020 & -0.01 & -0.04 & 303 \\
Exports of Goods/Services (\% GDP) & 2019--2020 & -0.08 & 0.08 & 331 \\
Inflation Rate (\%) & 2019--2020 & -0.01 & -0.12** & 321 \\
Internet Users (\%) & 2019--2020 & 0.10* & 0.61*** & 293 \\
Log GDP per Capita & 2019--2023 & 0.13*** & 0.41*** & 940 \\
Log Population & 2019--2023 & 0.14*** & 0.50*** & 940 \\
Oil Rents (\% GDP) & 2019--2020 & -0.03 & -0.05 & 356 \\
Tariff Rate (Simple Mean) & 2019--2020 & -0.04 & -0.47*** & 283 \\
Urban Population (\%) & 2019--2023 & 0.08** & 0.36*** & 975 \\
\bottomrule
\end{tabular}

\end{adjustbox}
\tabnote{This table shows pairwise correlations with related measures of state capacity, informational capacity, and e-government for 2019–2023.}
\label{tab:08raw_govip2}
\end{table}

\clearpage

\begin{table}[ht]
\centering
\small
\caption{Pearson Correlations with Gov IP Count and Log(Gov IP Count), 2019–2023}
\begin{adjustbox}{max width=\linewidth,max totalheight=0.78\textheight}

\begin{tabular}{lcccc}
\toprule
\textbf{Variable} & \textbf{Year} & \textbf{Gov IP Count} & \textbf{Log(Gov IP Count)} & \textbf{N} \\
\midrule
\multicolumn{5}{l}{\textbf{Statistical Capacity WB}} \\
Data Infrastructure & 2019--2023 & 0.13*** & 0.65*** & 877 \\
Data Products & 2019--2023 & 0.05* & 0.47*** & 979 \\
Data Services & 2019--2023 & 0.10*** & 0.59*** & 861 \\
Data Sources & 2019--2023 & 0.12*** & 0.64*** & 857 \\
Data Use & 2019--2023 & 0.08** & 0.54*** & 979 \\
Statistical Index & 2019--2023 & 0.12*** & 0.68*** & 857 \\
\addlinespace
\multicolumn{5}{l}{\textbf{UN E-Gov Indicators}} \\
E-Gov Index & 2020--2022 & 0.13** & 0.68*** & 355 \\
E-Gov Rank & 2020--2022 & -0.13** & -0.68*** & 355 \\
E-Participation Index & 2020--2022 & 0.14*** & 0.67*** & 355 \\
Human Capital Index & 2020--2022 & 0.10* & 0.58*** & 355 \\
Online Service Index & 2020--2022 & 0.13** & 0.69*** & 355 \\
Telecom Infra Index & 2020--2022 & 0.11** & 0.58*** & 355 \\
\addlinespace
\multicolumn{5}{l}{\textbf{Worldwide Governance}} \\
Ctrl Corruption & 2019--2023 & 0.10*** & 0.37*** & 916 \\
Gov Effectiveness & 2019--2023 & 0.11*** & 0.45*** & 916 \\
Political Stability & 2019--2023 & 0.01 & 0.16*** & 916 \\
Regulatory Quality & 2019--2023 & 0.12*** & 0.47*** & 916 \\
Rule of Law & 2019--2023 & 0.12*** & 0.39*** & 916 \\
Voice \& Accountability & 2019--2023 & 0.08** & 0.31*** & 916 \\
\bottomrule
\end{tabular}

\end{adjustbox}
\tabnote{This table shows pairwise correlations with related measures of state capacity, informational capacity, and e-government for 2019–2023.}
\label{tab:19raw_govip2}
\end{table}

\clearpage
\section{Additional Subnational, Level, and Function Descriptives}
\label{app:additional_level_function_descriptives}

\begin{figure}[ht]
    \centering
    \caption{Top 10 Democracies (Federal): Regional Government IP Allocations per Capita}
    \includegraphics[width=\linewidth]{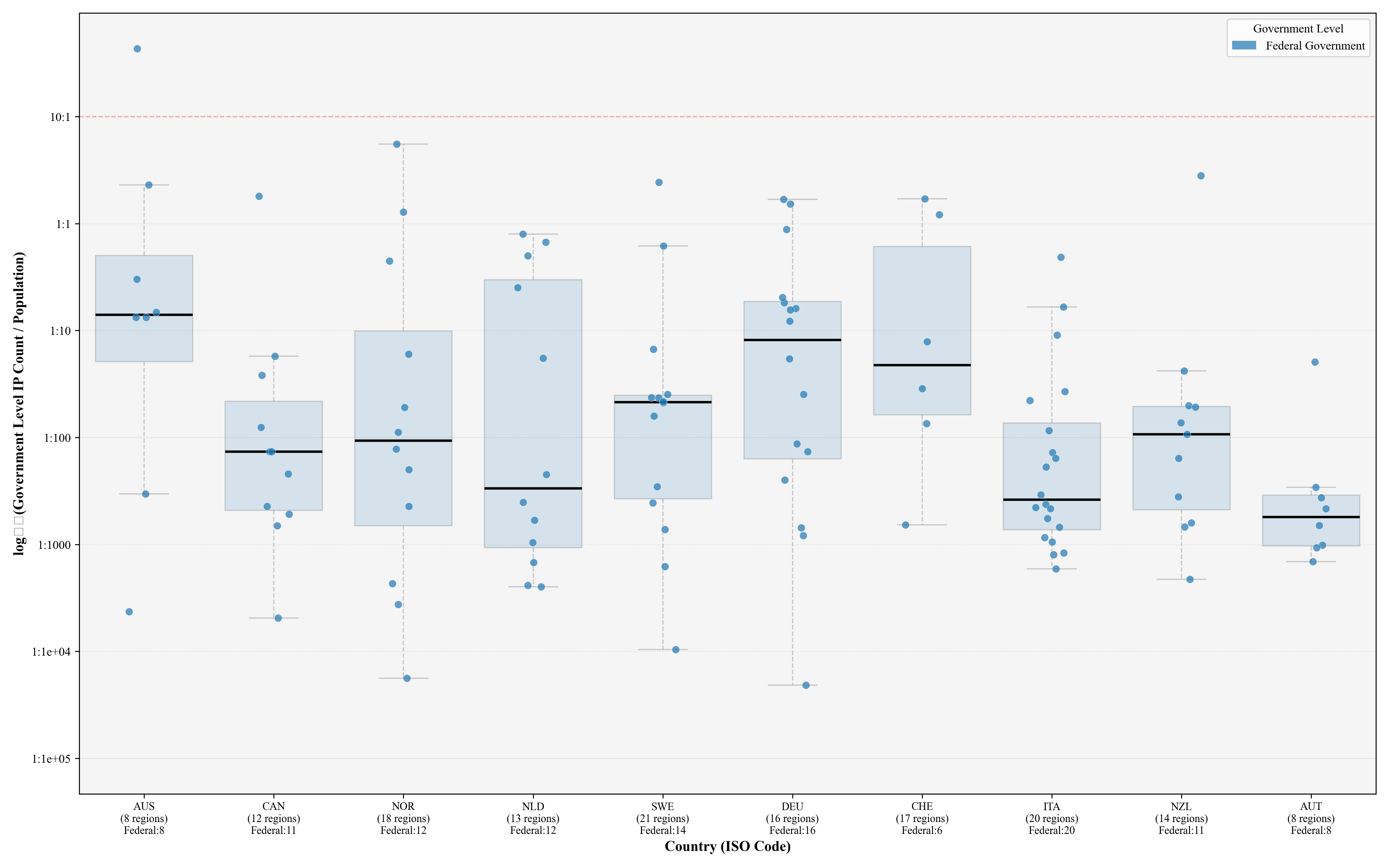}
    \fignote{Government IP addresses per capita are plotted on a base-10 logarithmic scale for federal-level regional governments in the top ten democracies by total government IP count. Box plots summarise countries; coloured dots represent regions.}
    \label{fig:levels_federal_democracy_logpop}
\end{figure}

\clearpage

\begin{figure}[ht]
    \centering
    \caption{Top 10 Democracies (Local): Regional Government IP Allocations per Capita}
    \includegraphics[width=\linewidth]{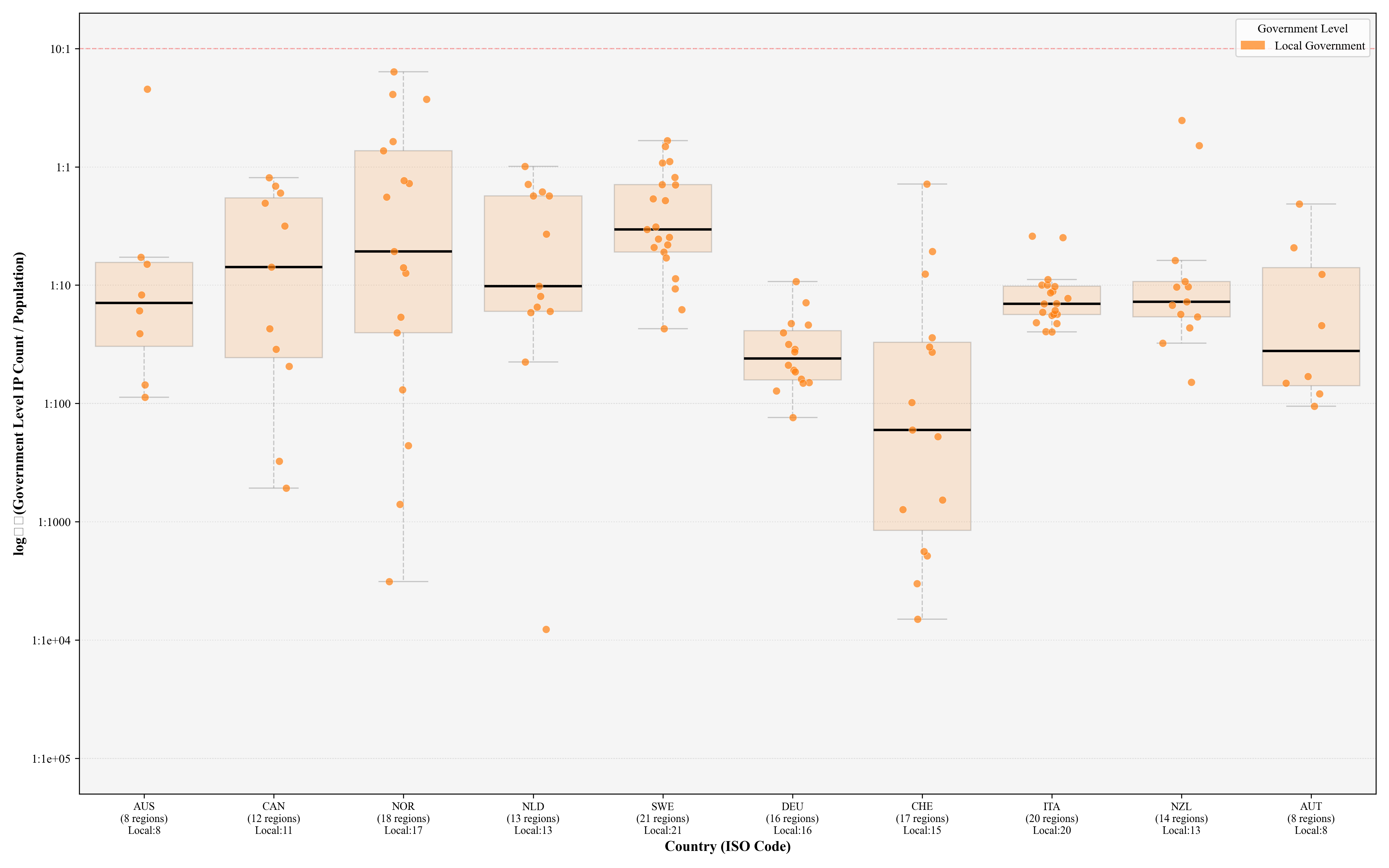}
    \fignote{Government IP addresses per capita are plotted on a base-10 logarithmic scale for local-level regional governments in the top ten democracies. Box plots summarise countries; coloured dots represent regions.}
    \label{fig:levels_local_democracy_logpop}
\end{figure}

\clearpage

\begin{figure}[ht]
    \centering
    \caption{Top 10 Non-Democracies (Federal and Local): Regional Government IP Allocations per Capita}
    \includegraphics[width=\linewidth]{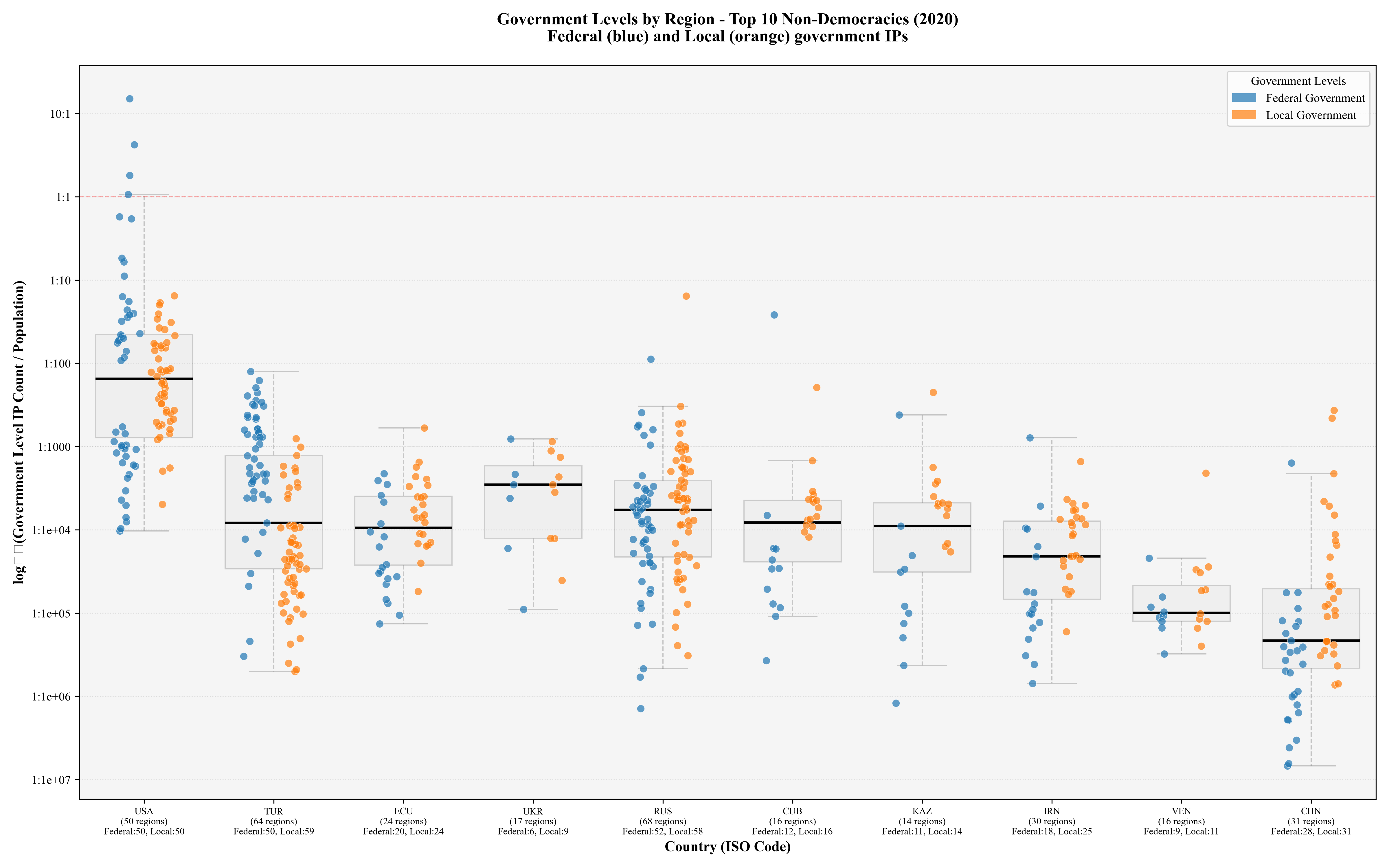}
    \fignote{Government IP addresses per capita are plotted on a base-10 logarithmic scale for federal- and local-level regional governments in the top ten non-democracies. Box plots summarise countries; coloured dots represent regions.}
    \label{fig:levels_nodemocracy_logpop}
\end{figure}

\clearpage

\begin{figure}[htbp]
    \centering
    \includegraphics[width=\linewidth,height=0.76\textheight,keepaspectratio]{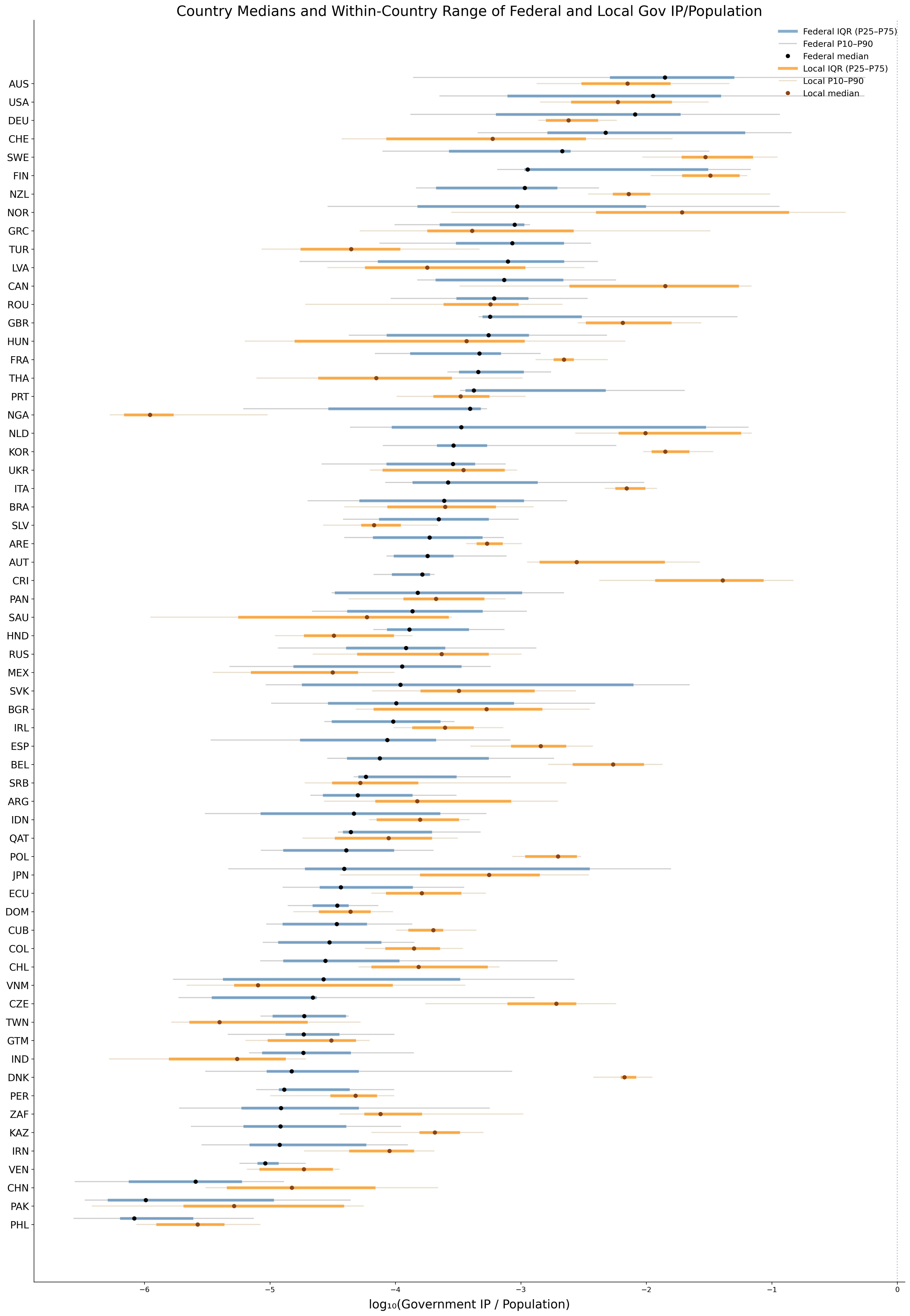}
    \caption{Within-country distribution of federal and local government IP addresses per capita. Each row shows a country's median (dots), interquartile range (thick bars), and 10th–90th percentile range (thin lines) of $\log_{10}(\text{Gov IP}/\text{Population})$ across Admin-1 regions. Federal values are shown in blue and local values in orange. Countries are sorted by ascending federal median. Coverage: 63 countries with $\geq$3 regions; 718 Admin-1 regions (federal), 1,121 regions (local).}
    \label{fig:caterpillar_joint}
\end{figure}

\clearpage

\begin{figure}[htbp]
    \centering
    \includegraphics[width=\linewidth,height=0.76\textheight,keepaspectratio]{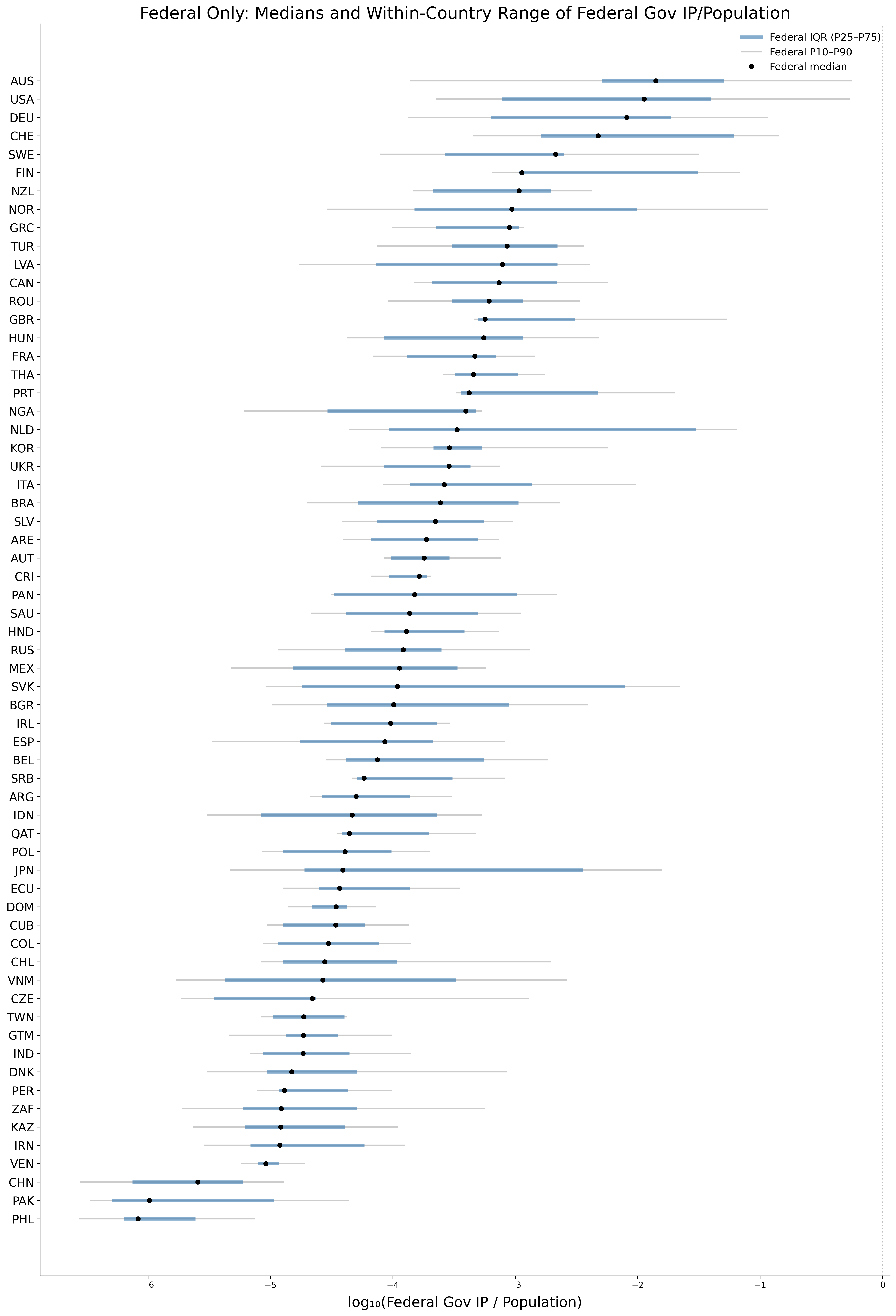}
    \caption{Within-country distribution of federal government IP addresses per capita. Each row shows a country's median (dots), interquartile range (thick bars), and 10th–90th percentile range (thin lines) of $\log_{10}(\text{Federal Gov IP}/\text{Population})$ across Admin-1 regions. Countries are sorted by ascending median. Coverage: 146 countries with federal data; 63 countries with $\geq$3 regions plotted; 718 Admin-1 regions.}
    \label{fig:caterpillar_federal}
\end{figure}

\clearpage

\begin{figure}[htbp]
    \centering
    \includegraphics[width=\linewidth,height=0.76\textheight,keepaspectratio]{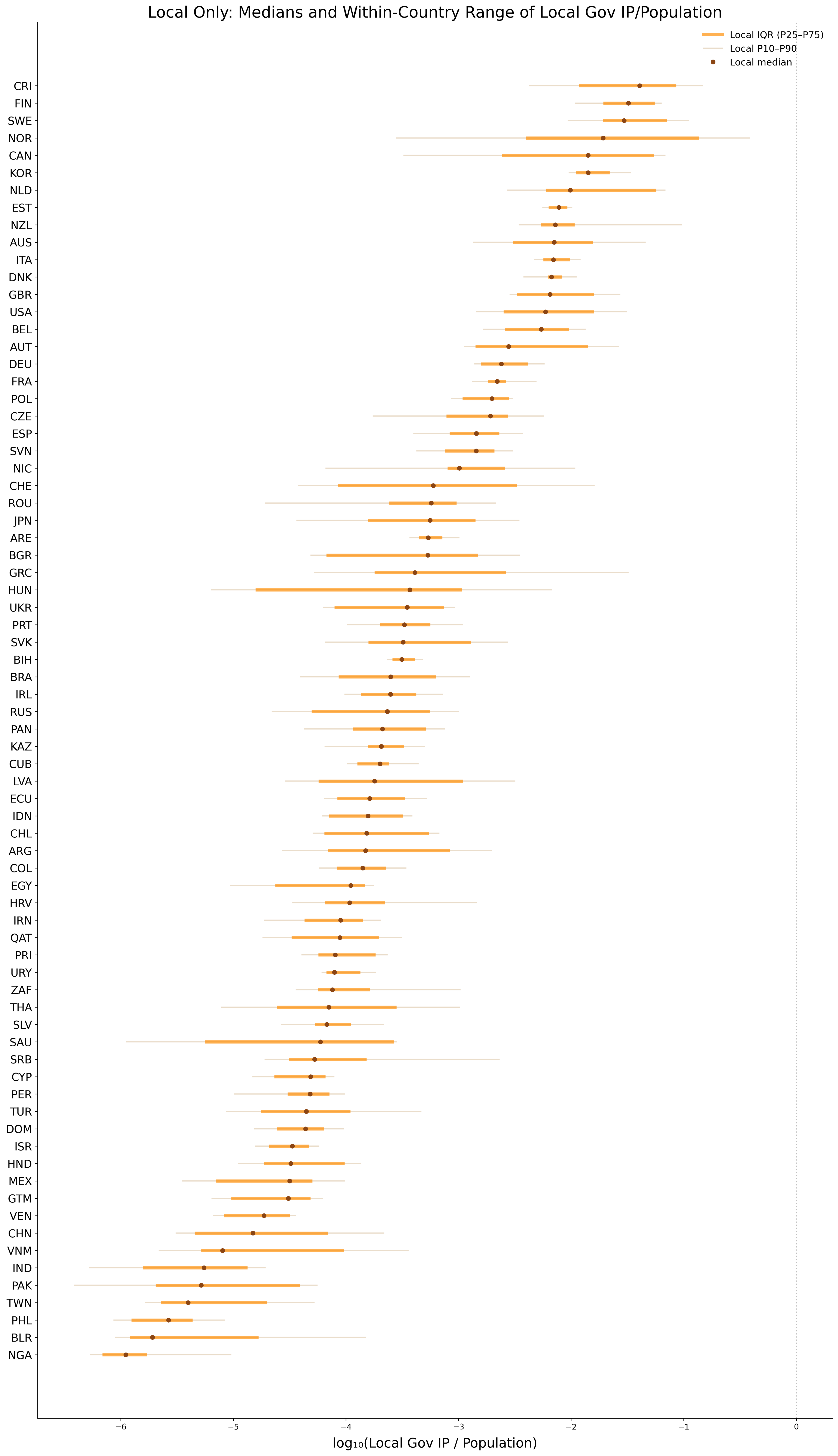}
    \caption{Within-country distribution of local government IP addresses per capita. Each row shows a country's median (dots), interquartile range (thick bars), and 10th–90th percentile range (thin lines) of $\log_{10}(\text{Local Gov IP}/\text{Population})$ across Admin-1 regions. Countries are sorted by ascending median. Coverage: 144 countries with local data; 74 countries with $\geq$3 regions plotted; 1,121 Admin-1 regions.}
    \label{fig:caterpillar_local}
\end{figure}

\clearpage

\begin{figure}[ht]
    \centering
    \caption{Top 10 Democracies: Regional Government IPs per Capita by COFOG Function}
    \includegraphics[width=\linewidth]{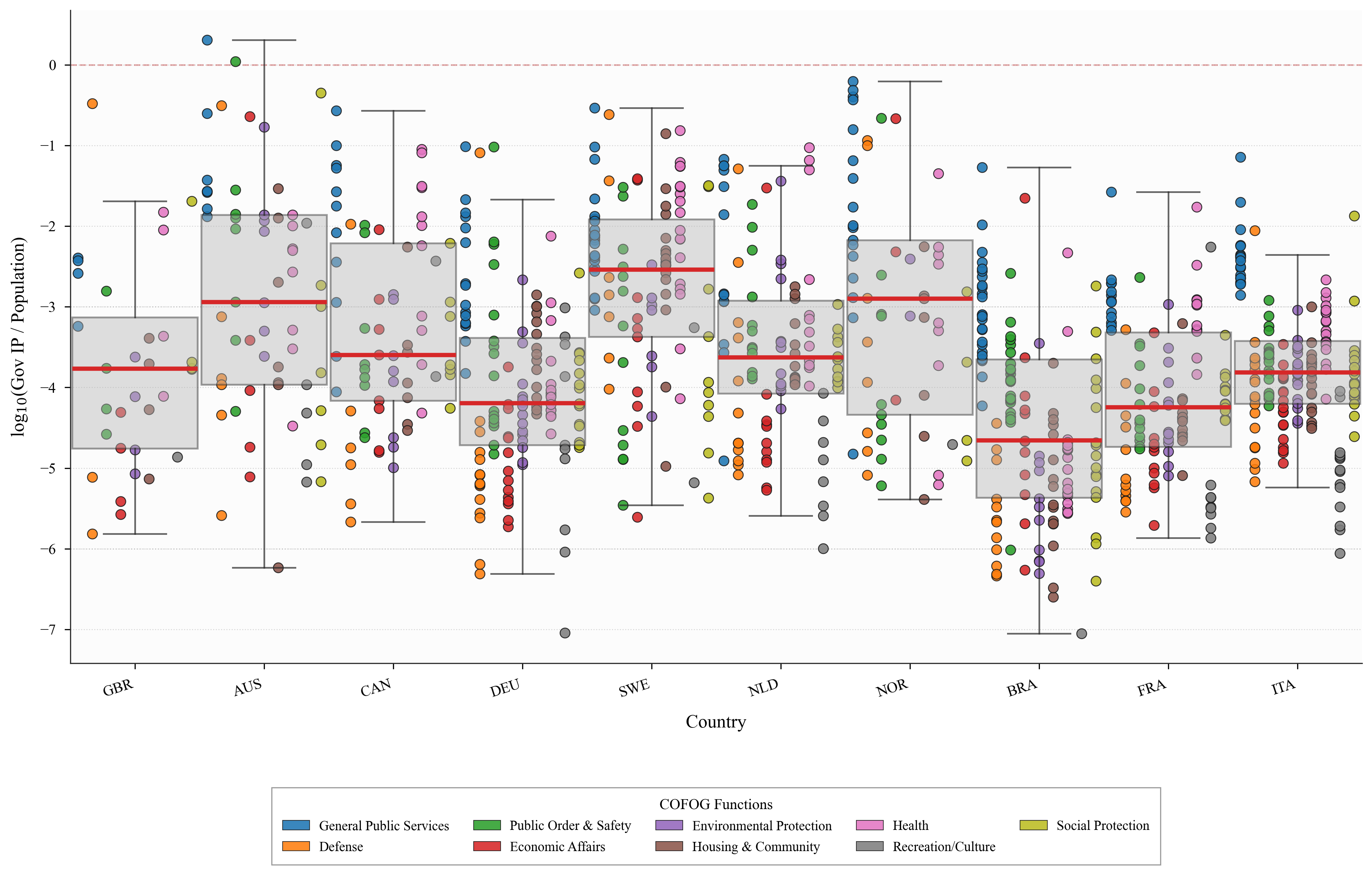}
    \fignote{Each country's subnational governments are plotted using $\log_{10}(\text{Gov IP}/\text{Population})$ for each COFOG function (coloured points), with overlaid box plots summarising all functions per country. The legend displays functions.}
    \label{fig:cofog_strip_box_democracy_logpop}
\end{figure}

\clearpage

\begin{figure}[ht]
    \centering
    \caption{Top 10 Non-Democracies: Regional Government IPs per Capita by COFOG Function}
    \includegraphics[width=\linewidth]{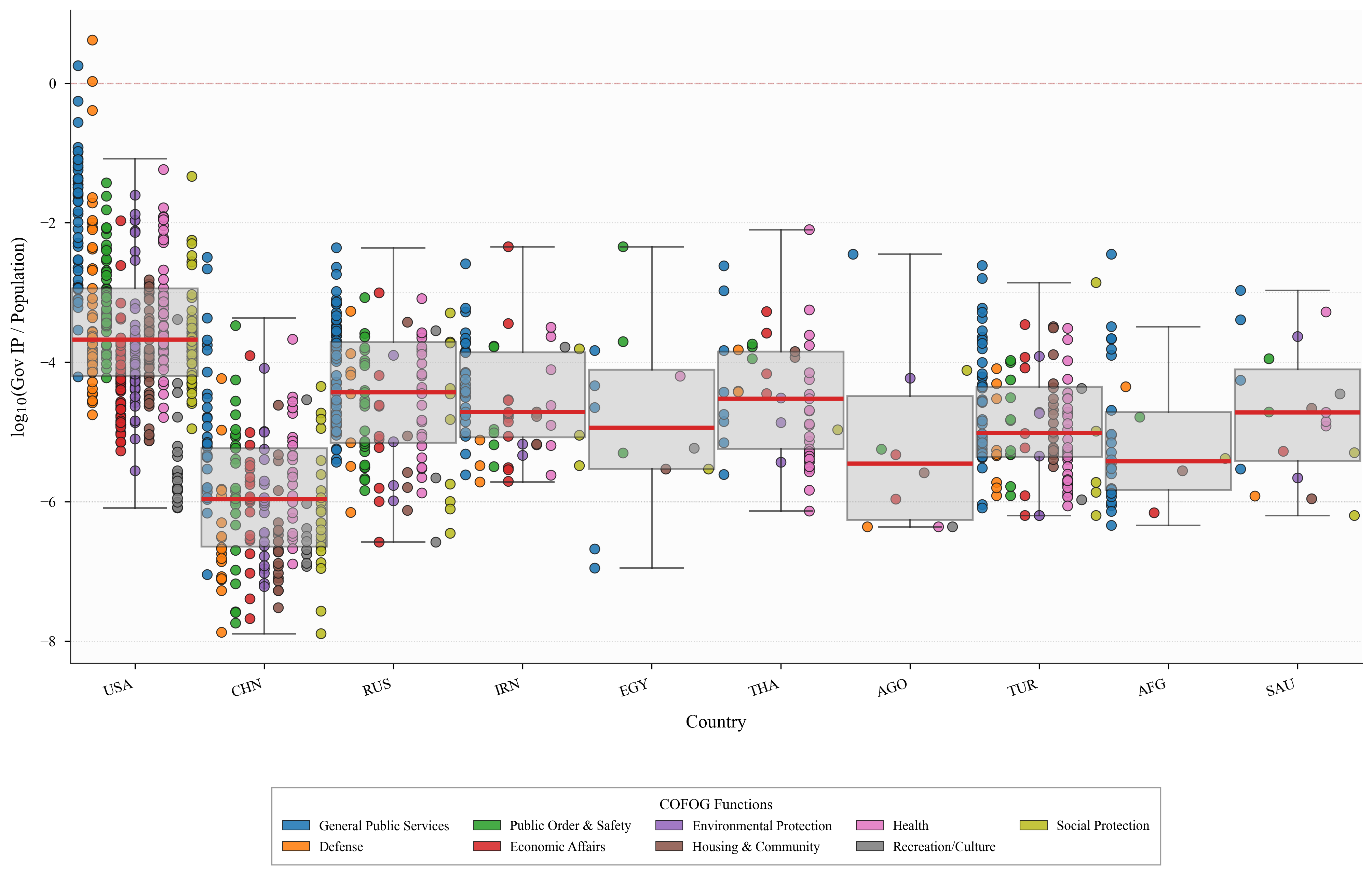}
    \fignote{Each country's subnational governments are plotted using $\log_{10}(\text{Gov IP}/\text{Population})$ for each COFOG function in the top ten non-democracies. Coloured points represent functions, with overlaid box plots summarising all functions per country.}
    \label{fig:cofog_strip_box_nondemocracy_logpop}
\end{figure}

\clearpage

\begin{figure}[ht]
    \centering
    \caption{Top 10 Democracies: Regional Government Total IP Address Allocations by COFOG Function}
    \includegraphics[width=\linewidth]{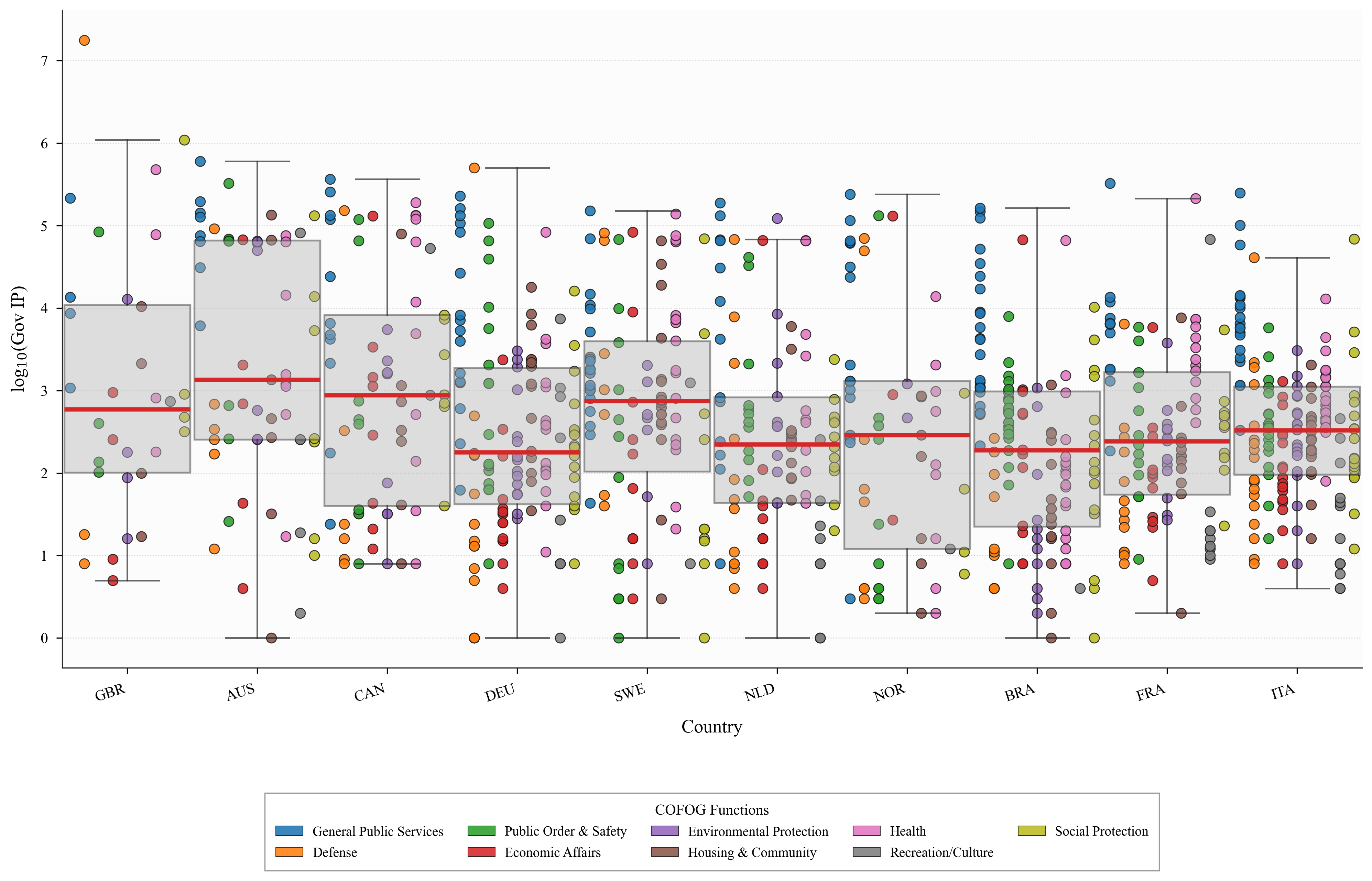}
    \fignote{The figure reports the distribution of $\log_{10}(\text{Gov IP})$, the absolute government IP block holdings per function and region. Colours represent COFOG functions; box plots summarise all regional values for each country.}
    \label{fig:cofog_strip_box_democracy_logtotal}
\end{figure}

\clearpage

\begin{figure}[ht]
    \centering
    \caption{Top 10 Non-Democracies: Regional Government Total IP Address Allocations by COFOG Function}
    \includegraphics[width=\linewidth]{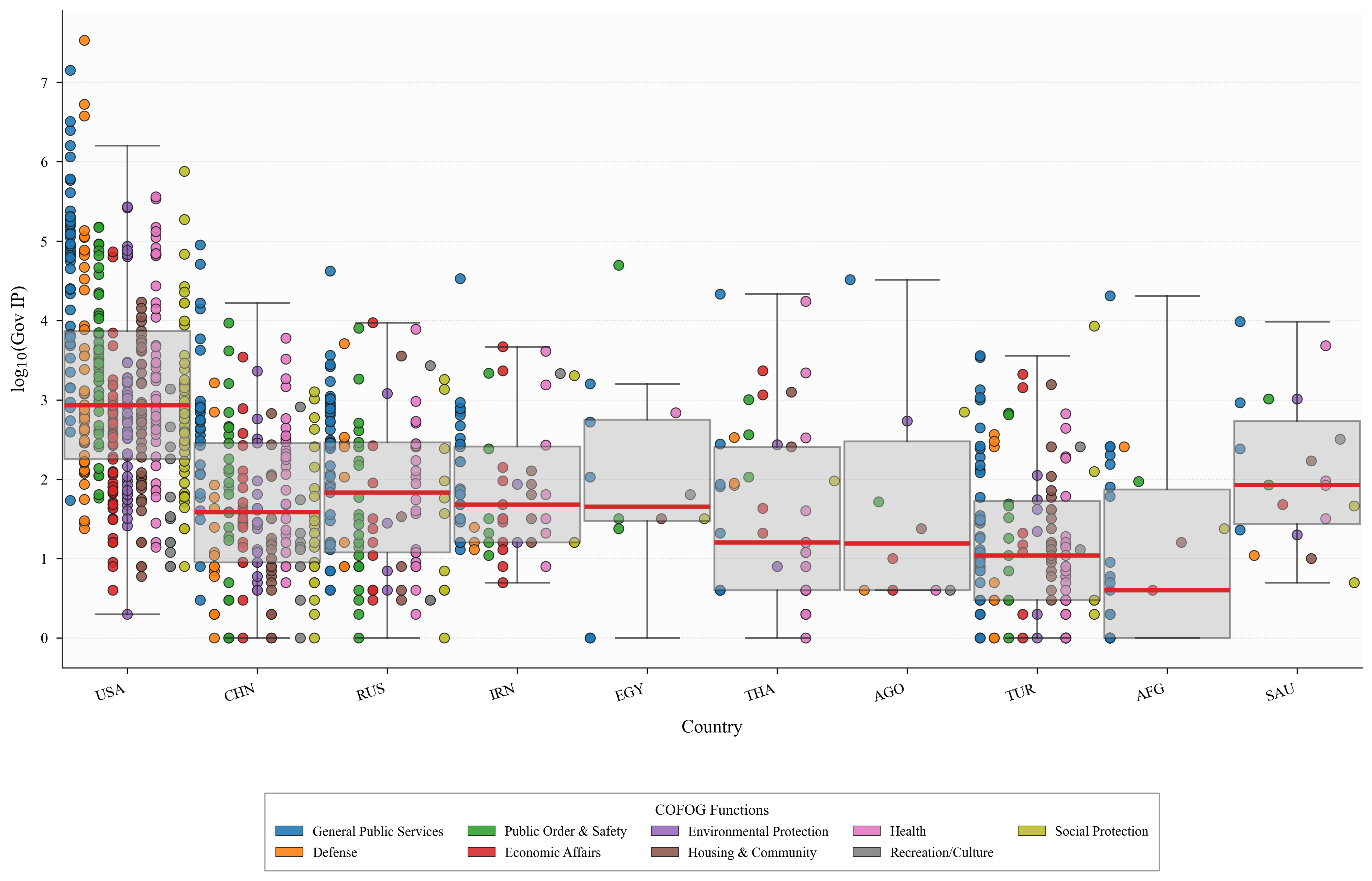}
    \fignote{The figure reports the distribution of $\log_{10}(\text{Gov IP})$ for non-democracies, using absolute government IP counts by function and region. Colours represent COFOG functions; box plots summarise all regional values for each country.}
    \label{fig:cofog_strip_box_nondemocracy_logtotal}
\end{figure}

\clearpage

\section{Specifications Without Country-Level Controls}
\label{app:bad_controls}

This section reports specifications without country-level controls as descriptive comparisons with the controlled estimates in the main text. The validation tables omit log GDP per capita, log population, and population internet access while retaining region-by-year fixed effects. The outcome tables omit the country-year controls and include separate region and year fixed effects. The estimation samples and data coverage also differ from the main tables. These comparisons therefore show whether the broad patterns are present without the controls; they do not isolate the effect of removing any single control. The controlled specifications in the main text remain the primary estimates.

\subsection{Validation Associations}

\begin{table}[H]
    \centering
    \caption{DSC per Capita and Existing State-Capacity Measures: Specifications Without Country-Level Controls}
        \resizebox{\linewidth}{!}{{
\def\sym#1{\ifmmode^{#1}\else\(^{#1}\)\fi}
\begin{tabular}{l*{5}{c}}
\toprule
                    &\\ Dep. var: \\ Measures of State Capacity         &                     &                     &                     &                     \\
                    &\multicolumn{1}{c}{(1)}&\multicolumn{1}{c}{(2)}&\multicolumn{1}{c}{(3)}&\multicolumn{1}{c}{(4)}&\multicolumn{1}{c}{(5)}\\
                    &\multicolumn{1}{c}{E-Gov}&\multicolumn{1}{c}{Statistical}&\multicolumn{1}{c}{Log (Ave Days)}&\multicolumn{1}{c}{Returned}&\multicolumn{1}{c}{Capacity}\\
\midrule
Log Total Gov IPs p.c.&       0.029\sym{***}&       0.021\sym{***}&      -0.059\sym{*}  &       0.020\sym{*}  &       0.026\sym{***}\\
                    &     (0.004)         &     (0.005)         &     (0.027)         &     (0.008)         &     (0.006)         \\
\midrule
Observations        &         922         &        1128         &          82         &          82         &         655         \\
Dep. Var. Mean      &       0.605         &       0.682         &       4.977         &       4.977         &       0.536         \\
Clusters            &          40         &          40         &           5         &           5         &          40         \\
Adjusted \(R^2\)        &       0.519         &       0.454         &       0.262         &       0.223         &       0.462         \\
Region-year FE      &         Yes         &         Yes         &         Yes         &         Yes         &         Yes         \\
Years               &   2014-2022         &   2016-2023         &        2010         &        2010         &   2008-2015         \\
\bottomrule
\multicolumn{6}{l}{\footnotesize Standard errors in parentheses}\\
\multicolumn{6}{l}{\footnotesize \sym{*} \(p<0.1\), \sym{**} \(p<0.05\), \sym{***} \(p<0.01\)}\\
\end{tabular}
}
}
    \label{tab:app_bad_controls_percapita_capacity_cal}
    \tabnote{This table reports associations between DSC per capita and existing state-capacity measures after omitting log GDP per capita, log population, and population internet access. Region-by-year fixed effects are retained. The years and estimation samples are reported in the table and differ from the main validation table. The estimates therefore show the pattern without these controls but do not isolate the contribution of any one control. Standard errors and significance levels are reported in the table.}
\end{table}

\clearpage
\subsection{Per Public-Sector Worker Validation Associations}

\begin{table}[H]
    \centering
    \caption{DSC per Public-Sector Worker and Existing State-Capacity Measures: Specifications Without Country-Level Controls}
    \label{tab:app_bad_controls_perworker_capacity_cal}
    \resizebox{\linewidth}{!}{{
\def\sym#1{\ifmmode^{#1}\else\(^{#1}\)\fi}
\begin{tabular}{l*{5}{c}}
\toprule
                    &\\ Dep. var: \\ Measures of State Capacity         &                     &                     &                     &                     \\
                    &\multicolumn{1}{c}{(1)}&\multicolumn{1}{c}{(2)}&\multicolumn{1}{c}{(3)}&\multicolumn{1}{c}{(4)}&\multicolumn{1}{c}{(5)}\\
                    &\multicolumn{1}{c}{E-Gov}&\multicolumn{1}{c}{Statistical}&\multicolumn{1}{c}{Log (Ave Days)}&\multicolumn{1}{c}{Returned}&\multicolumn{1}{c}{Capacity}\\
\midrule
Log Total Gov IPs Per Worker&       0.027\sym{***}&       0.023\sym{***}&      -0.060         &       0.020\sym{*}  &       0.027\sym{***}\\
                    &     (0.004)         &     (0.005)         &     (0.031)         &     (0.009)         &     (0.006)         \\
\midrule
Observations        &         858         &        1076         &          78         &          78         &         624         \\
Dep. Var. Mean      &       0.603         &       0.684         &       4.953         &       0.746         &       0.539         \\
Clusters            &          40         &          40         &           5         &           5         &          40         \\
Adjusted \(R^2\)        &       0.512         &       0.479         &       0.253         &       0.217         &       0.464         \\
Region-year FE      &         Yes         &         Yes         &         Yes         &         Yes         &         Yes         \\
Years               &   2014-2022         &   2016-2023         &        2010         &        2010         &   2008-2015         \\
\bottomrule
\multicolumn{6}{l}{\footnotesize Standard errors in parentheses}\\
\multicolumn{6}{l}{\footnotesize \sym{*} \(p<0.1\), \sym{**} \(p<0.05\), \sym{***} \(p<0.01\)}\\
\end{tabular}
}
}
    \tabnote{This table reports associations between DSC per public-sector worker and existing state-capacity measures after omitting log GDP per capita, log population, and population internet access. Region-by-year fixed effects are retained. The years and estimation samples are reported in the table and differ from the main validation table. The estimates therefore show the pattern without these controls but do not isolate the contribution of any one control. Standard errors and significance levels are reported in the table.}
\end{table}

\clearpage
\subsection{Per Capita Outcome Associations}

\subsubsection*{General Outcomes}

\begin{table}[H]
    \centering
    \caption{DSC per Capita and General Government Outcomes: Specifications Without Country-Level Controls}
    \resizebox{\linewidth}{!}{%
        {
\def\sym#1{\ifmmode^{#1}\else\(^{#1}\)\fi}
\begin{tabular}{l*{4}{c}}
\toprule
                    &\\ Dependent Variables: \\ General Outcomes         &                     &                     &                     \\
                    &\multicolumn{1}{c}{(1)}&\multicolumn{1}{c}{(2)}&\multicolumn{1}{c}{(3)}&\multicolumn{1}{c}{(4)}\\
                    &\multicolumn{1}{c}{CPI}&\multicolumn{1}{c}{Bribery Incidence}&\multicolumn{1}{c}{Tax/GDP}&\multicolumn{1}{c}{Soc. Ins. Cov.}\\
\midrule
Log Total Gov IPs p.c.&       0.041\sym{***}&       0.000         &       0.007\sym{***}&       0.018\sym{**} \\
                    &     (0.004)         &     (0.001)         &     (0.002)         &     (0.008)         \\
\midrule
Observations        &        1469         &         435         &        1070         &         224         \\
Dep. Var. Mean      &       0.464         &       0.457         &       0.176         &       0.227         \\
Clusters            &         164         &         162         &         129         &          56         \\
Adjusted \(R^2\)        &       0.474         &       0.994         &       0.240         &       0.625         \\
Year FE             &         Yes         &         Yes         &         Yes         &         Yes         \\
Region FE           &         Yes         &         Yes         &         Yes         &         Yes         \\
Years               &   2012-2023         &   2018-2020         &   2008-2020         &   2008-2019         \\
\bottomrule
\multicolumn{5}{l}{\footnotesize Standard errors in parentheses}\\
\multicolumn{5}{l}{\footnotesize \sym{*} \(p<0.1\), \sym{**} \(p<0.05\), \sym{***} \(p<0.01\)}\\
\end{tabular}
}

    }
    \tabnote{This table reports associations between DSC per capita and general government outcomes without country-year controls and with separate region and year fixed effects. The years and estimation samples are reported in the table and differ from the main specification. The estimates therefore show the pattern without the controls but do not isolate the contribution of any one control. Standard errors and significance levels are reported in the table.}
    \label{tab:app_bad_controls_pc_general_baseline}
\end{table}

\clearpage

\subsubsection*{Health Outcomes}

\begin{table}[H]
    \centering
    \caption{DSC per Capita and Health Outcomes: Specifications Without Country-Level Controls}
    \resizebox{\linewidth}{!}{%
        {
\def\sym#1{\ifmmode^{#1}\else\(^{#1}\)\fi}
\begin{tabular}{l*{5}{c}}
\toprule
                    &\\  Dependent Variables: \\ Health Outcomes         &                     &                     &                     &                     \\
                    &\multicolumn{1}{c}{(1)}&\multicolumn{1}{c}{(2)}&\multicolumn{1}{c}{(3)}&\multicolumn{1}{c}{(4)}&\multicolumn{1}{c}{(5)}\\
                    &\multicolumn{1}{c}{Vacc. Meas.}&\multicolumn{1}{c}{Vacc. Hep.}&\multicolumn{1}{c}{Vacc. Dip.}&\multicolumn{1}{c}{Neonatal Mort.}&\multicolumn{1}{c}{Infant Mort.}\\
\midrule
Log Total Gov IPs p.c.&       0.015\sym{***}&       0.018\sym{***}&       0.014\sym{**} &      -0.143\sym{***}&      -0.153\sym{***}\\
                    &     (0.003)         &     (0.004)         &     (0.006)         &     (0.025)         &     (0.021)         \\
\midrule
Observations        &        1391         &        1391         &        1288         &        1402         &        1402         \\
Dep. Var. Mean      &       0.902         &       0.905         &       0.887         &       1.794         &       2.431         \\
Countries           &          65         &          65         &          65         &          65         &          65         \\
Adjusted \(R^2\)        &       0.257         &       0.244         &       0.102         &       0.611         &       0.697         \\
Year FE             &         Yes         &         Yes         &         Yes         &         Yes         &         Yes         \\
Region FE           &         Yes         &         Yes         &         Yes         &         Yes         &         Yes         \\
Years               &   2008-2020         &   2008-2020         &   2008-2020         &   2008-2020         &   2008-2020         \\
\bottomrule
\multicolumn{6}{l}{\footnotesize Standard errors in parentheses}\\
\multicolumn{6}{l}{\footnotesize \sym{*} \(p<0.1\), \sym{**} \(p<0.05\), \sym{***} \(p<0.01\)}\\
\end{tabular}
}

    }
    \tabnote{This table reports associations between DSC per capita and health outcomes without country-year controls and with separate region and year fixed effects. The years and estimation samples are reported in the table and differ from the main specification. The estimates therefore show the pattern without the controls but do not isolate the contribution of any one control. Standard errors and significance levels are reported in the table.}
    \label{tab:app_bad_controls_pc_health_baseline}
\end{table}

\clearpage
\subsection{Per Public-Sector Worker Outcome Associations}

\subsubsection*{General Outcomes}

\begin{table}[H]
    \centering
    \caption{DSC per Public-Sector Worker and General Government Outcomes: Specifications Without Country-Level Controls}
    \resizebox{\linewidth}{!}{%
        {
\def\sym#1{\ifmmode^{#1}\else\(^{#1}\)\fi}
\begin{tabular}{l*{4}{c}}
\toprule
                    &\\ Dependent Variables: \\ General Outcomes         &                     &                     &                     \\
                    &\multicolumn{1}{c}{(1)}&\multicolumn{1}{c}{(2)}&\multicolumn{1}{c}{(3)}&\multicolumn{1}{c}{(4)}\\
                    &\multicolumn{1}{c}{CPI}&\multicolumn{1}{c}{Bribery Incidence}&\multicolumn{1}{c}{Tax/GDP}&\multicolumn{1}{c}{Soc. Ins. Cov.}\\
\midrule
Log Total Gov IPs Per Worker&       0.037\sym{***}&      -0.021\sym{**} &       0.007\sym{***}&       0.015\sym{*}  \\
                    &     (0.005)         &     (0.006)         &     (0.002)         &     (0.008)         \\
\midrule
Observations        &        1419         &         134         &        1016         &         213         \\
Dep. Var. Mean      &       0.461         &       0.229         &       0.177         &       0.235         \\
Clusters            &         159         &          81         &         121         &          54         \\
Adjusted \(R^2\)        &       0.423         &       0.304         &       0.293         &       0.603         \\
Year FE             &         Yes         &         Yes         &         Yes         &         Yes         \\
Region FE           &         Yes         &         Yes         &         Yes         &         Yes         \\
Years               &   2012-2023         &   2008-2020         &   2008-2020         &   2008-2019         \\
\bottomrule
\multicolumn{5}{l}{\footnotesize Standard errors in parentheses}\\
\multicolumn{5}{l}{\footnotesize \sym{*} \(p<0.1\), \sym{**} \(p<0.05\), \sym{***} \(p<0.01\)}\\
\end{tabular}
}

    }
    \tabnote{This table reports associations between DSC per public-sector worker and general government outcomes without country-year controls and with separate region and year fixed effects. The years and estimation samples are reported in the table and differ from the main specification. The estimates therefore show the pattern without the controls but do not isolate the contribution of any one control. Standard errors and significance levels are reported in the table.}
    \label{tab:app_bad_controls_pw_general_baseline}
\end{table}

\clearpage
\subsubsection*{Health Outcomes}

\begin{table}[H]
    \centering
    \caption{DSC per Public-Sector Worker and Health Outcomes: Specifications Without Country-Level Controls}
    \resizebox{\linewidth}{!}{%
        {
\def\sym#1{\ifmmode^{#1}\else\(^{#1}\)\fi}
\begin{tabular}{l*{5}{c}}
\toprule
                    &\\  Dependent Variables: \\ Health Outcomes         &                     &                     &                     &                     \\
                    &\multicolumn{1}{c}{(1)}&\multicolumn{1}{c}{(2)}&\multicolumn{1}{c}{(3)}&\multicolumn{1}{c}{(4)}&\multicolumn{1}{c}{(5)}\\
                    &\multicolumn{1}{c}{Vacc. Meas.}&\multicolumn{1}{c}{Vacc. Hep.}&\multicolumn{1}{c}{Vacc. Dip.}&\multicolumn{1}{c}{Neonatal Mort.}&\multicolumn{1}{c}{Infant Mort.}\\
\midrule
Log Total Gov IPs Per Worker&       0.014\sym{***}&       0.017\sym{***}&       0.014\sym{**} &      -0.132\sym{***}&      -0.139\sym{***}\\
                    &     (0.004)         &     (0.004)         &     (0.006)         &     (0.027)         &     (0.023)         \\
\midrule
Observations        &        1294         &        1294         &        1191         &        1304         &        1304         \\
Dep. Var. Mean      &       0.898         &       0.903         &       0.883         &       1.793         &       2.434         \\
Clusters            &         161         &         161         &         156         &         161         &         161         \\
Adjusted \(R^2\)        &       0.264         &       0.245         &       0.118         &       0.612         &       0.690         \\
Year FE             &         Yes         &         Yes         &         Yes         &         Yes         &         Yes         \\
Region FE           &         Yes         &         Yes         &         Yes         &         Yes         &         Yes         \\
Years               &   2008-2020         &   2008-2020         &   2008-2020         &   2008-2020         &   2008-2020         \\
\bottomrule
\multicolumn{6}{l}{\footnotesize Standard errors in parentheses}\\
\multicolumn{6}{l}{\footnotesize \sym{*} \(p<0.1\), \sym{**} \(p<0.05\), \sym{***} \(p<0.01\)}\\
\end{tabular}
}

    }
    \tabnote{This table reports associations between DSC per public-sector worker and health outcomes without country-year controls and with separate region and year fixed effects. The years and estimation samples are reported in the table and differ from the main specification. The estimates therefore show the pattern without the controls but do not isolate the contribution of any one control. Standard errors and significance levels are reported in the table.}
    \label{tab:app_bad_controls_pw_health_baseline}
\end{table}

\end{document}